\documentclass[journal=chreay,manuscript=review]{achemso}
\usepackage{graphicx}
\usepackage{amssymb}
\usepackage{amsmath}
\usepackage{amsfonts}
\usepackage{amssymb}
\usepackage{graphics}
\usepackage{color}
\usepackage{color}
\usepackage{xcolor}
\usepackage{hyperref}

\title{ Condensed Matter Theory of Dipolar Quantum Gases}
\date{\today}
\author{M. A. Baranov$^{1,2,3}$, M. Dalmonte$^{1,4}$, G. Pupillo$^{1,2,5}$ and P. Zoller$^{1,2}$}
\affiliation{$^1$ Institute for Quantum Optics and Quantum Information of the Austrian Academy
of Sciences, A-6020 Innsbruck, Austria\\
$^{2}$ Institute for Theoretical Physics, University of Innsbruck, A-6020 Innsbruck, Austria\\
$^{3}$ RRC "Kurchatov Institute", Kurchatov Square 1, 123182, Moscow, Russia\\
$^{4}$ Dipartimento di Fisica dell'Universit\`a di Bologna, via Irnerio 46, 40126
Bologna, Italy\\
$^{5}$ISIS (UMR 7006) and IPCMS (UMR 7504), Universit\'e de Strasbourg and CNRS,
Strasbourg, France \\}
\begin{document}

\maketitle
\tableofcontents

\section{Introduction.}

\label{s1}

The realization of Bose Einstein condensates (BEC) and quantum degenerate
Fermi gases with cold atoms have been highlights of quantum physics during the
last decade. Cold atoms in the tens of nanokelvin range are routinely obtained via combined laser- and evaporative-cooling techniques \cite{Metcalf1999}. For high-enough densities ($\gtrsim 10^{12}$ ${\rm cm}^{-3}$), the atomic de Broglie wavelength becomes larger than the typical
interparticle distance and thus quantum statistics governs the
many-body dynamics of these systems. Characteristic features of the physics of cold atomic gases are
the microscopic knowledge of the many-body Hamiltonians which are realized in
the experiments and the possibility of controlling and tuning system
parameters via external fields. External field control of contact inter-particle interactions
can be achieved, for example, by varying the scattering length
via Feshbach resonances \cite{Chi08}, while trapping of ultracold gases is obtained with magnetic, electric and optical fields \cite{Bloch2008}. In particular, optical lattices, which are artificial crystals made of light obtained via the interference of optical laser beams, can realize perfect arrays of hundreds of thousands of microtraps~\cite{BlochNatPhysRev2005, GreinerNatureRev2008}, allowing for the confinement of quantum gases to one-dimensional (1D), 2D and 3D geometries and even the manipulation of individual particles~\cite{bakr2009,sherson2010single}. This control over interactions and confinement is the key for the experimental realization of fundamental quantum
phases and phase transitions as illustrated by the BEC-BCS crossover in atomic
Fermi gases \cite{Ing08}, and the Berezinskii-Kosterlitz-Thouless transition
\cite{Had06} for cold bosonic atoms confined to 2D.

Breakthroughs in the experimental realization of BEC and degenerate
Fermi gases of atoms with a
comparatively large magnetic dipole moment, such as $^{52}\mathrm{Cr}$
\cite{Pfau0,Stuhler2005,Pfau2,Pfau3,Pfau4,Pfau5,Pfau6},  $^{168}\mathrm{Er}$
\cite{Aikawa2012} and $^{164}$Dy atoms
\cite{Lu2011,Lu2012b} (dipole moment $6\mu_{B}, 7\mu_{B} $ and 10$\mu_{B}$, respectively, with
$\mu_{B}$ Bohr's magneton), and the recent astounding progress in experiments
with ultracold polar molecules
\cite{Weinstein1998,doyle1999,PM3,PM4,bethlem2000,PM6,PM7,PM8,PM9,PM10,PM11,PM12}
have now stimulated great interest in the properties of low temperature
systems with dominant dipolar interactions
(see reviews Refs.~\cite{Baranov200871,Carr2009rev,Lahaye2009,Kre09,trefzger2011rev}
for discussions of various aspects of the problem).
The latter have a long-range and
anisotropic character, and their relative strength compared to, e.g.,
short-range interactions can be often controlled by tuning external fields, or
else by adjusting the strength and geometry of confining trapping potentials.
For example, in experiments with polarized atoms, magnetic dipolar
interactions can be made to overcome short-range interactions by tuning the
effective $s$-wave scattering length to zero using Feshbach resonances
\cite{Pfau0,Stuhler2005,Pfau2,Pfau3}. This has already led to the observation
of fundamental phenomena at the mean-field level, such as, the anisotropic
deformation during expansion and the directional
stability~\cite{Mueller2011,Lu2011} of dipolar BECs. Heteronuclear polar
molecules in a low vibrational and rotational state, on the other hand, can
have large permanent dipole moments along the internuclear axis with strength
ranging between one tenth and ten Debye (1Debye $\simeq3.335 \times10^{-30}$
C
$\cdot$%
m). In the presence of an external electric field (with a typical value of
$10^{3} - 10^{4}$ V/cm) mixing rotational excitations, the molecules can be
oriented in the laboratory frame and the induced dipole moment can approach
its asymptotic value, corresponding to the permanent dipole moment. This
effect can be used to tune the strength of the dipole-dipole interaction
\cite{Kre09}. Additional microwave fields allow for advanced tailoring of the
interactions between the molecules, where even the shape of interaction
potentials can be tuned with external fields, in addition to the strength.
This tunability of interactions forms the basis for the realization of novel
quantum phenomena in these systems, in the strongly interacting limit.

As a result of this progress, in recent years dipolar gases have become the
subject of intensive theoretical efforts, and there is now an extensive body
of literature predicting novel properties for these systems\cite{Baranov200871,Carr2009rev,Lahaye2009,Kre09,trefzger2011rev}. It is the purpose
of this review to provide a summary of these recent theoretical studies with a focus
on the many-body quantum properties, to demonstrate the connections and
differences between dipolar gaseous systems and traditional condensed-matter
systems, and to stress the inherent interdisciplinary nature of these studies.
This work covers spatially homogeneous as well as trapped systems, and
includes the analysis of the properties of dipolar gases in both the
mean-field (dipolar Bose-Einstein condensates and superfluid BCS pairing
transition) and in the strongly correlated (dipolar gases in optical lattices
and low-dimensional geometries) regimes.

We tried our best to include all relevant works of this exciting, ever
expanding field. We apologize in advance if some papers (hopefully, not many)
are not appearing below.

\section{The dipole-dipole interaction}

\label{s2}

\begin{figure}[t]
\begin{center}
\includegraphics[width=\columnwidth]{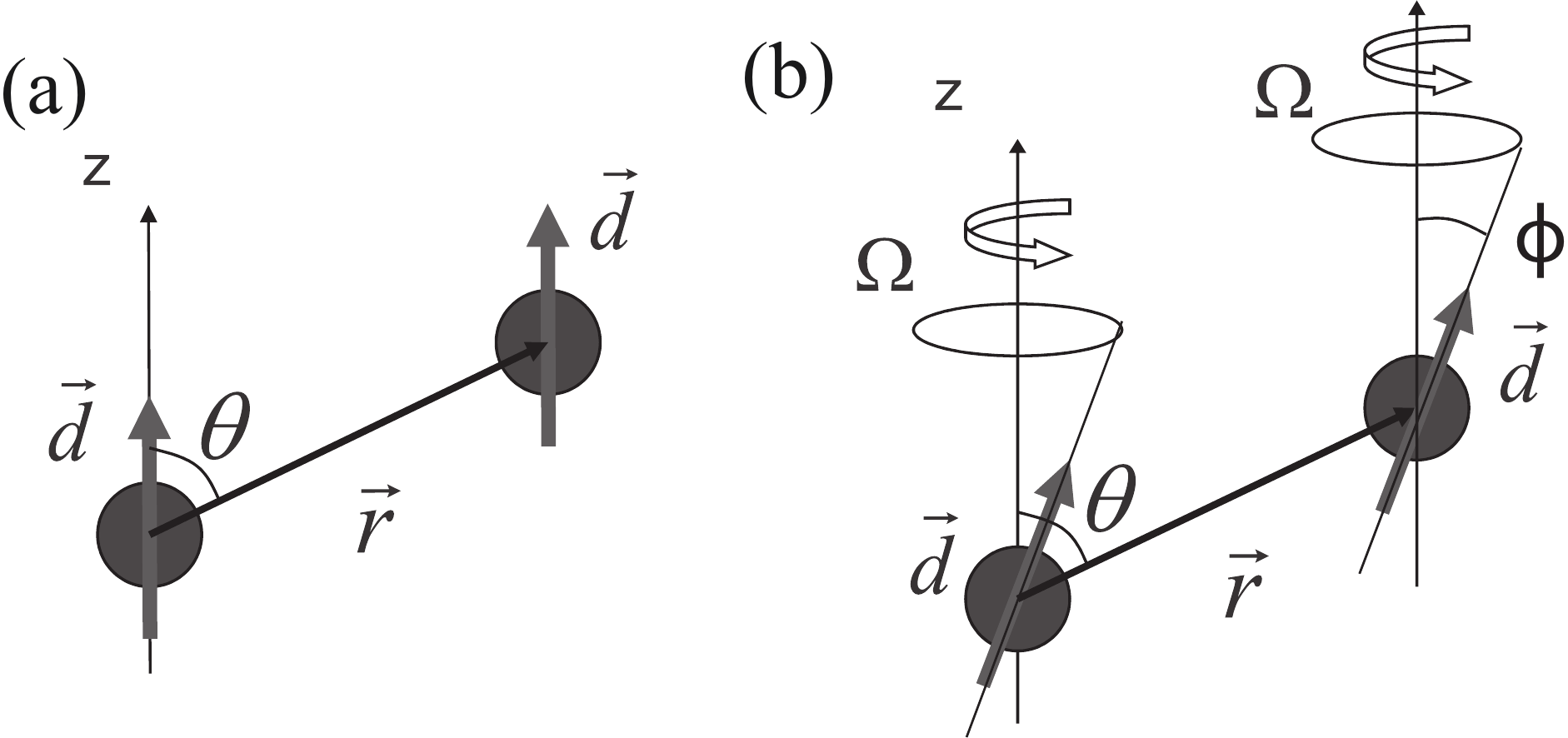}
\end{center}
\caption{(a) Geometry for the interaction of two aligned dipoles. (b)
Tunability of the dipole-dipole interaction by using a time-varying aligning
field. The angle $\phi$ between the dipole orientation and the $z$-axis
determines the strength and the sign of the effective interaction.}%
\label{Fig1}%
\end{figure}

For \textit{polarized} dipolar particles, interparticle interactions 
include both a short-range Van der Waals (vdW) part and a long-range
dipole-dipole one. The latter is dominant at large interparticle separations and assuming a polarization along the $z$-axis as in Fig.~\ref{Fig1}(a) the interparticle interaction reads
\begin{eqnarray}
V_{dd}(\mathbf{r})=\frac{d^{2}}{r^{3}}(1-3\cos^{2}\theta). \label{Vd}%
\end{eqnarray}
Here $d$ is the electric dipole moment (for magnetic dipoles $d^{2}$ should
be replaced with $\mu^{2}$, with $\mu$ the magnetic dipole moment),
$\mathbf{r}$ is the vector connecting two dipolar particles, and $\theta$ is
the angle between $\mathbf{r}$ and the dipole orientation (the $z$-axis). The
potential $V_{dd}(\mathbf{r})$ is both
\textit{long-range} and \textit{anisotropic}, that is, partially repulsive and
partially attractive. As discussed in the sections below, these features have important consequences 
for the scattering properties in the ultracold gas, for the stability of the system as well as 
for a variety of its properties. 

\subsection{Scattering of two dipoles}

\label{s2.1}

The long-range character ($\sim r^{-3}$) of the dipole-dipole interaction
results in all partial waves contributing to the scattering at low energies,
and not only, e.g., the $s$-wave, as is often the case for short-range
interactions. In fact, for dipole-dipole interactions the phase shift
$\delta_{l}$ in a scattering channel with angular momentum $l$ behaves as
$\delta_{l}\sim k$ for $l\geq 0$ and small $k$ (see, e.g., Refs. 
\cite{MarineskuYou1} and \cite{DebYou}).

The effect of the anisotropy of the interaction is instead that the angular
momentum is not conserved during scattering: for bosons and fermions the dipole-dipole interaction
mixes all even and odd angular momenta scattering channels, respectively. Due to the coupling between the various
scattering channels, the potential $V_{dd}$ then generates a short-range
contribution to the total effective potential in the $s$-wave channel ($l=0$).
This has the general effect to reduce the strength of the short-range part of the interaction.

Thus, for two bosonic dipolar particles (even angular momenta)
the scattering at low energies is determined by both the long-range and the short-range parts of
the interaction. This is in contrast to the low energy scattering of two
fermionic dipoles (odd angular momenta), which is universal in the sense
that it is determined only by the long-range dipolar part of the
interaction, and is insensitive to the short-range details.

For a dilute weakly interacting gas the above results allow a parametrization of the realistic interparticle
interaction between two particles of mass $m$  in terms of the following pseudopotential \cite{YiYou1}-\cite{Blume2}
(see also Refs. \cite{Derevianko} and \cite{DereviankoErr})%
\begin{eqnarray}
V(\emph{r})=g\delta(\mathbf{r})+\frac{d^{2}}{r^{3}}(1-3\cos^{2}\theta),
\label{V}%
\end{eqnarray}
with%
\begin{eqnarray}
g=\frac{4\pi\hbar^{2}a(d)}{m} \label{g}%
\end{eqnarray}
parametrizing the short-range part of the interaction. We note that the long-range part of the
pseudopotential $V(\emph{r})$ is identical to the long-range part of the original potential
and the scattering length $a(d)$ controlling the short-range part depends on the dipole moment. 
This dependence is important \cite{Blume1} when one changes the dipole moment,
using external, e.g. electric, fields, as explained below. \newline

The strength of the dipole-dipole interaction can be characterized by the
quantity
\[
a_{d}=\frac{md^{2}}{\hbar^{2}},
\]
which has the dimension of length and can be considered as a characteristic
range of the dipole-dipole interaction, or dipolar length. This length determines
the low energy limit of the scattering amplitudes, and, in this
sense, $a_{d}$ is analogous to the scattering length for the dipole-dipole
interaction. For chromium atoms with a comparatively large magnetic moment of
$6\mu_{B}$ (equivalent dipole moment $d=0.056$ Debye) we have
$a_{d}\approx2.4$ nm. For most polar molecules the electric dipole moment
ranges in between $0.1$ and $1$ Debye, while $a_{d}$ ranges from
$1$ to $10^{3}$ nm. For example, the dipole moment of fermionic
ammonia molecules $^{15}\mathrm{ND}_{3}$ is $d=1.5$ Debye with
$a_{d}=712$ nm, while for $\mathrm{H}%
^{12}\mathrm{C}^{14}\mathrm{N}$ it increases to $d=2.98$ Debye
and $a_{d}=3620$ nm. This latter value of the effective scattering length is an order of magnitude larger than, for example, the one for the
intercomponent interaction in the widely discussed case of a two-species
fermionic gas of $^{6}\mathrm{Li}$, where $a_{\mathrm{Li}}=-114$ nm. Thus, the
strength of the dipole-dipole interaction between polar molecules can be not
only comparable with but even much larger than the strength of the short-range
interatomic interaction.

\subsection{Tunability of the dipole-dipole interaction}

\label{s2.2}

One spectacular feature of the dipole-dipole interaction is its tunability. In
Sect.~\ref{sec:TunaAtoms} we first review methods for tuning the
\textit{strength} and \textit{sign} of dipolar interactions with an eye to
cold atoms, and then in Sect.~\ref{sec:TunaPolMol} we discuss tunability for
the specific case of polar molecules, where both the \textit{strength} as well
as the \textit{shape} of interactions can be engineered.

\subsubsection{Tunability of interactions in cold atoms}

\label{sec:TunaAtoms}

In Ref.~\cite{tuning} a technique has been developed to tune the strength as
well as the sign of dipolar interactions in atomic systems with a finite
permanent magnetic dipole moment. This technique uses a combination of a
static (e.g., magnetic) field along the $z$-axis and a fast rotating field in
the perpendicular $xy$-plane such that the resulting time dependent dipole
moment\ is [see Fig.~\ref{Fig1}(b)]%
\[
\mathbf{d}(t)=d\left\{  \mathbf{e}_{z}\cos\phi+\left[  \mathbf{e}_{x}%
\cos(\Omega t)+\mathbf{e}_{y}\sin(\Omega t)\right]  \sin\phi\right\} .
\]
Here $\Omega$ is the rotating frequency of the field and the angle $\phi$,
$0\leq\phi<\pi/2$ is determined by the ratio of the amplitudes of the static
and rotating fields. The above expression implies that the dipoles
follow the time-dependent external field adiabatically. This in turn sets an upper
limit on the values of the rotating frequency $\Omega$, which should be (much) smaller
than the level splitting in the field. However, if the frequency
$\Omega$ is much larger that the typical frequencies of the particle motion,
over the period $2\pi /\Omega$ the particles feel an average interaction $V_{d}$
\[
\left\langle V_{d}(\mathbf{r})\right\rangle =\frac{d^{2}}{r^{3}}(1-3\cos
^{2}\theta)\alpha(\phi).
\]
The latter differs from the interaction for aligned dipoles, Eq.~(\ref{Vd}), by
a factor $\alpha(\phi)=(3\cos^{2}\phi-1)/2$, which can be changed continuously 
from $1$ to $-1/2$ by varying the angle $\phi$. Thus this method allows
to "reverse" the sign of the dipole-dipole interaction and even cancel it
completely for $\phi=a\cos1/\sqrt{3}=54.7$, similar to NMR
techniques~\cite{mehring1983}. We note that an analogous technique can be also
applied for the electric dipole moments of, e.g., polar molecules. We will
review applications of this method below.

\subsubsection{Effective Hamiltonians for polar molecules}

\label{sec:TunaPolMol}

In the following we will be often interested in manipulating interactions for
polar molecules in the \textit{strongly interacting} regime. In particular, we
will aim at modifying not only the strength but also the \textit{shape} of
interaction potentials, as a basis to investigate new condensed matter
phenomena. This usually entails a combination of the following two steps: (i)
manipulating the internal (electronic, vibrational, rotational, ...) structure
of the molecules, and thus their mutual interactions, using external static
(DC) electric and microwave (AC) fields, and (ii) confining molecules to a
lower-dimensional geometry, using, e.g., optical potentials, as exemplified in
Fig.~\ref{fig:Set2D}. Under appropriate conditions, the resulting effective
interactions can be made \textit{purely repulsive} at large distances (e.g.,
at characteristic distances of 10nm or more), as in the 2D example of
Fig.~\ref{fig:circles}(a). On one hand, this has the effect to suppress
possible inelastic collisions and chemical reactions occurring at short-range
(i.e., at characteristic distances of $a_{c} \lesssim1$nm), and on the other
hand it allows to study interesting condensed matter phenomena originating
from the often-unusual form of the two-body (or many-body-) interaction
potentials. In the next few subsections we review techniques for the
engineering of the interaction potentials which will be used in the many-body
context in Sect.~\ref{sec:StronglyInteractingGases} below.\newline

Our starting point is the Hamiltonian for a gas of cold heteronuclear
molecules prepared in their electronic and vibrational ground-state,
\begin{eqnarray}
H(t)    =  \sum_{i}^{N}\left[ \frac{\mathbf{p}_{i}^{2}}{2m}+V_{\mathrm{trap}%
}(\mathbf{r}_{i})+H_{\mathrm{in}}^{(i)}-\mathbf{d}_{i}\mathbf{E}(t)\right]
+  \sum_{i<j}^{N}V_{\mathrm{dd}}(\mathbf{r}_{i}-\mathbf{r}_{j}%
).\label{eq:eqGeneral}%
\end{eqnarray}
Here the first term in the single particle Hamiltonian corresponds to the
kinetic energy of the molecules, while $V_{\mathrm{trap}}(\mathbf{r}_{i})$
represents a trapping potential, as provided, for example, by an optical
lattice, or an electric or magnetic trap. The term $H_{\mathrm{in}}^{(i)}$
describes the internal low energy excitations of the molecule, which for a
molecule with a closed electronic shell ${}^{1}\Sigma(\nu=0)$ (e.g. SrO, RbCs
or LiCs) correspond to the rotational degree of freedom of the molecular axis.
This term is well described by a rigid rotor $H_{\mathrm{in}}^{(i)}\equiv
H_{\mathrm{rot}}^{(i)}=B\mathbf{J}_{i}^{2}$ with $B$ the rotational constant
(in the few to tens of GHz regime) and $\mathbf{J}_{i}$ the dimensionless
angular momentum. The rotational eigenstates $|J,M\rangle$ for a quantization
axis $z$, and with eigenenergies $BJ(J+1)$ can be coupled by a static (DC) or
microwave (AC) field $\mathbf{E}$ via the \emph{electric} dipole moment
$\mathbf{d}_{i}$, which is typically of the order of a few Debye.

In the absence of electric fields, the molecules prepared in a ground
rotational state $J=0$ have no net dipole moment, and interact via a
van-der-Waals attraction $V_{\mathrm{vdW}}\sim-C_{6}/r^{6}$, reminiscent of
the interactions of cold alkali metal atoms in the electronic ground-states.
Electric fields admix excited rotational states and induce static or
oscillating dipoles, which interact via strong dipole-dipole interactions
$V_{dd}$ with the characteristic $1/r^{3}$ dependence given in Eq.~\eqref{Vd}.
For example, a static DC field couples the spherically symmetric rotational
ground state of the molecule to excited rotational states with different
parity, thus creating a non-zero average dipole moment. The field strength
therefore determines the degree of polarization and the magnitude of the
dipole moment. As a result, the effective dipole-dipole interaction may be
tuned by the competition between an orienting, e.g., DC electric field and the
quantum (or thermal) rotation of the molecule. This method effectively works
for the values of the field up to the saturation limit, at which the molecule
is completely polarized (typically $10^{4}$ \textrm{V/cm}). \newline

The many body dynamics of cold polar molecules is thus governed by an
interplay between dressing and manipulating the rotational states with DC and
AC fields, and strong dipole-dipole interactions. In condensed matter physics
one is often interested in effective theories for the low-energy dynamics of
the many-body system, after the high-energy degrees of freedom have been
traced out. The connection between the full molecular $N$-particle Hamiltonian
(\ref{eq:eqGeneral}) including rotational excitations and dressing fields, and
an effective low-energy theory can be made using the following
Born-Oppenheimer approximation: The diagonalization of the Hamiltonian
$H_{BO}=\sum_{i}^{N}\left[ H_{\mathrm{in}}^{(i)}-\mathbf{d}_{i}\mathbf{E}%
\right] +\sum_{i<j}^{N}V_{\mathrm{dd}}(\mathbf{r}_{i}-\mathbf{r}_{j})$ for
frozen spatial positions $\{\mathbf{r}_{i}\}$ of the $N$ molecules yields a
set of energy eigenvalues $V_{\text{eff}}^{\mathrm{3D}}\left( \{\mathbf{r}%
_{i}\}\right) $, which can be interpreted as the effective interaction
potential in the single-channel many-body Hamiltonian
\begin{eqnarray}
H_{\text{eff}}=\sum_{i=1}^{N}\left[ \frac{\mathbf{p}_{i}^{2}}{2m}%
+V_{\mathrm{trap}}(\mathbf{r}_{i})\right] +V_{\text{eff}}^{\mathrm{3D}}\left(
\{\mathbf{r}_{i}\}\right) .\label{eq:NbodyHamiltonian}%
\end{eqnarray}
The term $V_{\mathrm{\scriptscriptstyle
eff}}^{\mathrm{3D}}\left( \{\mathbf{r}_{i}\}\right) $ represents an effective
$N$-body interaction, which can be expanded as a sum of two-body and many-body
interactions
\begin{eqnarray}
V_{\mathrm{\scriptscriptstyle eff}}^{\mathrm{3D}}\left( \{\mathbf{r}%
_{i}\}\right) =\sum_{i<j}^{N}V^{\mathrm{3D}}\left( \mathbf{r}_{i}%
\!-\!\mathbf{r}_{j}\right) +\!\sum_{i<j<k}^{N}W^{\mathrm{3D}}\left(
\mathbf{r}_{i},\mathbf{r}_{j},\mathbf{r}_{k}\right) +\ldots,\label{effint1}%
\end{eqnarray}
where in most cases only two-body interactions are considered. The dependence
of $V_{\text{eff}}^{\mathrm{3D}}\left( \{\mathbf{r}_{i}\}\right) $ on the
electric fields $\mathbf{E}$ provides the basis for the engineering of the
many body interactions, as described below. \newline

We note that the attractive part of the interaction potential can induce
instabilities in a dipolar gas at the few body level as well as at the
many-body level (this latter case will be discussed in Sect.~\ref{s3} below).
For example, for several experimentally relevant mixed alkali-metal diatomic
species such as KRb, LiNa, LiK, LiRb, and LiCs~\cite{Zuchowski2010} there
exist chemically reactive channels that are energetically favorable, leading
to particle recombination and two-body losses in the gas. The rate of chemical
reactions can be strongly enhanced by dipole-dipole interactions which can
attract molecules in a \textit{head-to-tail} configuration [e.g., $\theta=0$
in Fig.~\ref{Fig1}(a)] to distances on the order of typical chemical
interaction distances, $a_{c}\lesssim1$
nm~\cite{Idziaszek2010,Idziaszek2010b,Ni2010,deMiranda2011,Ticknor2010,
Quemener2010b, Petrov2001, Li2008, Li2009}. One aim of interaction engineering
is to control these interactions in order to stabilize the gas against
particle losses. This will enable the study of complex condensed matter
phenomena in these systems.

\subsubsection{Stabilization of dipolar interactions in 2D}

\label{sec:Stable2D}

\begin{figure}[tb]
\includegraphics[width=0.9\columnwidth]{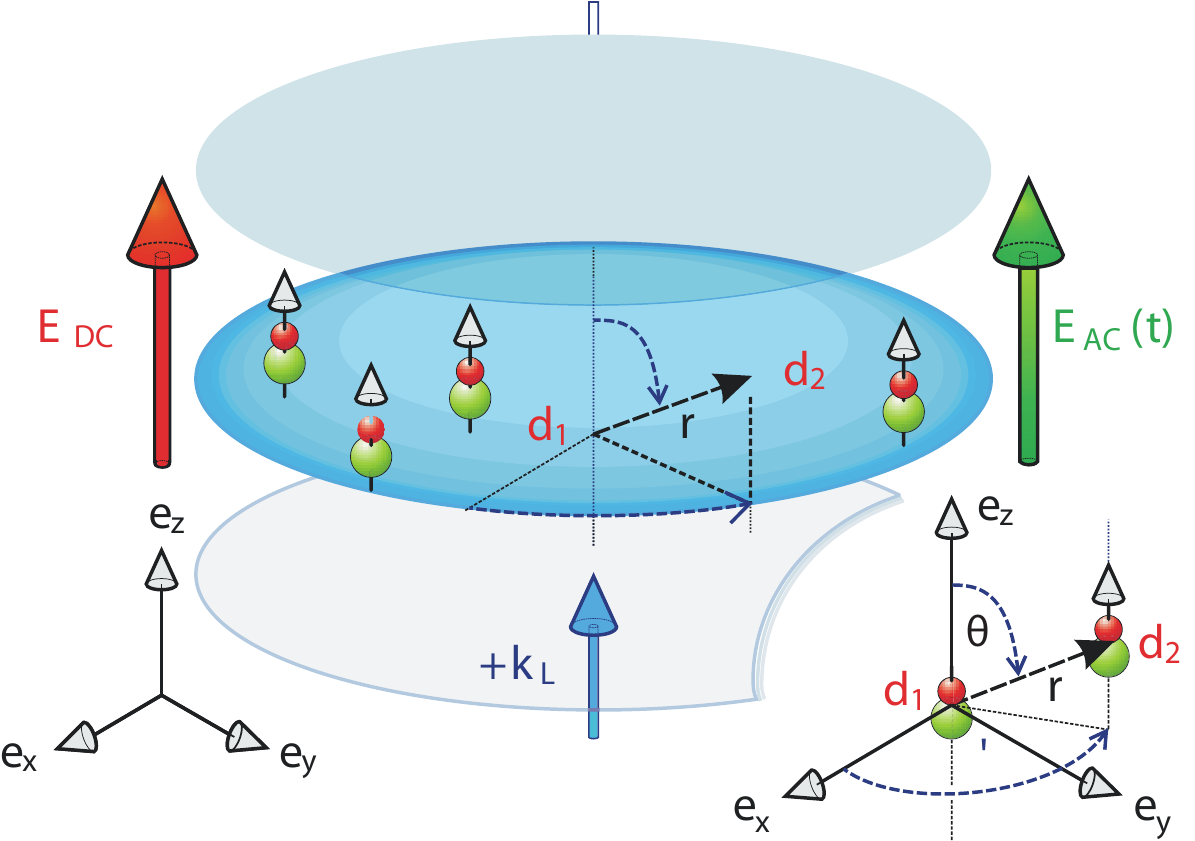}
\caption{System setup: Polar molecules are trapped in the ($x,y$)-plane by an
optical lattice made of two counter-propagating laser beams with wavevectors
$\pm\mathbf{k}_{\mathbf{L}}=\pm{k}_{\mathbf{L}}\mathbf{e}_{z}$, (blue arrows).
The dipoles are aligned in the $z$-direction by a DC electric field
$\mathbf{E}_{\mathrm{DC}}\equiv E_{\mathrm{DC}}\mathbf{e}_{z}$ (red arrow). An
AC microwave field is indicated (green arrow). Inset: Definition of polar
($\vartheta$) and azimuthal ($\varphi$) angles for the relative orientation of
the inter-molecular collision axis $\mathbf{r}_{12}$ with respect to a
space-fixed frame, with axis along $z$. [Adapted from Ref.~\cite{Micheli2007}]
}%
\label{fig:Set2D}%
\end{figure}

The simplest example of stabilization of dipolar interactions against
inelastic collisions is sketched in Fig.~\ref{fig:Set2D} and consists of a
system of cold polar molecules in the presence of a polarizing DC electric
field oriented in the $z$-direction, and of a strong harmonic transverse
confinement $V_{\mathrm{trap}}= \mu\Omega^{2}z^{2}/2$ with frequency $\Omega$
and characteristic length $a_{h}=\sqrt{\hbar/(m \Omega)}$. The latter is
provided, e.g., by an optical lattice along $z$.

\begin{figure}[tb]
\includegraphics[width=\columnwidth]{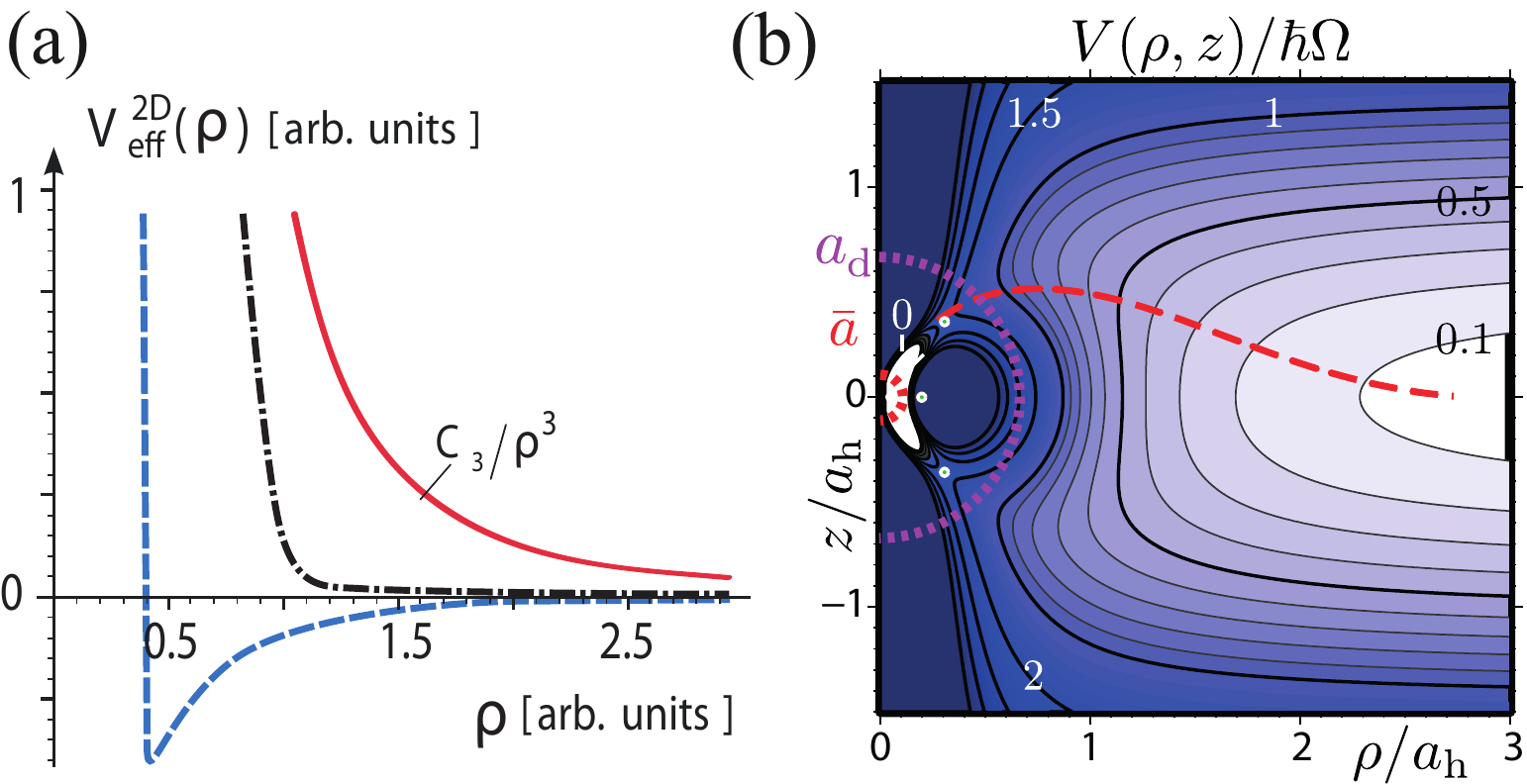}
\caption{(a) Effective repulsive interaction potentials in 2D. Solid line:
Dipolar potential 
$V_{\mathrm{eff}}^{\mathrm{2D}} (\rho)=D/\rho^{3}$ 
induced by a DC electric field in the
configuration of Fig.~\ref{fig:Set2D} [$\mathbf{E}_{\mathrm{DC}} > 0$ and
$\mathbf{E}_{\mathrm{AC}}(t) =0$ ].  Dash-dotted line: "Step-like" potential
induced by a single AC microwave field and a weak DC field [see also
Fig.~\ref{fig20000}]. Dashed line: Attractive potential induced by the
combination of several AC fields and a weak DC field. The potentials
$V_{\mathrm{eff}}^{\mathrm{2D}}(\rho)$ and the separation
$\rho$ are given in arbitrary units. (b) Potential $V(\rho,z)/(\hbar\Omega)$
versus $\rho/a_{h}$ and $z/a_{h}$ for $|m|=1$ for $^{40}$K$^{87}$Rb in a
$\Omega=2\pi(50\,\mathrm{kHz})$ trap, where $a_{h}=56.4$ nm, $\bar{a}=$ 6.25
nm~\cite{Idziaszek2010} and $d=0.2$ D. White circles: saddle points. Dashed
lines: semiclassical trajectory for the collision of two molecules. The dashed
half-circles show $\bar{a}$ and $a_{d}$. [Panels (a) and (b) adapted from
Refs.~\cite{Micheli2007} and \cite{Micheli2010}, respectively.]}%
\label{fig:circles}%
\end{figure}

Figure~\ref{fig:circles}(b) shows a countour plot of the interaction potential
$V(\rho,z)$ for two dipoles in this quasi-2D geometry, where
\begin{eqnarray}
\label{eq:Coll2D}V(\mathbf{r})= V^{\mathrm{3D}}_{\mathrm{eff}}(\mathbf{r}) +
\frac{m\Omega^{2}z^{2}}{4}= -\frac{C_{6}}{r^{6}} +\frac{d^{2}(1-3z^{2}/r^{2}%
)}{4\pi\epsilon_{0} r^{3}} + \frac{m \Omega^{2}z^{2}}{4} \,.\nonumber
\end{eqnarray}
Here $\vec{\rho}=(\rho,\phi,z)$ represents the distance between the two
molecules in cylindrical coordinates, and $r\equiv|\vec{\rho}|$.
The first term is the isotropic vdW potential, assuming the molecules are in
their rotational ground state, with a vdW length $\bar{a} = \left( 2
\pi/\Gamma(\frac14)^{2}\right)  \left( 2 \mu C_{6}/\hbar^{2}\right) ^{1/4}%
$~\cite{Gribakin1993,Julienne2009}. The second term is the anisotropic dipolar
potential, with induced dipole moment $d$ and dipolar length $a_{d}$.


Figure~\ref{fig:circles}(b) illustrates essential features of reduced
dimensional collisions: for finite $d$, the repulsive dipole-dipole
interaction overcomes the attractive van-der-Waals potential in the
($z=0$)-plane at distances $r > \bar a \gg a_{c}$, realizing a repulsive
in-plane potential barrier (blue dark region). In addition, the harmonic
potential confines the particles's motion in the $z$ direction. The
combination of the dipole-dipole interaction and of the harmonic confinement
thus yields a \textit{three-dimensional} potential barrier separating the
long-distance, where interactions are repulsive, from the short-distance one,
where interactions are attractive and inelastic processes can occur. If the
collision energy is smaller than the height of the barrier at the saddle point
(white circles), the particles' motion is confined to the long-distance
region, where particles scatter elastically. Particle losses are due to
tunneling through the potential barrier at a rate $K^{\mathrm{(re)}}_{j}$.
Within a semiclassical (\emph{instanton}) approximation valid for $a_{d}
\gg\max\{a_{h}, \bar a \}$, the tunneling rate $K^{\mathrm{(re)}}_{j}$ is well
approximated by the exponential form
\begin{eqnarray}
\label{eq:ExpInstanton}K^{\mathrm{(re)}}_{j} \propto\omega_{p} \exp\left[ -c
(a_{d}/a_{h})^{2/5}\right] .
\end{eqnarray}
The constant $c$ has been recently computed numerically by Julienne, Hanna and
Idziaszek~\cite{Julienne2011} to be $c\approx3.03$, while the "attempt rate"
$\omega_{p}$~\cite{Coleman1977} for the scattering of two isolated dipoles
reads $\omega_{p} \propto(\hbar\kappa^{4} a_{h}^{4}/\mu)$,
\textit{independent} of particles' statistics. Here $\kappa$ is the momentum
for a collision with relative kinetic energy $E_{\kappa}= \hbar^{2} \kappa
^{2}/(2\mu)$, with $a_{\kappa}= 2 \pi/ \kappa$ the DeBroglie wavelength. For
particles in a crystalline configuration (see Sect.~\ref{sec:secSelfAssembled}
below), $\omega_{p}$ will be proportional to the frequency of phonon
oscillations around the mean particle positions $\omega_{p} \sim\sqrt
{d^{2}/(\mu a^{5})}$, with $a$ the mean inter-particle distance. The
expression Eq.~\eqref{eq:ExpInstanton} shows that collisional losses may be
strongly suppressed for \textit{any} molecular species for a large enough
dipole moment or strength of transverse confinement.\newline

\begin{figure}[tb]
\includegraphics[width=0.65\columnwidth]{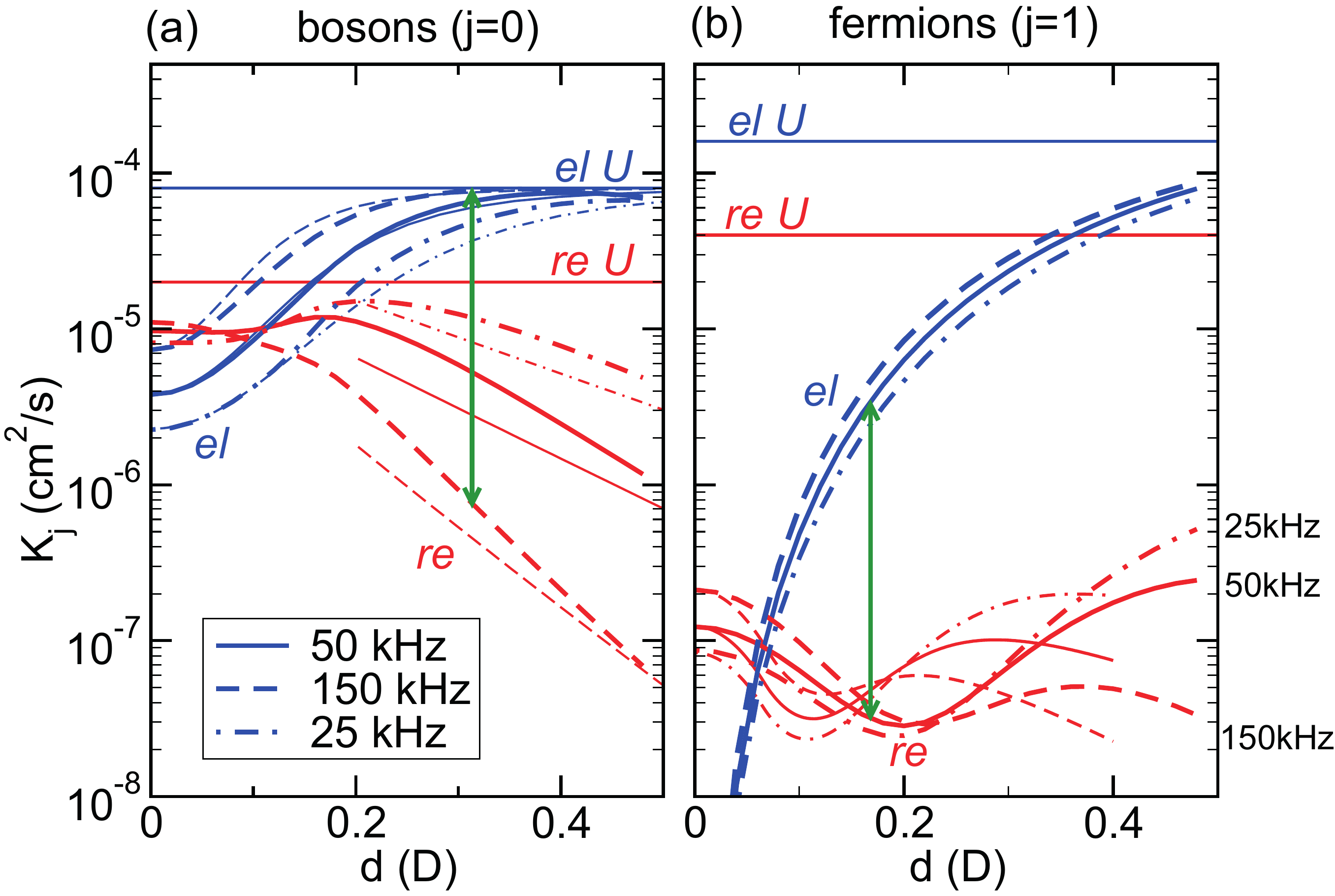} \caption{Quasi-2D elastic
(el) and reaction (re) rate constants $K_{j}$ for (a) identical KRb bosons ($j
= 0$) and (b) identical KRb fermions ($j=1$) at a collision energy of
$E=k_{B}(240\,\mathrm{nK})=h(5\,\mathrm{kHz})$ for three different trap
frequencies $\Omega/(2\pi)=$ 25 kHz, 50 kHz, and 150 kHz. Heavy lines: CC.
Light lines: UBA or instanton. Horizontal lines show the unitarity limits.
Vertical arrows show where $\eta_{j}=100$. [From Ref.~\cite{Micheli2010}] }%
\label{fig:Rates}%
\end{figure}

In ultracold collisions one often has the following separation of length
scales: $a_{\kappa}\gg a_{h} \gg\bar a \gg a_{c}$, and $a_{d}$ can be tuned
by, e.g, increasing the external DC field. Figure~\ref{fig:Rates}(a) and (b)
show numerical results for reactive and elastic collision rates of bosonic and
fermionic KRb molecules, respectively, and for several strengths of transverse
confinement. Here $a_{\kappa}$, $a_{h}$ and $\bar{a}$ are on the order of
hundreds of nm, tens of nm, and less than 10 nm, respectively. Because of the
moderate $d_{\mathrm{max}}=0.5$D of KRb molecules, here $a_{d} \lesssim a_{h}$
and the semiclassical regime of large $a_{d}$ of Eq.~\eqref{eq:ExpInstanton}
is not reached. Nevertheless, in stark contrast to collisions in
3D~\cite{Ni2010}, the figure shows that the ratio between elastic and
inelastic collision rates increases rapidly with $d$, signaling an increased
stability with increasing $d$. For bosons, the exact numerical results (thick
lines) approach rapidly the semiclassical instanton limit (thin lines) with
increasing $d$. The behavior of the inelastic rates for fermions is explained
in detail in Refs.~\cite{Quemener2010,Micheli2010}.

Recent landmark experimental results from the JILA group with fermionic KRb
molecules show a strong suppression of inelastic collisions and increase of
elastic ones with $d$, in excellent agreement with the predictions of
Fig.~\ref{fig:circles}. This opens the way to the study of strongly correlated
phenomena in these systems.

\subsubsection{Advanced interaction designing: Blue-shielding}

\label{sec:secBlueThree}

\begin{figure}[ptb]
\par
\begin{center}
\includegraphics[width=0.65\columnwidth]{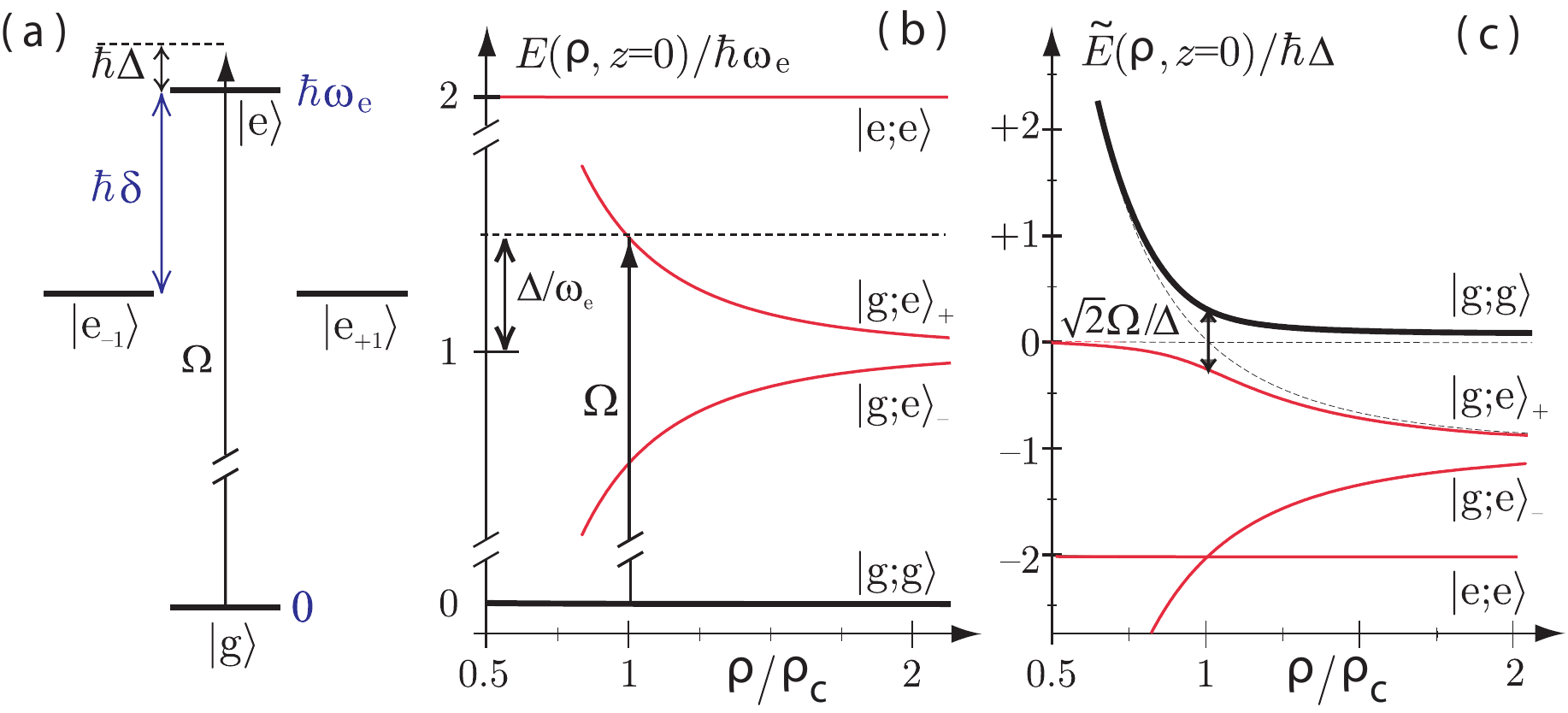}
\end{center}
\caption{ Design of the step-like potential of Fig.~\ref{fig:circles}(a)
(black dashed-dotted line): (a) Rotational spectrum of a molecule in a weak DC
field. The DC field splits the $(J=1)$-manifold by an amount $\hbar\delta$.
The linearly-polarized microwave transition with detuning $\Delta$ and Rabi
frequency $\Omega$ is shown as an arrow. (b) BO-potentials for the internal
states for $\Omega=0$ (bare potentials), where $|g;e\rangle_{\pm}%
\equiv(|g;e\rangle\pm|e;g\rangle)/\sqrt{2}$. The resonant Condon point
$\rho_{C}$ is indicated by an arrow. (c) AC-field-dressed BO-potentials. The
dressed groundstate potential has the largest energy. }%
\label{fig20000}%
\end{figure}

By combining DC and AC fields to dress the manifold of rotational energy
levels it is possible to design effective interaction potentials
$V^{\mathrm{3D}}_{\mathrm{eff}}\left( \mathbf{r}_{i}\!-\!\mathbf{r}_{j}\right)
$ with (essentially) any shape as a function of distance. For example, the
addition of a single linearly-polarized AC field to the configuration of
Sect.~\ref{sec:Stable2D} leads to the realization of the 2D ``step-like''
potential of Fig.~\ref{fig:circles}(a) (black dashed-dotted line), where the
character of the repulsive potential varies considerably in a small region of
space. The derivation of this effective 2D interaction is sketched in
Fig.~\ref{fig20000}(a-c)~\cite{Buechler07,Micheli2007}. The (weak) DC-field
splits the first-excited rotational ($J=1$)-manifold of each molecule by an
amount $\hbar\delta$, while a linearly polarized AC-field with Rabi frequency
$\Omega$ is blue-detuned from the ($|g\rangle-|e\rangle$)-transition by
$\hbar\Delta$, see Fig.~\ref{fig20000}(a). Because of $\hbar\delta$ and the
choice of polarization, for distances $\rho\gg(d^{2}/\hbar\delta)^{1/3}$ the
relevant single-particle states for the two-body interaction reduce to the
states $|g\rangle$ and $|e\rangle$ of each molecule. Figure~\ref{fig20000}(b)
shows that the dipole-dipole interaction splits the excited state manifold of
the two-body rotational spectrum, making the detuning $\Delta$
\emph{position-dependent}. As a consequence, the combined energies of the bare
groundstate of the two-particle spectrum and of a microwave photon become
degenerate with the energy of a (symmetric) excited state at a characteristic
resonant (Condon) point $\rho_{C}=(d^{2}/\hbar\Delta)^{1/3}$, which is
represented by an arrow in Fig.~\ref{fig20000}(b). At this Condon point, an
avoided crossing occurs in a \emph{field-dressed} picture, and the new
(dressed) groundstate potential inherits the character of the bare ground and
excited potentials for distances $\rho\gg\rho_{C}$ and $\rho\ll\rho_{C}$,
respectively. Fig.~\ref{fig20000}(d) shows that the dressed groundstate
potential (which has the \emph{largest energy}) is almost flat for $\rho
\gg\rho_{C}$ and it is strongly repulsive as $1/\rho^{3}$ for $\rho\ll\rho
_{C}$, which corresponds to the realization of the step-like potential of
Fig.~\ref{fig:circles}(a). We remark that, due to the choice of polarization,
this strong repulsion is present only in the plane $z=0$, while for $z\neq0$
the groundstate potential can become attractive. The optical confinement along
$z$ of Sect.~\ref{sec:Stable2D} is therefore necessary to ensure the stability
of the system.

The interactions in the presence of a single AC field are described in detail
in Ref.~\cite{Micheli2007}, where it is shown that in the absence of external
confinement this case is analogous to the (3D) optical blue-shielding
developed in the context of ultracold collisions of neutral
atoms~\cite{Zilio96,Napolitano97,Weiner99}, however with the advantage of the
long lifetime of the excited rotational states of the molecules, as opposed to
the electronic states of cold atoms. The strong inelastic losses observed in
3D collisions with cold atoms~\cite{Zilio96,Napolitano97,Weiner99} can be
avoided via a judicious choice of the field's polarization, eventually
combined with a tight confinement to ensure a 2D geometry (as e.g. in
Fig.~\ref{fig:Set2D} above). For example, in Ref.~\cite{Gorshkov08} it is
shown that in the presence of a DC field and of a circularly polarized AC
field the \emph{attractive} time-averaged interaction due to the rotating
(AC-induced) dipole moments of the molecules allows for the cancelation of the
total dipole-dipole interaction. The residual interactions remaining after
this cancelation are \emph{purely repulsive 3D interactions} with a
characteristic van-der-Waals behavior $V_{\mathrm{eff}}^{\mathrm{3D}%
}(\mathbf{r})\sim(d^{4}/\hbar\Delta)/r^{6}$. This 3D repulsion provides for a
shielding of the inner part of the interaction potential and thus it will
strongly suppress inelastic collisions in experiments.

Recent works~\cite{Aldegunde2009,Ran2010} have considered the microwave
spectra of alkali-metal dimers including hyperfine interactions. It is an
important open question to determine the effects that the presence of internal
states such as, e.g. hyperfine states, have on the broad class of shielding
techniques described above.\newline

\section{Weakly interacting dipolar Bose gas}

\label{s3}

\subsection{BEC in a spatially homogeneous gas.}

\label{s3.1}

Let us discuss now the influence of the dipole-dipole interaction on the
properties of a homogeneous single-component dipolar Bose gas \footnote{%
This and the next Sections are substantially revised and updated version of
the corresponding part of Ref. \cite{Baranov200871}}. This can be most
conveniently done in the language of second quantization. For this purpose
we introduce particle creation and annihilation field operators $\hat{\psi}%
^{\dagger }(\mathbf{r})$ and $\hat{\psi}(\mathbf{r})$ satisfying standard
bosonic commutation relation%
\begin{align*}
\left[ \hat{\psi}(\mathbf{r}),\hat{\psi}^{\dagger }(\mathbf{r}^{\prime })%
\right] & \equiv \hat{\psi}(\mathbf{r})\hat{\psi}^{\dagger }(\mathbf{r}%
^{\prime })-\hat{\psi}^{\dagger }(\mathbf{r}^{\prime })\hat{\psi}(\mathbf{r}%
)=\delta (\mathbf{r}-\mathbf{r}^{\prime }), \\
\left[ \hat{\psi}(\mathbf{r}),\hat{\psi}(\mathbf{r}^{\prime })\right] & =%
\left[ \hat{\psi}^{\dagger }(\mathbf{r}),\hat{\psi}^{\dagger }(\mathbf{r}%
^{\prime })\right] =0.
\end{align*}%
The corresponding second quantized Hamiltonian of the system then reads%
\begin{eqnarray}
\hat{H} &=&\int d\mathbf{r}\hat{\psi}^{\dagger }(\mathbf{r})\left[ -\frac{%
\hbar ^{2}}{2m}\nabla ^{2}-\mu \right] \hat{\psi}(\mathbf{r})+  \notag \\
&+&\frac{1}{2}\int d\mathbf{r}d\mathbf{r}^{\prime }\hat{\psi}^{\dagger }(%
\mathbf{r})\hat{\psi}^{\dagger }(\mathbf{r}^{\prime })V(\mathbf{r}-\mathbf{r}%
^{\prime })\hat{\psi}(\mathbf{r}^{\prime })\hat{\psi}(\mathbf{r}),
\label{HamBosehom}
\end{eqnarray}%
%
%
%
%
%
where $m$\ is the mass of the particles, $V(\mathbf{r})$ is the
interparticle interaction, and the chemical potential $\mu $ fixes the
average density $n$ of the gas. We consider the case when the system is away
from any "shape" resonances \cite{MarineskuYou1},\cite{DebYou},\cite%
{ShaperesBohn} and, therefore, replace the original interparticle
interaction with the pseudopotential (\ref{V}). Assuming that the system is
dilute, $na^{3}\ll 1$, we can write the Hamiltonian as
\begin{eqnarray}
\hat{H} &=&\int d\mathbf{r}\hat{\psi}^{\dagger }(\mathbf{r})\left[ -\frac{%
\hbar ^{2}}{2m}\nabla ^{2}-\mu +\frac{1}{2}g\left\vert \hat{\psi}(\mathbf{r}%
)\right\vert ^{2}\right] \hat{\psi}(\mathbf{r})+  \notag \\
&+&\frac{1}{2}\int d\mathbf{r}d\mathbf{r}^{\prime }\hat{\psi}^{\dagger }(%
\mathbf{r})\hat{\psi}^{\dagger }(\mathbf{r}^{\prime })V_{d}(\mathbf{r}-%
\mathbf{r}^{\prime })\hat{\psi}(\mathbf{r}^{\prime })\hat{\psi}(\mathbf{r}),
\label{HamBosehom1}
\end{eqnarray}%
%
%
%
%
%
where $V_{d}$ is given by Eq.~(\ref{Vd}) and $g=4\pi \hbar ^{2}a/m$ [as
compared to Eq.~(\ref{g}), we omit the dependence of the scattering length $a
$ on the dipole moment $d$]. Note that the scattering length has to be
positive, $a>0$, to avoid an absolute instability due to local collapses 
\cite{Pitaevskii}.

As we will see below, an important parameter that determines the properties
of the system described by the Hamiltonian (\ref{HamBosehom1}) is%
\begin{eqnarray}
\varepsilon _{dd}=\frac{4\pi }{3}\frac{d^{2}}{g}=\frac{md^{2}}{3\hbar ^{2}a}=%
\frac{1}{3}\frac{a_{d}}{a}.  \label{epsilondd}
\end{eqnarray}%
It measures the strength of the dipole-dipole interaction relative to the
short-range repulsion. In the case $\varepsilon _{dd}<1$, the short-range
part of the interparticle interaction is dominant while the dipole-dipole
interaction results in only small corrections. For a positive scattering
length $a$, the system is stable and exhibits BEC at low temperatures. This
case corresponds to earlier experiments \cite{Pfau0,Pfau3} with \textrm{Cr }%
BEC ($\varepsilon _{dd}\approx 0.16$ \cite{Pfau3}). It was found that the
corrections due to magnetic dipole-dipole interaction between $^{52}$\textrm{%
Cr} atoms are of the order of $10\%$.

For the opposite case $\varepsilon _{dd}>1$, the anisotropic dipole-dipole
interaction plays the dominant role resulting in instability of a spatially
homogeneous system \cite{YiYou2},\cite{Gora}, \cite{Lushnikov}. This
instability can be seen in the dispersion relation $E(\mathbf{p})$ between
the energy $E$ and the momentum $\mathbf{p}$ of excitations in the
Bose-condensed gas, which can be easily obtained within the standard
Bogoliubov approach:
\begin{eqnarray}
E(\mathbf{p}) &=&\sqrt{\frac{p^{2}}{2m}\left[ \frac{p^{2}}{2m}+2nV(\mathbf{p}%
)\right] } \\
&=&\frac{p}{2m}\sqrt{p^{2}+4mng\left[ 1+\varepsilon _{dd}(3\cos ^{2}\theta
-1)\right] }.  \notag
\end{eqnarray}%
%
%
%
%
%
Here $\theta $ is the angle between the excitation momentum $\mathbf{p}$ and
the direction of dipoles, and $V(\mathbf{p})=g+(4\pi d^{2}/3)(3\cos
^{2}\theta -1)$ is the Fourier transform of $V(\mathbf{r})=g\delta (\mathbf{r%
})+V_{d}(\mathbf{r})$. For $\varepsilon _{dd}>1$, the excitation energies $E(%
\mathbf{p})$ at small $p$ and $\theta $ close to $\pi /2$ become imaginary
signalling the instability (collapse). This instability of a spatially
homogeneous dipolar Bose gas with dominant dipole-dipole interaction is a
result of a partially attractive nature of the dipole-dipole interaction.

\subsection{BEC in a trapped gas.}

\label{s3.2}

The above consideration shows that the behavior of a spatially homogeneous
Bose gas with a strong dipole-dipole interaction is similar to that of a
Bose gas with an attractive short-range interaction characterized by a
negative scattering length $a<0$. In the latter case, however, the collapse
of the gas can be prevented by confining the gas in a trap provided the
number of particles $N$ \ in the gas is smaller than some critical value $%
N_{c}$, $N<N_{c}$ (see, e.g., \cite{Pitaevskii}). This is due to the finite
energy difference between the ground and the first excited states in a
confined gas. For a small number of particle this creates an effective
energy barrier preventing the collapse and, therefore, results in a
metastable condensate. The same arguments are also applicable to a dipolar
BEC in a trap, see Refs. \cite{Pfau5}\ and \cite{Pfau6}, with one very
important difference: The sign and the value of the dipole-dipole
interaction energy in a trapped dipolar BEC depends on by the trapping
geometry and, therefore, the stability diagram contains the trap anisotropy
as a crucial parameter.

\subsubsection{Ground state.}

\label{s3.2.1}

The Hamiltonian for a trapped dipolar Bose gas reads
\begin{eqnarray}
\hat{H} &=&\int d\mathbf{r}\hat{\psi}^{\dagger }(\mathbf{r})\left[ -\frac{%
\hbar ^{2}}{2m}\nabla ^{2}-\mu +U_{\mathrm{tr}}(\mathbf{r})+\right.\nonumber\\
&+&\left.\frac{1}{2}%
g\left\vert \hat{\psi}(\mathbf{r})\right\vert ^{2}\right] \hat{\psi}(\mathbf{%
r})+   \\
&+&\frac{1}{2}\int d\mathbf{r}d\mathbf{r}^{\prime }\hat{\psi}^{\dagger }(%
\mathbf{r})\hat{\psi}^{\dagger }(\mathbf{r}^{\prime })V_{d}(\mathbf{r}-%
\mathbf{r}^{\prime })\hat{\psi}(\mathbf{r}^{\prime })\hat{\psi}(\mathbf{r}),\nonumber
\label{HamBosetrap}
\end{eqnarray}%
%
%
%
%
%
where%
\begin{eqnarray}
U_{\mathrm{tr}}(\mathbf{r})=\frac{m}{2}\left[ \omega _{\rho
}^{2}(x^{2}+y^{2})+\omega _{z}^{2}z^{2}\right]  \label{Utrap}
\end{eqnarray}%
is the trapping potential and we again use the pseudopotential (\ref{V}) for
the short-range part of the interparticle interaction assuming that the
system is away from "shape" resonances. For the trapping potential we
consider the experimentally most common case of an axially symmetric
harmonic trap characterized by the axial $\omega _{z}$ and radial $\omega
_{\rho }$ trap frequencies. The aspect ratio of the trap $l$ is defined
through the ratio of the frequencies: $l=\sqrt{\omega _{\rho }/\omega _{z}}%
=l_{z}/l_{\rho }$, where $l_{z}=\sqrt{\hbar /m\omega _{z}}$ and $l_{\rho }=%
\sqrt{\hbar /m\omega _{\rho }}$ are the axial and radial sizes of the ground
state wave function in the harmonic oscillator potential (\ref{Utrap}),
respectively. For $l<1$ one has a pancake-form (oblate) trap, while the
opposite case $l>1$ corresponds to a cigar-form (prolate) trap. Taking into
account the anisotropy of the dipole-dipole interaction, one can easily \
see that the aspect ratio $l$ should play a very important role in the
behavior of the system.

The standard mean-field approximation corresponds to taking the many-body
wave function in the form of a product of single-particle wave functions:%
\begin{eqnarray}
\Psi (\mathbf{r}_{1},\ldots ,\mathbf{r}_{N};t)=\prod\limits_{j=1}^{N}\psi
_{1}(\mathbf{r}_{j},t).  \label{GPwavefunc}
\end{eqnarray}%
The condensate is then described by the condensate wave function $\psi (%
\mathbf{r},t)=\sqrt{N}\psi _{1}(\mathbf{r},t)$ normalized to the total
number of particles, $\int d\mathbf{r}\left\vert \psi (\mathbf{r}%
,t)\right\vert ^{2}=N$, and governed by the time-dependent Gross-Pitaevskii
(GP) equation
\begin{eqnarray}
&&i\hbar \frac{\partial }{\partial t}\psi (\mathbf{r},t)=\left[ -\frac{\hbar
^{2}}{2m}\nabla ^{2}-\mu +U_{\mathrm{tr}}(\mathbf{r})+g\left\vert \psi (%
\mathbf{r},t)\right\vert ^{2}+\right.  \notag \\
&+&\left. d^{2}\int d\mathbf{r}^{\prime }\frac{1-3\cos ^{2}\theta }{%
\left\vert \mathbf{r-r}^{\prime }\right\vert ^{3}}\left\vert \psi (\mathbf{r}%
^{\prime },t)\right\vert ^{2}\right] \psi (\mathbf{r},t).  \label{GPgen}
\end{eqnarray}%
%
%
%
%
%
The validity of this approach was tested in Refs. \cite{Blume1} and \cite%
{Blume2} by using many-body diffusion Monte-Carlo calculations with the
conclusion that a GP equation with the pseudopotential (\ref{V}) provides a
correct description of the gas in the dilute limit $na^{3}\ll 1$. Note that,
being the product of single-particle wave functions, the many-body wave
function (\ref{GPwavefunc}) does not take into account interparticle
correlations at short distances due to their interaction, which takes place
at interparticle distances $\left\vert \mathbf{r}-\mathbf{r}^{\prime
}\right\vert \lesssim a_{d}=md^{2}/\hbar ^{2}$. This change of the wave
function is taken into account in Eq.~(\ref{GPgen}) by the contact part of
the pseudopotential (\ref{V}) [the fourth term in the right-hand-side in
Eq.~(\ref{GPgen})] but ignored in the last term of Eq.~(\ref{GPgen}) because
the main contribution to the integral comes from large interparticle
distances (of order the spatial size of the condensate).

Let us first consider stationary solutions of Eq.~(\ref{GPgen}), for which $%
\psi (\mathbf{r},t)=\psi _{0}(\mathbf{r})$ and $\psi _{0}(\mathbf{r})$ obeys
the stationary GP equation
\begin{eqnarray}
&&\left[ -\frac{\hbar ^{2}}{2m}\nabla ^{2}+\frac{m}{2}(\omega _{\rho
}^{2}\rho ^{2}+\omega _{z}^{2}z^{2})+g\left\vert \psi (\mathbf{r}%
,t)\right\vert ^{2}+\right.  \notag \\
&&+\left. d^{2}\int d\mathbf{r}^{\prime }\frac{1-3\cos ^{2}\theta }{%
\left\vert \mathbf{r-r}^{\prime }\right\vert ^{3}}\left\vert \psi _{0}(%
\mathbf{r}^{\prime })\right\vert ^{2}\right] \psi _{0}(\mathbf{r})=\mu \psi
_{0}(\mathbf{r}),  \label{GPstat}
\end{eqnarray}%
%
%
%
%
%
where $\rho ^{2}=x^{2}+y^{2}$. Numerical analysis of Eqs. (\ref{GPgen}) and (%
\ref{GPstat}) was performed in Refs. \cite{YiYou1},\cite{YiYou2}, \cite{Gora}%
,\cite{Goraerr}-\cite{3Dtrap} on the basis of numerical solutions of the
non-linear Schr\"{o}dinger equation (\ref{GPstat}) together with variational
considerations with the Gaussian ansatz for the condensate wave function. In
Refs. \cite{Blume1} and \cite{Blume2} the problem was treated using
diffusive Monte-Carlo calculations, while the authors of Ref. \cite{Eberl1}
apply the Thomas-Fermi approximation that neglects the kinetic energy and
allows to obtain analytical results.

We begin the discussion of the results with the case of a dominant
dipole-dipole interaction, $\varepsilon _{dd}\gg 1$, such that the third
term in the left-hand-side of Eq.~(\ref{GPstat}) can be neglected. This case
demonstrates already all important features of the behavior of dipolar
condensates. The general case will be briefly discussed at the end of this
section.

Let us introduce the mean-field dipole-dipole interaction energy per particle%
\begin{eqnarray}
V=\frac{1}{N}\int d\mathbf{r}d\mathbf{r}^{\prime }\left\vert \psi _{0}(%
\mathbf{r})\right\vert ^{2}\frac{1-3\cos ^{2}\theta }{\left\vert \mathbf{r-r}%
^{\prime }\right\vert ^{3}}\left\vert \psi _{0}(\mathbf{r}^{\prime
})\right\vert ^{2},  \label{dipenergy}
\end{eqnarray}%
which together with the trap frequencies $\omega _{z}$ and $\omega _{\rho }$
are important energy scales of the problem. One can easily see that the
value of the chemical potential $\mu $ and the behavior of the dipolar
condensate are determined by the aspect ratio of the trap $l$, the quantity $%
V/\hbar \omega _{\rho }$, and the parameter $\sigma =Na_{d}/l_{\rho }$.
Notice also that the anisotropy of the dipole-dipole interaction results in
squeezing the cloud in the radial direction and stretches it in the axial
one (along the direction of dipoles) in order to low the interaction energy.
For this reason the aspect ratio of the cloud $L=L_{z}/L_{\rho }$is always
larger than the aspect ratio $l$ of the trap. Here $L_{z}$ and $L_{\rho }$
are the axial and the radial sizes of the cloud, respectively.

We \ now summarize the results of the stability analysis of the dipolar
condensate with $\varepsilon _{dd}\gg 1$ (Eq.~(\ref{GPstat}) with $g=0$) 
\cite{Gora},\cite{Goraerr},\cite{radialrotons} (see also Ref. \cite{3Dtrap}
for the stability analysis in a general harmonic trap). The mean-field
dipole-dipole interaction is always attractive, $V<0$, for a cigar shaped
trap $l\geq 1$ causing instability (collapse) of the gas if the particle
number $N$ exceeds a critical value $N_{c}$. This critical value depends
only on the trap aspect ratio $l$. It was found that the shape of the cloud
with $N$ close to $N_{c}$ is approximately Gaussian with the aspect ratio $%
L\approx 2.1$ for a spherical trap ($l=1$), and $L\approx 3.0$ for an
elongated trap with $l\gg 1$.

For a pancake shaped trap with $l\leq 1$, the situation is more subtle. In
this case there exists a critical trap aspect ratio $l_{\star }\approx 0.43$%
, which splits the pancake shaped traps into soft pancake traps ($l_{\star
}<l\leq 1$) and hard pancake traps ($l<l_{\star }$). For soft pancake traps
one has again a critical number of particles $N_{c}$ such that the
condensates with $N>N_{c}$ are unstable. For $N$ close to $N_{c}$ and $%
l\rightarrow l_{\star }$, the aspect ratio of the cloud $L_{c}$ approaches
the aspect ratio of the trap, $L_{c}\rightarrow l_{\star }$. Note that in
this case the collapse occurs even in a pancake shaped cloud with positive
mean dipole-dipole interaction $V$ due to the behavior of the lowest
quadrupole and monopole excitations (see Section \ref{s3.2.2}).

For hard pancake traps, it was argued in Refs. \cite{Gora} and \cite{Goraerr}
that the dipolar condensate is stable for any $N$ because the dipole-dipole
interaction energy $V$ is always positive. On the other hand, by using more
advanced numerical analysis and larger set of possible trial condensate wave
functions, the authors of Ref. \cite{radialrotons} found that the dipolar
condensate in a hard pancake trap is also unstable for sufficiently large
number of particles. Similar conclusions were drawn in Ref. \cite{3Dtrap}.
It was found that the critical values of the parameter $\sigma $ for the
instability to occur are orders of magnitude larger than in soft pancake and
cigar shaped traps. In addition, the regions in parameter space were
discovered where the maximum density of the condensate is not in the center
of the cloud such that the condensate has a biconcave shape. (Analogous
behavior of the condensate in a general three-dimensional harmonic trap were
found in Ref. \cite{3Dtrap}, see also \cite{GoralRzPfau} and \cite%
{Martikainen}.) These regions exist also in the presence of\ a small contact
interaction with $\left\vert a\right\vert \lesssim 0.2a_{d}$, but their
exact position and size depend on $a$. It is important to mention that
condensates with normal and biconcave shapes behave differently when the
instability boundary is crossed. The condensate with a normal shape develops
a modulation of the condensate density in the radial direction, so-called
"radial roton" instability similar to the roton instability for the
infinite-pancake trap ($l\rightarrow \infty $) \cite{roton}, see Section \ref%
{s3.2.3}. On the other hand, it is the density modulations in the angular
coordinate that lead to the collapse of biconcave condensates - a kind of
\textquotedblleft angular roton\textquotedblright\ instability in the trap.
In the latter case one has spontaneously broken cylindrical symmetry.

The behavior of the trapped dipolar condensate can be simply captured by
means of a Gaussian variational ansatz for the condensate wave function $%
\psi _{0}(\mathbf{r})$:%
\begin{eqnarray}
\psi _{0}(\mathbf{r})=\sqrt{\frac{N}{\pi ^{3/2}L_{\rho }^{2}L_{z}}}\exp
\left( -\frac{\rho ^{2}}{2L_{\rho }^{2}}-\frac{z^{2}}{2L_{z}^{2}}\right) ,
\label{gausanzatz}
\end{eqnarray}%
where the equilibrium radial size $L_{\rho }$ and the cloud aspect ratio $L$
can be found by minimizing the energy. Note that in order to describe
biconcave shaped condensates, one has to consider (see Ref. \cite%
{radialrotons}) a linear combination of two wave functions: the first one is
a Gaussian (\ref{gausanzatz}) and the second one is the same Gaussian
multiplied by $H_{2}(x/L_{\rho })+H_{2}(y/L_{\rho })$, where $H_{2}$ is the
Hermite polynomial of the second order.

For large values of the parameters $Na/l_{i}$, where $i=\rho $ or $z$, are
large (but still $na^{3}\ll 1$), one can use the Thomas-Fermi approximation
to find the chemical potential and the shape of the cloud \cite{Eberl1}.
This case corresponds to the small\ the kinetic energy, as compared to other
energies, and, therefore, we can neglect the corresponding term with
derivatives in Eq.~(\ref{GPstat}). The GP equation then becomes%
\begin{eqnarray}
&&\left[ \frac{m}{2}(\omega _{\rho }^{2}\rho ^{2}+\omega _{z}^{2}z^{2})+%
\frac{1}{2}g\left\vert \psi (\mathbf{r},t)\right\vert ^{2}+\right.  \\
&&+\left. d^{2}\int d\mathbf{r}^{\prime }\frac{1-3\cos ^{2}\theta }{%
\left\vert \mathbf{r-r}^{\prime }\right\vert ^{3}}\left\vert \psi _{0}(%
\mathbf{r}^{\prime })\right\vert ^{2}\right] \psi _{0}(\mathbf{r})=\mu \psi
_{0}(\mathbf{r}).  \notag
\end{eqnarray}%
%
The solution of this equation reads
\begin{equation*}
\psi _{0}^{2}(\mathbf{r})=n(\mathbf{r})=n_{0}\left( 1-\frac{\rho ^{2}}{%
R_{\rho }^{2}}-\frac{z^{2}}{R_{z}^{2}}\right) 
\end{equation*}%
with the chemical potential 
\begin{equation*}
\mu =gn_{0}[1-3\varepsilon _{dd}F(L)],
\end{equation*}%
where $n_{0}$ is the density of the condensate in the center of the trap and%
\begin{eqnarray}
F\left( L\right) =\frac{1}{3}-\frac{G(L)-1}{L^{2}-1}.  \label{funcF}
\end{eqnarray}%
The energy of the condensate is%
\begin{eqnarray}
E=\frac{1}{14}Nm\omega _{\rho }^{2}R_{\rho }^{2}\left( 2+\frac{L^{2}}{l^{4}}%
\right) +\frac{15}{28\pi }\frac{N^{2}}{R_{\rho }^{2}R_{z}}g[1-3\varepsilon
_{dd}F(L)],  \label{EnrgyTF}
\end{eqnarray}%
and the radii of the condensate in the radial and axial directions $R_{\rho }
$ and $R_{z}$ are%
\begin{eqnarray}
R_{\rho }& =&\left\{ \frac{15gN}{4\pi m\omega _{\rho }^{2}L}\left[
1+\varepsilon _{dd}\left( \frac{9}{2}\frac{F(L)}{L^{2}-1}-1\right) \right]
\right\} ^{1/5},  \label{Rradial} \\
R_{z}& =&LR_{\rho },  \label{Raxial}
\end{eqnarray}%
and the corresponding aspect ratio of the cloud $L$ can be found from the
equation%
\begin{eqnarray}
3\varepsilon _{dd}\left[ 3\left( \frac{1}{2l^{4}}+1\right) \frac{L^{2}F(L)}{%
L^{2}-1}-1\right] +(\varepsilon _{dd}-1)\left( 1-\frac{L^{2}}{l^{4}}\right)
=0.  \label{eqcloudar}
\end{eqnarray}

Note, that the above equation coincides with the equation on the aspect
ratio for the Gaussian variational ansatz (\ref{gausanzatz}) when the
kinetic energy contribution is neglected, as shown in Ref. \cite{YiYou2}. It
was also found that the Thomas-Fermi approximation agrees well with
numerical results when used to analyze the stability of the condensate.
However, the critical number of particles cannot be found in the
Thomas-Fermi approximation because both terms in the expression (\ref%
{EnrgyTF}) for the energy have the same dependence $N^{7/5}$ on the number
of particles $N$ after taking into account the expressions (\ref{Rradial})
and (\ref{Raxial}) for $R_{\rho }$ and $R_{z}$.

Let us now briefly discuss the stability of a dipolar condensate in the
general case with $g\neq 0$. It is obvious that for an attractive
short-range interaction with $g<0$ the condensate can only be (meta)stable
for a small number of particles. For a repulsive short-range interaction
with $g>0$ and weak dipole-dipole interaction $0\leq \varepsilon _{dd}<1$,
the condensate is always stable. For $\varepsilon _{dd}>1$ the dipolar
condensate can be only metastable for number of particles smaller than a
critical value, $N<N_{c}$, which depends on $\varepsilon _{dd}$ and the trap
aspect ratio $l$. This means that the (metastable)condensate solution
provides only a local minimum of the energy, while the global minimum
presumably corresponds to a collapsed state with $L\rightarrow \infty $ or,
for $l<1$, a kind of density modulated state.

\subsubsection{Collective excitations and instability.}

\label{s3.2.2}

We have already mentioned that collective excitations play an important role
in the stability analysis of a dipolar condensate. They also determine the
dynamics of the gas and, therefore, are of experimental interest.

For a trapped dipolar condensate, the analysis of excitations is usually
performed on the basis of the Bogoliubov-de Gennes equations which can be
obtained by linearizing the time-dependent GP equation (\ref{GPgen}) around
the stationary solution $\psi _{0}(\mathbf{r})$. This can be achieved by
writing a solution of Eq.~(\ref{GPgen}) in the form%
\begin{equation*}
\psi (\mathbf{r},t)=\psi _{0}(\mathbf{r})+\varepsilon \lbrack u(\mathbf{r}%
)\exp (-i\omega t)+v^{\ast }(\mathbf{r})\exp (-i\omega t)]
\end{equation*}%
where the second term describes small ($\varepsilon \ll 1$) oscillations of
the condensate around $\psi _{0}(\mathbf{r})$ with (complex) amplitudes $u(%
\mathbf{r})$ and $v(\mathbf{r})$. To the first order in $\varepsilon $ the
linearization of Eq.~(\ref{GPgen}) gives the Bogoliubov-de Gennes equations%
%
\begin{eqnarray}
&&\hbar \omega u(\mathbf{r})=[-\frac{\hbar ^{2}}{2m}\nabla ^{2}-\mu +U_{%
\mathrm{tr}}(\mathbf{r})+  \notag \\
&+&2\int d\mathbf{r}^{\prime }V(\mathbf{r}-\mathbf{r}^{\prime })\left\vert
\psi _{0}(\mathbf{r}^{\prime })\right\vert ^{2}]u(\mathbf{r})+  \notag \\
&+&\int d\mathbf{r}^{\prime }V(\mathbf{r}-\mathbf{r}^{\prime })\left\vert
\psi _{0}(\mathbf{r}^{\prime })\right\vert ^{2}v(\mathbf{r}), \label{BdGu}
\end{eqnarray}%
\begin{eqnarray}
&&  \nonumber \\
&&-\hbar \omega v(\mathbf{r})=[-\frac{\hbar ^{2}}{2m}\nabla ^{2}-\mu +U_{%
\mathrm{tr}}(\mathbf{r})+  \notag \\
&&+2\int d\mathbf{r}^{\prime }V(\mathbf{r}-\mathbf{r}^{\prime })\left\vert
\psi _{0}(\mathbf{r}^{\prime })\right\vert ^{2}]v(\mathbf{r})+  \notag \\
&&+\int d\mathbf{r}^{\prime }V(\mathbf{r}-\mathbf{r}^{\prime })\left\vert
\psi _{0}(\mathbf{r}^{\prime })\right\vert ^{2}u(\mathbf{r}),  \label{BdGv}
\end{eqnarray}%
%
%
%
%
%
where $V(\mathbf{r}-\mathbf{r}^{\prime })$ is given by Eq.~(\ref{V}). The
solution of these linear equations provides the eigenfunctions ($u_{j}(%
\mathbf{r})$,$v_{j}(\mathbf{r})$) with the amplitudes $u_{j}(\mathbf{r})$
and $v_{j}(\mathbf{r})$ obeying the normalization condition%
\begin{equation*}
\int d\mathbf{r}[u_{i}^{\ast }(\mathbf{r})u_{j}(\mathbf{r})-v_{i}^{\ast }(%
\mathbf{r})v_{j}(\mathbf{r})]=\delta _{ij},
\end{equation*}%
and the corresponding eigenfrequencies $\omega _{j}$ of the collective
modes. The Bogoliubov-de Gennes equations (\ref{BdGu}) and (\ref{BdGv}) can
also be obtained by diagonalizing the Hamiltonian (\ref{HamBosetrap}) in the
Bogoliubov approximation, which corresponds to splitting the field operator $%
\hat{\psi}(\mathbf{r})$ into its mean-field value $\psi _{0}(\mathbf{r})$
and the fluctuating quantum part expressed in terms of annihilation and
creation operators $\hat{\alpha}_{j}$ and $\hat{\alpha}_{j}^{\dagger }$ of
bosonic quasiparticles (quanta of excitations):%
\begin{equation*}
\hat{\psi}(\mathbf{r})=\psi _{0}(\mathbf{r})+\sum_{j}[u_{j}(\mathbf{r})\hat{%
\alpha}_{j}+v_{j}^{\ast }(\mathbf{r})\hat{\alpha}_{j}^{\dagger }].
\end{equation*}%
The normalization condition for the amplitudes $u_{j}(\mathbf{r})$ and $%
v_{j}(\mathbf{r})$ ensures the bosonic nature of the excitations: The
operators $\hat{\alpha}_{j}$ and $\hat{\alpha}_{j}^{\dagger }$ obey the
canonical Bose commutation relations.

Nonlocality of the dipole-dipole interaction results in an
integrodifferential character of the Bogoliubov-de Gennes equations (\ref%
{BdGu}) and (\ref{BdGv}), making it hard to analyze them both analytically
and numerically. A simpler way is to study the spectrum of small
perturbations around the ground state solution of the time-dependent GP
equation (\ref{GPgen}) (see Ref. \cite{nobel28} for this approach to atomic
condensates). Using this approach in combination with the Gaussian
variational ansatz \cite{YiYou2}, \cite{GoralSantos} or the Thomas-Fermi
approximation \cite{Eberl2}, it is possible to obtain analytic results for
several low energy excitation modes.

As an illustration, let us consider a Gaussian variational wave function%
\begin{eqnarray}
&&\psi (\mathbf{r},t)= \\
&=&A(t)\prod\limits_{\eta =x,y,z}\exp \left\{ -\frac{\left[ \eta -\eta
_{0}(t)\right] ^{2}}{2R_{\eta }^{2}(t)}+i\eta \alpha _{\eta }(t)+i\eta
^{2}\beta _{\eta }(t)\right\} .  \notag  \label{timegaussanzatz}
\end{eqnarray}%
%
%
%
%
%
The variational parameters here are the complex amplitude $A$, the widths $%
R_{\eta }$, the coordinates of the center of the cloud $\eta _{0}$, and the
quantities $\alpha _{\eta }$ and $\beta _{\eta }$ related to the slope and
the curvature, respectively. The normalization of the wave function to the
total number of particles $N$ provides the constraint%
\begin{eqnarray}
N=\pi ^{3/2}\left\vert A(t)\right\vert ^{2}R_{x}R_{y}R_{z}=\mathrm{const}.
\label{norm}
\end{eqnarray}

\begin{figure}[tbp]
\begin{center}
\includegraphics[width=0.5\columnwidth]{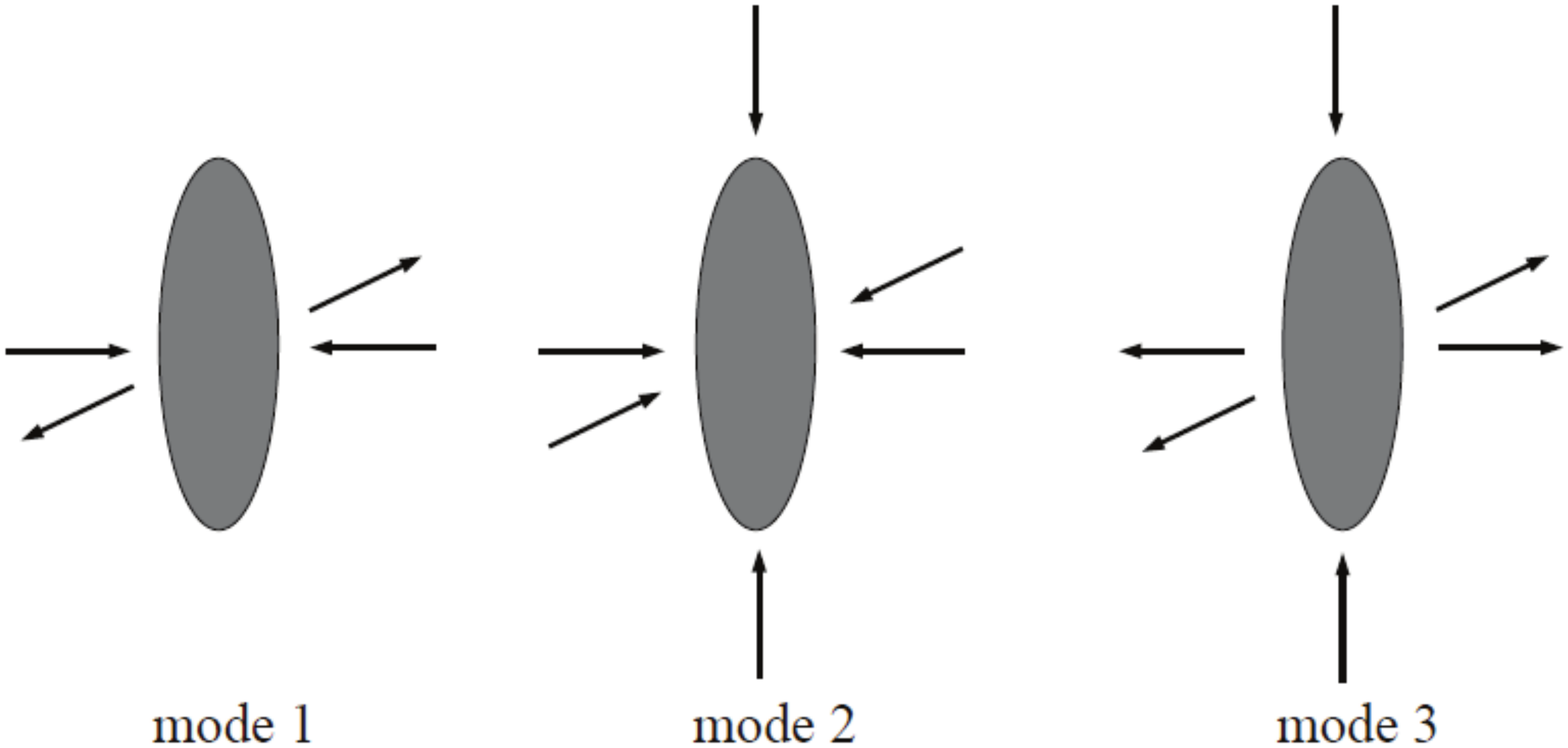}
\end{center}
\caption{Low-energy excitation modes of a dipolar condensate.}
\label{fig3}
\end{figure}

To find the equations governing the variational parameters, we notice that
the time-dependent GP equation (\ref{GPgen}) are equivalent to the
Euler-Lagrange equations for the action%
\begin{eqnarray}
S=\int dtL  \label{S}
\end{eqnarray}%
with the Lagrangian
\begin{eqnarray}
L &=&\int d\mathbf{r}\left\{ \frac{i}{2}\hbar \left[ \psi ^{\ast }\frac{%
\partial \psi }{\partial t}-\psi \frac{\partial \psi ^{\ast }}{\partial t}%
\right] -\frac{\hbar ^{2}}{2m}\left\vert \nabla \psi \right\vert ^{2}+\left[
\mu +\right. \right.   \notag \\
&-&\left. \left. U_{\mathrm{tr}}(\mathbf{r})\right] \left\vert \psi
\right\vert ^{2}\right\} -\frac{1}{2}\int d\mathbf{r}d\mathbf{r}^{\prime
}\left\vert \psi (\mathbf{r})\right\vert ^{2}V(\mathbf{r}-\mathbf{r}^{\prime
})\left\vert \psi (\mathbf{r}^{\prime })\right\vert ^{2}.  \notag  \label{L}
\end{eqnarray}%
%
%
%
%
%
%
We therefore can obtain an effective Lagrangian $L_{\mathrm{eff}}$ that
depends on the variational parameters by inserting Eq.~(\ref{timegaussanzatz}%
) into Eq.~(\ref{L}) and integrating over space coordinates. We obtain 
\begin{eqnarray}
L_{\mathrm{eff}} &=&-\hbar N\overset{.}{\varphi }+  \notag \\
&-&\frac{N}{2}\sum_{\eta }\left[ \frac{\hbar ^{2}}{2mR_{\eta }^{2}}+\left(
\hbar \overset{.}{\beta }_{\eta }+\frac{2\hbar ^{2}\beta _{\eta }^{2}}{m}+%
\frac{m\omega _{\eta }^{2}}{2}\right) R_{\eta }^{2}\right]   \notag \\
&-&\frac{N^{2}}{4\sqrt{2}\pi ^{3/2}R_{x}R_{y}R_{z}}\left[ g+\right.   \notag
\\
&+&\left. d^{2}\int d\mathbf{r}\exp \left( -\sum_{\eta }\frac{\eta ^{2}}{%
2R_{\eta }^{2}}\right) \frac{1-3\cos ^{2}\theta }{r^{3}}\right] ,
\label{Leff}
\end{eqnarray}%
%
%
%
%
%
%
where $\varphi $ is the phase of $A$ [the modulus of $A$ was excluded by
using Eq.~(\ref{norm})] and we set $\eta _{0}(t)=0$ and $\alpha _{\eta }(t)=0
$ for simplicity (this corresponds to ignoring the so-called sloshing motion
of the condensate). The standard Euler-Lagrange variational procedure%
\begin{equation*}
\frac{d}{dt}\left( \frac{\partial L}{\partial \overset{.}{q}_{j}}\right) -%
\frac{\partial L}{\partial q_{j}}=0
\end{equation*}%
with $q_{j}=$($R_{\eta }$, $\beta _{\eta }$) provides equations of motion
for the parameters $R_{\eta }$, and $\beta _{\eta }$:%
\begin{equation*}
\overset{.}{\beta }_{\eta }=\frac{m\overset{.}{R_{\eta }}}{2\hbar R_{\eta }},
\end{equation*}%
and%
\begin{eqnarray}
\overset{..}{R_{\eta }}+\frac{\partial }{\partial R_{\eta }}U(R)=0.
\label{Req}
\end{eqnarray}%
The above equation describes the motion of a particle with a unit mass in
the potential %
\begin{eqnarray}
U(R) &=&\sum_{\eta }\left( \frac{\hbar ^{2}}{2m^{2}R_{\eta }^{2}}+\frac{%
\omega _{\eta }^{2}R_{\eta }^{2}}{2}\right) +\frac{N}{4\sqrt{2}\pi
^{3/2}R_{x}R_{y}R_{z}}\left[ g+\right.   \notag \\
&+&\left. d^{2}\int d\mathbf{r}\exp \left( -\sum_{\eta }\frac{\eta ^{2}}{%
2R_{\eta }^{2}}\right) \frac{1-3\cos ^{2}\theta }{r^{3}}\right] .
\end{eqnarray}%
%
%
%
%
%
%
Therefore, the frequencies of small amplitude oscillations around the
stationary solution can be read from the second derivatives of the potential 
$U(R)$ at its minimum. In this way one can obtain the frequencies for the
first three compressional excitation modes. In Refs. \cite{YiYou2} and \cite%
{GoralSantos}, these frequencies and the corresponding shapes of the cloud
oscillations were found for a cylindrical symmetric trap, $%
R_{x}=R_{y}=R_{\rho }$, $R_{z}=LR_{\rho }$, see Fig.~\ref{fig3}. 
In the considered cylindrical geometry with dipoles are oriented along the $z
$-axis, the projection $M$ of the angular momentum on the $z$-axis is a good
quantum number that can characterize the mode: One has $M=0$ for modes $2$
and $3$ and $M=1$ for mode $1$. The modes $2$ and $3$ as often called
breathing \ and quadrupole modes, respectively, and we will follow this
convention here. (In the Thomas-Fermi approximation, one can find analytical
expressions for these modes, see Ref. \cite{Eberl2}.) Important is that with
increasing the strength of the dipole-dipole interaction, the quadrupole
mode 3 demonstrates the tendency towards instability, and becomes unstable
when $\varepsilon _{dd}$ reaches some critical value. This character of
instability via softening of the mode 2 is similar to that in a Bose gas
with a short-range attractive interaction ($a<0$).

The situation for a dipolar gas with dominant dipole interactions is more
complicated \cite{radialrotons},\cite{GoralSantos}, \cite{nobel}. It was
found (see Ref. \cite{GoralSantos}) that the instability of collective modes
of a dipolar BEC reminds that of a gas with an attractive short-range
interaction only if the trap aspect ratio is larger than the critical one, $%
l\gg l_{\star }$ (numerically was found $l>1.29$): The lowest frequency
\textquotedblright breathing\textquotedblright\ mode 2 becomes unstable when
the parameter $\sigma =Na_{d}/l_{\rho }\rightarrow \sigma _{c}$. The
variational approach discussed above provides the scaling behavior of its
frequency $\omega _{2}$ near the critical point (see Ref. \cite{GoralSantos}%
): $\omega _{2}\sim (\sigma _{c}-\sigma )^{\beta }$, with $\beta =1/4$,
which is very close to the experimental value $\beta \approx 0.2$ for
Chromium BEC \cite{Pfau5}.

For intermediate values of $l$ above $l_{\star }$ ($0.75<l<1.29$), the mode
which drives the instability (the lowest frequency mode) is a superposition
of breathing and quadrupole modes with the exponent $\beta $ still close to $%
1/4$. The mode has the breathing symmetry (mode 2) for $\sigma $ far below $%
\sigma _{c}$, while it changes and becomes quadrupole-like (mode 3) as $%
\sigma $ approaches the critical value $\sigma _{c}$.

For $l$ close to $l_{\star }$ ($l<0.75$) the lowest frequency is the
quadrupole mode 3. The frequency of this mode $\omega _{3}$ tends to zero as 
$\sigma $ approaches the critical value, $\omega _{3}\sim (\sigma
_{c}-\sigma )^{\beta }$,  with the exponent $\beta \approx 1/4$ if $l$ is
not too close to $l_{\star }$. When $l$ approaches $l_{\star }$, one has $%
\sigma _{c}\rightarrow \infty $, and $\beta >1/4$. Finally, when $l=l_{\star
}$, the frequency of the lowest frequency quadrupole mode $\omega _{3}$ can
be zero only for $\sigma =\infty $ \cite{nobel}. (Note that this result
cannot be reproduced within the Gaussian variational ansatz, which in
general does not provide reliable results close to the instability, see Ref. 
\cite{GoralSantos}.)

Collective modes for the case $l<l_{\star }$ were analyzed in Ref. \cite%
{radialrotons} on the basis of the Bogoliubov-de Gennes equations (\ref{BdGu}%
) and (\ref{BdGv}). The two possible type of solutions for the stable
condensate were already mentioned above: A pancake (normal) shaped
condensate (the maximum condensate density is in the center of the trap),
and a biconcave shaped condensate (the maximum condensate density is at some
distance from the center of the trap). It was found that in the case of a
pancake condensate, the mode which drives the instability has zero
projection of angular momentum on the $z$-axis, $M=0$, and consists of a
radial nodal pattern. The number of the nodal surfaces increases with
decreasing $l$ (flattening of the condensate). This \textquotedblleft radial
roton\textquotedblright\ mode in a confined gas can be viewed as an analog
of the roton mode in an infinite-pancake trap from Ref. \cite{roton}, see
below. In a biconcave condensate near the instability, the lowest frequency
mode has non-zero projection of the angular momentum on the $z$-axis, $M\neq
0$. This mode is an \textquotedblleft angular roton\textquotedblright\ in
the trap: For a biconcave-shaped condensate, the maximum density is along
the ring, and an angular roton corresponds to density modulation along this
ring. The instability in this case corresponds to the collapse of the
condensate due to buckling of the density in the angular coordinate, and,
therefore, breaks the cylindrical symmetry spontaneously (see Ref. \cite%
{radialrotons} for more details).

\subsubsection{Roton instability of a quasi 2D dipolar condensate.}

\label{s3.2.3}

Let us now discuss the effects of the long-range and anisotropic character
of dipole-dipole forces in the physically simpler case of an infinite
pancake shaped trap, with the dipoles perpendicular to the trap plane \cite%
{roton}. It was found that a condensate with a large density $n_{0}$ can be
dynamically stable only when a sufficiently strong short-range repulsive
interaction is present. Otherwise, excitations with the certain in-plane
momenta $q$ become unstable when the condensate density $n_{0}$ exceeds the
critical value $n_{c}$. Interestingly, the excitation spectrum of a stable
condensate with the density $n_{0}<n_{c}$ has a roton-maxon form similar to
that in the superfluid helium (see also Ref. \cite{UFischer} for the
quasi-2D version of this problem).

The time-dependent GP equation for the condensate wave function $\psi (%
\mathbf{r},t)$ of dipolar particles harmonically confined in the direction
of the dipoles ($z$-axis) reads%
\begin{eqnarray}
i\hbar \frac{\partial }{\partial t}\psi (\mathbf{r},t) &=&\left[ -\frac{%
\hbar ^{2}}{2m}\nabla ^{2}-\mu +\frac{m\omega _{z}^{2}z^{2}}{2}+g\left\vert
\psi (\mathbf{r},t)\right\vert ^{2}+\right.   \notag \\
&+&\left. \int d\mathbf{r}^{\prime }V_{d}(\mathbf{r}-\mathbf{r}^{\prime
})\left\vert \psi (\mathbf{r}^{\prime },t)\right\vert ^{2}\right] \psi (%
\mathbf{r},t),  \label{GP2D}
\end{eqnarray}%
%
%
%
%
%
where $\omega _{z}$ is the confining frequency. Let us assume the ground
state to be uniform in the in-plane directions such that the ground state
wave function $\psi _{0}(z)$ is independent of the in-plane coordinate $%
\mathbf{\rho }=(x,y)$. We can\ then integrate over $\mathbf{\rho }^{\prime }$
in the dipole-dipole term of Eq.~(\ref{GP2D}) with the result%
\begin{eqnarray}
\left[ -\frac{\hbar ^{2}}{2m}\frac{d^{2}}{dz^{2}}+\frac{m\omega _{z}^{2}z^{2}%
}{2}+(g+g_{d})\psi _{0}^{2}(z)-\mu \right] \psi _{0}(z)=0,  \label{groundwf}
\end{eqnarray}%
where $g_{d}=8\pi d^{2}/3$. This one-dimensional equation is similar to the
GP equation for short-range interactions. The simplest case corresponds to $%
g+g_{d}>0$, where the chemical potential $\mu $ is always positive. Let us
consider the case $\mu \gg \hbar \omega _{z}$ (large condensate density)
such that we can use the Thomas-Fermi approximation to find the condensate
density profile in the confined direction:%
\begin{equation*}
n_{0}(z)=\psi _{0}^{2}(z)=n_{0}(1-z^{2}/L^{2}),
\end{equation*}%
where $n_{0}=\mu /(g+g_{d})$ is the condensate maximum density and $L=(2\mu
/m\omega _{z}^{2})^{1/2}$ is the Thomas-Fermi size.

Eq.~(\ref{GP2D}) can now be linearized around the ground state solution $%
\psi _{0}(z)$ to obtain the Bogoliubov--de Gennes equations for the
excitations. These equation are Eqs. (\ref{BdGu}) and (\ref{BdGv}) with $U_{%
\mathrm{tr}}(\mathbf{r})=m\omega _{z}^{2}z^{2}/2$ and $\psi _{0}(\mathbf{r}%
)=\psi _{0}(z)$. Having translational symmetry in the in-plane directions,
we can characterized the solutions of these equations by the momentum $%
\mathbf{q}$ of the in-plane free motion. In addition to $\mathbf{q}$, we
also have an integer quantum number $j\geq 0$ related to the motion in the $z
$-direction such that the amplitudes $\left\{ u(\mathbf{r}),v(\mathbf{r}%
)\right\} $ have the form $\left\{ u(z),v(z)\right\} \exp (i\mathbf{q\rho })$%
. After introducing the new functions $f_{\pm }=u\pm v$, the Bogoliubov--de
Gennes equations read \cite{roton}%
\begin{eqnarray}
\omega f_{-}& =&H_{\mathrm{kin}}f_{+},  \label{eqfminus} \\
\omega f_{+}& =&H_{\mathrm{kin}}f_{-}+H_{\mathrm{int}}[f_{-}],
\label{eqfplus}
\end{eqnarray}%
where%
\begin{equation*}
H_{\mathrm{kin}}=\frac{\hbar ^{2}}{2m}\left[ -\frac{d^{2}}{dz^{2}}+q^{2}+%
\frac{\Delta \psi _{0}}{\psi _{0}}\right] 
\end{equation*}%
is the kinetic energy operator and
\begin{eqnarray}
H_{\mathrm{int}}[f_{-}] &=&2(g+g_{d})\psi _{0}^{2}(z)f_{-}(z)+ \\
&-&\frac{3}{2}qg_{d}\psi _{0}(z)\int\limits_{-\infty }^{\infty }dz^{\prime
}\psi _{0}(z^{\prime })\exp (-q\left\vert z-z^{\prime }\right\vert
)f_{-}(z^{\prime })  \notag  \label{Hint}
\end{eqnarray}%
%
%
%
%
%
is the interaction operator. The solution of the above equations provides
excitation frequencies $\omega _{j}(q)$ which depend on both $j$ and $%
\mathbf{q}$. The most relevant for the stability analysis is the lowest
frequency branch $\omega _{0}(q)$ for which the confined motion is not
excited in the limit $q\rightarrow 0$.

Because of nonlocality of the dipole-dipole interaction, an effective
coupling [the second term in the right-hand-side of Eq.~(\ref{Hint})]
becomes momentum dependent. For small in-plane momenta $qL\ll 1$,
excitations of the lowest branch are essentially two-dimensional with a
repulsive effective coupling, and their spectrum has been found in Ref. \cite%
{roton19}. These excitations are phonons propagating in the $xy$-plane with
the sound velocity $c_{s}$:

\begin{equation*}
\omega _{0}(q)=c_{s}q,\quad c_{s}=(2\mu /3m)^{1/2}.
\end{equation*}

In the opposite limit of large in-plane momenta $qL\gg 1$, the excitations
are three-dimensional and the interaction term is then reduced to%
\begin{equation*}
H_{\mathrm{int}}[f_{-}]=(2g-g_{d})\psi _{0}^{2}(z)f_{-}(z).
\end{equation*}%
Eqs. (\ref{eqfminus}) and (\ref{eqfplus}) \ are then equivalent to the
Bogoliubov-de Gennes equations for the excitations in a condensate with a
short-range interaction with the strength $2g-g_{d}$. This interaction is
repulsive if the parameter $\beta \equiv g/g_{d}>1/2$, and all excitation
frequencies in this case are real and positive for any in-plane momentum $q$
and condensate density $n_{0}$. In the other case $\beta <1/2$, the
interaction is attractive resulting in dynamical instability of a condensate
with regard to high momentum excitations at a sufficiently large density.

\begin{figure}[tbp]
\begin{center}
\includegraphics[width=.9\columnwidth]{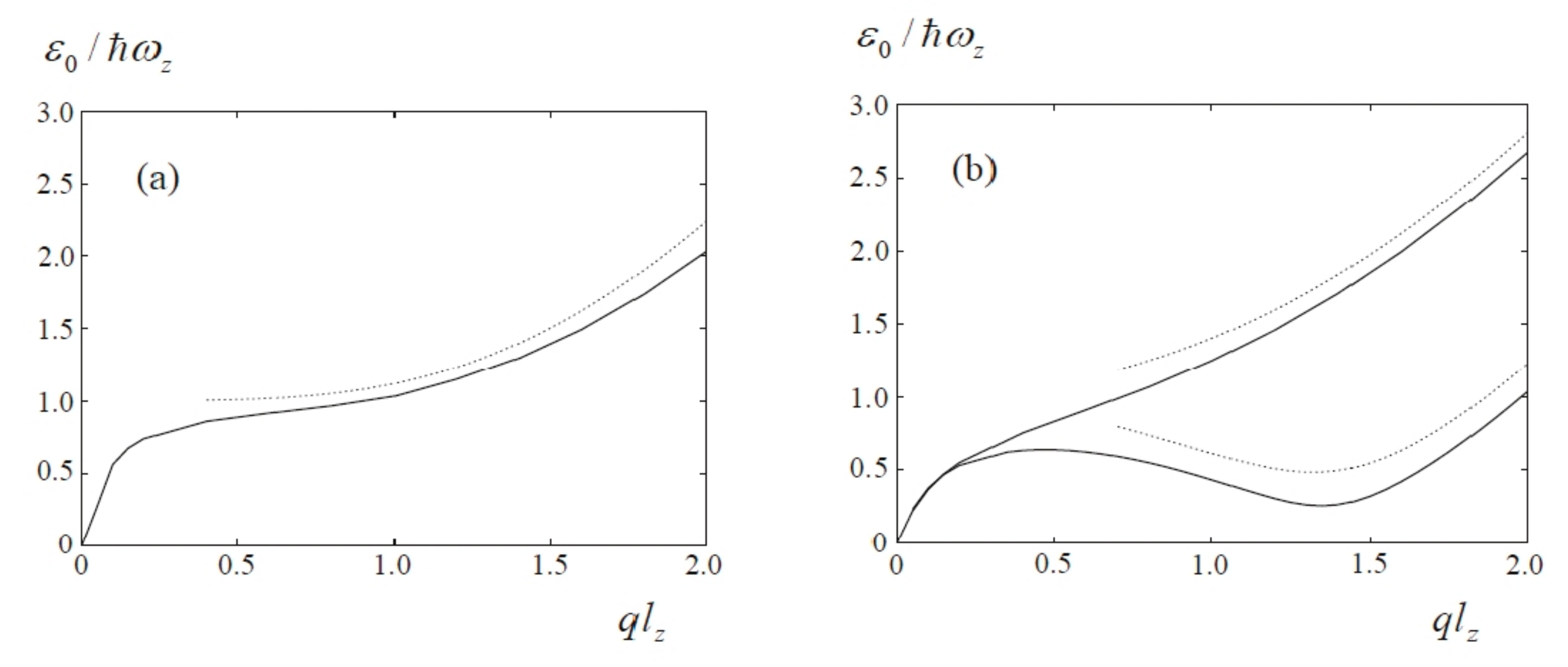}
\end{center}
\caption{Dispersion law $\protect\omega _{0}(q)$ for (a) $\protect\beta =1/2$%
, $\protect\mu /\hbar \protect\omega _{z}=343$; (b) $\protect\beta =0.53$, $%
\protect\mu /\hbar \protect\omega _{z}=46$ (upper curve) and $\protect\beta %
=0.47$, $\protect\mu /\hbar \protect\omega _{z}=54$ (lower curve). The solid
curves show numerical results. (Taken from Ref. \protect\cite{roton}.)}
\label{fig4}
\end{figure}

The analysis in the Thomas-Fermi regime of the system of equations (\ref%
{eqfminus}) and (\ref{eqfplus}) in the most interesting case $qL\gg 1$ and $%
\beta $ close to the critical value $1/2$ was performed in Ref. \cite{roton}%
. It was found that at the critical value $\beta =1/2$ the momentum
dependence of the excitation frequencies is characterized by a plateau [see
Fig.~\ref{fig4}(a)], and the $j$-th branch reads%
\begin{equation*}
\omega _{j}^{2}(q)=\varepsilon _{q}^{2}+\hbar ^{2}\omega
_{z}^{2}[1+j(j+3)/2],\quad qL\gg 1,
\end{equation*}%
where $\varepsilon _{q}=\hbar ^{2}q^{2}/2m$. 

For $\beta \neq 1/2$, the lowest branch of the spectrum is%
\begin{eqnarray}
\omega _{0}^{2}(q)=\varepsilon _{q}^{2}+\frac{(2\beta -1)(5+2\beta )}{%
3(1+\beta )(2+\beta )}\mu \varepsilon _{q}+\hbar ^{2}\omega _{z}^{2},\quad
qL\gg 1,  \label{lowestbranch}
\end{eqnarray}%
where the condition $\mu \varepsilon _{q}\left\vert 2\beta -1\right\vert
/(1+\beta )\ll \hbar ^{2}\omega _{z}^{2}$ was assumed. Eq.~(\ref%
{lowestbranch}) provides us with two types of behavior of the
lowest-frequency mode $\omega _{0}(q)$. It is either monotonously increases
with $q$ [see Fig.~\ref{fig4}(b)] when $\beta >1/2$, or has a minimum if $%
\beta <1/2$.  
Being combined it with the fact that $\omega _{0}(q)$ grows with $q$ for $%
qL\ll 1$ , the existence of this minimum results in a roton-maxon character
of the spectrum as a whole [see Fig.~\ref{fig4}(b)]. This type of the
excitation spectrum in an infinite pancake trap can be understood as
follows: For small in-plane momenta $q\ll L^{-1}$ excitations have
two-dimensional character and are phonons because dipoles being oriented
perpendicular to the plane of the trap, repel each. On the other hand,
excitations with large momenta $q\gg L^{-1}$ have three-dimensional
character and, hence, the repulsion between them is reduced. The excitation
frequency therefore decreases with an increase of $q$,reaches a minimum, and
starts to increase again as the excitations continuously enter the
single-particle regime.

The roton minimum for $\beta $ close to $1/2$ found from Eq.~(\ref%
{lowestbranch}) is located at $q=(16\mu \delta /15\hbar \omega
_{z})^{1/2}/l_{z}$, where $\delta =1/2-\beta $, and $l_{z}=\sqrt{\hbar
/m\omega _{z}}$ is the harmonic oscillator length for the confined motion,
and correspond to the excitation frequency%
\begin{equation*}
\omega _{0\mathrm{\min }}=\sqrt{\hbar ^{2}\omega _{z}^{2}-(8\mu \delta
/15)^{2}}.
\end{equation*}%
This minimum becomes deeper with increasing the density (chemical potential)
or $\delta $, and reaches zero at $q=\sqrt{2}/l_{z}$ for $\mu \delta /\hbar
\omega _{z}=15/8$. Excitations for larger values of $\mu \delta /\hbar
\omega _{z}$ have imaginary frequencies for $q\sim l_{z}^{-1}$, and,
therefore, the condensate becomes unstable. 

Eqs. (\ref{eqfminus}) and (\ref{eqfplus}) for various values of $\beta $ and 
$\mu /\hbar \omega _{z}$ were solved numerically in Ref. \cite{roton}. The
results for the excitation spectrum in the Thomas-Fermi regime are shown in
Fig.~\ref{fig4}, demonstrating a good agreement between numerical and
analytical approaches. 

For non-Thomas-Fermi condensates, the stability does not require as strong
short-range repulsive interaction as in the Thomas-Fermi regime because of a
large kinetic energy in the confined direction. The spectrum of excitations
in this case also has a roton-maxon character, although the appearance of
the roton minimum and the instability take place at smaller values of $\beta 
$, see Ref. \cite{roton} for details.

Up to now, a roton-maxon dispersion was observed only in liquid \textrm{He}
with strong interparticle interactions. Dipolar condensate provides the
first example of a weakly interacting system with a roton-maxon excitation
spectrum. This spectrum can be controlled and manipulated by changing the
density, the strength of the confinement, and the short-range interaction,
starting from the Bogoliubov-type spectrum, then creating the roton minimum,
and finally reach the point of instability.

It is important to point out that the existence of the roton minimum with $%
q\neq 0$ at a given $\beta <1/2$ for $\mu /\hbar \omega _{z}$ just below the
point of instability is likely to indicate the existence of a new ground
state presumably with a periodic density modulation. This is in contrast to
the instability of condensates with attractive short-range interaction,
which is driven by unstable long wavelength excitations resulting in local
collapses. In this case the chemical potential is negative and not bounded
from below such that no new ground state exists. In the Section \ref{sec:StronglyInteractingGases}
we will show that the excitation spectrum of a two-dimensional dipolar gas
in the strongly interacting regime $n_{2}a_{d}^{2}\gtrsim 1$ also has the
roton minimum, and the system undergoes a liquid to solid quantum phase
transition.

\section{Weakly interacting dipolar Fermi gas}

\label{s4}

In this Section we discuss fermionic dipolar gases in the weakly interacting
regime. Most of the discussion will be devoted to a single-component
(polarized) dipolar gas with only brief mentioning some results available
for two- and more component dipolar systems.

The crucial differences in the behavior of many-body fermionic systems as
compared to bosonic ones are related to the Pauli principle: identical
fermions are not allowed to be in the same quantum state. As a result, the
many-body wave function\ of a single component Fermi gas should be
antisymmetric with respect to permutations of the positions of any two
particles. In the second quantization, this requires that the field
operators $\hat{\psi}(\mathbf{r})$ and $\hat{\psi}^{\dagger }$ obey the
canonical anticommutation relations%
\begin{align*}
\left\{ \hat{\psi}(\mathbf{r}),\hat{\psi}^{\dagger }(\mathbf{r}^{\prime
})\right\} & \equiv \hat{\psi}(\mathbf{r})\hat{\psi}^{\dagger }(\mathbf{r}%
^{\prime })+\hat{\psi}^{\dagger }(\mathbf{r}^{\prime })\hat{\psi}(\mathbf{r}%
)=\delta (\mathbf{r}-\mathbf{r}^{\prime }), \\
\left\{ \hat{\psi}(\mathbf{r}),\hat{\psi}(\mathbf{r}^{\prime })\right\} &
=\left\{ \hat{\psi}^{\dagger }(\mathbf{r}),\hat{\psi}^{\dagger }(\mathbf{r}%
^{\prime })\right\} =0.
\end{align*}%
As a direct consequence, the wave function of a relative motion of two
identical fermions is allowed to have components with only odd values, $%
L=1,3,\ldots $, of the angular momentum, and vanishes when the interparticle
distance tends to zero. Therefore, the low-energy scattering of two
identical fermions is insensitive to the short-range part of their
interaction and is solely determined by the long-range dipole-dipole part $%
V_{d}$. As a result, for a single-component polarized dipolar Fermi gas we
can omit the contact term in Eq.~(\ref{V}), and the corresponding
Hamiltonian then reads%
\begin{eqnarray}
\hat{H} &=&\int d\mathbf{r}\hat{\psi}^{\dagger }(\mathbf{r})\left[ -\frac{%
\hbar ^{2}}{2m}\nabla ^{2}-\mu +U_{\mathrm{tr}}(\mathbf{r})\right] \hat{\psi}%
(\mathbf{r})+  \notag \\
&+&\frac{1}{2}\int d\mathbf{r}d\mathbf{r}^{\prime }\hat{\psi}^{\dagger }(%
\mathbf{r})\hat{\psi}^{\dagger }(\mathbf{r}^{\prime })V_{d}(\mathbf{r}-%
\mathbf{r}^{\prime })\hat{\psi}(\mathbf{r}^{\prime })\hat{\psi}(\mathbf{r}),
\label{Hdiptrap}
\end{eqnarray}%
%
%
%
%
%
where $V_{d}$ is given by Eq.~(\ref{Vd}) and $U_{\mathrm{tr}}(\mathbf{r})$
is the trapping potential (if present).

Another consequence of the Pauli principle is that the state of a many-body
system of fermions at a low temperature $T$ is completely different from
that for bosons. The average number of ideal fermions in\ a quantum state $i$
with the energy $\varepsilon _{i}$ is given by the Fermi-Dirac distribution%
\begin{equation*}
n(\varepsilon _{i})=f_{\mathrm{FD}}(\varepsilon _{i})=\frac{1}{\exp
[(\varepsilon _{i}-\mu )/T]+1},
\end{equation*}%
where $\mu $ is the chemical potential, which depends on $T$ and, as usual,
ensures the fixed total number of particles $N=\sum_{i}n(\varepsilon _{i})$.
The ground state ($T=0$) therefore corresponds to all quantum state with $%
\varepsilon _{i}\leq \mu (T=0)\equiv \varepsilon _{F}$ being completely
completely occupied [$n(\varepsilon _{i})=1$], while the states with $%
\varepsilon _{i}>\varepsilon _{F}$ are are being empty [$n(\varepsilon
_{i})=0$]. The energy $\varepsilon _{F}$ is called the Fermi energy and sets
the typical energy scale  a many-body system of fermions. 

The ground state of an ideal homogeneous Fermi gas with $\varepsilon
_{p}=p^{2}/2m$ corresponds to the so-called Fermi sphere: All quantum states
with momenta $p$ below the Fermi momentum $p_{F}=\sqrt{2m\varepsilon _{F}}$
are occupied and the states with $p>p_{F}$ are empty. The states with
momentum $p=p_{F}$ form a surface in the momentum space called the Fermi
surface, which separates the filled and empty states. Semiclassical state
counting provides the relation between the Fermi momentum $p_{F}$ and the
density $n$ of a single-component homogeneous gas:%
\begin{eqnarray}
n=\frac{p_{F}^{3}}{6\pi ^{2}\hbar ^{3}}  \label{npF}
\end{eqnarray}

For a trapped Fermi gas we can establish the similar relation but between
the local Fermi momentum $p_{F}(\mathbf{r})$ and the local density of the
gas $n(\mathbf{r})$,%
\begin{eqnarray}
n(\mathbf{r})=\frac{p_{F}(\mathbf{r})^{3}}{6\pi ^{2}\hbar ^{3}},
\label{npFlocal}
\end{eqnarray}%
where%
\begin{eqnarray}
p_{F}(\mathbf{r})=\sqrt{2m\left[ \mu -U_{\mathrm{tr}}(\mathbf{r})\right] },
\label{pFlocal}
\end{eqnarray}%
provided the chemical potential $\mu $ is much larger than the level spacing
in the trapping potential $U_{\mathrm{tr}}(\mathbf{r})$. This condition
corresponds to a large number of particle $N$ in the trap, most of them
occupying high energy states of the trapping potential. The wave functions
of these states are quasiclassical (see, for example, \cite{LL3}), and the
calculation of the gas density results in Eq. \ref{npFlocal}, which is the
essence of the local-density (Thomas-Fermi) approximation. This
approximation is legitimate when the trapping potential changes slowly over
the distances of the order of the average interparticle separation $%
n^{-1/3}\sim \hbar /p_{F}$. For an ideal Fermi gas in a harmonic potential
expression (\ref{pFlocal}) gives%
\begin{align*}
p_{F}(\mathbf{r})& =\sqrt{2m\left[ \mu -U_{\mathrm{tr}}(\mathbf{r})\right] }
\\
& =p_{F}(0)\sqrt{1-\frac{x^{2}}{R_{TFx}^{2}}-\frac{y^{2}}{R_{TFy}^{2}}-\frac{%
z^{2}}{R_{TFz}^{2}}},
\end{align*}%
where $p_{F}(0)=\sqrt{2m\mu }$ is the Fermi momentum in the center of the
trap and $R_{TF\alpha }=\sqrt{2\mu /m\omega _{\alpha }^{2}}$ is the
Thomas-Fermi size of the gas cloud in the $\alpha $-direction. The density
of the gas in this approximation, according to Eq. (\ref{npFlocal}), reads%
\begin{eqnarray}
n_{TF}(\mathbf{r})=n_{0}\left( 1-\frac{x^{2}}{R_{TFx}^{2}}-\frac{y^{2}}{%
R_{TFy}^{2}}-\frac{z^{2}}{R_{TFz}^{2}}\right) ^{3/2},  \label{nTF}
\end{eqnarray}%
where $n_{0}=(2m\mu )^{3/2}/(6\pi ^{2}\hbar ^{3})$ is the density in the
trap center. The calculation of the total number of particle with the use of
the above density distribution relates the chemical potential $\mu $ to the
total number of particle $N$ and the parameters of the trap:

\begin{equation*}
\mu =\hbar \overline{\omega }(6N)^{1/3},
\end{equation*}%
where $\overline{\omega }=(\omega _{x}\omega _{y}\omega _{z})^{1/3}$.

For understanding the properties of the fermionic systems it is important to
keep in mind that the ground state in the form of a filled Fermi sphere
stores a large amount of kinetic energy. This guaranties applicability of
the perturbation theory for dilute dipolar systems with $p_{F}a_{d}/\hbar
\ll 1$. Another consequence is the improved stability of fermionic dipolar
gases, as compared to the bosonic ones, against collapse due to the
attractive part of the dipole-dipole interaction. This can be understood as
follows: The energy per volume for a homogenous dipolar Fermi gas with
density $n$ effectively attractive two-body interaction can be written as%
\begin{equation*}
\frac{E(n)}{V}=\frac{3}{5}\varepsilon _{F}n-Ad^{2}n^{2},
\end{equation*}%
where the first term is the kinetic energy of the filled Fermi sphere and
the second term is the interaction energy with some numerical coefficient $A$
of the order unity. The first term scales as $n^{2/3}$ [see Eq.~(\ref{npF})]
and provides an energy barrier between states with small $n$ and positive
energy and collapsing states with $n\rightarrow \infty $ and negative
energy. Therefore, one expects the stability against collapse when the
system is dilute: $n^{-1/3}\gg a_{d}$ or, equivalently, $p_{F}a_{d}/\hbar
\ll 1$, and instability in the dense system with  $n^{-1/3}\lesssim a_{d}$
when the interaction energy becomes comparable or larger that the kinetic
energy. Applying this arguments to a trapped single-component dipolar Fermi
gas, one expects to have a stable gas for $p_{F}(0)a_{d}/\hbar <1$ or $%
N^{1/6}a_{d}/l<1$, where $l$ is the oscillator length and we use Eq.~(\ref%
{nTF}) to obtain $N\sim \lbrack p_{F}(0)l/\hbar ]^{6}$. We provide more
details on the issue of stability later.

\subsection{Effects of dipole-dipole interactions.}

When considering effects of interparticle interactions in Fermi systems, one
has to keep in mind two possible scenarios depending on whether the
properties of the system (the ground state and excitations) change
continuously or abruptly when interactions are switched on. In the first
case an interacting system is called normal Fermi liquid (in other words,
belongs to Fermi liquid universality class) and has properties that are very
much similar to those of an ideal Fermi gas. Of course, the interaction
leads to appearance of new features (collective modes, for example), which
are absent in a non-interacting gas, but for many applications the system
can be considered as an ideal gas of fermionic non-interacting
quasiparticles. For weak interparticle interactions, the properties of the
interacting system can be obtained with the help of perturbation theory
starting from the non-interacting Fermi gas. In the second scenario, the
ground state and excitations of interacting system are qualitatively
different from those of the non-interacting Fermi gas, and a system is in
non-Fermi liquid universality class. This scenario is usually associated
with breaking\ of some symmetries of an ideal gas: phase (or gauge) symmetry
in a superfluid Fermi liquid or translational symmetry in a charge-density
wave or crystal state. The new ground state cannot be continuously connected
with the filled Fermi sphere (ground state of a non-interacting Fermi gas),
and, therefore, one has to go beyond simple perturbative expansions to
describe those states. It is important to mention that one does not
necessarily need a strong interaction for the second scenario. For example,
even an infinitesimally small attractive interaction results in a superfluid
ground state. The smallness of the interaction in this case manifests itself
in low (much smaller that $T_{F}$) critical temperature $T_{c}$ - the
temperature above which the superfluid properties disappear and the system
returns to normal Fermi liquid. In contrast, the charge-density wave state
requires strong interaction, and this state disappears at temperature
comparable or larger that $T_{F}$ when the gas is essentially classical.

As we will discuss below, depending on an experimental setup, both scenarios
are possible in a polarized dipolar gas: A 3D polarized dipolar gas is in
the superfluid state for low temperatures, $T<T_{c}\ll T_{F}$, and in the
normal state (Fermi-liquid) for $T>T_{c}$. The state of a monolayer of
polarized dipoles depends on temperature and relative angle between the
dipole moments and the motion plane of molecules. For the perpendicular
orientation of dipoles, the gas is in the normal state, but, starting from
some critical tilting angle, becomes a superfluid at small enough
temperatures $T<T_{c}$. In both cases, the increase of the strength of the
dipole-dipole interaction leads to the instability of the homogeneous state
resulting to a collapse or formation of density-wave state with broken
translational symmetry.

\subsection{Normal (anisotropic) Fermi liquid state.}

We begin with discussion of a normal Fermi liquid state of a dipolar fermi
gas, which is a generic state for a fermionic dipolar gas at finite ($%
T_{c}<T<T_{F}$) temperatures, as well as for a purely repulsive (in a
monolayer, for example) dipolar gas, in a weakly interacting regime $%
k_{F}a_{d}<1$. Following the original idea of Landau, an interacting normal
Fermi system (Fermi liquid) can be described in terms of fermionic
quasiparticles, which can be viewed as particles together with disturbances
they produce in the system due to interactions with another particles
(particles surrounded by particle-hole excitations) - dressed particles. In
the ground state, the quasiparticles occupy all states with energies smaller
or equal than the chemical potential $\mu\approx\varepsilon_{F}$ forming a
filled Fermi sphere (in a spatially uniform dipolar gas this corresponds to
a deformed Fermi sphere in momentum space due to anisotropy of the
dipole-dipole interaction, see below). Excited states are obtained by moving
some quasiparticle from occupied states below $\mu$ to empty ones above -
creation particle-hole excitations. The advantage of this description is
that weakly excited states correspond to small number of particle-hole
excitations near the Fermi surface and, hence, can be described using the
dilute gas approximation. Note that, although we are talking about filled
quasiparticle states inside the Fermi sphere, quasiparticles in the Fermi
liquid are well-defined only in the vicinity of the Fermi surface where
their energies $\varepsilon(\mathbf{p})$ are much larger than the inverse of
their life-times $\tau_{\mathbf{p}}$ due to decay via creation of
particle-hole pairs. (In a weakly interacting gas the quasiparticles are
well-defined for all momenta.) This is because the presence of occupied
states below the Fermi energy strongly reduces the phase space volume for
such processes, and, as a result, the life-time $\tau_{\mathbf{p}}$ of
quasiparticle near the fermi surface in a Fermi liquid are much larger than
the corresponding time $\tau_{c}$ in a classical gas with the same
interparticle interactions and density, $\tau_{\mathbf{p}}\sim\lbrack%
\varepsilon_{F}/\varepsilon(\mathbf{p})]^{2}\tau_{c}\gg\tau_{c}$. But those
are quasiparticles we actually need to describe low-energy excitations of
the Fermi system and its behavior at low temperatures and under weak
external perturbations.

The change of the quasiparticle distribution $\delta n_{\mathbf{p}}$ (we
assume here a spatially homogeneous gas) results in the change of the energy
of the system%
\begin{eqnarray}
\delta E=\sum_{\mathbf{p}}[\varepsilon(\mathbf{p})+\mu]\delta n_{\mathbf{p}}+%
\frac{1}{2V}\sum_{\mathbf{p},\mathbf{p}^{\prime}}f(\mathbf{p},\mathbf{p}%
^{\prime})\delta n_{\mathbf{p}}\delta n_{\mathbf{p}^{\prime}},
\label{Energy_f-func}
\end{eqnarray}
where $\varepsilon(\mathbf{p})=\delta E/\delta n_{\mathbf{p}}|_{\delta n_{%
\mathbf{p}}=0}$ is the quasiparticles energy (counted from the chemical
potential $\mu$) and the second term describes the interaction between
quasiparticles with $f(\mathbf{p},\mathbf{p}^{\prime})=\delta^{2}E/\delta n_{%
\mathbf{p}}\delta n_{\mathbf{p}^{\prime}}|_{\delta n_{\mathbf{p}}=0}$ being
the Landau $f$-function, which plays a crucial role in the Fermi-liquid
theory, and can be either calculated perturbatively (if the gas is weakly
interacting) or measured experimentally. Note that the function $f$
describes the change of the quasiparticles energy under the change of
quasiparticle distribution as a result of their interaction,%
\begin{equation*}
\delta\varepsilon(\mathbf{p})=\sum_{\mathbf{p}^{\prime}}f(\mathbf{p},\mathbf{%
p}^{\prime})\delta n_{\mathbf{p}^{\prime}},
\end{equation*}
which gives rise to Fermi-liquid corrections and make possible collective
motion of quasiparticles (collective modes) even when collisions between
quasiparticles can be neglected.

For states close to the Fermi surface (the boundary between occupied and
empty states), the quasiparticle energy has the form%
\begin{equation*}
\varepsilon(\mathbf{p})\approx\frac{p_{F}}{m^{\ast}}(p-p_{F}),
\end{equation*}
where $p_{F}$ is the Fermi momentum specifying the Fermi surface in momentum
space, and $m^{\ast}$ is the effective mass. The Fermi momentum $p_{F}$ is
related to the density in the same way as in the ideal gas, Eq.~(\ref{npF}),
reflecting the fact that numbers of particles and quasiparticles are equal,
while the effective mass $m^{\ast}$ can be expressed in terms of the $f$%
-function (see, for example Ref. \cite{LL9}). The compressibility $%
\kappa=n^{-2}d\mu/dn$, where $\mu=dE/dn$ is the chemical potential, is
another important quantity, which can also be expressed in terms of $f$%
-function. For a stable system one must have $\kappa>0$.$\,$Therefore,\ the
knowledge of the compressibility as a function of system parameters provides
us with stability conditions of the system against collapse. The stability
of the system against possible deformations $\delta n_{\mathbf{p}}$ of the
Fermi surface around its equilibrium form (Pomeranchuk criterion\textbf{\ }%
\cite{Pomeranchuk1958}, \cite{LL9}]) can be obtained from the requirement
that the change of the energy caused by this deformation, Eq.~(\ref%
{Energy_f-func}), is positive. In this way one can detect instabilities
different from collapse, related to the nonuniform change of the Fermi
surface.

The $f$-function determines also collective modes in the Fermi liquid
(Landau zero sound), which correspond to a collisionless coherent dynamics
of particle-hole excitations. The simplest way to describe zero sound is to
use a semiclassical (or Wigner) quasiparticle distribution function $n(%
\mathbf{r},\mathbf{p},t)$, which is the Fourier transform of a
single-particle density matrix with respect to the relative coordinate,%
\begin{equation*}
n(\mathbf{r},\mathbf{p},t)=\int d\mathbf{r}^{\prime}\left\langle \hat{\psi }%
^{\dagger}(\mathbf{r+r}^{\prime}/2,t)\hat{\psi}(\mathbf{r}-\mathbf{r}%
^{\prime}/2,t)\right\rangle \exp(-i\mathbf{pr}^{\prime}/\hbar),
\end{equation*}
and describes the local momentum distribution of particles at position $%
\mathbf{r}$. In the ground state of a spatially homogeneous system, $n_{0}(%
\mathbf{r},\mathbf{p},t)=\theta(p_{F}-p)$ corresponds to a filled Fermi
sphere. For a thermal equilibrium state, the step function $\theta(p_{F}-p)$
has to be replaced with the Fermi-Dirac distribution, $n_{T}(\mathbf{r},%
\mathbf{p},t)=[\exp(\varepsilon(\mathbf{p})/T)+1]^{-1}$.Time evolution of
non-equilibrium distributions $n(\mathbf{r},\mathbf{p},t)=n_{\mathrm{eq}}(%
\mathbf{p})+\delta n(\mathbf{r},\mathbf{p},t)$ are described by the
quasiparticle kinetic equation%
\begin{eqnarray}
(\frac{\partial}{\partial t}+\frac{\partial\varepsilon}{\partial\mathbf{p}}%
\frac{\partial}{\partial\mathbf{r}}+\frac{\partial\varepsilon}{\partial 
\mathbf{r}}\frac{\partial}{\partial\mathbf{p}})n(\mathbf{r},\mathbf{p},t)=0,
\label{KinEq}
\end{eqnarray}
where $\varepsilon=\varepsilon(\mathbf{p})+\sum_{\mathbf{p}^{\prime}}f(%
\mathbf{p},\mathbf{p}^{\prime})\delta n(\mathbf{r},\mathbf{p},t)$ and the
collision integral, which normally appears on the right-hand side, is set to
zero assuming low temperatures, as discussed above. The solutions of this
equation of the form $\delta n(\mathbf{r},\mathbf{p},t)\sim\varkappa (%
\mathbf{p})\exp[i(\mathbf{kr}-\omega t)]\ll n_{\mathrm{eq}}(\mathbf{p})$
with $\omega=ck$ and $c>v_{F}$ are called Landau zero sound and described
coherent motion of particle-hole pairs - propagation of a deformation of the
Fermi surface. Generically, the solutions of this kind exist when $f(\mathbf{%
p},\mathbf{p}^{\prime})$ is positive (for more details and exact criterion
see, for example, Refs. \cite{LL9}). Note that the condition $c>v_{F}$
separates the zero-sound from the continuum of particle-hole excitations and
ensures its long life-time. In the opposite case the energy of zero-sound
waves would be inside the continuum of particle-hole excitations and, hence,
the waves would rapidly decay into incoherent particle-hole excitations
(Landau damping).

\subsubsection{Anisotropic Fermi surface and single-particle excitations}

Due to anisotropy of the dipole-dipole interaction, the Fermi surface in a
dipolar gas is not a sphere any more and the modulus of the Fermi momentum
depends on the direction. The effective mass becomes a tensor that can be
defined from the relation between the Fermi momentum $\mathbf{p}_{F}$ and
fermi velocity $\mathbf{v}_{F}=\partial\varepsilon(\mathbf{p})/\partial 
\mathbf{p}|_{\mathbf{p}=\mathbf{p}_{F}}$, $p_{Fi}=m_{ij}^{\ast}v_{Fj}$. This
can easily be seen by using the following variational ansatz \cite%
{PhysRevA.77.061603}, \cite{SogoNJP}%
\begin{eqnarray}
n(\mathbf{p})=\theta\lbrack p_{F}^{2}-\frac{1}{\alpha}(p_{x}^{2}+p_{y}^{2})-%
\alpha^{2}p_{z}^{2}]  \label{deformation_ansatz}
\end{eqnarray}
and find the variational parameter $\alpha$ by minimizing the total energy
of the system with the interaction energy calculated in the Hartree-Foch
approximation:%
\begin{eqnarray}
\frac{E}{V}=\int\frac{d\mathbf{p}}{(2\pi\hbar)^{3}}\frac{p^{2}}{2m}n(\mathbf{%
p})-\frac{1}{2}\int\frac{d\mathbf{p}d\mathbf{p}^{\prime}}{(2\pi \hbar)^{6}}n(%
\mathbf{p})V_{\mathrm{d}}(\mathbf{p}-\mathbf{p}^{\prime })n(\mathbf{p}%
^{\prime}),  \label{EdipHF}
\end{eqnarray}
where $V$ is the volume of the system and only exchange (Fock) term
contribute to the dipole-dipole interaction energy because the direct
Hartree contribution%
\begin{equation*}
E_{d}=\frac{1}{2}\int d\mathbf{r}d\mathbf{r}^{\prime}n(\mathbf{r})V_{\mathrm{%
d}}(\mathbf{r}-\mathbf{r}^{\prime})n(\mathbf{r}^{\prime}),
\end{equation*}
where $n(\mathbf{r})$ is the gas density, vanishes in a homogeneous gas as a
result of angular integrations. It was found that $\beta<1$ so that the
Fermi surface is deformed into a spheroid stretched along the direction of
dipoles (prolate spheroid). These findings were supported by microscopic
calculations in the spirit of Landau liquid theory in Refs. \cite%
{PhysRevA.81.033601} and \cite{PhysRevA.81.023602}. The quasiparticle energy
calculated from Eq.~(\ref{EdipHF}) reads%
\begin{equation*}
\varepsilon(\mathbf{p})=\frac{p^{2}}{2m}-\int\frac{d\mathbf{p}^{\prime}}{%
(2\pi\hbar)^{3}}V_{\mathrm{d}}(\mathbf{p}-\mathbf{p}^{\prime})n(\mathbf{p}%
^{\prime})-\mu,
\end{equation*}
which corresponds to the following Landau $f$-function for a spatially
homogeneous gas 
\begin{eqnarray}
f^{\mathrm{\hom}}(\mathbf{p},\mathbf{p}^{\prime})=-V_{\mathrm{d}}(\mathbf{p}-%
\mathbf{p}^{\prime})  \label{f_func}
\end{eqnarray}
with only exchange contribution. Note that the condition of spatial
homogeneity of the gas is essential for validity of Eq.~(\ref{f_func}). This
is because the Fourier component of the dipole-dipole interaction $V_{%
\mathrm{d}}(\mathbf{q})$ in non-analytic for $\mathbf{q}\rightarrow0$ (the
limit depends on the direction $\mathbf{q}$ approaches zero). As a result,
the direct (Hartree) contribution vanishes only in the spatial homogeneity
gas, in which one has $V_{\mathrm{d}}(\mathbf{q})$ averaged over the
direction of $\mathbf{q}$, which is zero. In an inhomogeneous gas, this is
not the case and one also has the contribution of the direct dipole-dipole
interaction, see, for example, Eq.~(\ref{zero-sound}) describing spatially
inhomogeneous variations of the quasiparticle distribution.

Setting $\varepsilon(\mathbf{p})$ to zero gives the position of the Fermi
surface in momentum space $\mathbf{p}_{F}=\mathbf{n}p_{F}(\mathbf{n})$,
where $\mathbf{n}$ is a (radial) unit vector (direction) in momentum space.
The chemical potential $\mu$ then has to be defined selfconsistently from
assuming a fixed gas density $n=k_{F}^{3}/6\pi^{2}$: 
\begin{equation*}
n=\int\frac{d\mathbf{p}}{(2\pi\hbar)^{3}}\theta\lbrack-\varepsilon (\mathbf{p%
})]=\frac{1}{(2\pi\hbar)^{3}}\int d\mathbf{n}\int_{0}^{p_{F}(\mathbf{n}%
)}p^{2}dp\text{.}
\end{equation*}
For weak interaction one finds \cite{PhysRevA.81.033601}%
\begin{equation*}
p_{F}(\mathbf{n})=\hbar k_{F}\left[ 1+\frac{1}{9\pi}a_{d}k_{F}(3\cos
^{2}\theta_{\mathbf{n}}-1)\right] ,
\end{equation*}
where $\theta_{\mathbf{n}}$ is the angle between $\mathbf{n}$ and the $z$%
-axis. This gives $\beta=1-2a_{d}k_{F}/9\pi$. The energy and the chemical
potential are%
\begin{equation*}
\frac{E}{V}=\frac{3}{5}\frac{\hbar^{2}k_{F}^{2}}{2m}\left[ 1-\frac{4}{%
81\pi^{2}}(a_{d}k_{F})^{2}\right]
\end{equation*}
and%
\begin{equation*}
\mu=\frac{\hbar^{2}k_{F}^{2}}{2m}\left[ 1-\frac{28}{405\pi^{2}}%
(a_{d}k_{F})^{2}\right] .
\end{equation*}
After expanding the quasiparticle energy in the vicinity of the Fermi
surface, $\varepsilon(\mathbf{p})\approx\mathbf{v}_{F}(\mathbf{p}-\mathbf{p}%
_{F})$, one finds \cite{PhysRevA.81.023602} that the tensor of the effective
mass has only longitudinal $m_{L}^{\ast}(\mathbf{n})$ and and transverse $%
m_{T}^{\ast }(\mathbf{n})$ components:%
\begin{equation*}
\mathbf{v}_{F}=\mathbf{n}\frac{p_{F}(\mathbf{n})}{m_{L}^{\ast}(\mathbf{n})}+%
\mathbf{e}_{\theta}\frac{p_{F}(\mathbf{n})}{m_{L}^{\ast}(\mathbf{n})},
\end{equation*}
where $\mathbf{e}_{\theta}$ is the polar angle unit vector and 
\begin{align*}
\frac{m}{m_{L}^{\ast}(\mathbf{n})} & =\left[ 1-\frac{1}{6\pi}%
a_{d}k_{F}(3\cos^{2}\theta_{\mathbf{n}}-1)\right] , \\
\frac{m}{m_{T}^{\ast}(\mathbf{n})} & =\frac{1}{3\pi}a_{d}k_{F}\sin 2\theta_{%
\mathbf{n}}.
\end{align*}
Calculations for moderate strengths of the dipole-dipole interactions and
finite temperatures were performed in Refs. \cite{PhysRevA.77.061603},%
\textbf{\ }\cite{PhysRevA.81.033601}, \cite{PhysRevA.82.033608}\textbf{\ }
including the trapped case (Ref. \cite{PhysRevA.77.061603}), as well as 2D
(monolayer) and 1D (tube) gases, and two-component dipolar gas (Ref. \cite%
{PhysRevA.81.023602}). We mention here only some details for a monolayer and
refer to these references for more details.

In a 2D dipolar gas (monolayer), when the chemical potential $\mu$ is much
smaller than the frequency $\omega_{z}$ of the transverse confining
potential in the $z$-direction, $\mu\ll\omega_{z}$, the transverse wave
function of particles is limited to the ground state wave function $%
\phi_{0}(z)$ of the harmonic oscillator, such that $\psi(\mathbf{r})=\psi(%
\mathbf{\rho})\phi _{0}(z)$, where $\mathbf{\rho}=(x,y)=(\rho\cos\varphi_{%
\mathbf{\rho}},\rho \sin\varphi_{\mathbf{\rho}})$ is the in-plane vector.
The corresponding effective 2D dipole-dipole interaction for the in-plane
motion
\begin{eqnarray}
V_{\mathrm{d}}^{2D}(\mathbf{\rho })& =&\int dzdz^{\prime }\phi _{0}(z)^{2}V_{%
\mathrm{d}}(\mathbf{\rho },z-z^{\prime })\phi _{0}(z^{\prime })^{2}  \notag
\\
& =&-\frac{d^{2}}{\sqrt{2}l_{z}}\frac{1}{\rho ^{2}}\left\{ 2\mathrm{P}%
_{2}(\cos \theta )\Psi (1/2,0;\rho ^{2}/2l_{z}^{2})+\right.  \notag \\
&-&3\left[ \mathrm{P}_{2}(\cos \theta )-\frac{1}{2}\sin ^{2}\theta \cos
2\varphi _{\mathbf{\rho }}\right] \times  \notag \\
&\times&\left.\Psi (1/2,-1;\rho ^{2}/2l_{z}^{2})\right\} ,  \label{V2D}
\end{eqnarray}%
%
%
%
%
%
%
where $\Psi(a,b;z)$ is the confluent hypergeometric function and $\theta$ is
the angle between the direction of dipoles (in the $(x,z)$-plane) and the
motion $(x,y)$-plane, has the following Fourier transform%
\begin{eqnarray}
V_{\mathrm{d}}^{2D}(\mathbf{p})=-\sqrt{2}\pi\frac{d^{2}}{l_{z}}w(pl_{z}/\hbar%
\sqrt{2})[2\mathrm{P}_{2}(\cos\theta)-\sin^{2}\theta\cos2\varphi _{\mathbf{p}%
}],  \label{V2Dgeneral}
\end{eqnarray}
where $w(x)=x\exp(x^{2})\mathrm{erfc}(x)$ with $\mathrm{erfc}(x)$ being the
error function and $\phi_{\mathbf{p}}$ is the angle between $\mathbf{p}$ and
the $x$-axis. For $p\sim p_{F}\ll l_{z}$ one has%
\begin{eqnarray}
V_{\mathrm{d}}^{2D}(\mathbf{p})\approx-\pi\frac{d^{2}}{\hbar}p\,[2\mathrm{P}%
_{2}(\cos\theta)-\sin^{2}\theta\cos2\varphi_{\mathbf{p}}],  \label{Vdip2D}
\end{eqnarray}
which is linear in $p$. (Strictly speaking, expression (\ref{Vdip2D})
contains also a constant which depends on the regularization of the Fourier
integral at the origin. This constant corresponds to a short-range
inerparticle interaction and, hence, has no physical effect in a single
component Fermi gas because all its contributions should vanish upon proper
antisymmetrization. We therefore set this constant to zero.) Within the
Hartree-Fock approximation one then obtains (assuming $a_{d}k_{F}\ll1$)%
\begin{equation*}
p_{F}(\mathbf{n})=\hbar k_{F}\left[ 1+\frac{8}{15\pi}a_{d}k_{F}\sin^{2}%
\theta\cos2\varphi_{\mathbf{p}}\right]
\end{equation*}
for the position $\mathbf{p}=\mathbf{n}p_{F}(\mathbf{n})$ of the Fermi
surface, where now $\mathbf{n}=(\cos\varphi_{\mathbf{n}},\sin\varphi _{%
\mathbf{n}})$ is the unit vector of direction in the $(x,y)$-plane, and%
\begin{align*}
\frac{m}{m_{2DL}^{\ast}(\mathbf{n})} & =1+\frac{4a_{d}k_{F}}{3\pi}[\mathrm{P}%
_{2}(\cos\theta)-\frac{7}{10}\sin^{2}\theta\cos2\varphi _{\mathbf{p}}], \\
\frac{m}{m_{2DT}^{\ast}(\mathbf{n})} & =\frac{16}{15\pi}a_{d}k_{F}\sin
^{2}\theta\cos2\varphi_{\mathbf{p}}
\end{align*}
for the longitudinal (along $\mathbf{n}$) and transverse (perpendicular to $%
\mathbf{n}$) components of the effective mass, respectively \cite%
{PhysRevA.81.023602}. Note that for $\theta=0$ (dipoles are perpendicular to
the plane and, therefore, the system has rotational symmetry around the $z $%
-axis) the deformation of the Fermi surface disappears and $\mathbf{v}_{F}=%
\mathbf{p}_{F}/m^{\ast}$ with $m^{\ast}/m\approx1-4a_{d}k_{F}/3\pi$.

\subsubsection{Collective modes (Landau zero sound)}

\begin{figure}[ptb]
\begin{center}
\includegraphics[width=.9\columnwidth]{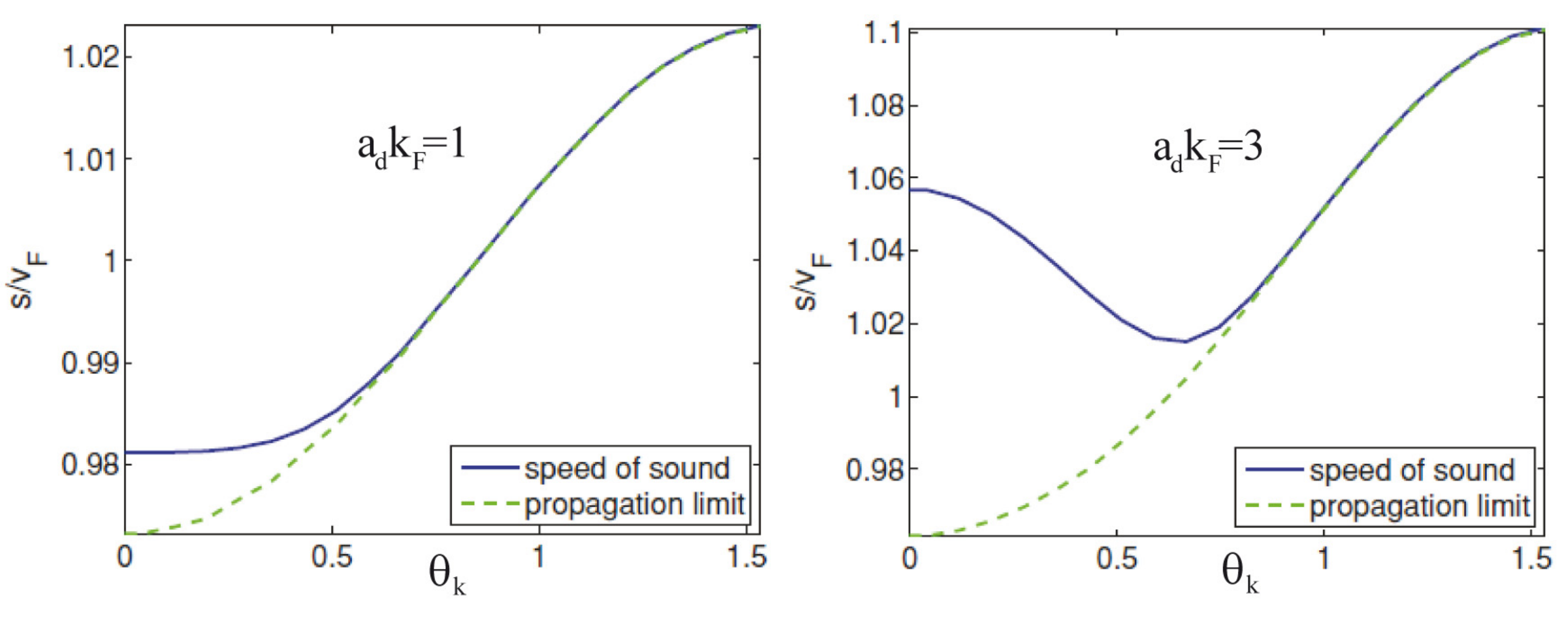}
\end{center}
\caption{Speed of sound (solid curve) in a 3D dipolar Fermi gas as a
function of the angle of propagation $\protect\theta_{\mathbf{k}}$ relative
to the direction of polarization, for dipolar interaction strengths $%
a_{d}k_{F}=1$ (left panel) and $a_{d}k_{F}=3$ (right panel). The speed is
measured in units of $v_{F}$. The dashed curve represents a lower bound on
the speed of any undamped mode. (Taken from Ref. \protect\cite%
{PhysRevA.81.033601}.)}
\label{XXX}
\end{figure}

Collective modes in a dipolar gas can be studied on the basis of the kinetic
equation, Eq.~(\ref{KinEq}). For small deviation of the quasiparticle
distribution from equilibrium of the form 
\begin{equation*}
\delta n(\mathbf{r},\mathbf{p},t)\sim\varkappa(\mathbf{p})\exp[i(\mathbf{kr}%
-\omega t)]\ll n_{\mathrm{eq}}(\mathbf{p}),
\end{equation*}
the kinetic equation reduces to the following equation on the unknown
function $\varkappa(\mathbf{p})$
\begin{eqnarray}
&&\lbrack \omega -\mathbf{k}\nabla _{\mathbf{p}}\varepsilon (\mathbf{p}%
)]\varkappa (\mathbf{p})=  \notag \\
&&\mathbf{k}\nabla _{\mathbf{p}}\varepsilon (\mathbf{p})\frac{\partial n_{%
\mathrm{eq}}(\mathbf{p})}{\partial \varepsilon (\mathbf{p})}\int \frac{d%
\mathbf{p}^{\prime }}{(2\pi \hbar )^{D}}f(\mathbf{k},\mathbf{p},\mathbf{p}%
^{\prime })\mathbf{\varkappa }(\mathbf{p}^{\prime }),  \label{zero-sound}
\end{eqnarray}%
%
%
%
%
%
%
where $\nabla_{\mathbf{p}}=\partial/\partial\mathbf{p}$ and the wave vector $%
k$ is assumed to be much smaller than $k_{F}=p_{F}/\hbar$, $k\ll k_{F}$. In
the Hartree-Fock approximation, the $f$-function is $f(\mathbf{k},\mathbf{p},%
\mathbf{p}^{\prime})=V_{\mathrm{d}}(\mathbf{k})-V_{\mathrm{d}}(\mathbf{p}-%
\mathbf{p}^{\prime})$, see comments below Eq.~(\ref{f_func}) and Refs. \cite%
{PhysRevA.81.033601},\textbf{\ }\cite{PhysRevA.81.023602}. Eq.~(\ref%
{zero-sound}) was analyzed numerically for a 3D gas in Refs. \cite%
{PhysRevA.81.033601} and \cite{PhysRevA.81.023602}\textbf{, }and in 2D gas
in Refs. \cite{PhysRevA.81.023602}, \cite{SiebererBaranov}. (Note that in a
2D gas -- monolayer -- the first term in the $f$-function can be omitted
following the arguments from the end of the previous section.) The results
for the sound velocity in a 3D gas together with the propagation limit due
to particle-hole continuum are shown in Fig.~\ref{XXX} as a function of the
propagation angle (the angle between $\mathbf{k}$ and the $z$-axis). 
They show that the collective mode can propagate only in the certain cone of
directions around the direction of dipoles polarization. This is
counterintuitive to some extend because this is the direction in which two
dipoles attract each other; on the other hand in the perpendicular
direction, in which dipoles repel each other and, hence, one would expect
the existence of the collective mode, no zero sound is possible. This can be
understood by noting that the zero-sound propagation is dominated by the
exchange contribution, not direct one, and, therefore, the intuition based
on the direct interaction does not work, see Ref. \cite{PhysRevA.81.033601}
for more discussions of this issue. The sound velocity depends strongly on
both the propagation direction $\theta_{\mathbf{k}}$ and interactions
strength $a_{d}k_{F}$: It increases monotonically with $\theta_{\mathbf{k}}$
for $a_{d}k_{F}\lesssim1$ and becomes a non-monotonic in $\theta_{\mathbf{k}%
} $ for $a_{d}k_{F}>1$, see Fig.~\ref{XXX}.

\begin{figure}[tbp]
\begin{center}
\includegraphics[width=0.4\columnwidth]{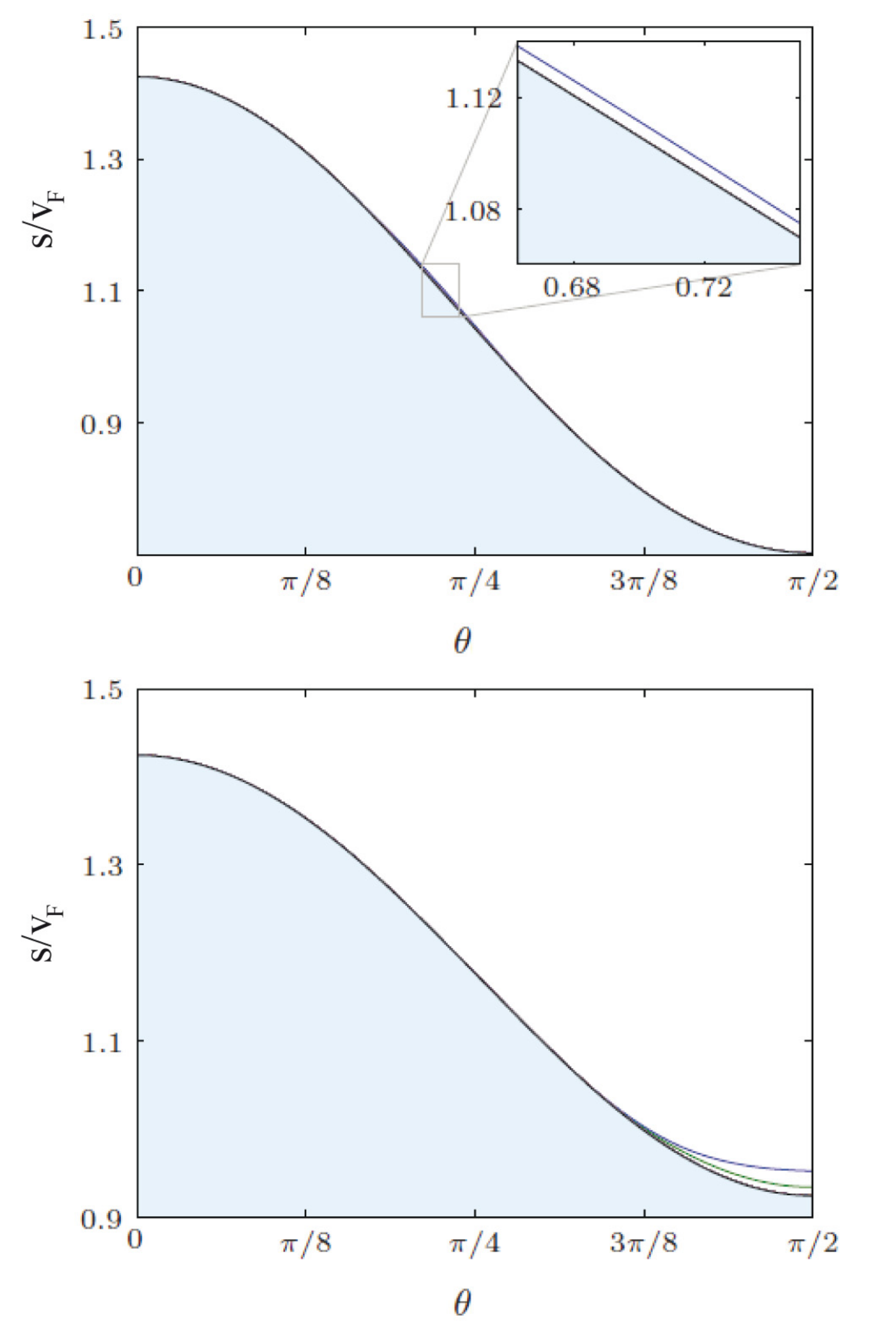}
\end{center}
\caption{The speed of zero sound in a 2D dipolar Fermi gas as a function of
the tilting angle $\protect\theta $ for $a_{d}k_{F}=1$ and the propagation
angles $\protect\phi _{\mathbf{k}}=0$ (upper panel) and $\protect\phi _{%
\mathbf{k}}=\protect\pi /2$ (lower panel). The shaded region correspond to
strong damping. (Taken from Ref. \protect\cite{SiebererBaranov}.)}
\label{YYY1}
\end{figure}

\begin{figure}[tbp]
\begin{center}
\includegraphics[width=.6\columnwidth]{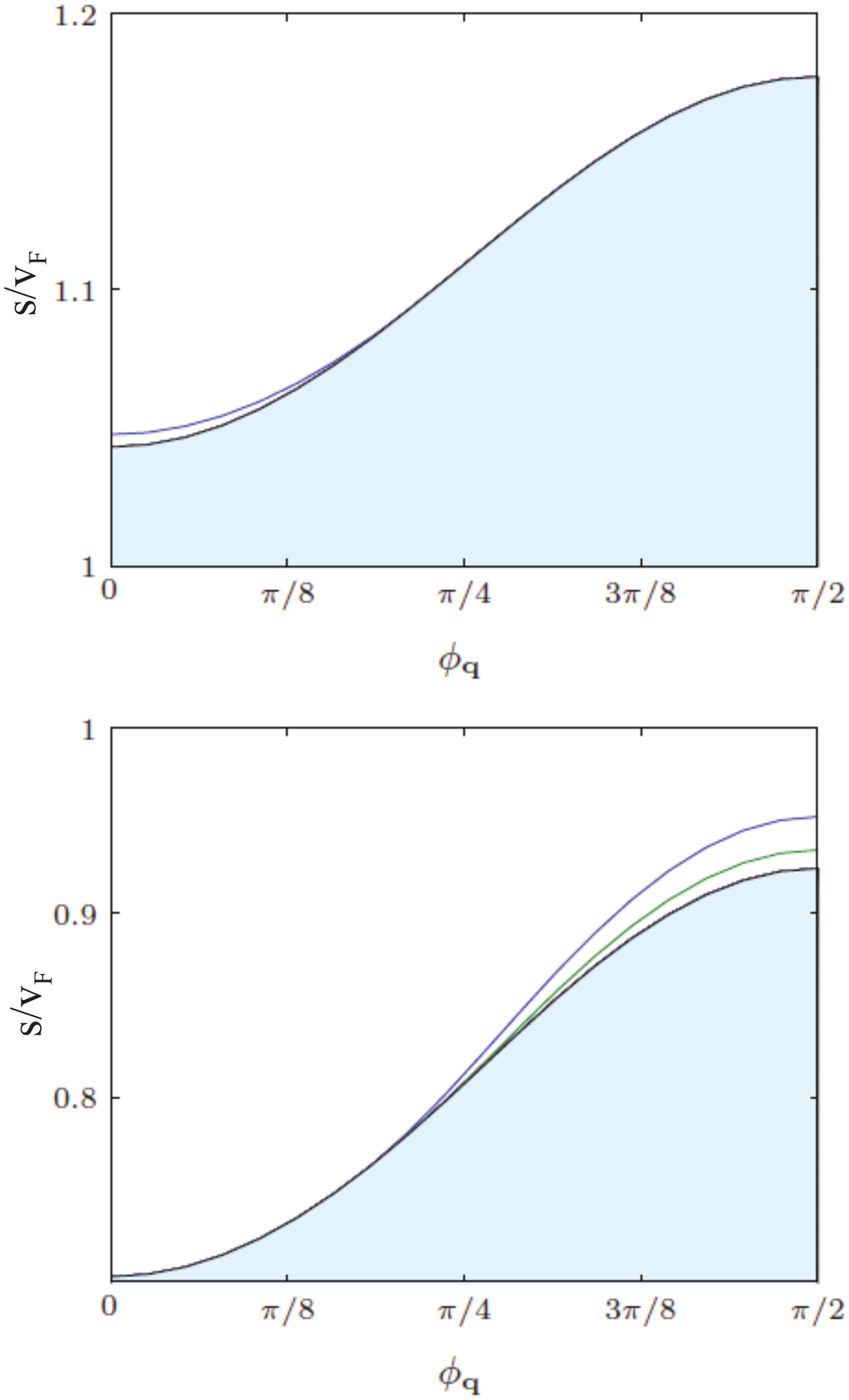}
\end{center}
\caption{The speed of zero sound in a 2D dipolar Fermi gas as a function of
the propagation angle $\protect\phi _{\mathbf{k}}$ for $a_{d}k_{F}=1$ and $%
\protect\theta =\protect\pi /4$ (upper panel) and $\protect\theta =\protect%
\pi /2$ (lower panel). The shaded region correspond to strong damping.
(Taken from Ref. \protect\cite{SiebererBaranov}.)}
\label{YYY2}
\end{figure}

In a dipolar monolayer (quasi-2D gas), the situation is even more intriguing
because the existence of zero-sound and the value of the sound velocity
strongly depend on the propagation direction ( $\phi _{\mathbf{k}}$ is the
angle between $\mathbf{k}$ and the $x$-axis), on the tilting angle $\theta $%
, and on the strength of the interaction. In this case, there is no
collective modes if the tilting angle is smaller that some critical value
that depends on the interaction strength, see\textbf{\ }Fig.~\ref{YYY1}. 
This is again counterintuitive because the direct interaction for small
tilting angles is purely repulsive (dipoles are almost perpendicular to the
motion plane), and one would expect stable collective zero-sound modes.
However, similar to the 3D case, the $f$-function contains only exchange
contribution, and this explains such peculiar behavior of collective modes.
(The collective modes without exchange contribution were considered in Ref.%
\textbf{\ }\cite{PhysRevA.82.033608}.) Note also that with increasing the
tilting angles from the critical one, the directions in plane, in which one
has propagating zero-sound, changes from those around the projection $%
\mathbf{d}_{\parallel }$ of the dipole moment $\mathbf{d}$ on the plane to
those around the direction perpendicular to $\mathbf{d}_{\parallel }$ (when
the polarization of dipoles approaches the plain), see Fig.~\ref{YYY2}. 

Note, however, that the above results for collective modes were obtained
with the $f$-function in the lowest-order Hartree-Fock approximation, $f(%
\mathbf{k},\mathbf{p},\mathbf{p}^{\prime })=V_{\mathrm{d}}(\mathbf{k})-V_{%
\mathrm{d}}(\mathbf{p}-\mathbf{p}^{\prime })$. Taking higher order terms
into account can change the situation: As it was shown in Ref. %
\cite{lu2012} for the case of a 2D dipolar Fermi gas polarized
perpendicular to the motion plane, $\theta =0$, the inclusion of second
order contributions to the $f$-function results is the appearance of a
stable zero-sound mode with the velocity $s=v_{F}[1+2(a_{d}k_{F})^{4}]$.

\subsection{BCS pairing in a homogeneous single-component dipolar Fermi gas.}

\label{s4.2}

The partial attractiveness of the dipole-dipole interaction opens the
possibility for BCS pairing in a fermionic many-body dipolar system at
sufficiently low temperatures. As we will see in this section, the pairing
in dipolar systems has generically an unconventional character (different
from a singlet isotropic $s$-wave pairing as in a two-component fermionic
system with an isotropic attractive interaction), and a superfluid state has
many peculiar properties that are different from those of conventional
superconductors. Indeed, the $s$-wave (together with other even angular
momentum) two-particle interaction channel is forbidden in a
single-component Fermi gas by the Pauli principle. On the other hand, the
angular part of the matrix element for the dipole-dipole interaction between
the states with the angular momentum $L=1$ ($p$-wave channel) and its
projection on the $z$-axis $M=0$ is negative (i.e. corresponding to an
attractive interaction):%
\begin{eqnarray}
\left\langle L=1,M=0\left\vert 1-3\cos ^{2}\theta \right\vert
L=1,M=0\right\rangle =-\frac{4\pi }{5}<0,  \label{attraction}
\end{eqnarray}%
and, therefore, can lead to BCS pairing. (The matrix elements between the
states with $M=\pm 1$ are positive.) It easy to see that this pairing should
be anisotropic, reaching its maximum amplitude in the direction of dipolar
polarization when two dipoles attract each other, and being zero in the
perpendicular directions corresponding to repulsive dipole-dipole
interaction. As we will see, the dominant contribution has $p$-wave symmetry.

The Cooper pairing in a polarized single-component dipolar Fermi gas has
been discussed in Refs. \cite{YouMarinescu} and \cite{dipstoof} within the
BCS approach with the restriction to purely $p$-wave pairing. An exact value
of the critical temperature and the angular dependence of the order
parameter for a dilute gas were found in Ref. \cite{BCShom}.

After omitting the contribution of the short-range part of the interparticle
interaction, as discussed above, the Hamiltonian of a homogeneous
single-component polarized dipolar Fermi gas reads 
\begin{eqnarray}
H &=&\int d\mathbf{r}\hat{\psi}^{\dagger }(\mathbf{r})\left[ -\frac{\hbar
^{2}}{2m}\nabla ^{2}-\mu \right] \hat{\psi}(\mathbf{r})+  \notag \\
&+&\frac{1}{2}\int d\mathbf{r}d\mathbf{r}^{\prime }\hat{\psi}^{\dagger }(%
\mathbf{r})\hat{\psi}^{\dagger }(\mathbf{r}^{\prime })V_{d}(\mathbf{r}-%
\mathbf{r}^{\prime })\hat{\psi}(\mathbf{r}^{\prime })\hat{\psi}(\mathbf{r}).
\label{1}
\end{eqnarray}%
We considered the property of the system with this Hamiltonian in the dilute
limit $na_{d}^{3}\ll 1$ and at temperatures $T$ much smaller than the
chemical potential $\mu $ (or the Fermi energy $\varepsilon _{F}$), $T\ll
\mu =\varepsilon _{F}\approx (\hbar ^{2}/2m)(6\pi ^{3}n)^{2/3}$, relevant
for Cooper pairing. In this case one can neglect the corrections $\delta \mu 
$ to the chemical potential due to the dipole-dipole interaction because $%
\delta \mu \sim $ $d^{2}n\sim \varepsilon _{F}(na_{a}^{3})^{1/3}\ll
\varepsilon _{F}$. 

The BCS pairing corresponds to a nonzero value of the order parameter 
\begin{equation*}
\Delta (\mathbf{r}_{1}-\mathbf{r}_{2})=V_{\mathrm{d}}(\mathbf{r}_{1}\mathbf{%
-r}_{2}\mathbf{)}\left\langle \hat{\psi}(\mathbf{r}_{1})\hat{\psi}(\mathbf{r}%
_{2})\right\rangle ,
\end{equation*}%
which can be viewed as a wave function of Cooper pairs. because of
anticommutativity of the fermionic field operators, $\Delta (\mathbf{r}_{1}-%
\mathbf{r}_{2})$ changes sign under the exchange of particles $\mathbf{r}%
_{1}\longleftrightarrow \mathbf{r}_{2}$ forming a pair. As a consequence,
the order parameter in momentum space%
\begin{equation*}
\Delta (\mathbf{p})=\int d\mathbf{r}\exp (-i\mathbf{pr}/\hbar )\Delta (%
\mathbf{r})
\end{equation*}%
is also antisymmetric, $\Delta (-\mathbf{p})=-\Delta (\mathbf{p})$. Because
of the anisotropy of the dipole-dipole interaction, the angular momentum $L$
of the relative motion of two particles is not a conserved quantum number,
but its projection on the $z$-axis (in the considered geometry) does. We can
therefore write $\Delta (\mathbf{p})$ in the form%
\begin{eqnarray}
\Delta (\mathbf{p})=\sum_{\mathrm{odd}\,L}\Delta _{L}(p)\mathrm{Y}_{L0}(\hat{%
\mathbf{p}}),  \label{Deltasum}
\end{eqnarray}%
where $\mathrm{Y}_{LM}(\hat{\mathbf{p}})$ are the spherical harmonics and $%
\hat{\mathbf{p}}$ is the unit vector in the direction of the momentum $%
\mathbf{p}$. We keep in the sum only odd angular momentum $L$ following the
discussion above and set $M=0$ in every term. This is because $M$ is a
conserved and, following Eq.~(\ref{attraction}), only for $M=0$ one has an
attractive interaction.

A nonzero order parameter and, therefore, the superfluid properties in the
system appear for temperatures below some temperature $T_{c}$ which is the
critical temperature of the superfluid transition. This critical temperature
and the order parameter $\Delta $ for temperatures below $T_{c}$ can be
found from the gap equation \cite{BCS},\cite{leggett} (we use the momentum
representation and assume the order parameter to be real a real function of
momentum $\mathbf{p}$)

\begin{eqnarray}
\Delta (\mathbf{p})=-\int \frac{d\mathbf{p}^{\prime }}{(2\pi \hbar )^{3}}V(%
\mathbf{p},\mathbf{p}^{\prime })\frac{\tanh (E(\mathbf{p}^{\prime })/2T)}{2E(%
\mathbf{p}^{\prime })}\Delta (\mathbf{p}^{\prime }),  \label{1.1}
\end{eqnarray}%
where $E(\mathbf{p})=\sqrt{\Delta ^{2}(\mathbf{p})+(p^{2}/2m-\mu )^{2}}$ is
the energy of single-particle excitations in the superfluid gas. The
effective interparticle interaction is described by the function $V(\mathbf{p%
},\mathbf{p}^{\prime })=V_{d}(\mathbf{p-p}^{\prime })+\delta V(\mathbf{p},%
\mathbf{p}^{\prime })$. Here $V_{d}(\mathbf{q})$ is the Fourier transform of
the bare dipole-dipole interaction potential $V_{d}(\mathbf{r})$:

\begin{eqnarray}
V_{d}(\mathbf{q})=\frac{4\pi }{3}d^{2}(3\cos ^{2}(\theta _{\mathbf{q}})-1),
\label{2'}
\end{eqnarray}%
with $\theta _{\mathbf{q}}$ being the angle between the momentum $\mathbf{q}$
and the $z$-axis, and $\delta V(\mathbf{p},\mathbf{p}^{\prime })$
corresponds to corrections to the bare interparticle interaction $V_{d}$
resulted from many-body effects. The effective interaction $V(\mathbf{p},%
\mathbf{p}^{\prime })$ describes all scattering processes in the system
which transform a pair of particles with momenta $\mathbf{p}^{\prime }$ and $%
-\mathbf{p}^{\prime }$ into a pair with momenta $\mathbf{p}$ and $-\mathbf{p}
$. The leading process here is the direct scattering of the two particles on
each other [the term $V_{d}(\mathbf{p-p}^{\prime })$], while the many-body
corrections $\delta V(\mathbf{p},\mathbf{p}^{\prime })$ describe processes
of higher order in $V_{d}$. The leading terms in $\delta V$ are second order
in $V_{d}$ (see Ref. \cite{GM}) and correspond to scattering processes in
which the two colliding particles interact with each other indirectly with
an involvement of particle-hole excitations which they create in the system
(see more details in Ref. \cite{BCShom}). These processes are important even
in the weakly interacting case (although they are of the second order in the
small parameter) because they result in the pre-exponential factor in the
expression for the critical temperature, see Eq.~(\ref{Tc}).

The gap equation (\ref{1.1}) can be simplified for temperatures just below $%
T_{c}$ because for such temperatures the order parameter $\Delta (\mathbf{p})
$ is small and, hence, the right-hand-sides of Eq.(\ref{1.1}) can be
expanded in powers of $\Delta (\mathbf{p})$. The resulting equation

\begin{eqnarray}
\Delta (\mathbf{p})=-\int \frac{d\mathbf{p}^{\prime }}{(2\pi \hbar )^{3}}V(%
\mathbf{p},\mathbf{p}^{\prime })\left[ K(p^{\prime })\Delta (\mathbf{p}%
^{\prime })+\,\frac{\partial K(p^{\prime })}{\partial \xi ^{\prime }}\frac{%
\Delta ^{3}(\mathbf{p}^{\prime })}{2\xi ^{\prime }}\right] ,  \label{2}
\end{eqnarray}%
where $K(p)=\tanh (\xi /2T)/2\xi $ and $\xi =p^{2}/2m-\mu $, is equivalent
to the Ginzburg-Landau equation for a spatially homogeneous order parameter.
.

Note that Eq. (\ref{2}) always has the trivial solution $\Delta =0$ which
corresponds to a normal phase of the Fermi gas. The Cooper pairing is
associated with a nontrivial solution of the gap equation (\ref{2}) which
exists for temperatures $T\leq T_{c}$. To find the critical temperature $%
T_{c}$, it is sufficient to keep only the linear term in the square brackets
in the right-hand-side of Eq.(\ref{2}) because $\Delta \rightarrow 0$ for $%
T\rightarrow T_{c}$. The corresponding linearized gap equation allows also
finding the momentum dependence of the order parameter. The absolute
temperature dependent value of $\Delta $ is determined by the nonlinear term
in the right-hand-side of Eq.(\ref{2}).

The result for the critical temperature reads (see Ref\textbf{.} \cite%
{BCShom} for details)%
\begin{eqnarray}
T_{c}=1.44\varepsilon _{F}\exp \left( -\frac{\pi \varepsilon _{F}}{12nd^{2}}%
\right) .  \label{Tc}
\end{eqnarray}%
For temperatures $T$ close to $T_{c}$, the anisotropic order parameter $%
\Delta (\mathbf{p})$ on the Fermi surface, $p=p_{F}$, has the form%
\begin{eqnarray}
\Delta (p_{F}\hat{\mathbf{p}})\approx 2.5T_{c}\sqrt{1-\frac{T}{T_{c}}}\phi
_{0}(\hat{\mathbf{p}}),  \label{DeltapF}
\end{eqnarray}%
where%
\begin{eqnarray}
\phi _{0}(\hat{\mathbf{p}})=\sqrt{2}\sin \left( \frac{\pi }{2}\cos \theta _{%
\mathbf{p}}\right)   \label{phi0}
\end{eqnarray}%
with $\theta _{\mathbf{p}}$ being the angle between the vector $\mathbf{p}$
and the $z$-axis.

\begin{figure}[tbp]
\begin{center}
\includegraphics[width=.9\columnwidth]{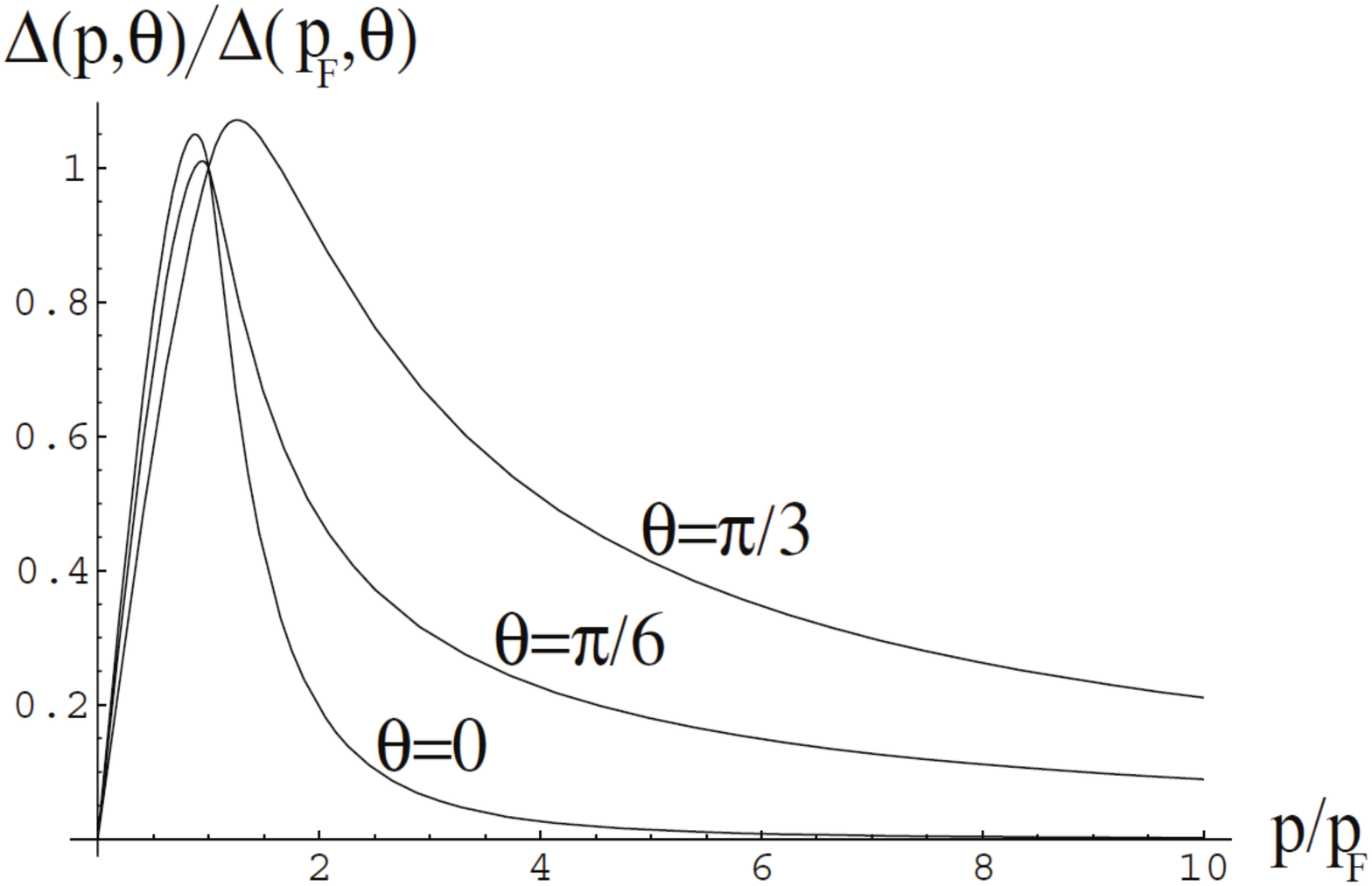}
\end{center}
\caption{The order parameter $\Delta (p,\protect\theta )$ [in units of $%
\Delta (p_{F},\protect\theta )$] as a function of the momentum $p$ (in units
of $p_{F}$ ) for various values of the polar angle $\protect\theta $. (Taken
from Ref. \protect\cite{BCShom}.)}
\label{fig9}
\end{figure}

For momenta away from the Fermi surface the order parameter is%
\begin{equation*}
\Delta (\mathbf{p})\approx -\frac{\pi }{8d^{2}}\int \frac{d\hat{\mathbf{p}}%
^{\prime }}{4\pi }V_{d}(\mathbf{p}-p_{F}\hat{\mathbf{p}}^{\prime })\Delta
(p_{F}\hat{\mathbf{p}}^{\prime }).
\end{equation*}%
The dependence of the order parameter $\Delta (\mathbf{p})$ on the modulus $p
$ of the momentum $\mathbf{p}$ for several values of the angle $\theta _{%
\mathbf{p}}$ is shown in Fig.~\ref{fig9}.  
This momentum dependence of the order parameter in a dipolar Fermi gas is in
contrast to that for pairing (both $s$- and $p$-wave) due to a short-range
interparticle interaction, in which case the order parameter is a constant
for momenta $p\lesssim \hbar /r_{0}$, where $r_{0}$ is the range of the
interparticle interaction, and rapidly decays $p>\hbar /r_{0}$.

The anisotropy of the order parameter in the momentum space described by the
function $\phi _{0}(\mathbf{n})=\sqrt{2}\sin {[(\pi /2)\cos \theta ]}$, see
Eq.~(\ref{DeltapF}), provides another difference from the conventional $s$%
-wave pairing, say in a two-component Fermi gas with a short-range
intercomponent attractive interaction. As a result of this anisotropy, the
gap $\left\vert \Delta (p_{F}\hat{\mathbf{p}})\right\vert $ in the spectrum
of single-particle excitations in a dipolar superfluid gas depends on the
direction of momentum $\hat{\mathbf{p}}$: The gap reaches its maximum in the
directions parallel to the direction of the dipoles ($\theta _{\mathbf{p}}=0$%
, $\pi $), while it vanishes in the direction perpendicular to the dipoles ($%
\theta _{\mathbf{p}}=\pi /2$). Similar anisotropic is expected in the
properties of collective excitations, and, as a result, in the response of
the dipolar superfluid dipolar Fermi gas.

The vanishing of the single-particle gap at $\theta _{\mathbf{p}}=\pi /2$
(and arbitrary azimuthal angle $\varphi $, i.e. on the line on the Fermi
surface $p=p_{F}$) results in the $T^{2}$ dependence of the specific heat of
the gas at low temperatures $T\ll \Delta _{0}\sim T_{c}$ (the contribution
of collective excitations is proportional to $T^{3}$). Note that for the
conventional $s$-wave pairing, the low-temperature specific heat is
determined by the contribution of collective modes ($\propto T^{3}$), while
the contribution of single-particle excitations is exponentially suppressed.

The above mentioned properties of a superfluid dipolar Fermi gas is similar
to those of the polar phase of superfluid liquid $^{3}$\textrm{He}. This
phase, however, cannot be realized in experiments because it has higher
energy than experimentally observed $A$ and $B$ phases (see, e.g. Ref. \cite%
{3He}). Note that several heavy-fermion compounds in a superconducting state
(for a review of superconducting phases of heavy-fermion compounds see, e.g.
Refs. \cite{heavy-fermions},\cite{heavy-fermions1}, and \cite%
{heavy-fermions2}) also have lines of zeros of the order parameter on the
Fermi surface and, as a consequence, the $T^{2}$-dependence of the
low-temperature specific heat (see, e.g. Ref. \cite{T2}).

\subsection{BCS pairing in a trapped single-component dipolar Fermi gas.}

\label{s4.3}

Similar to the bosonic case, the trap geometry strongly influences the BCS
pairing in a polarized dipolar Fermi gas as a result of the dipole-dipole
interaction. It is natural to expect that cigar-shaped traps are more
favorable for pairing then pancake-shaped ones because the dipole-dipole
interaction is on average attractive in the former case and repulsive in the
latter one. As a result, the critical temperatures in cigar-shaped traps
should be higher. This question was addressed in Refs. \cite{njp} and \cite%
{nobel}.

In a trapped gas, the BCS order parameter 
\begin{eqnarray}
\Delta (\mathbf{r}_{1},\mathbf{r}_{2})=V_{\mathrm{d}}(\mathbf{r}_{1}\mathbf{%
-r}_{2}\mathbf{)}\left\langle \hat{\psi}(\mathbf{r}_{1})\hat{\psi}(\mathbf{r}%
_{2})\right\rangle   \label{DeltaTrap}
\end{eqnarray}%
depends on both coordinates $\mathbf{r}_{1}$ and $\mathbf{r}_{2}$ and not
only on their difference $\mathbf{r}_{1}\mathbf{-r}_{2}$ as in the spatially
homogeneous case, because the translational symmetry is broken by the
trapping potential. To find the critical temperature $T_{c\mathrm{trap}}$ in
the trap, it is sufficient\ to consider the linearized gap equation: 
\begin{eqnarray}
\Delta (\mathbf{r}_{1}\mathbf{,r}_{2})=-V_{\mathrm{d}}(\mathbf{r}_{1}\mathbf{%
-r}_{2}\mathbf{)}\int d\mathbf{r}_{3}d\mathbf{r}_{4}K(\mathbf{r}_{1}\mathbf{%
,r}_{2}\mathbf{;r}_{3}\mathbf{,r}_{4})\Delta (\mathbf{r}_{3}\mathbf{,r}_{4})
\label{lingapeq}
\end{eqnarray}%
with the kernel%
\begin{equation*}
K(\mathbf{r}_{1}\mathbf{,r}_{2}\mathbf{;r}_{3}\mathbf{,r}_{4})=\sum_{\nu
_{1},\nu _{2}}\frac{\tanh (\xi _{\nu _{1}}/2T)+\tanh (\xi _{\nu _{2}}/2T)}{%
\xi _{\nu _{1}}+\xi _{\nu _{1}}}\phi _{\nu _{1}}(\mathbf{r}_{1})\phi _{\nu
_{2}}(\mathbf{r}_{2})\phi _{\nu _{1}}^{\ast }(\mathbf{r}_{3})\phi _{\nu
_{2}}^{\ast }(\mathbf{r}_{4}),
\end{equation*}%
%
where $\xi _{\nu }=\varepsilon _{\nu }-\mu $ and $\phi _{\nu }(%
\mathbf{r})$ are the eigenenergies (shifted by the chemical potential $\mu $%
) and the eigenfunctions of the single-particle Schr\"{o}dinger equation in
the trap 
\begin{equation*}
\left\{ -\frac{\hbar ^{2}}{2m}\Delta +U_{\mathrm{tr}}(\mathbf{r)}\right\}
\phi _{\nu }(\mathbf{r})=\varepsilon _{\nu }\phi _{\nu }(\mathbf{r}),
\end{equation*}%
where $U_{\mathrm{tr}}(\mathbf{r})$ is given by Eq.~(\ref{Utrap}). Note that
the gap equation~(\ref{lingapeq}) does not contain the mean-field
Hartree-Fock corrections. They lead to unimportant change of parameters of
the Hamiltonian (\ref{1}) and, therefore are not relevant for
pairing. The many-body contributions $\delta V$ to the effective
interparticle interaction which are also absent in Eq.~(\ref{lingapeq})
[compare with Eq.~(\ref{1.1})], will be taken into account later.

The results of the analysis of Eq.~(\ref{lingapeq}) in Ref. \cite{njp} are
the following. When $\omega _{\rho }\sim \omega _{z}\ll T_{c}$ (shallow
nearly spherical trap), where $T_{c}$ is the critical temperature of the BCS
transition in a spatially homogeneous gas with the density $n$ equals to the
central density $n_{0}$ in the trap, it is convenient to perform the Fourier
transformation with respect to the relative coordinate $\mathbf{r}=\mathbf{r}%
_{1}-\mathbf{r}_{2}$,

\begin{equation*}
\widetilde{\Delta }(\mathbf{R},\mathbf{p})=\int d\mathbf{r}\exp (-i\mathbf{pr%
}/\hbar )\Delta (\mathbf{R}+\mathbf{r}/2,\mathbf{R}-\mathbf{r}/2),
\end{equation*}%
where $\mathbf{R}=(\mathbf{r}_{1}+\mathbf{r}_{2})/2$ is the coordinate of
the center of mass. The characteristic scale of the $p$-dependence of $%
\widetilde{\Delta }(\mathbf{R},\mathbf{p})$ is of the order of the Fermi
momentum $p_{F}$, while for the $R$-dependence it is of the order of the
size of the cloud $R_{TF}$, which is much larger than the typical size $\xi
_{0}=p_{F}/mT_{c}$ of pairing correlations (coherence length), $R_{TF}\gg
\xi _{0}$. We therefor can write the order parameter on the local Fermi
surface as $\widetilde{\Delta }[\mathbf{R},\mathbf{p}=\mathbf{n}p_{F}(%
\mathbf{R})]=\widetilde{\Delta }(\mathbf{R})\varphi _{0}(\mathbf{n})$, where
the function $\widetilde{\Delta }(\mathbf{R})$ obeys the equation (see Ref. 
\cite{njp} for details) 
\begin{eqnarray}
\left\{ -\frac{7\zeta(3)}{48\pi^{2}}\left( \frac{p_{F}}{mT_{c}}\right)
^{2}\sum_{\alpha=x,y,z}f_{\alpha}\,\nabla_{\mathbf{R}_{\alpha}}^{2}+\frac{U_{%
\mathrm{trap}}(\mathbf{R})}{\mu}\left( 1+\frac{\pi\varepsilon_{F}}{24nd^{2}}%
\right) \right\} \widetilde{\Delta}(\mathbf{R})=\ln\frac{T_{c}}{T_{c\mathrm{%
trap}}}\widetilde{\Delta}(\mathbf{R}),  \label{18}
\end{eqnarray}
with $f_{x}=f_{y}=1-3/\pi ^{2}$, $f_{z}=1+6/\pi ^{2}$. \ The
solution of this equation which is formally equivalent to the Schr\"{o}%
dinger equation for a three-dimensional anisotropic harmonic oscillator, for
the lowest eigenvalue gives the following expression%
\begin{eqnarray}
\frac{T_{c}-T_{c\mathrm{trap}}}{T_{c}}\approx\ln\frac{T_{c}}{T_{c\mathrm{trap%
}}}=\sqrt{\frac{7\zeta(3)}{48\pi^{2}}\left( 1+\frac{\pi\varepsilon_{F}}{%
24nd^{2}}\right) }\left[ 2\frac{\omega_{\rho}}{T_{c}}\sqrt{1-\frac{3}{\pi^{2}%
}}+\frac{\omega_{z}}{T_{c}}\sqrt{1+\frac{6}{\pi^{2}}}\right] .
\label{trapTc}
\end{eqnarray}
for the change of the critical temperature due to the
presence of the trapping potential. According to this expression, the
critical temperature in the trap is always smaller than that in the
homogeneous gas. In the considered case we have $\omega _{\alpha }/T_{c}\ll 1
$ but $\pi \varepsilon _{F}/24nd^{2}>1$ (weakly interacting gas). Taking
into account that $7\zeta (3)/(48\pi ^{2})\approx 0.018$ we see that the
difference between $T_{c\mathrm{trap}}$ and $T_{c}$ is small if $\pi
\varepsilon _{F}/24nd^{2}$ is not very large.

The order parameter just below $T_{c}$ is given by the corresponding
eigenfunction and has the Gaussian form 
\begin{equation*}
\widetilde{\Delta }(\mathbf{R})\propto \exp (-\sum_{\alpha =x,y,z}R_{\alpha
}^{2}/2l_{\Delta \alpha }^{2}),
\end{equation*}%
where%
\begin{equation*}
l_{\Delta \alpha }=\frac{p_{F}}{m\omega _{\alpha }}\sqrt{\frac{\omega
_{\alpha }}{T_{c}}}\left[ \frac{7\zeta (3)f_{\alpha }}{48\pi ^{2}}\left( 1+%
\frac{\pi \varepsilon _{F}}{24nd^{2}}\right) ^{-1}\right] ^{1/4}
\end{equation*}%
is the characteristic size in the $\alpha $-th direction. If again the
quantity $\pi \varepsilon _{F}/24nd^{2}$ is not very large, we have $%
l_{\Delta \alpha }\ll R_{TF}^{(\alpha )}$, where $R_{TF}^{(\alpha
)}=p_{F}/m\omega _{\alpha }$ is the Thomas-Fermi radius of the trapped gas
cloud in the $\alpha $-th direction. This means that the pairing takes place
only in the central part of the gas.

\begin{figure}[tbp]
\begin{center}
\includegraphics[width=.6\columnwidth]{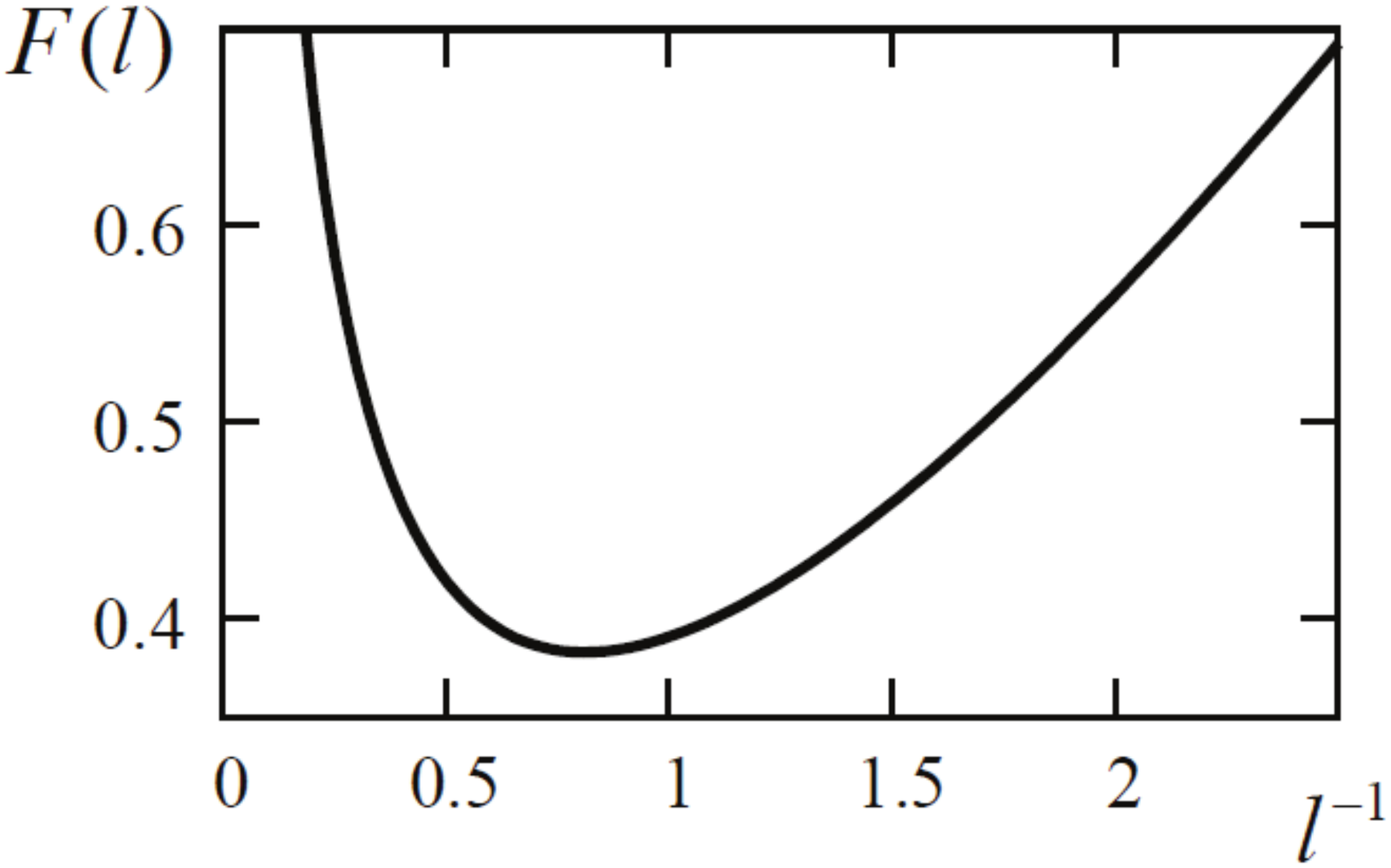}
\end{center}
\caption{The function $F$ versus the inverse trap aspect ratio $l$. (Taken
from Ref. \protect\cite{njp}.)}
\label{fig10}
\end{figure}

For a large number of particles the gas is in the Thomas-Fermi regime [Eq.~(%
\ref{nTF})], and we have the following relation 
\begin{equation*}
N=\mu (n_{0})^{3}/6\omega _{z}\omega _{\rho }^{2},
\end{equation*}%
between the total number of particle $N$ in the gas and the density $n_{0}$
in the center, where $\mu (n_{0})=(6\pi ^{2}\hbar ^{3}n_{0})^{2/3}/2m$ is
the chemical potential. \thinspace For a fixed $N$ and $n_{0}$, this gives

\begin{eqnarray}
\frac{T_{c\mathrm{trap}}-T_{c}}{T_{c}}=-\frac{\overline{\omega }}{T_{c}}%
\sqrt{\left( 1+\frac{\pi \varepsilon _{F}}{24nd^{2}}\right) }F(l),
\label{Tctrapaspr1}
\end{eqnarray}%
where $\overline{\omega }=(\omega _{z}\omega _{\rho }^{2})^{1/3}$ and $l=%
\sqrt{\omega _{\rho }/\omega _{z}}$ is the trap aspect ratio. The function \\$
F(l)= \sqrt{7\zeta (3)/48\pi ^{2}} [ 2 \sqrt{1-3/\pi ^{2}}l^{2/3}+\sqrt{1+6/\pi
^{2}}l^{-4/3}]$ is shown in Fig.~\ref{fig10}. 
The minimum of $F(l)$ provides the optimal trap aspect ration $l^{\ast }=1.23
$ (cigar shaped trap, as expected) for the highest critical temperature in
the trapped gas with the fixed total number of particles $N$ and the density
in the center $n_{0}$ [the critical temperature $T_{c}$ of the homogeneous
gas as a function of $n_{0}$ is given by Eq.~(\ref{Tc})]. The optimal value
is a result of the competition between the anisotropic dipole-dipole
interparticle interaction and the finite-size effects: The former favours
larger $l$ (cigar shaped traps) while the latter, due to the zero boundary
condition on the order parameter, acts on pairing destructively imposing an
upper limit on $l$.

Numerical solution of the linearized gap equation (\ref{lingapeq}) for the
case $\mu \gg \omega _{z},\omega _{\rho }\sim T_{c}$ shows the existence of
a critical trap aspect ratio $l_{c}$ for a given interaction strength $%
\Gamma =36n(0)d^{2}/\pi \mu $ and a number of particles $N$. This critical
aspect ratio corresponds to zero critical temperature, $T_{c\mathrm{trap}}=0$%
, such that no pairing is possible in traps with $l<l_{c}$. Alternatively,
for fixed values of the trap aspect ratio $l$ and the number of particles $N$%
, BCS pairing takes place if the interaction parameter $\Gamma $ is large
enough, $\Gamma >\Gamma _{c}$. The existence of the critical values $l_{c}$
and $\Gamma _{c}$ can be understood as follows: Due to the fact that the
order parameter changes sign when the direction of the $z$-axis is reversed,
single-particle states involved in forming the order parameter should have
different (by an odd integer) quantum numbers $n_{z}$. As a result, these
states have different energies (the minimum difference is $\omega _{z}$) and
can be paired only if the energy gain due to pairing (which is of the order
of the critical temperature $T_{c}$) exceeds this difference. In the
limiting case of an infinite pancake trap with the confinement only in the $z
$-direction, this results in the appearance of a critical trap frequency $%
\omega _{zc}=1.8T_{c}$ \cite{nobel} with no pairing possible in traps with $%
\omega _{z}>\omega _{zc}$.

\begin{figure}[tbp]
\begin{center}
\includegraphics[width=.9\columnwidth]{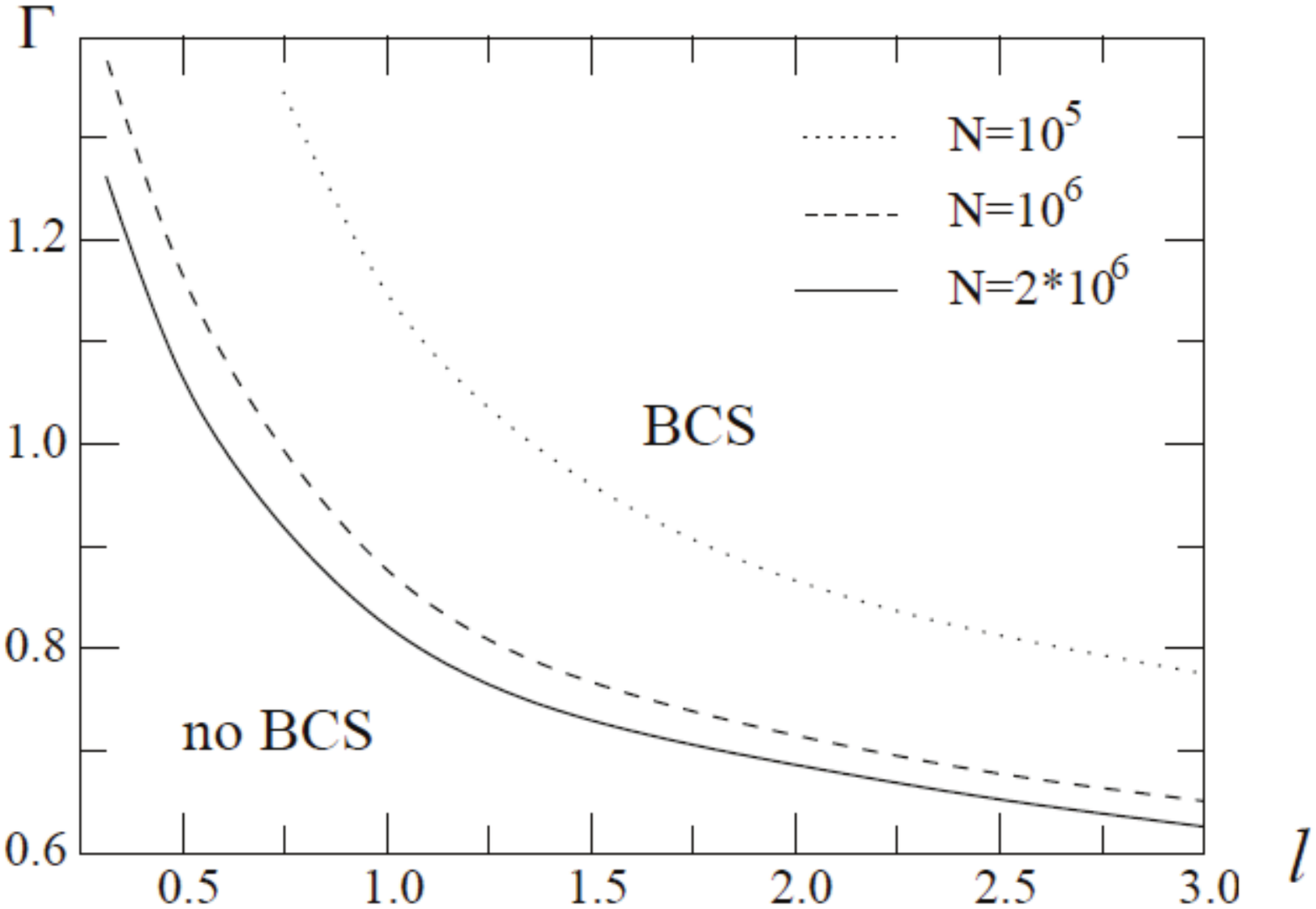}
\end{center}
\caption{Critical lines $\Gamma _{c}$ versus the aspect ratio $l$ for
different numbers of particles $N$. The BCS pairing takes place above the
depicted curves. (Taken from Ref. \protect\cite{njp}.)}
\label{fig11}
\end{figure}

The dependence of the critical interaction strength $\Gamma _{c}$ on the
trap aspect ratio $l$ for different $N$ is shown in Fig.~\ref{fig11}. 
As expected, the critical interaction strength $\Gamma _{c}$ decreases with
increasing $l$. On the other hand, with increasing the number of particles $N
$, the interaction parameter becomes larger, $\Gamma \sim N^{1/6}$, and the
critical aspect ratio $l_{c}$ decreases.

\begin{figure}[tbp]
\begin{center}
\includegraphics[width=.9\columnwidth]{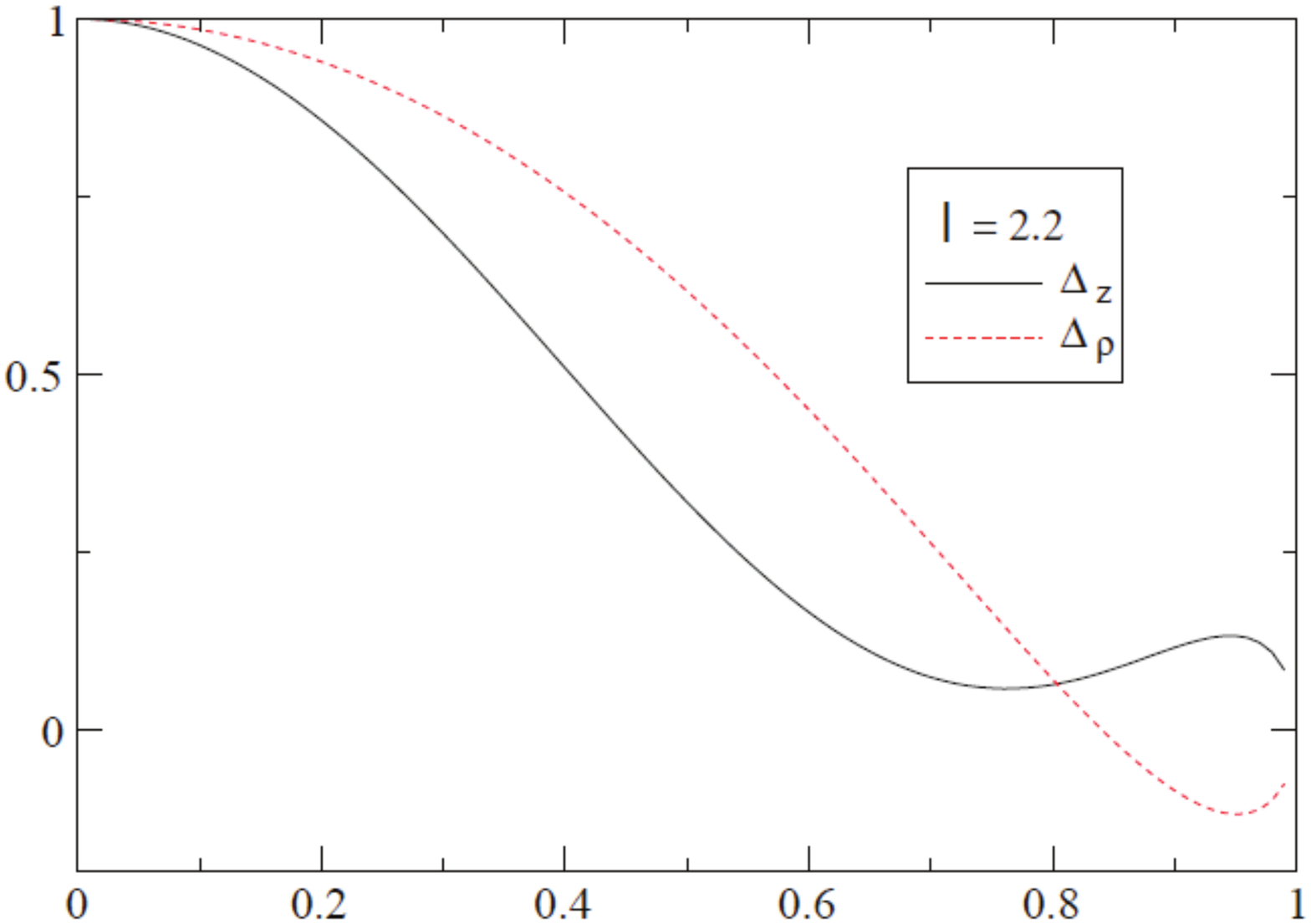}
\end{center}
\caption{The order parameter for the aspect ratio $l=2.2$ (cigar shaped
trap). The solid line shows $\Delta (z,\protect\rho =0)$ and the dotted line
corresponds to $\Delta (z=0,\protect\rho )$. (Taken from Ref. \protect\cite%
{njp}.)}
\label{fig12}
\end{figure}

The order parameter $\Delta _{0}(\mathbf{R})$ for a cigar shaped trap with $%
l=2.2$ (see Fig.~\ref{fig12}) exhibits a non-monotonic behavior with the
distance from the trap center. This is to be compared with a monotonic
behavior of the BCS order parameter in a two component Fermi gas with a
short-range attractive interaction~\cite{monotonicDelta}-\cite%
{monotonicDelta2} under similar conditions. (It should be noted, however,
that an oscillating and highly non-monotonic behaviors of the order
parameter in a two component Fermi gas was obtained in Ref. \cite{intrashell}
in the regime of an intershell pairing $T_{c}\ll \omega $.) 

\subsection{BCS pairing in a two component dipolar Fermi gas}

Adding the second component in a dipolar Fermi gas opens the possibility for
a singlet interspecies pairing (equivalent to the $s$-wave pairing in a two
component atomic Fermi gas) in addition to the triplet intraspecies pairing
considered in Sec. \ref{s4.2}, and competition between them. Assuming
chemical stability of a mixture of two species of polarized dipolar
particles with equal masses, concentrations, and dipole moments (a mixture
of fermionic polar molecules with two different hyperfine states, for
example), this problem was considered in Ref. \cite{PhysRevA.82.033623}. The
corresponding Hamiltonian reads%
\begin{equation*}
H=\sum_{\alpha }\int d\mathbf{r}\hat{\psi }_{\alpha }^{\dagger }(\mathbf{r})%
\left[ -\frac{\hbar ^{2}}{2m}\nabla ^{2}-\mu \right] \hat{\psi }_{\alpha }(%
\mathbf{r})+\frac{1}{2}\sum_{\alpha ,\alpha ^{\prime }}\int d\mathbf{r}d%
\mathbf{r}^{\prime }\hat{\psi }_{\alpha }^{\dagger }(\mathbf{r})\hat{\psi }%
_{\alpha ^{\prime }}^{\dagger }(\mathbf{r}^{\prime })V_{\alpha \alpha
^{\prime }}(\mathbf{r}-\mathbf{r}^{\prime })\hat{\psi }_{\alpha ^{\prime }}(%
\mathbf{r}^{\prime })\hat{\psi }_{\alpha }(\mathbf{r}),
\end{equation*}%
%
%
%
%
where $\alpha=\pm$ denotes two different species, the intraspecies
interaction $V_{\alpha\alpha}(\mathbf{r})$ is the dipole-dipole one, $%
V_{\alpha\alpha }(\mathbf{r})=V_{d}(\mathbf{r})$, while the interspecies
interaction $V_{\alpha\alpha^{\prime}}(\mathbf{r})$ for $\alpha^{\prime}\neq%
\alpha$ contains a short-range part $V_{s}(\mathbf{r})=(4\pi a_{s}/m)\delta (%
\mathbf{r})$ in addition to the dipole-dipole interaction, $V_{\alpha
\alpha^{\prime}}(\mathbf{r})=V_{d}(\mathbf{r})+V_{s}(\mathbf{r})$,
parametrized by the $s$-wave scattering amplitude $a_{s}$.

The order parameter of singlet interspecies pairing%
\begin{equation*}
\Delta_{s}(\mathbf{r}_{1}-\mathbf{r}_{2})=V_{-+}(\mathbf{r}_{1}\mathbf{-r}%
_{2}\mathbf{)}\left\langle \hat{\psi}_{-}(\mathbf{r}_{1})\hat{\psi}_{+}(%
\mathbf{r}_{2})\right\rangle
\end{equation*}
in momentum space is now a sum of all partial waves with even angular
momentum $L$ and zero azimuthal quantum number $M$,%
\begin{eqnarray}
\Delta_{s}(\mathbf{p})=\sum_{\mathrm{even}\,L}\Delta_{sL}(p)\mathrm{Y}_{L0}(%
\hat{\mathbf{p}}).  \label{Deltasums}
\end{eqnarray}
The corresponding gap equation is similar to Eq.~(\ref{1.1}) with $V(\mathbf{%
p},\mathbf{p}^{\prime})=V_{d}(\mathbf{p-p}^{\prime})+4\pi a_{s}/m+\delta V(%
\mathbf{p},\mathbf{p}^{\prime})$.

In the BCS approach, when one neglects many-body contributions including the
deformation of the Fermi-sphere in momentum space, the critical temperature
of the singlet superfluid transition is (see Ref. \cite{PhysRevA.82.033623})%
\begin{eqnarray}
T_{c}^{(s)}\sim\varepsilon_{F}\exp\left( -\frac{\varepsilon_{F}}{\pi nd^{2}}%
\frac{1}{\left\vert \lambda_{s}\right\vert }\right)  \label{Tcs}
\end{eqnarray}
and the corresponding order parameter parameter on the Fermi surface reads
[compare with Eq.~(\ref{phi0})]%
\begin{eqnarray}
\Delta_{s}(p_{F}\hat{\mathbf{p}})\sim\cos[\cos\theta_{\mathbf{p}}\sqrt{%
3/\left\vert \lambda_{s}\right\vert }],  \label{delta_s}
\end{eqnarray}
where $\left\vert \lambda_{s}\right\vert $ is the largest positive root of
the equation%
\begin{equation*}
t+\left( 1+\frac{2a_{s}k_{F}}{\pi}\frac{\varepsilon_{F}}{\pi nd^{2}}\right) 
\sqrt{\frac{t}{3}}\tan\sqrt{\frac{3}{t}}=0.
\end{equation*}
The order parameter (\ref{delta_s}) is now a symmetric function under $%
\mathbf{p}\rightarrow-\mathbf{p}$ as it should be for a singlet pairing.

\begin{figure}[ptb]
\begin{center}
\includegraphics[width=.9\columnwidth]{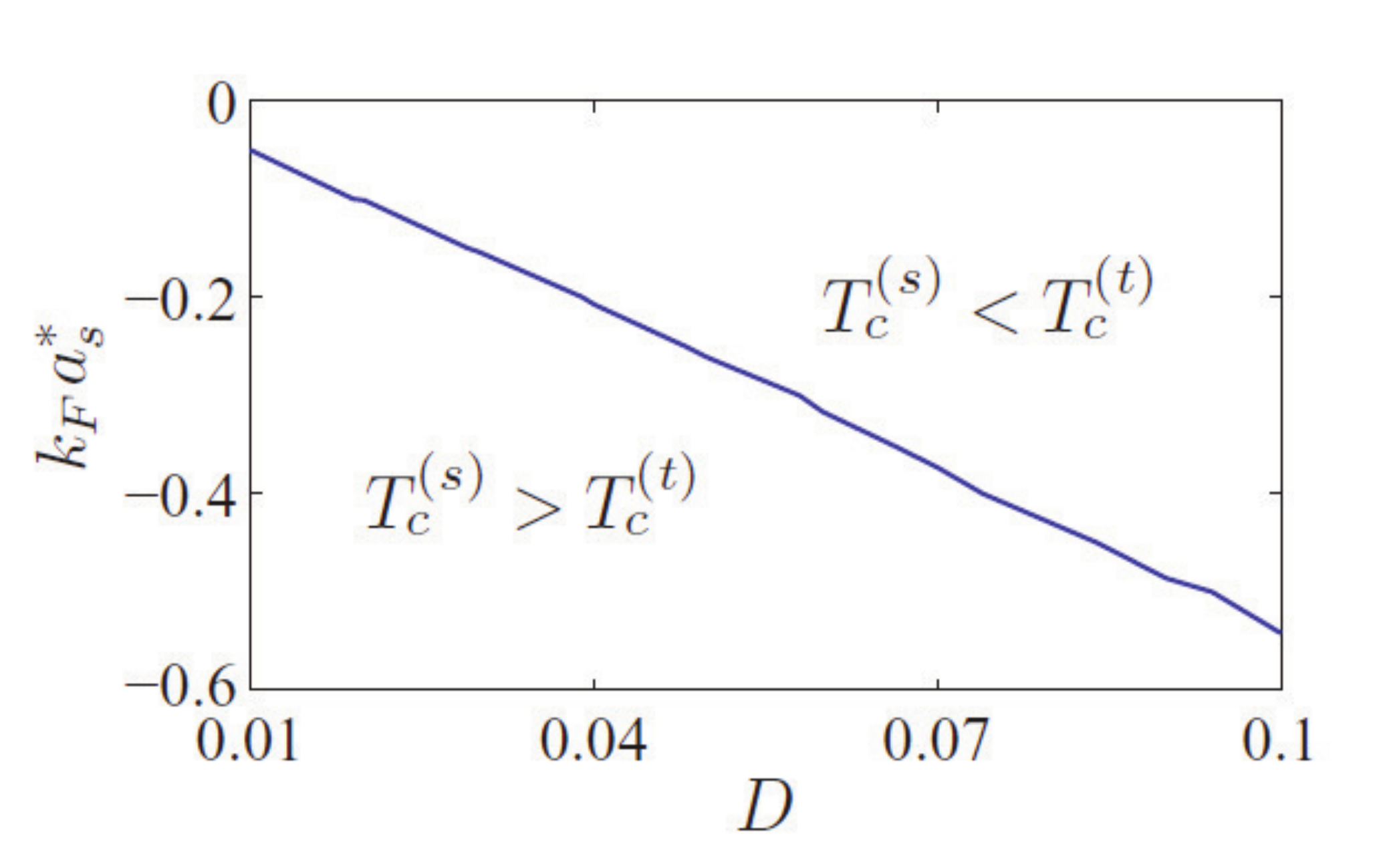}
\end{center}
\caption{The dependence of the critical scattering length $a_{s}^{\ast}$ on
the strength of the dipole interaction $D=nd^{2}/\protect\varepsilon_{F}$,
and the resulting superfluid phase diagram. (Taken from Ref. \protect\cite%
{PhysRevA.82.033623}.)}
\label{ZZZ}
\end{figure}

With the deformation of the Fermi surface taken into account, the problem
was solved numerically (see Ref. \cite{PhysRevA.82.033623} for details). It
turns out that without contact interaction ($a_{s}=0$) the critical
temperature of the triplet intraspecies pairing is always higher and,
therefore, the system undergoes the transition into the triplet intraspecies
BCS state, see Sec. \ref{s4.2}. The ground state in this case is a mixture
of two intraspecies triplet superfluids. However, switching on an attractive
contact interaction ($a_{s}<0$) one can increase the critical temperature of
the singlet pairing and make it larger than for the triplet one such that
the ground state corresponds to an interspecies singlet superfluid. The
critical value of $a_{s}k_{F}$ as a function of $D=nd^{2}/\varepsilon_{F}$
and the corresponding phase diagram are shown in Fig.~\ref{ZZZ}. 

\subsection{BCS pairing in a dipolar monolayer}

\begin{figure}[ptb]
\begin{center}
\includegraphics[width=.7\columnwidth]{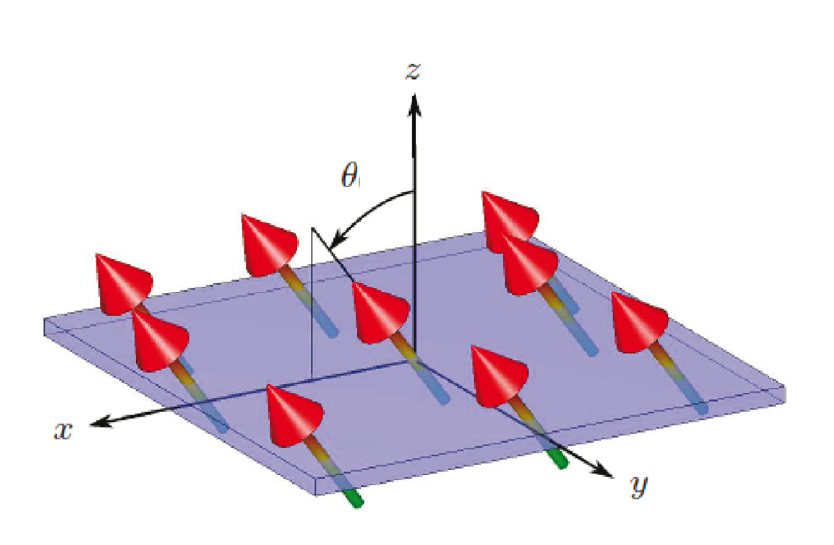}
\end{center}
\caption{Fermionic monolayer: Fermionic dipoles are confined to the $xy$%
-plane. The dipoles are aligned in the $xz$-plane and form an angle $\protect%
\theta$ with the $z$-axis.}
\label{ZZZ1}
\end{figure}

Let us now consider the possibility for superfluid pairing in a polarized
dipolar monolayer -- a polarized single-component dipolar gas confined to a
(quasi)2D geometry by a harmonic trapping potential $V(z)=m%
\omega_{z}^{2}z^{2}/2$, assuming $\hbar\omega_{z}\gg\varepsilon_{F}$, where $%
\varepsilon _{F}=p_{F}^{2}/2m$ is the Fermi energy of a Fermi gas with the
2D density $n_{2D}=p_{F}^{2}/4\pi\hbar^{2}$. An important parameter of the
problem is the angle $\theta$ between the $z$-axis (normal to the 2D motion
plane) and the direction of the dipole polarization, see Fig.~\ref{ZZZ1}. 
The effective 2D interaction between dipoles is given by Eqs. (\ref{V2D}), (%
\ref{V2Dgeneral}), and (\ref{Vdip2D}). To see the possibility of BCS
pairing, one has to look at the component in the $p$-wave channel.
Straightforward but lengthy calculations \cite{PhysRevLett.101.245301} show
that it becomes negative when $\sin\theta>2/3$, and, therefore, the system
becomes unstable against BCS pairing. Note that the critical value $\theta
_{c}=\arcsin2/3=0.73$ ($42^{\circ}$) is larger than the value $\arcsin 1/%
\sqrt{3}=0.62$ ($35^{\circ}$) of the angle $\theta$ above which the
dipole-dipole interaction has attractive directions in the $(x,y)$-plane.

Due to the anisotropy of the effective interaction (for $\theta>0$), the
azimuthal quantum number is no longer a conserving quantity and, therefore,
the superfluid order parameter contains all odd harmonics in the azimuthal
angle $\varphi$. In momentum space,%
\begin{equation*}
\Delta(\mathbf{k})=\sum_{n=1}^{\infty}\Delta_{n}(k)\cos[(2n-1)\varphi _{%
\mathbf{k}}]
\end{equation*}
with the components $\Delta_{n}(k)$ coupled to each other through the gap
equation. With only $n=1$ component taken into account, the problem was
solved in Ref.\cite{PhysRevLett.101.245301}, in which is was also found that
the Fermi surface deformation due to anisotropy of the effective interaction
does not play any significant role in the pairing problem and only slightly
decreases the critical angle $\theta_{c}$ when the strength of the
interaction increases. This result was confirmed by the analysis of Ref. 
\cite{SiebererBaranov}, in which all components of $\Delta(\mathbf{k})$ were
taken into account. It appears that the critical angle remains practically
the same, and higher components of $\Delta(\mathbf{k})$ with $n>1$ are
visible only at angles very close to $\theta_{c}$. For larger values of $%
\theta$, one has $\Delta_{n>1}/\Delta_{1}\lesssim10^{-2}$ such that the
pairing has indeed the $p$-wave character, $\Delta(\mathbf{k}%
)\approx\Delta_{1}(k)\cos \varphi_{\mathbf{k}}$.

The critical temperature for the BCS pairing reads \cite{SiebererBaranov}%
\begin{eqnarray}
T_{c}=\frac{2e^{\gamma }}{\pi }\varepsilon _{F}\exp \left[ \frac{3\pi }{%
4k_{F}a_{d}}\frac{1}{(9/4)\sin ^{2}\theta -1}\right] F_{1}(\theta
)F_{2}(k_{F}l_{z},\theta ),\quad \theta >\theta _{c}=\arcsin 2/3,
\label{Tcmonolayer}
\end{eqnarray}%
%
%
%
%
where%
\begin{equation*}
F_{1}(\theta)=\frac{0.52-2.47\sin^{2}\theta+2.83\sin^{4}\theta}{%
0.18-0.81\sin ^{2}\theta+0.91\sin^{4}\theta}
\end{equation*}
and%
\begin{equation*}
F_{2}(\eta,\theta)=\eta^{f(\theta)},\quad f(\theta)=\frac{0.25-1.13\sin
^{2}\theta+1.28\sin^{4}\theta}{0.18-0.81\sin^{2}\theta+0.91\sin^{4}\theta}.
\end{equation*}
The two functions $F_{1}(\theta)$ and $F_{2}(k_{F}l_{z},\theta)$ in Eq.~(\ref%
{Tcmonolayer}) describe deviations from the simplest BCS approach (with no
account of the Fermi surface deformation): The exponent in Eq.~(\ref%
{Tcmonolayer}) is simply the $p$-wave component of the effective
dipole-dipole interaction on the undeformed Fermi surface. These deviations
result from many-body corrections to the interparticle interaction and
Fermi-liquid effects, as well as from the second order contribution to the
two-body quasi-2D scattering amplitude. The latter includes virtual
transitions to intermediate states which are not necessarily limited to the
ground state of the transverse confinement. The virtual transitions to
excited states of the transverse confinement, together with many-body
contributions (those include corrections to the interparticle interaction,
Fermi surface deformation, and Fermi-liquid effects -- effective mass),
contribute to the function $F_{1}(\theta)$. The virtual transitions to the
states with ground state motion in the transverse confining potential
contribute to both functions $F_{1}(\theta)$ and $F_{2}(k_{F}l_{z},\theta)$
(see Ref. \cite{SiebererBaranov}\ for details). The latter transitions
provide the second order Born term to purely 2D scattering amplitude on the
effective potential (\ref{V2D}), which has the so-called anomalous
scattering contribution $\sim k^{2}\ln k$ (see more discussions below) due
to long-range power decay of the potential. This anomalous contribution
gives rise to the function $F_{2}(k_{F}l_{z},\theta)$. Eq.~(\ref{Tcmonolayer}%
) predicts a rapid growth of $T_{c}$ with increasing of the angle $\theta$
from the critical value $\theta_{c}=\arcsin2/3$, to the values of the order
of tens of $\mathrm{nK}$ for realistic experimental parameters; see Fig.~\ref%
{TcND3}. 
\begin{figure}[ptb]
\begin{center}
\includegraphics[width=0.5\columnwidth]{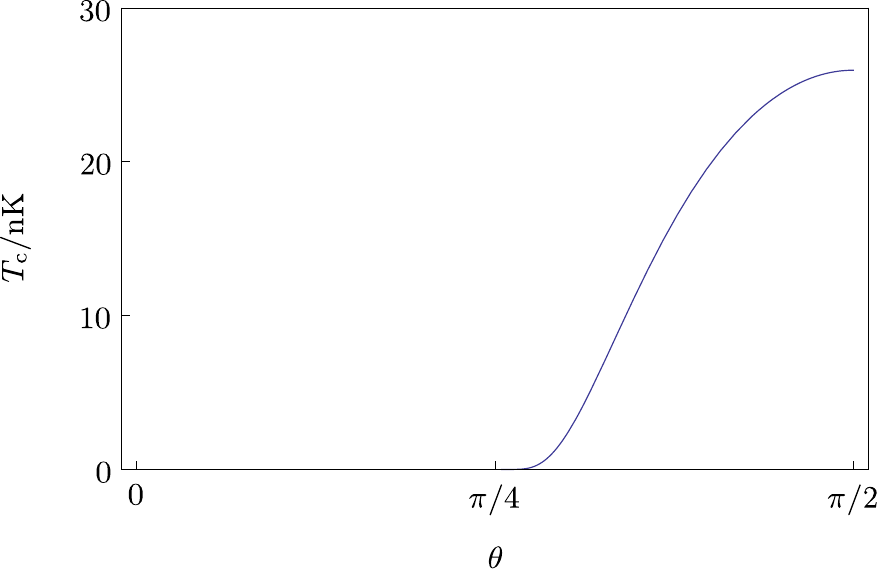}
\end{center}
\caption{The critical temperature as a function of the tilting angle $%
\protect\theta$ for $a_{d}k_{F}=2.5$ and $k_{F}l\approx0.2$. These values
correspond to a gas of fermionic $^{15}\mathrm{ND}_{3}$ molecules with the
density $n_{2D}=10^{8}\mathrm{cm}^{-2}$ and $\protect\omega_{z}=2\protect\pi%
\times100\mathrm{KHz}$. (Taken from Ref. \protect\cite{SiebererBaranov}.)}
\label{TcND3}
\end{figure}

We obtain our results for the superfluid critical temperature using the
mean-field approach. However, as it is well-known, this approach in two
dimensions is only applicable at zero temperature, while at finite
temperature the long-range order is destroyed by phase fluctuations and,
therefore, the mean-field order parameter is zero. In this case, the
transition into the superfluid phase follows the
Berezinskii-Kosterlitz-Thouless (BKT) scenario \cite{berezinskii1972}, \cite%
{kosterlitz1973}, and \cite{Kosterlitz1974}. In the weak coupling limit,
however, as it was pointed out by Miyake \cite{Miyake}, the difference
between the critical temperature calculated within the mean-field approach $%
T_{c}$ and the critical temperature of the BKT transition $T_{\mathrm{BKT}}$
can be estimated as $T_{c}-T_{\mathrm{BKT}}\sim T_{c}^{2}/\mu$ and,
therefore, small as compared to $T_{c}$. As a result, our mean-field
calculations provide a reliable answer for the critical temperature in the
considered weak coupling regime $a_{d}k_{F}<1$.

Another interesting possibility to create a topologically nontrivial
superfluid state in a monolayer was considered in Refs. \cite%
{PhysRevLett.103.155302} and \cite{PhysRevA.84.013603}, in which the authors
consider a monolayer with fermionic dipoles oriented perpendicular to the
plane ($\theta=0$) and use the RF-dressing technique discussed in Sec.~\ref%
{sec:TunaAtoms} to create an effective (time-averaged) attractive potential $%
V_{\mathrm{eff}}(\rho\rightarrow\infty)\approx-d_{\mathrm{eff}}^{2}/\rho^{3}$%
. At short distances, the potential $V_{\mathrm{eff}}(\rho)$ has a repulsive
core that prevents low-energy particle from approaching each other and,
therefore, suppresses inelastic collisions resulting in losses. The leading $%
p$-wave 2D scattering amplitude for identical fermions was found \cite%
{PhysRevA.84.013603} to be%
\begin{eqnarray}
f_{1}(k)=\int_{0}^{\theta }\mathrm{J}_{1}(k\rho )V_{\mathrm{eff}}(\rho )\psi
_{k}^{(+)}(\rho )2\pi \rho d\rho \approx -\frac{8}{3}\frac{\hbar ^{2}}{m}%
kr_{\ast }+\frac{\pi }{2}\frac{\hbar ^{2}}{m}(kr_{\ast })^{2}\ln (Ckr_{\ast
}),  \label{f1monolayer}
\end{eqnarray}%
%
%
%
%
where $\mathrm{J}_{1}(z)$ is the Bessel function, $\psi_{k}^{(+)}(\rho)$ is
the radial wave function of the $p$-wave relative motion, the constant $C$
is determined by the behavior of the potential at short distances, and the
length scale $r_{\ast}=md_{\mathrm{eff}}^{2}/\hbar^{2}$ depends on the
details of the RF-dressing (for BCS pairing $r_{\ast}$ has to be positive,
for more details see in Ref. \cite{PhysRevA.84.013603}). The first term in
Eq.~(\ref{f1monolayer}) corresponds to the $p$-wave Born amplitude for the
potential $V_{\mathrm{eff}}(\rho)=-(\hbar^{2}/m)r_{\ast}\rho^{-3}$ while the
second term contains both the anomalous scattering due to the dipole-dipole
tail of the interparticle interaction and the short-range contribution.

For $kr_{\ast}\ll1$, the first term in Eq.~(\ref{f1monolayer}) is negative
and, therefore, leads to $p$-wave superfluid transition. The most stable
low-temperature $p$-wave superfluid phase in 2D has $p_{x}+ip_{y}$ symmetry, 
$\Delta_{\mathbf{k}}=\Delta(k)\exp(i\varphi_{\mathbf{k}})$, because this is
the only $p$-wave superfluid phase with a non-zero energy gap on the entire
Fermi sphere. The numerical solution of the gap equation \cite%
{PhysRevLett.103.155302}, \cite{PhysRevA.84.013603} shows that $\Delta(k)$
raises linearly for $k\lesssim k_{F}$ and approaches a constant $\pi
e^{-\gamma}T_{c}$ for $k\gtrsim k_{F}$. The critical temperature $T_{c}$
reads%
\begin{equation*}
T_{c}=\varepsilon_{F}\frac{\kappa}{(k_{F}r_{\ast})^{9\pi^{2}/64}}\exp\left[ -%
\frac{3\pi}{4k_{F}r_{\ast}}\right] ,
\end{equation*}
where%
\begin{equation*}
\kappa\simeq0.16\exp\left( -\frac{9\pi^{2}}{64}A\right)
\end{equation*}
with the numerical coefficient $A$ determined by short-distance behavior of
the interparticle interaction. Note that the value of the critical
temperature is very sensitive to the short-range part of the effective
potential $V_{\mathrm{eff}}(\rho)$ and, by modifying it, can be varied
within a few orders of magnitude. reaching the values of the order of tens
of $\mathrm{nK}$ for realistic experimental parameters that correspond to
the life-time of the system of the order of seconds; see Ref. \cite%
{PhysRevA.84.013603} for details and discussions.

The resulting $p_{x}+ip_{y}$ superfluid pairing spontaneously breaks
time-reversal invariance (the degenerate time-reversal partner is the $%
p_{x}-ip_{y}$ state). This phase belongs to the class of the so-called
topological superconductors and can exist in one of two topologically
distinct phases, depending on the sign of the chemical potential $\mu$ \cite%
{Volovik1992}. The phase with $\mu<0$ is topologically trivial (may be
continuously deformed to the vacuum state), while the phase with $\mu>0$ is
topologically non-trivial ( cannot be continuously deformed to the vacuum)
and has several very interesting properties. One of the most interesting of
them is that the vortices in this superfluid carry localized zero-energy
states, described by a Majorana fermion. These Majorana states obey
non-Abelian exchange statistics \cite{PhysRevB.61.10267}, \cite{Stern2008}
and can possibly be used for topologically protected quantum information
processing \cite{RevModPhys.80.1083}. In the considered case of superfluid
state of dipoles in a monolayer, the chemical potential is positive, $\mu>0$%
, and the resulting superfluid phase is topologically non-trivial.

\subsection{BCS\ pairing in a bilayer dipolar system}

\begin{figure}[ptb]
\begin{center}
\includegraphics[width=.9\columnwidth]{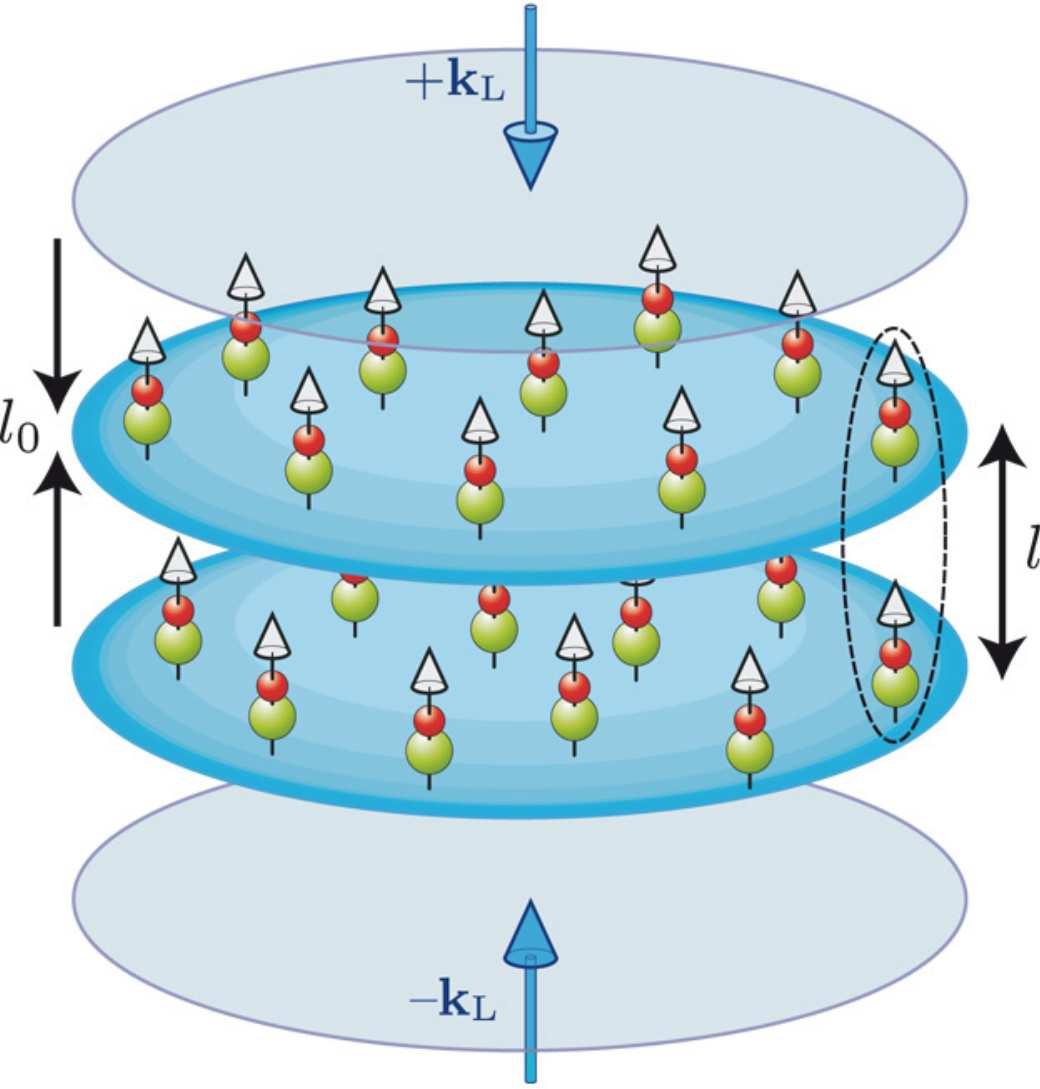}
\end{center}
\caption{The setup of the dipolar bilayer system: Two layers with the
thickness $l_{0}$ of a 1D optical lattice formed by two counterpropagating
laser waves with wavevectors $\mathbf{k}_{L}$ and $-\mathbf{k}_{L}$ are
filled with dipoles oriented perpendicular to the layers. The interlayer
distance $l$ is $\protect\pi/k_{L}$. An interlayer Cooper pair or molecule}
\label{Setup2}
\end{figure}

The single-component fermionic bilayer dipolar system (see Fig.~\ref{Setup2}%
) 
provides an example of a relatively simple many-body system in which an
entire range of nontrivial many-body phenomena are solely tied to the
dipole-dipole interparticle interaction with its unique properties:
long-range and anisotropy. The long-range character provides an
interparticle interaction in single-component Fermi gases inside each layer
that otherwise would remain essentially noninteracting. For the considered
setup, this intralayer interaction is always repulsive and gives rise to the
crystalline phase for a large density of particles (see Sec.~\ref%
{sec:Stable2D}). More important, the long-range dipole-dipole interaction
couples particles from different layers allowing them interact in the $s$%
-wave channel that is dominant at low energies, allowing formation of bound
states and BCS pairing \cite{PhysRevLett.105.215302}, \cite%
{PhysRevA.83.043602}. The Hamiltonian of the system reads%
\begin{eqnarray}
H  =\sum_{\alpha=\pm}\int d\mathbf{r}\hat{\psi}_{\alpha}^{\dagger }(\mathbf{%
r})\left\{ -\frac{\hbar^{2}}{2m}\Delta+\frac{1}{2}m\omega_{z}^{2}z_{%
\alpha}^{2}-\mu^{\prime}\right\} \hat{\psi}_{\alpha}(\mathbf{r})
\label{H0bilayer}
 +\frac{1}{2}\sum_{\alpha,\beta}\int d\mathbf{r}d\mathbf{r}^{\prime}\hat{%
\psi}_{\alpha}^{\dagger}(\mathbf{r})\hat{\psi}_{\beta}^{\dagger }(\mathbf{r}%
^{\prime})V_{\mathrm{d}}(\mathbf{r}-\mathbf{r}^{\prime})\hat{\psi }_{\beta}(%
\mathbf{r}^{\prime})\hat{\psi}_{\alpha}(\mathbf{r}), 
\end{eqnarray}
where $\alpha=\pm$ is the layer index, $z_{\pm}\equiv z\pm l/2$, with $%
l$ being the layer separation, $\hat{\psi}_{\alpha}(\mathbf{r})$ with $%
\mathbf{r}=(\boldsymbol{\rho},z)$ is the field operator for fermionic
dipolar particles ($\boldsymbol{\rho}=x\mathbf{e}_{x}+y\mathbf{e}_{y}$) on
the corresponding layer $\alpha$, $\Delta=\Delta_{\boldsymbol{\rho}%
}+\partial ^{2}/\partial_{z}^{2}$ is the Laplace operator, $\omega_{z}$ is
the confining frequency in each layer such that $l_{z}=\sqrt{%
\hbar/m\omega_{z}}$, and $\mu^{\prime}$ is the chemical potential. The last
term describes the intra- ($\alpha=\beta$) and interlayer ($\alpha\neq\beta$%
) dipole-dipole interparticle interactions. Assuming a strong confinement, $%
\hbar\omega_{z}\gg\mu^{\prime},T$, where $T$ is the temperature, we can
write $\hat{\psi }_{\alpha}(\mathbf{r})=\hat{\psi}_{\alpha}(\boldsymbol{\rho}%
)\phi _{0}(z_{\alpha})$ and, therefore, reduce the Hamiltonian (\ref{H0bilayer}) to%
\begin{eqnarray}
H_{2D} & =&\sum_{\alpha=\pm}\int d\boldsymbol{\rho}\hat{\psi}_{\alpha
}^{\dagger}(\boldsymbol{\rho})\left\{ -\frac{\hbar^{2}}{2m}\Delta _{%
\boldsymbol{\rho}}-\mu\right\} \hat{\psi}_{\alpha}(\boldsymbol{\rho }) 
\notag \\
& +&\frac{1}{2}\sum_{\alpha,\beta}\int d\boldsymbol{\rho}d\boldsymbol{\rho }%
^{\prime}\hat{\psi}_{\alpha}^{\dagger}(\boldsymbol{\rho})\hat{\psi}_{\beta
}^{\dagger}(\boldsymbol{\rho}^{\prime})V_{\alpha\beta}(\boldsymbol{\rho }-%
\boldsymbol{\rho}^{\prime})\hat{\psi}_{\beta}(\boldsymbol{\rho}^{\prime })%
\hat{\psi}_{\alpha}(\boldsymbol{\rho}),  \label{Hbilayer}
\end{eqnarray}
for a two-component fermionic field $\hat{\psi}_{\alpha}(\boldsymbol{\rho}) $%
, $\alpha=\pm$, with shifted chemical potential $\mu=\mu^{\prime}-\hbar
\omega_{z}/2$. The intracomponent (intralayer) interaction $V_{\alpha\alpha
}(\boldsymbol{\rho})$ coincides with $V_{\mathrm{d}}^{2D}(\mathbf{\rho})$ in
Eq.~(\ref{V2D}) for $\theta=0$, and the intercomponent (interlayer)
interaction is%
\begin{eqnarray}
V_{+-}(\boldsymbol{\rho})=V_{-+}(\boldsymbol{\rho})\equiv V_{2D}(\boldsymbol{%
\rho})\approx d^{2}\frac{\rho^{2}-2l^{2}}{(\rho^{2}+l^{2})^{5/2}}.
\label{V2Dinterlayer}
\end{eqnarray}

The considered system is characterized by three characteristic lengths: the
dipolar length $a_{d}=md^{2}/\hbar^{2}$, the interlayer separation $l$, and
the mean interparticle separation inside each layer $\sim k_{F}^{-1}$ with $%
k_{F}=\sqrt{4\pi n}$ being the Fermi wave vector for a 2D single-component
fermionic gas with the density $n$. Therefore, the physics of the system is
completely determined by two dimensionless parameters which are independent
ratios of the above lengths. The first parameter $g=a_{d}/l$ (the ratio of
the dipolar length and the interlayer separation) is a measure of the
interlayer interaction strength relevant for pairing. In experiments with
polar molecules, the values of the dipolar length $a_{d}$ is of the order of 
$10^{2}\div10^{4}\,\mathrm{nm}$: for a $^{40}\mathrm{K}^{87}\mathrm{Rb}$
with currently available $d\approx0.3\,\mathrm{D}$ one has $a_{d}\approx
170\,\mathrm{nm}$ (with $a_{d}\approx600\,\mathrm{nm}$ for the maximum value 
$d\approx0.566\,\mathrm{D}$), and for $^{6}\mathrm{Li}^{133}\mathrm{Cs}$
with the tunable dipole moment from $d=0.35\,\mathrm{D}$ to $d=1.3\,\mathrm{D%
}$ in an external electric field $\sim1\,\mathrm{kV/cm}$ the value of $a_{d}$
varies from $a_{d}\approx260\,\mathrm{nm}$ to $a_{d}\approx3500\,\mathrm{nm}$%
. For the interlayer separation $l=500\,\mathrm{nm}$ these values of $a_{d}$
corresponds to $g\lesssim10$. The second parameter $k_{F}l$ measures the
interlayer separation in units of the mean interparticle distance in each
layer. This parameter can also be both smaller (dilute regime) and of the
order or larger (dense regime) than unity for densities $n=10^{6}\div
10^{9}\,\mathrm{cm}^{-2}$ (for, example, for $l=500\,\mathrm{nm}$ one has $%
k_{F}l=1$ for $n\approx3\cdot10^{7}\,\mathrm{cm}^{-2}$). The two parameters $%
g$ and $k_{F}l$ determine the regime of interlayer scattering at typical
energies of particles ($\sim$ Fermi energy $\varepsilon_{F}=%
\hbar^{2}k_{F}^{2}/2m$), and their product, $gk_{F}l=a_{d}k_{F}$, as usual,
controls the perturbative expansion in the system and, therefore, many-body
effects.

The interlayer interaction has a very specific form resulting from the
anisotropy of the interaction: Two particles from different layers attract
each other at short and repel each other at large distances, respectively,
as a result of different mutual orientations of their relative coordinate
and of their dipole moments. A peculiar property of $V_{2D}(\boldsymbol{\rho}%
)$ is%
\begin{equation*}
\int d\boldsymbol{\rho}V_{2D}(\boldsymbol{\rho})=0.
\end{equation*}
This means that its Fourier transform%
\begin{eqnarray}
\tilde{V}_{2D}(\mathbf{q})=\int d\boldsymbol{\rho}V_{2D}(\boldsymbol{\rho }%
)e^{-i\mathbf{q}\boldsymbol{\rho}}=-\frac{2\pi\hbar^{2}}{m}gqle^{-ql},
\label{V2DFourier}
\end{eqnarray}
vanishes for small $q$,%
\begin{equation*}
\tilde{V}_{2D}(\mathbf{q\rightarrow0})\approx-\frac{2\pi\hbar^{2}}{m}%
gql\rightarrow0.
\end{equation*}
The potential well at short distances is strong enough to support at least
one bound state for any strength of interlayer coupling. For a weak coupling
between layers ($g\ll1$), the bound state is extremely shallow and has an
exponentially large size \cite{PhysRevA.83.043602} (see also \cite%
{PhysRevA.82.044701}):%
\begin{eqnarray}
E_{b}\approx\frac{4\hbar^{2}}{ml^{2}}\exp\left[ -\frac{8}{g^{2}}+\frac {128}{%
15g}-\frac{2521}{450}-2\gamma+\mathcal{O}(g)\right] ,  \label{Eb}
\end{eqnarray}
for the binding energy and%
\begin{equation*}
R_{b}=\sqrt{\hbar^{2}/mE_{b}}\sim l\exp(4/g^{2})\gg l
\end{equation*}
for the size, respectively. However, in the intermediate and strong coupling
cases ($g\gtrsim1$) the size of the deepest bound state becomes comparable
with the interlayer separation:%
\begin{equation*}
E_{b}=(\hbar^{2}/ml^{2})2g(1-\sqrt{6/g}
\end{equation*}
and%
\begin{equation*}
R_{b}\sim l(6g)^{-1/4},
\end{equation*}
respectively.

The specific properties of the interlayer potential (\ref{V2DFourier}): It
decays exponentially for large momenta $k\gg l^{-1}$, while it is
proportional to $k$ for $k\ll l^{-1}$, lead to different regimes of
scattering and, therefore, of the BCS pairing, depending on the relation
between $g$ and $k_{F}l$. This can be conveniently formulated in terms of
the vertex function $\Gamma(E,\mathbf{k},\mathbf{k}^{\prime})$, where the
arguments $E$, $\mathbf{k}$,and $\mathbf{k}^{\prime}$ are independent of
each other. This function satisfies the following integral equation \cite%
{Taylor1972}%
\begin{eqnarray}
\Gamma(E,\mathbf{k},\mathbf{k}^{\prime})  =V_{2D}(\mathbf{k}-\mathbf{k}%
^{\prime})+\int\frac{d\mathbf{q}}{(2\pi)^{2}}\tilde{V}_{2D}(\mathbf{k}-%
\mathbf{q}) 
 \times\frac{1}{E-\hbar^{2}q^{2}/m+i0}\Gamma(E,\mathbf{q},%
\mathbf{k}^{\prime}).  \label{ScatteringIntegralEquation}
\end{eqnarray}
The 2D scattering amplitude $f_{k}(\varphi)$, where $\varphi$ is the angle
between $\mathbf{k}$ and $\mathbf{k}^{\prime}$, corresponds to $(m/\hbar
^{2})\Gamma(E,\mathbf{k},\mathbf{k}^{\prime})$ with $E=\hbar^{2}k^{2}/m=%
\hbar^{2}k^{\prime2}/m$. The solution of this equation \cite%
{PhysRevA.83.043602}, \cite{PhysRevA.82.044701} reads:

For $g<k_{F}l\lesssim1$, the leading contribution to scattering is given by
the first Born term 
\begin{eqnarray}
\Gamma(E,\mathbf{k},\mathbf{k}^{\prime})\approx-\frac{2\pi\hbar^{2}}{m}%
g\left\vert \mathbf{k}-\mathbf{k}^{\prime}\right\vert l;  \label{Gamma_a}
\end{eqnarray}

For $\exp(-1/g^{2})\ll kl<g<1$, the scattering is dominated by the second
order Born contribution%
\begin{eqnarray}
\Gamma(E,\mathbf{k},\mathbf{k}^{\prime})\approx-\frac{2\pi\hbar^{2}}{m}\frac{%
g^{2}}{4},  \label{Gamma_b}
\end{eqnarray}
which is momentum and energy independent and, hence, is equivalent to a
pseudopotential $V_{0}(\boldsymbol{\rho})=-(2\pi\hbar^{2}/m)(g^{2}/4)\delta(%
\boldsymbol{\rho})$;

For $\exp(-1/g^{2})\lesssim kl\ll g<1$, higher order contributions become
important and one has to sum leading contributions from the entire Born
series. The result of this summation is%
\begin{eqnarray}
\Gamma(E,\mathbf{k},\mathbf{k}^{\prime})\approx\frac{2\pi\hbar^{2}}{m}\frac {%
2}{\ln(E_{b}/E)+i\pi},  \label{Gamma_c}
\end{eqnarray}
where $E_{b}$ is the energy of the bound state from Eq.~(\ref{Eb}). This
expression recovers the standard energy dependence of the 2D low-energy
scattering and has a pole at $E=-E_{b}$, as it should be. The real part of
the scattering amplitude, being zero at $E=E_{b}$, changes from negative to
positive values for $E>E_{b}$ and $E<E_{b}$, respectively. Note that within
the lowest order terms, a unique expression for the scattering amplitude can
be written as%
\begin{equation*}
\Gamma(E,\mathbf{k},\mathbf{k}^{\prime})\approx-\frac{2\pi\hbar^{2}}{m}\left[
g\left\vert \mathbf{k}-\mathbf{k}^{\prime}\right\vert l-\frac{2}{\ln
(E_{b}/E)+i\pi}\right] .
\end{equation*}

As one can see, the interlayer scattering amplitude is negative in the $s$%
-wave channel [for the case $\exp(-1/g^{2})\lesssim kl\ll g<1$ this requires 
$E\sim\varepsilon_{F}\gg E_{b}$, which is realistic in the limit $g<1$].
This means that at sufficiently low temperatures, the bilayer fermionic
dipolar system undergoes a BCS pairing transition into a superfluid state
with interlayer $s$-wave Cooper pairs, characterized by an order parameter $%
\Delta(\mathbf{p})\sim\left\langle \hat{\psi}_{-}(\mathbf{p})\hat{\psi}_{+}(-%
\mathbf{p})\right\rangle $ with $\hat{\psi}_{\alpha }(\mathbf{p})$ being the
field operator in the momentum space, which is independent of the azimuthal
angle $\varphi$, $\Delta(\mathbf{p})=\Delta(p)$.

\begin{figure}[ptb]
\begin{center}
\includegraphics[width=0.5\columnwidth]{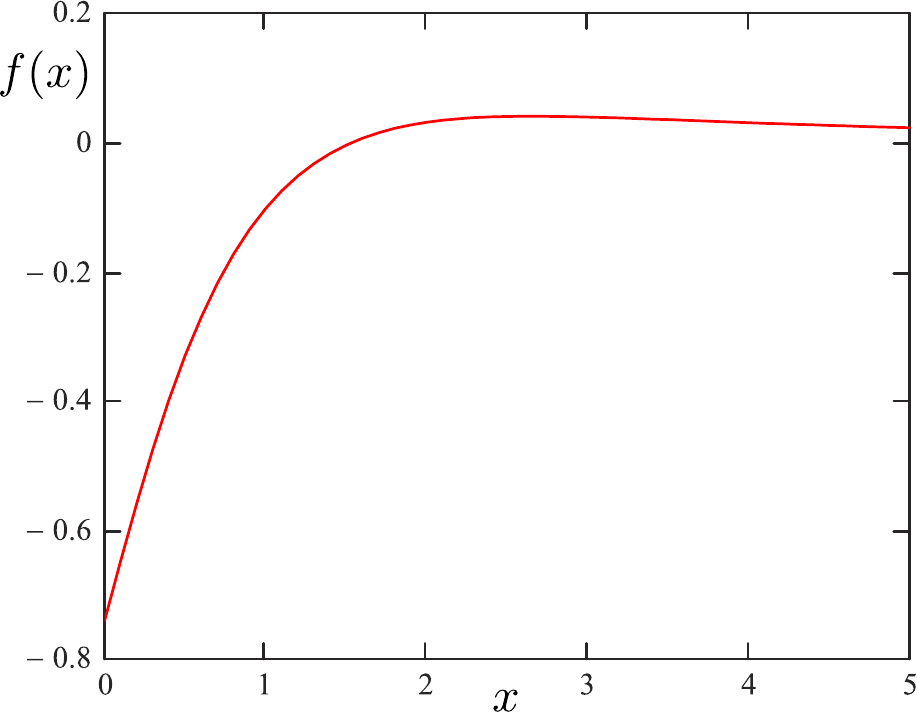}
\end{center}
\caption{The function $f(x)$. (Taken from Ref. \protect\cite%
{PhysRevA.83.043602}.)}
\label{Functionf}
\end{figure}

\begin{figure}[ptbp]
\begin{center}
\includegraphics[width=0.5\columnwidth]{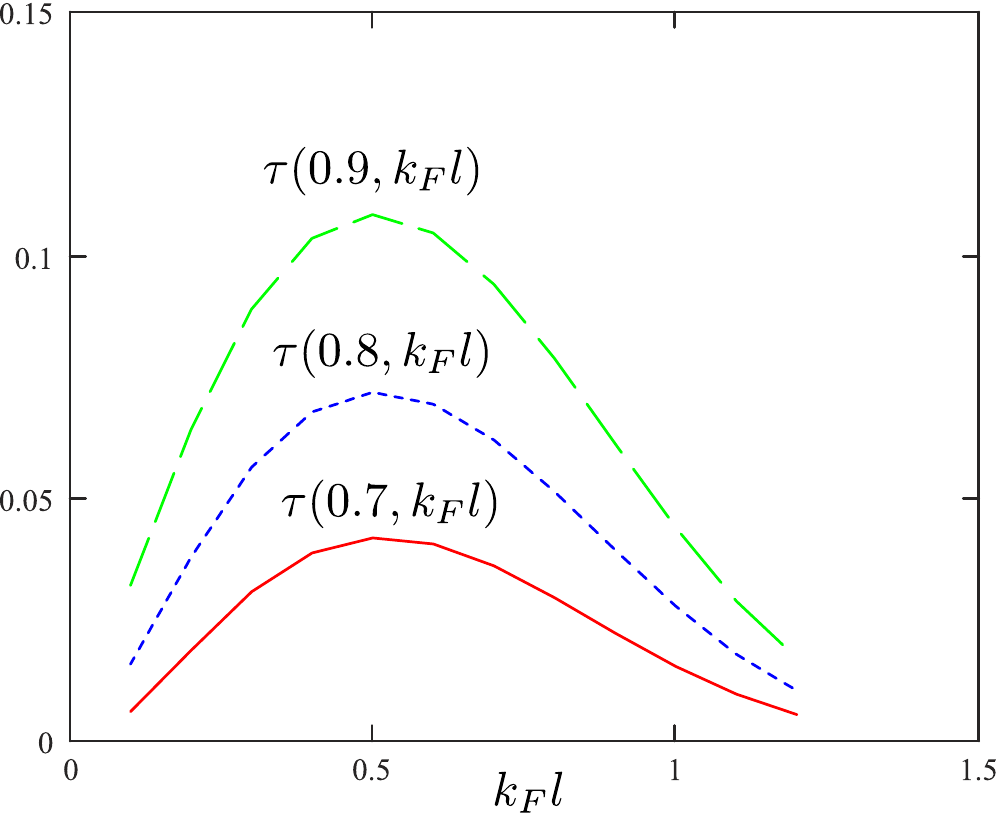}
\end{center}
\caption{The function $\protect\tau(g,x)$ for $g=0.7$ (solid line), $g=0.8$
(short-dashed line), and $g=0.9$ (long-dashed line). (Taken from Ref. 
\protect\cite{PhysRevA.83.043602}.)}
\label{tau}
\end{figure}

The analysis of the corresponding gap equation was performed in Refs. \cite%
{PhysRevLett.105.215302} and \cite{PhysRevA.83.043602} (the latter includes
many-body effects), and we present here only the results for the critical
temperature $T_{c}$ in the experimentally most interesting case $k_{F}l\sim1 
$ and $g<1$ such that $gk_{F}l=a_{d}k_{F}<1$ (results for other cases scan
be found in Ref. \cite{PhysRevA.83.043602}:%
\begin{align*}
T_{c} & =\frac{2e^{\gamma}}{\pi}\varepsilon_{F}\exp\left[ -\frac{1}{%
3\gamma(k_{F}l\,)}-\left( \frac{\pi}{4}\right) ^{2}\frac{f(k_{F}l\,)}{%
\gamma(k_{F}l\,)^{2}}\right] \\
& \times\exp\left[ -\frac{\pi}{4gk_{F}l\,\gamma(k_{F}l\,)}\frac{1}{%
1-(4/\pi)gk_{F}l\,\gamma(k_{F}l\,)\Omega(k_{F}l)}\right] \\
& \equiv\frac{2e^{\gamma}\mu}{\pi}\tau(g,k_{F}l),
\end{align*}
where 
\begin{equation*}
\gamma(x)=\frac{1}{2}\int_{0}^{\pi}d\varphi\sin(\varphi)e^{-x\sin(\varphi )}=%
\frac{\pi}{4}\left[ \mathbf{L}_{-1}(2x)-\mathrm{I}_{1}(2x)\right] ,
\end{equation*}
$\mathbf{L}_{n}(z)$ and $\mathrm{I}_{n}(z)$ being the modified Struve and
Bessel functions, respectively,

\begin{equation*}
\Omega(x)=\frac{1}{2}\int_{0}^{\infty}ds\ln\left\vert 1-s^{2}\right\vert 
\mathrm{sign}(s-1)\frac{d}{dx}\left[ V(s,x)^{2}\right]
\end{equation*}
with%
\begin{equation*}
V(s,x)=\int_{0}^{\pi}d\varphi\sqrt{1+s^{2}-2s\cos\varphi}e^{-x\sqrt {%
1+s^{2}-2s\cos\varphi}},
\end{equation*}
and the function $\ f(x)$ is shown in Fig.~\ref{Functionf}. 
The dependence of the function $\tau(g,k_{F}l)$ on $k_{F}l$ for several
values of $g$ is shown in Fig.~\ref{tau}. 
We see that the critical temperature decreases very rapidly for $k_{F}l>1$
due to the fast decay of the scattering amplitude. The optimal value of $%
k_{F}l$ is around $0.5$ with the critical temperature reaching values of the
order of $0.1\mu$ for $g\approx0.9$ that corresponds to $gk_{F}l%
\approx0.45<1 $.

This BCS state with interlayer Cooper pairs occurs in the weak (interlayer)
coupling regime when the size of the bound state is larger than the
interparticle separation (in other words, the Fermi energy is larger than
the binding energy). With increasing interlayer coupling, the BCS state
smoothly transforms into a BEC state of tightly bound interlayer molecules
when the interparticle separation is larger than the size of the bound
state; see, for example, Refs. \cite{PhysRevLett.105.215302}\textbf{\ }and 
\cite{ZinnerEprint}. Of course, the BEC regime and BEC-BCS crossover are
possible only when the mean interparticle separation in each layer is larger
than the distance between the layers.

Let us now discuss possible physical realizations of the interlayer pairing.
In the experiments with polar molecules, the values of the dipolar length $%
a_{d}$ are of the order of $10^{2}\div10^{4}\,\mathrm{nm}$: for a $^{40}%
\mathrm{K}^{87}\mathrm{Rb}$ with currently available $d\approx 0.3\,\mathrm{D%
}$ one has $a_{d}\approx170\,\mathrm{nm}$ (with $a_{d}\approx600\,\mathrm{nm}
$ for the maximum value $d\approx0.566\,\mathrm{D}$), and for $^{6}\mathrm{Li%
}^{133}\mathrm{Cs}$ with a tunable dipole moment from $d=0.35\,\mathrm{D}$
to $d=1.3\,\mathrm{D}$ (in an external electric field $\sim1\,\mathrm{kV/cm}$%
) the value of $a_{d}$ varies from $a_{d}\approx260\,\mathrm{nm}$ to $%
a_{d}\approx3500\,\mathrm{nm}$. For the interlayer separation $l$ of the
order of few hundreds nanometers, the corresponding values of the parameter $%
g$ can be both smaller and larger than unity ($g\lesssim10$).

The values of the parameter $k_{F}l$ are also within this range for
densities $n=10^{6}\div10^{9}\,\mathrm{cm}^{-2}$ (for example, one has $%
k_{F}l=1$ for $l=500\,\mathrm{nm}$ and $n\approx3\cdot10^{7}\,\mathrm{cm}%
^{-2}$). Note, however, that the optimal values of this parameter are around 
$k_{F}l\sim0.5$ (see Fig.~\ref{tau}) and, hence, the optimum value of the
interlayer separation is related to the density, which, in turn, should be
large enough to provide a substantial value for the Fermi energy. For$^{40}%
\mathrm{K}^{87}\mathrm{Rb}$ molecules at the density $n\approx4\cdot10^{8}\,%
\mathrm{cm}^{-2}$ in each layer one has $\varepsilon_{F}\approx 100\,\mathrm{%
nK}$ and $k_{F}=$. Therefore, the interlayer separation $l$ should be
relatively small, $l\lesssim150\,\mathrm{nm}$, to meet the optimal
conditions. For $l=150\,\mathrm{nm}$ one then has $g\approx1.1$ (with
current $d\approx0.3\,\mathrm{D}$), $k_{F}l\approx1$, and $%
T_{c}\approx0.1\varepsilon _{F}\approx10\,\mathrm{nK}$. Note that strictly
speaking these values of parameters $g$ and $k_{F}l$ do not correspond to
the weak coupling regime considered in this paper, rather to the
intermediate regime of the BCS-BEC crossover. However, based on the
experience with the BEC-BCS crossover in two-component atomic fermionic
mixtures, in which the critical temperature continues to grow when
approaching the crossover region from the BCS side, we could expect that the
above value of the critical temperature provides a good estimate for the
onset of the superfluidity in the intermediate coupling regime.

\subsection{Stability of fermionic dipolar systems}

As we have already mentioned, a polarized homogeneous fermionic dipolar gas
becomes unstable for strong dipole-dipole interaction ($a_{d}k_{F}>1$).
Unfortunately, for this strongly interacting regime the usage of the
Hartree-Fock approximation could not be rigorously justified. It is commonly
accepted, however, that the Hartree-Fock method is able to provide a correct
qualitative picture even in this regime, although its quantitative results
should be taken with care.

The simplest way to get a quantitative insight into the instability region
for a 3D dipolar gas is to use the variational ansatz (\ref%
{deformation_ansatz}) to calculate the compressibility. It turns out \cite%
{SogoNJP}, \cite{PhysRevA.81.033601} that the compressibility if a
homogeneous gas becomes negative for $a_{d}k_{F}>9.5$ signalling the
instability of the gas leading to a collapse. 

\begin{figure}[tbp]
\begin{center}
\includegraphics[width=0.5\columnwidth]{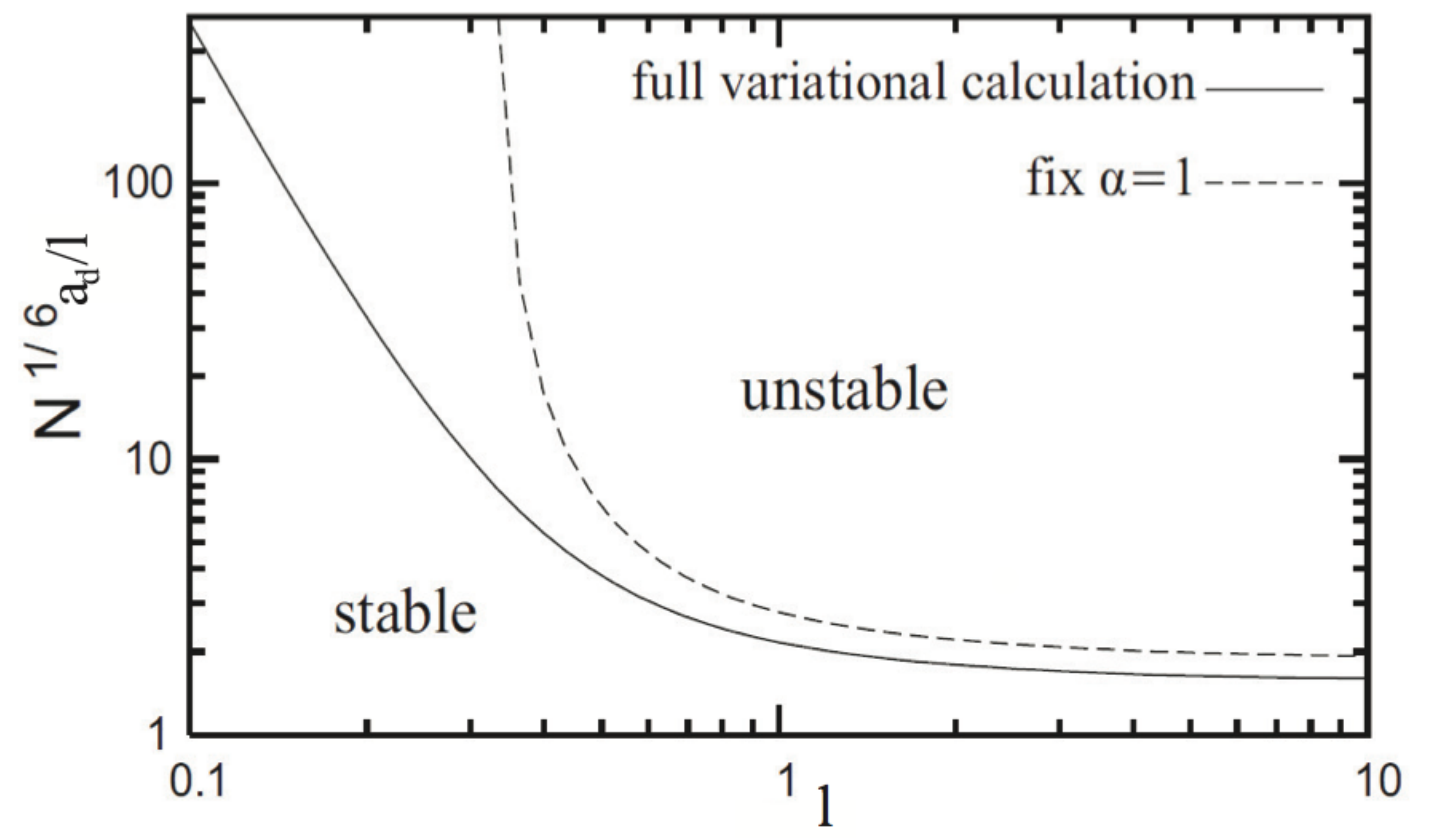}
\end{center}
\caption{Critical value of $N^{1/6}a_{d}k_{F}$ as a function of the trap
aspect ratio $l$. The solid line represents the full variational
calculation, while the dashed line is obtained by forcing $\protect\alpha =1$%
. (Taken from Ref. \protect\cite{PhysRevA.77.061603}.)}
\label{XXX2}
\end{figure}

For a trapped gas with $N$ particles \cite{PhysRevA.77.061603}, the ansatz (%
\ref{deformation_ansatz}) for the momentum distribution has to be
generalized to the ansatz for the Wigner distribution function%
\begin{equation*}
n(\mathbf{r},\mathbf{p})=\theta \lbrack p_{F}(\mathbf{r})^{2}-\frac{1}{%
\alpha }(p_{x}^{2}+p_{y}^{2})-\alpha ^{2}p_{z}^{2}],
\end{equation*}%
where $p_{F}(\mathbf{r})^{2}=p_{F}^{2}-\lambda ^{2}l_{\omega }^{-4}[\beta
(x^{2}+y^{2})-\beta ^{-2}z^{2}]$ is the position-dependent square of the
Fermi momentum, $\lambda $ and $\beta $ are variational parameters, $%
p_{F}=(48N)^{1/6}\lambda ^{1/2}\hbar /l_{\omega }$, and $l_{\omega }=\sqrt{%
\hbar /m\omega }$ with $\omega =(\omega _{\rho }^{2}\omega _{z})^{1/3}$.
Minimization of the energy with respect to the variational parameters $%
\alpha $, $\beta $, and $\lambda $ gives equilibrium density and local
momentum distribution for a given value of the trap aspect ratio $l=\sqrt{%
\omega _{\rho }/\omega _{z}}$. The calculation of the compressibility shows
that it becomes negative for any value of the trap aspect ratio $l$ provided
the value of the parameter $N^{1/6}a_{d}/l_{\omega }$, which measures the
strength of interparticle interaction, is sufficiently large. The dependence
of the critical value of $N^{1/6}a_{d}/l_{\omega }$ on the trap aspect ratio 
$l$ is shown in\textbf{\ }Fig.~\ref{XXX2}. 
Note the found in Ref. \cite{WKBFdip} critical aspect ratio $l_{c}=0.19$,
below which the trapped gas is stable for any value of $N^{1/6}a_{d}/l_{%
\omega }$, is an artefact of the approximation used in this paper resulted
from neglecting the deformation of local momentum distribution, see
discussion in Ref. \cite{PhysRevA.77.061603}.

In a dipolar monolayer, the calculations of the compressibility in Ref. \cite%
{PhysRevLett.101.245301} shows that the collapse of the system takes place
for $a_{d}k_{F}\approx1$ for $\theta=\pi/2$ (dipoles are oriented parallel
to the plane) but for rapidly increasing value of $a_{d}k_{F}$ when the
angle between the dipole polarization and the plane increases. More careful
analysis of the both long wavelength (collapse) and finite wavelength
(density wave) instabilities of a dipolar monolayer can be performed by
looking at the stability of the collective modes: A collective mode becomes
unstable when its frequency tends to zero. The corresponding equation valid
for a general wave vector $\mathbf{k}$ of the collective mode reads%
\begin{eqnarray}
\gamma_{\mathbf{k}}(\mathbf{p})=\int\frac{d\mathbf{p}^{\prime}}{(2\pi
\hbar)^{2}}\widetilde{\Gamma}(\mathbf{k},\mathbf{p},\mathbf{p}^{\prime})%
\frac{n(\mathbf{p}^{\prime})-n(\mathbf{p}^{\prime}+\hbar\mathbf{k})}{%
\hbar\omega+\varepsilon(\mathbf{p}^{\prime})-\varepsilon(\mathbf{p}^{\prime
}+\hbar\mathbf{k})}\gamma_{\mathbf{k}}(\mathbf{p}^{\prime}),
\label{coll_modes}
\end{eqnarray}
where $\widetilde{\Gamma}(\mathbf{k},\mathbf{p},\mathbf{p}^{\prime })=V_{%
\mathrm{d}}^{2D}(\hbar\mathbf{k})-V_{\mathrm{d}}^{2D}(\mathbf{p}-\mathbf{p}%
^{\prime}+\hbar\mathbf{k})$ for our case. [Note that Eq.~(\ref{zero-sound})
follows from (\ref{coll_modes}) after taking the limit $k\rightarrow0$ and
using the relation $\varkappa(\mathbf{p})=[\omega -\mathbf{k}\nabla_{\mathbf{%
p}}\varepsilon(\mathbf{p})]\gamma_{\mathbf{k}}(\mathbf{p})$.] As a result,
an instability occurs when the equation%
\begin{eqnarray}
\gamma_{\mathbf{k}}(\mathbf{p})=\int\frac{d\mathbf{p}^{\prime}}{(2\pi
\hbar)^{2}}\widetilde{\Gamma}(\mathbf{k},\mathbf{p},\mathbf{p}^{\prime})%
\frac{n(\mathbf{p}^{\prime})-n(\mathbf{p}^{\prime}+\hbar\mathbf{k})}{%
\varepsilon(\mathbf{p}^{\prime})-\varepsilon(\mathbf{p}^{\prime}+\hbar%
\mathbf{k})}\gamma_{\mathbf{k}}(\mathbf{p}^{\prime})  \label{stability}
\end{eqnarray}
has a non-trivial solution for some value of $\mathbf{k}$. Note that for $%
\widetilde{\Gamma}(\mathbf{k},\mathbf{p},\mathbf{p}^{\prime})=V_{\mathrm{d}%
}^{2D}(\hbar\mathbf{k})$ (when only the direct interaction is taken into
account and the exchange one is neglected), $\gamma_{\mathbf{k}}(\mathbf{p}) 
$ is $\mathbf{p}$-independent and Eq.~(\ref{coll_modes}) reduces to 
\begin{eqnarray}
1-V_{\mathrm{d}}^{2D}(\hbar\mathbf{k})\Pi(\omega,k)=0,  \label{2Dplasmons}
\end{eqnarray}
where 
\begin{equation*}
\Pi(\omega,k)=\int\frac{d\mathbf{p}^{\prime}}{(2\pi\hbar)^{2}}\frac {n(%
\mathbf{p}^{\prime})-n(\mathbf{p}^{\prime}+\hbar\mathbf{k})}{\hbar
\omega+\varepsilon(\mathbf{p}^{\prime})-\varepsilon(\mathbf{p}^{\prime}+\hbar%
\mathbf{k})}
\end{equation*}
is the 2D polarization operator. Eq.~(\ref{2Dplasmons}) is used to study
long wavelength ($k\rightarrow0$) plasmon oscillations in electrically
charged systems (see, for example, \cite{PinesNozieresbook}). Although
keeping only direct interaction in the long wave-length limit is legitimate
for Coulomb systems (because of divergence of the Coulomb interaction, while
the exchange one is finite due to non-zero momentum transfer, $\left\vert 
\mathbf{p}-\mathbf{p}^{\prime}\right\vert \sim p_{F}$), this approximation
gives physically incorrect results in a Fermi system with a finite Fourier
component of the interparticle interaction for small momentum transfer (like
in the considered case of a dipolar monolayer). In this case, the direct and
the exchange contributions are of the same order and keeping only the former
results in unphysical results. Actually, for a short range interparticle
interaction (with a momentum-independent Fourier component), the two
contributions cancel each other, as it should be in a single-component Fermi
gas. Similar considerations are also applied to the analysis of
instabilities in a dipolar systems on the basis of Eq.~(\ref{stability}):
Keeping the exchange contribution in this equation is essential in order to
obtain correct results consistent with fermionic statistics of particles.

\begin{figure}[tbp]
\begin{center}
\includegraphics[width=0.5\columnwidth]{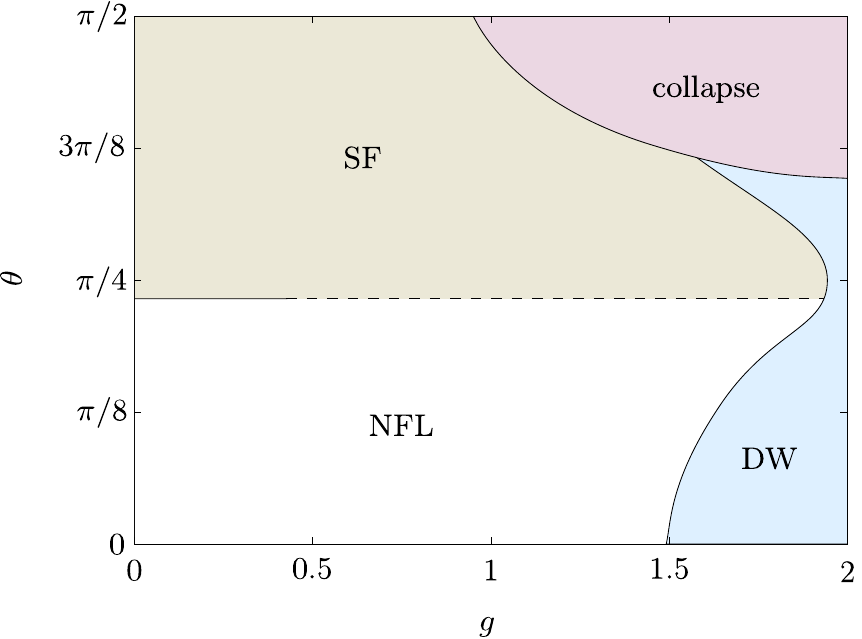}
\end{center}
\caption{Phase diagram of the 2D dipolar Fermi gas at $T=0$. For $0\leq
a_{d}k_{F}\lesssim 1.5$ and small tilting angles $\protect\theta $ the
system is a normal Fermi liquid (NFL). Thje transition to the supefluid
state (SF) occurs at the critical angle $\protect\theta _{c}=0.72$. At
moderately strong interactions, the system either collapses or undergoes the
transition into the density-wave phase (DW). (Taken from Ref. \protect\cite%
{SiebererBaranov}.)}
\label{Phase_diagram}
\end{figure}

For $k\rightarrow 0$, Eq.~(\ref{stability}) is equivalent to the Pomeranchuk
criterion\textbf{\ }\cite{Pomeranchuk1958}, \cite{LL9}] formulated in the
framework of Landau Fermi-liquid. Numerical solution of this equation (for $%
k=0$) \cite{SiebererBaranov} shows that the instability of the system for $%
\theta \gtrsim 3\pi /8$ corresponds to mostly isotropic with some addition
of the quadrupole ($\sim \cos 2\varphi $) deformation of the Fermi surface,
i.e. to the collapse, that takes place for $a_{d}k_{F}\gtrsim 1$ (see Fig.~%
\ref{Phase_diagram}). 
For $\theta \lesssim 3\pi /8$, the leading instability corresponds to the $p$%
-wave ($\sim \sin \varphi $) deformation of the Fermi surface. However, this
instability is unobservable because for these values of the tilting angle $%
\theta $ the system undergoes a density-wave instability at smaller values
of $a_{d}k_{F}$ \cite{PhysRevB.82.075105}\textbf{, } 
This instability corresponds to a non-trivial solution of Eq.~(\ref%
{stability}) with finite $k$ and takes place for $a_{d}k_{F}\gtrsim 1.5$.
For $0<\theta \lesssim 3\pi /8$, the corresponding vector $\mathbf{k}$ is
along the $y$-axis ($\varphi _{\mathbf{k}}=\pm \pi /2$) and has the modulus
that is twice larger than the Fermi wave-vector in the $y$-direction, $%
\mathbf{k}=\hbar ^{-1}2p_{F}(\pi /2)\mathbf{e}_{y}$. For $\theta =0$
(isotropic case), the instability vector $\mathbf{k}$ has no preferable
directions and the system is believed to become unstable against formation
of a crystalline state with the triangular lattice.

\section{Dipolar multilayer systems}

Let us now briefly discuss known results on dipolar multilayer systems. The
stability against formation of inhomogeneous (density wave) phases in
fermionic dipolar multilayer system was discussed in Ref. %
\cite{zinner2011} for a particular choice of the tilting angle $\theta
=\arccos 1/\sqrt{3}$ and in Ref. \cite{babadi2011} for an arbitrary $%
\theta $. The analysis of these papers show that inclusion of exchange
interactions tends to stabilize the homogeneous state resulting in higher
values of critical dipolar interaction strength as compared to the simple
random-phase-approximation (RPA) approach. On the other hand, for multiple
layers this critical dipolar interaction strength decreases with the number
of layers.

Another interesting feature of multilayer systems of dipoles is the
formation of many-body bound states in the form of a chain (or filament)
made of one dipole in each layer \cite%
{PhysRevLett.97.180413,PhysRevA.81.013604,volosniev2011, volosniev2011b} (It
was argued in Ref. \cite{volosniev2011} that bound states involving two
molecules from the same layer do not exist.) The binding energy of such
chains increases with the number of involved molecules (or layers). As a
result, the ground state contains chains of maximum length, while at finite
temperatures, the competition between entropy that favors shorter chains and
energy preferring longer ones results in a non-monotonic dependence of the
distribution on the length of the chains. For bosonic dipoles, quantum
fluids of such self-assembled chains (dipolar chains fluid) and
superfluidity of dipolar chains was considered in Ref. %
\cite{PhysRevLett.97.180413}. For fermionic dipoles, the situation is
even more interesting because chains with even number of dipoles are bosons,
while with odd number of dipoles are fermions. In this case, even\ at zero
temperature, there is an interplay between the Fermi statistics in the form
of a Pauli principle giving rise to a finite kinetic energy of a filled
Fermi sphere, and the binding energy, see Ref. %
\cite{PhysRevA.81.013604}, where a Bose-Fermi mixture of
self-assembled noninteracting. chains was considered for the simple case of
a three-layer system of fermionic polar molecules oriented perpendicular to
the layers. For a more general case for both bosonic and fermionic dipoles,
which also includes interactions between chains, see Ref. %
\cite{volosniev2011b}.

The superfluidity in fermionic dipolar multilayer systems was addressed in
Ref. \cite{PhysRevLett.105.220406}. The interlayer character of Cooper
pairs in this case leads to the competition for pairing among adjacent
layers resulting in a dimerized superfluid state as the ground state, in
which the system can be viewed as a stack of bilayers with interlayer
pairing correlations inside each bilayer and no such correlations between
layers belonging to different bilayers. This state is characterized by a
quasi-long-range superfluid order in every bilayer. At some finite critical
temperature, this phase undergoes a phase transition into a dimerized
"pseudogap" phase with only short-range superfluid correlations. These
correlations disappear above the second critical temperature, and the system
is in the normal phase (see details and proposals for experimental
detections of the phases and phase transitions in Ref. %
\cite{PhysRevLett.105.220406}).

\section{Strongly interacting dipolar gas}

\label{sec:StronglyInteractingGases}

Strong correlations are at the core of a number of fundamental phenomena in
many-body physics, ranging from the formation of self-assembled ionic crystals
to exotic phases such as high-Tc superconductivity and spin liquids. The
regime of strong correlations between particles in a gas is generally obtained
when the strength of the inter-particle interactions becomes comparable to, or
larger than, the average kinetic energy. There are two main avenues to achieve
this regime of strong correlations: (i) the first is to decrease the kinetic
energy by placing particles on a lattice - which in the case of a dilute gas
has the effect of increasing the effective mass - and the second (ii) is to
increase the relative strength of interactions. In gases of dipolar particles
strong correlations can be achieved either way. In the following we first
discuss the phase diagram for dipoles confined to 2D with tunable interactions
(Sect.~\ref{sec:secSelfAssembled}), and then we review several works on exotic
many-body phases for interacting dipoles trapped in optical lattice,
Sect.~\ref{sec:OpticalLattices}. \newline

\subsection{Two-dimensional dipoles: phase diagram.}

\label{sec:secSelfAssembled}

The conceptually simplest example, although remarkably rich from a physics
point of view, is a system of cold polar molecules in a DC electric field
under strong transverse confinement. The setup is illustrated in
Fig.~\ref{fig:Set2D}(a). A weak DC field along the $z$-direction induces a
dipole moment $d$ in the ground state of each molecule. These molecules
interact via the effective dipole-dipole interaction $V_{\mathrm{eff}%
}^{\mathrm{3D}}(\mathbf{r})=D(r^{2}-3z^{2})/r^{5}$ according to their induced
dipoles, with $D=d^{2}$. For molecules confined to the $x,y$-plane
perpendicular to the electric field this interaction is purely repulsive. For
molecules displaced by $z>r/\sqrt{3}$ the interaction becomes attractive,
resulting in few-body and many-body instabilities. As discussed in
Sect.~\ref{s2.2}, these instabilities can be suppressed by a sufficiently
strong 2D confinement with the potential $V_{\mathrm{trap}}(z_{i})$ along $z$,
due to, for example, an optical force induced by an off-resonant light
field~\cite{Buechler07}.

\begin{figure}[ptb]
\begin{centering}
\includegraphics[width=0.9 \columnwidth]{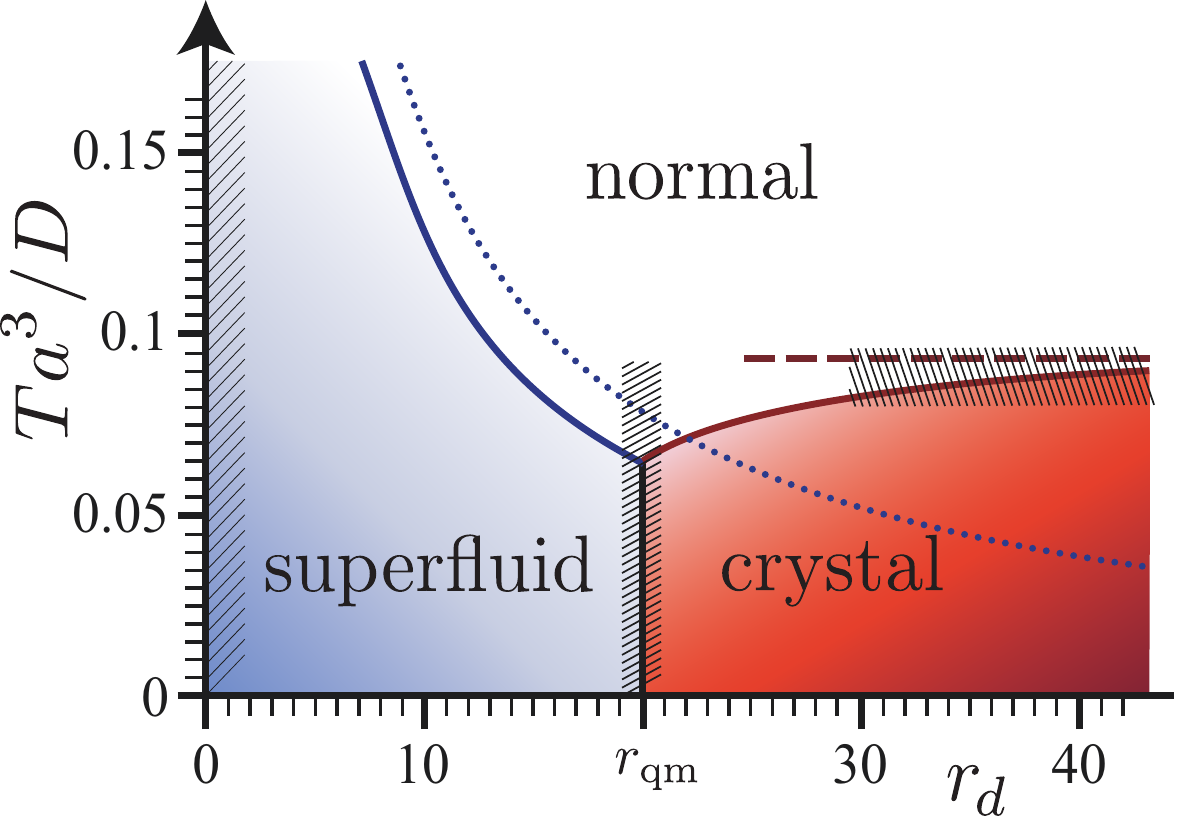}
\par\end{centering}
\caption{Tentative phase diagram for bosonic dipoles in 2D [see setup in
Fig.~\ref{fig:Set2D}] in the $T-r_{d}$ plane: crystalline phase for
interactions $r_{d}>r_{\mathrm{\scriptscriptstyle QM}}$ and temperatures below
the classical melting temperature $T_{m}$ (dashed line)~\cite{Kalia}. The
superfluid phase appears below the upper bound $T<\pi\hbar^{2}n/2 m $ (dotted
line)~\cite{berezinskii1972,kosterlitz1973}. The crossover to the unstable
regime for small repulsion and finite transverse confinement $\Omega$ for
polar molecules is indicated (hatched region). Hatched regions for $r_{d} \sim
r_{\mathrm{qm}}$ and at the crystal-normal phase transitions correspond to the
existence of possible exotic phases [see the \textit{Open Questions} section
in the text]. [Adapted from Ref.~\cite{Buechler07}] }%
\label{figs:fig1000}%
\end{figure}

The 2D dynamics in this pancake configuration is described by the Hamiltonian
\begin{eqnarray}
\label{hamilton1}H_{\mathrm{eff}}^{\mathrm{2D}}=\sum_{i}\frac{\mathbf{p}%
_{\mathbf{\rho}i}^{2}}{2m}+\sum_{i<j}V_{\mathrm{eff}}^{\mathrm{2D}%
}(\mbox{\boldmath$\rho$}_{ij}),
\end{eqnarray}
which is obtained by integrating out the fast $z$-motion.
Equation~(\ref{hamilton1}) is the sum of the 2D kinetic energy in the $x$%
,$y$-plane and the repulsive 2D dipolar interaction
\begin{eqnarray}
V_{\mathrm{eff}}^{\mathrm{2D}}(\mbox{\boldmath$\rho$})=d^{2}/\rho
^{3},\label{eq:eqIn}%
\end{eqnarray}
with $\mbox{\boldmath$\rho$}_{ij}\equiv(x_{j}-x_{i},y_{j}-y_{i})$ a vector in
the $x,y$-plane {[}solid line in Fig.~\ref{fig:circles}(a)]. The
distinguishing feature of the system described by the Hamiltonian
(\ref{hamilton1}) is that tuning the induced dipole moment $d$ drives the
system from a weakly interacting gas (a 2D superfluid in the case of bosons or
a 2D Fermi liquid~\cite{Bruun2008}), to a crystalline phase in the limit of
strong repulsive dipole-dipole interactions. This transition and the
crystalline phase have no analog in the atomic bose gases with short range
interactions modelled by a pseudopotential of a given scattering length.

A crystalline phase corresponds to the limit of strong repulsion where
particles undergo small oscillations around their equilibrium positions, which
is a result of the balance between the repulsive long-range dipole-dipole
forces and an additional (weak) confining potential in the $x,y$-plane. The
relevant parameter is
\begin{eqnarray}
r_{d}\equiv\frac{E_{\text{pot}}}{E_{\text{kin}}}=\frac{d^{2}/a^{3}}{\hbar
^{2}/ma^{2}}=\frac{d^{2}m}{\hbar^{2}a} =\frac{a_{d}}{a},\label{eq:eqrd}%
\end{eqnarray}
which is the ratio of the interaction energy and the kinetic energy at the
mean interparticle distance $a$. This parameter is tunable as a function of
$d$ from small $r_{d}$ to large. A crystal forms for $r_{d}\gg1$, when
interactions dominate, which for a dipolar crystal is the limit of large
densities. This density dependence is different from that in Wigner crystals
with $1/r$- Coulomb interactions, as realized e.g. with laser cooled trapped
ions~\cite{ions}. In the latter case $r_{c}=(e^{2}/a)/\hbar^{2}/ma^{2}\sim a$
and the crystal forms at low densities. In addition, the charge $e$ is a fixed
quantitiy, while $d$ can be varied as a function of the DC field.\newline

Figure~\ref{figs:fig1000} shows a schematic phase diagram for a dipolar gas of
bosonic molecules in 2D as a function of $r_{d}$ and temperature $T$. In the
limit of weak interactions $r_{d}<1$, the ground state is a superfluid (SF)
with a finite (quasi-)condensate. The SF is characterized by a superfluid
fraction $\rho_{s}(T)$, which depends on temperature $T$, with $\rho
_{s}(T=0)=1$. A Berezinskii--Kosterlitz--Thouless
transition~\cite{berezinskii1972,kosterlitz1973} from the superfluid to a
normal fluid occurs at a finite temperature
$T_{\mathrm{{\scriptscriptstyle KT}}}=\pi\rho_{s}\hbar^{2}n/2m$, as expected
in two dimensions. Recent numerical results in Ref.~\cite{Filinov2010}
obtained with an exact Path-Integral Monte-Carlo technique
(PIMC)~\cite{Boninsegni2006} have shown that the superfluid fraction $\rho
_{s}(T)$, and thus $T_{\mathrm{{\scriptscriptstyle KT}}}$, has a non-monotonic
behavior as a function of the interaction strength $r_{d}$, reaching a maximum
of about $\rho_{s}(T_{\mathrm{KT}})= 0.9$ at $r_{d}\simeq1$.

In the opposite limit of strong interactions $r_{d}\gg1$ the polar molecules
are in a crystalline phase for temperatures $T<T_{m}$ with $T_{m}%
\approx0.089D/a^{3}$, see Ref.~\cite{Kalia}. The configuration with minimal
energy is a triangular lattice with excitations given by acoustic phonons,
with characteristic Debye frequency $\hbar\omega_{\mathrm{D}}/(\hbar
^{2}/ma^{2})\sim7.9\sqrt{r_{d}}$. The intermediate strongly interacting regime
with $r_{d}\gtrsim1$ has been investigated using several numerical, especially
quantum Monte-Carlo, techniques in
Refs.~\cite{Buechler07,Astrakharchik07,Mora07,Mazzanti2009,Hufnagl2011}. Using
an exact PIMC technique, a quantum melting transition from the crystalline to
the superfluid phases has been determined to occur at a critical interaction
strength $r_{\mathrm{{\scriptscriptstyle QM}}}=18\pm4$,~\cite{Buechler07}.
\newline



\begin{figure}[ptb]
\includegraphics[width=\columnwidth]{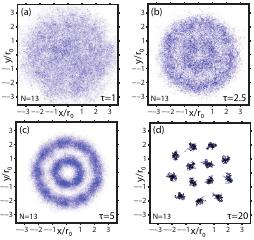}
\caption{(color online) (a-d) Monte Carlo snapshots of the density of
particles in all mesoscopic phases for $N=13$ dipoles, as a function of the
effective mass $\tau$. (a) superfluid; (b) supersolid; (c) ring-like crystals;
(d) classical crystal. [From Ref.~\cite{Pupillo2010}]}%
\label{fig:Mesoscopic}%
\end{figure}\emph{Confined geometries:} The addition of an in-plane parabolic
confinement of frequency $\omega_{\parallel}$ as realized in experiments with
magnetic or optical traps introduces a term $\sum_{i} m \omega_{\parallel}^{2}
\mathbf{\rho}_{i}^{2}/2$ in Eq.~\eqref{hamilton1}. The strength of
interactions is now characterized by the dimensionless ratio $\tau=
(a_{d}/\ell)^{2}$, with $\ell= \sqrt{\hbar/m \omega_{\parallel}}$ the harmonic
oscillator length. The physics of \textit{classical} mesoscopic crystals with
$\tau\gg1$ and dominant thermal fluctuations has been first discussed in the
context of excitonic
materials~\cite{belousov2000,Lozovik2004,Ludwig2008,bedanov1994}. It was found
that small dipolar clusters with $N \lesssim40$ particles confined to
parabolic potentials may not crystallize in a triangular lattice, but rather
arrange in shell-like structures. Finite-$T$ melting of these structures would
usually proceed through two separate (non-sharp) transitions, corresponding to
the loss of rotational and radial order.

Motivated by dipolar molecular and atomic gases, the focus has now shifted
towards the low-temperature regime of dominant quantum
fluctuations~\cite{Lozovik2004,Filinov2008,Pupillo2009,Pupillo2010,Jain2011,Golomedov2011,Cremon2010}%
. For bosons, it was recently determined~\cite{Pupillo2009,Pupillo2010} that
the quantum melting of mesoscopic crystals ($\tau\gg1$) into a superfluid
($\tau\lesssim1$) is a sharp crossover involving two intermediate phases:
these are ring-crystals, with vanishing superfluid fraction, and mesoscopic
superfluids with a modulated (e.g., non-homogeneous) density. Snapshots of
quantum Monte-Carlo simulations for $N=13$ particles are shown in
Fig.~\ref{fig:Mesoscopic} for all of these quantum phases. \newline

Having determined the low-temperature phase-diagram both in the homogeneous
situation and in confined geometries, the remaining question is whether these
phases, and in particular the crystalline phase emerging at strong
dipole-dipole interactions, are in fact accessible with polar molecules. As
discussed in Sec.~\ref{s2.2}, stable 2D configurations for the molecules exist
in the parameter region where the combination of strong dipole-dipole
interactions and transverse (optical) trapping confines the particles' motion
to the large distance region with $a > l_{\perp}$, with $l_{\perp} \sim(12
d^{2}/m \Omega^{2})^{1/5}$ the position of the saddle points in
Fig.~\ref{fig:circles}(b) (white circles). For a given induced dipole $d$ the
ground-state of an ensemble of polar molecules is thus a crystal for mean
interparticle distances $l_{\perp}\lesssim a \lesssim a_{\text{max}}$, where
$a_{\text{max}}\equiv d^{2} m/\hbar^{2} r_{\mathrm{QM}}$ corresponds to the
distance at which the crystal melts into a superfluid. For SrO (RbCs)
molecules with permanent dipole moment $d=8.9$D ($d=1.25$D), $a_{\text{min}%
}\sim200nm$($100$nm), while $a_{\text{max}}$ can be several $\mu$m. Since for
large enough interactions the melting temperature $T_{\mathrm{m}}$ can be of
the order of several $\mu$K, the self-assembled crystalline phase should be
accessible for reasonable experimental parameters using cold polar molecules.
\newline

\emph{Open questions:} Remarkably, there are still important open questions
concerning the phase diagram of two-dimensional dipoles, and in particular the
order of the phase transitions between the solid and liquid phases at zero and
finite temperature. While as often happens~\cite{Bloch2008} numerical results
are usually consistent with direct first order transitions, exotic
intermediate phases may occur in either case: (i) \emph{at finite $T \sim
T_{m}$} an intermediate \textit{hexatic} phase characterized by a short-range
positional and a quasi-long-range orientational (sixfold) order may exist
between the solid and the isotropic liquid phases. Evidence for this hexatic
phase, originally proposed by Kosterlitz, Thouless, Halperin, Nelson, and
Young~\cite{kosterlitz1973,nelson1979,young1979}, has been recently found
numerically in Ref.~\cite{Lin2006}. (ii) \emph{At low-temperature} theoretical
results suggest the presence of an intermediate \textit{microemulsion bubble
phase} between the superfluid and the solid. First introduced in the context
of 2D electron gases in Si MOSFETs by Spivak and Kivelson~\cite{Spivak2004},
and never observed so far (however, see below), a microemulsion should prevent
a first order transition in two dimensions. The observation of either phase
with cold dipolar gases would constitute a breakthrough for condensed matter theory.
\\
\\
Strong correlations can also occur in a weakly interacting dipolar gases
subject to rotation\cite{Eberl3,Cooper1,ZhangZhai,YiPu,Cooper2}, which,
as extensively reviewed in Ref.~\cite{Baranov200871}, represents
a key element to engineer effective magnetic
fields in ultracold atomic and molecular gases. 
Strongly correlated phases similar to the Laughlin quantum Hall
states have been proposed in Ref. \cite{Cooper1} for dipolar bosons and
for in Ref. \cite{FQHE} for dipolar fermions, while the transition from the Laughlin
liquid state to a dipolar crystal state was addressed in Ref. \cite{WignerL}.
Recent work\cite{baranov2011} has now provided quantitative estimates for the realization
of Abelian and non-Abelian gauge fields with polar molecules.

\subsection{Optical lattices}

\label{sec:OpticalLattices}

Hubbard Hamiltonians are model Hamiltonians describing the low-energy physics
of interacting fermionic and bosonic particles in a lattice~\cite{hubbard1963}%
. They have the general tight-binding form
\begin{eqnarray}
\label{Hubbard10}H= - \sum_{i, j, \sigma} J_{i j}^{\sigma} b^{\dag}_{i,\sigma}
b_{j,\sigma} + \sum_{i , j,\sigma,\sigma^{\prime}} \frac{U_{i j}^{\sigma
\sigma^{\prime}}}{2} n_{i,\sigma} n_{j,\sigma^{\prime}}.
\end{eqnarray}
Here $b_{i,\sigma}$ ($b^{\dag}_{i,\sigma}$) are the destruction (creation)
operators for a particle at site $i$ in the internal state $\sigma$, $J_{i
j}^{\sigma}$ describes coherent hopping of a particle from site $i$ to site
$j$ (typically the nearest neighbor), and $U_{i j}^{\sigma\sigma^{\prime}}$
describes the onsite ($i=j$) or offsite ($i\neq j$) two-body interactions
between particles, with $n_{i,\sigma}= b^{\dag}_{i,\sigma} b_{i,\sigma}$.
Hubbard models have a long history in condensed matter physics, where they
have been used as tight-binding approximations of strongly correlated systems.
For example, for a system of electrons in a crystal hopping from the orbital
of a given atom to that of its nearest neighbor, $\sigma$ represents the
electron spin. A (fermionic) Hubbard model comprising electrons in a 2D
lattice with interspecies onsite interactions is thought to be responsible for
the high-temperature superconductivity observed in cuprates~\cite{Cup1}%
.\newline

In recent years, Hubbard models have been shown to provide excellent
\textit{microscopic} descriptions of the low-energy physics of interacting
bosonic and fermionic atoms trapped at the bottom of an optical
lattice~\cite{jaksch98,Hofs02}. Since the interactions between cold atoms are
short-ranged, in these systems Hubbard Hamiltonian typically have
\emph{onsite} interactions only [$U_{i,i}^{\sigma,\sigma^{\prime}}$ in
Eq.~\eqref{Hubbard10}]. This is readily shown for the simple case of
single-species ($\sigma=\sigma^{\prime}$) bosonic atoms with contact
interactions, such as $^{133}$Cs atoms prepared in their absolute internal
(hyperfine, fine, ...) ground-state, and trapped in the lowest band of an
optical lattice. In the limit in which all energies involved in the system
dynamics are small compared to excitation energies to the second band and
neglecting the often-small overlap beyond nearest neighboring densities, the
microscopic many-body Hamiltonian reduces to one of the form of
Eq.~\eqref{Hubbard10} with nearest-neighbor hopping energy $J = J^{\sigma}_{i,
i+1}$ and on-site interactions $U=U^{\sigma\sigma}_{i i}$~\cite{jaksch98}
with
\begin{eqnarray}
\label{eq:JUHabbard}J & = - \int w_{i}^{*}({\boldsymbol{r}}) \left(  -
\frac{\hbar^{2} \Delta}{2m} + V_{\mathrm{0}}({\boldsymbol{r}}) \right)
w_{i+1}({\boldsymbol{r}}) \; \mathrm{d}^{3} { r} ,\nonumber\\
U & = \frac{4 \pi a_{s} \hbar^{2}}{m} \int|w_{i}({\boldsymbol{r}})|^{4} \;
\mathrm{d}^{3} { r}.
\end{eqnarray}
Here, $\{ w_{i}({\boldsymbol{r}}) \}$ is a complete set of single-particle
basis functions, known as \textit{Wannier} functions, which are linear
combinations of exact solutions of the Schr\"odinger equation in the periodic
optical potential $V_{\mathrm{0}}({\boldsymbol{r}}) = \sum_{\alpha=1,3}
V_{0,\alpha} \sin^{2}(k_{\alpha}x_{\alpha})$ (known as Bloch functions), and
are localized at individual sites $j$ (here we focus on the lowest lattice
band only). The optical lattice has a depth $V_{0,\alpha}$ proportional to the
intensity of the confining laser beams, with wavevector $k_{\alpha}$. In
atomic systems, the Hamiltonian parameters in Eq.~\eqref{eq:JUHabbard} can
then be accurately controlled independently using external (optical, magnetic)
fields: by increasing the intensity of the lattice laser light, $J$ decreases
exponentially, while $U$ can be broadly tuned, e.g., by varying the scattering
length $a_{s}$ using magnetic Feshbach resonances~\cite{Bloch2008}.

\begin{figure}[ptb]
\includegraphics[width=0.7\columnwidth]{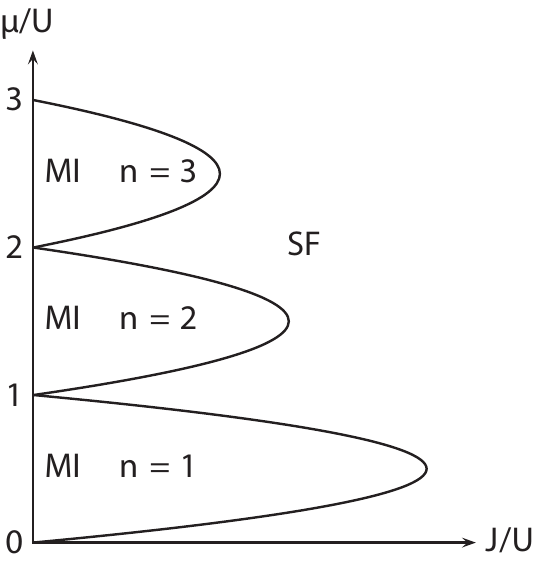}\caption{(color
online) (a) Sketch of the phase diagram of the Bose-Hubbard model at $T=0$ in
the plane $\mu/U$ vs $J/U$. The regions MI denote Mott Insulator phases, with
an integer average onsite density $n$. [Adapted from Ref.~\cite{Lahaye2009}] }%
\label{fig:GuidoBHM}%
\end{figure}

The resulting Bose-Hubbard Hamiltonian (BHH) has been extensively studied in
condensed matter physics~\cite{Fisher1989,Bloch2008}. When the atom number is
commensurate with the number of lattice sites, the BHH predicts a
zero-temperature phase transition from a superfluid (SF) phase to a Mott
insulator (MI) with an increasing ratio of the on site interaction $U$ (due to
repulsion of atoms) to the tunneling matrix element $J$. In the MI phase the
density (occupation number per site) is pinned at an integer value $n$ = 1, 2,
..., and the excitation spectrum shows a gap of the order of $U$,
corresponding to particle-hole excitations. When the density is not integer,
the low-energy phase is superfluid for all strengths of the ratio $J/U$. The
associated phase diagram is sketched in Fig.~\ref{fig:GuidoBHM} as a function
of the chemical potential $\mu$ and the ratio $J/U$, the lobes denoting MI
regions of constant density.

Spectacular experiments with ultracold atoms have lead to the first
observation and characterization of this superfluid/Mott-insulator quantum
phase transition for bosonic atoms~\cite{greiner2002,Spielman06, Haller2010},
by looking at the interference of the expanded cloud, the measurement of the
gapped excitations and the (lack of) conductivity in the Mott phase. Further
experimental work with fermions may resolve the phase diagram of the fermionic
Hubbard model in 2D by performing an \emph{analog quantum simulation} of
Eq.~\eqref{Hubbard10} with two-species cold
fermions~\cite{Ess1,Ess2,Schneider2008}.


\subsubsection{Dipoles on a 2D lattice monolayer}

\label{sec:2DLatticeMono}

\begin{figure}[t]
\vspace*{-0.5cm}
\center{\includegraphics[width=0.55\columnwidth]{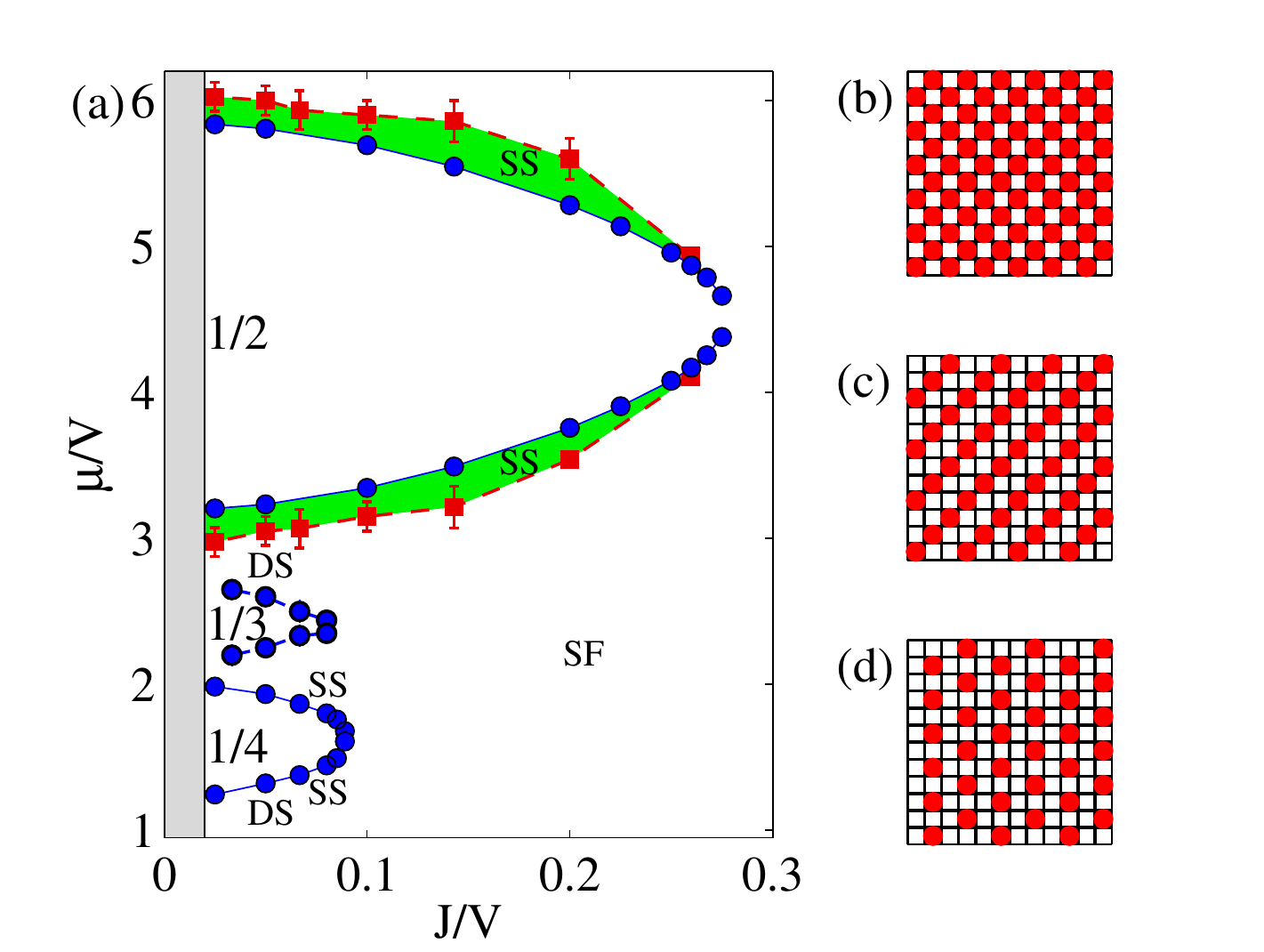}}
\caption{(color online) (a) Phase diagram of Hamiltonian~\eqref{H} at $T=0$.
Lobes: Mott solids (densities indicated); SS: supersolid phase; SF: superfluid
phase. DS: devil's staircase. Panels (b-d): sketches of the groundstate
configuration for the Mott solids in panel (a), with density $n=1/2, 1/3$ and
1/4, respectively. [Adapted from Ref.~\cite{Capogrosso2010}] }%
\label{fig_lobes}%
\end{figure}

The long-range and anisotropic character of dipole-dipole interactions add to
the Bose-Hubbard model new possibilities to observe quantum phases of
fundamental interest. The simplest case is that of an ensemble of
single-species dipoles ($\sigma= \sigma^{\prime}$) which are all polarized
perpendicular to the lattice plane, resulting in isotropic in-plane
interactions. This adds to the BH Hamiltonian terms of the kind $\sum_{i<j}
U^{\sigma\sigma}_{ij}n^{\sigma}_{i} n^{\sigma}_{j}$, with $U^{\sigma\sigma
}_{ij}=V/r^{3}_{ij}$.\newline

Extended Hubbard models have been extensively studied in literature. It has
been predicted that in 2D lattices the presence of finite range interactions
gives rise to novel quantum phases, like the charge-density wave
(checkerboard), which is an insulating phase with modulated density, and the
supersolid (SS) phase, with coexistence of superfluidity and of a periodic
spatial modulation of the density, different from the one of the lattice
\cite{bruder1993}. This latter phase has particularly interesting and storied
history. First proposed in the context of the search for the groundstate of
helium, it has been the subject of extended theoretical and experimental
investigations for almost 40 years. While recent experiments may have spotted
it in bulk solid helium, its very own existence in free-space, that is in the
absence of an underlying periodic potential, is a matter of active research
and debate. With few exact theoretical tools available in the
strongly-interacting regime, quantum Monte-Carlo (QMC) methods have so far
established SS behavior in free-space to be based on defects and disorder
mechanisms, such as the presence of superfluid dislocations and grain
boundaries~\cite{sasaki2006}.
Crucially, any commensurate bulk solid (including $^{4}$He) should be
insulating~\cite{Prokofev2007}, notwithstanding recent theoretical
proposals~\cite{Anderson2010}.

Several theoretical studies have demonstrated SS behavior in tight-binding
lattice
models~\cite{bruder1993,Batrouni1995,Yi2007,Danshita2009,Pollet2010,Capogrosso2010,Boninsegni2008,Dang:2008ij,
Boninsegni:2005hc, Wessel:qa, Sengupta:2005kl, Heidarian:2005mi,
Melko:2005pi}. Model systems with nearest-neighbor (NN) or
next-neighrest-neighbor (NNN) interactions have been generally considered. The
variety of theoretical models and techniques which have been used has resulted
in a zoo of predictions. A general conclusion is that SS behavior seems to be
favored by finite-range interactions as well as finite on-site interactions,
$U^{\sigma\sigma}_{ii} \gtrsim J^{\sigma}_{i, i+1}$. Quantum Monte-Carlo
methods have determined the SS to occur for the following models: \textit{i)}
\textit{hard-core bosons} (infinitely-large $U^{\sigma\sigma}_{ii}$) on a
triangular lattice with NN interactions only, for densities comprised between
$1/3 < n < 2/3$; \textit{ii)} \textit{hard-core bosons} on a square lattice
with NN and NNN interactions for $n < 0.25$ and $0.25 < n < 0.5$ between a
"star" and a "stripe" solid at half filling; \textit{iii)} \textit{soft-core
bosons} (finite value of the ratio $U^{\sigma\sigma}_{ii}/J^{\sigma}_{i, i+1}%
$) on a square lattice with NN and $n > 0.5$. \textit{Phase separation},
characterized by a negative compressibility, has been predicted to occur in
several models, for example in the latter case (iii) with NN interactions and
$n < 0.5$. However, its origin can be traced back to the finite-range
character of the interactions (NN) considered in that specific model, and thus
phase-separation may be expected to disappear in the case of the
infinite-range interactions considered below.\newline

Dipolar atomic and molecular gases trapped in optical lattices can provide
physical systems where the dynamics is \textit{microscopically} described by
extended-Hubbard Hamiltonians with long-range, anisotropic,
interactions~\cite{goral2002b}, with Hamiltonian parameters are tunable with
external fields. In particular, polar molecules in optical lattices can
provide for strong offsite dipolar interactions, of the order of hundreds of
kHz, decaying with distance as $1/r^{3}$. Due to these strong interactions,
two molecules cannot hop onto the same site, and thus the particles may be
treated as effectively "hard-core" (soft-core particles may be realized with
dipolar magnetic atoms). For a lattice of 2D polarized hard-core dipoles, the
microscopic extended Bose-Hubbard Hamiltonian in the presence of long-range
interactions is
\begin{eqnarray}
H = -J \sum_{<i,j>} b^{\dag}_{i} b^{}_{j} + V\sum_{i<j}\frac{n_{i} n_{j}%
}{r^{3}_{ij}} - \mu\sum_{i} n_{i}\;. \label{H}%
\end{eqnarray}


\begin{figure}[t]
\includegraphics[width=0.55\columnwidth]{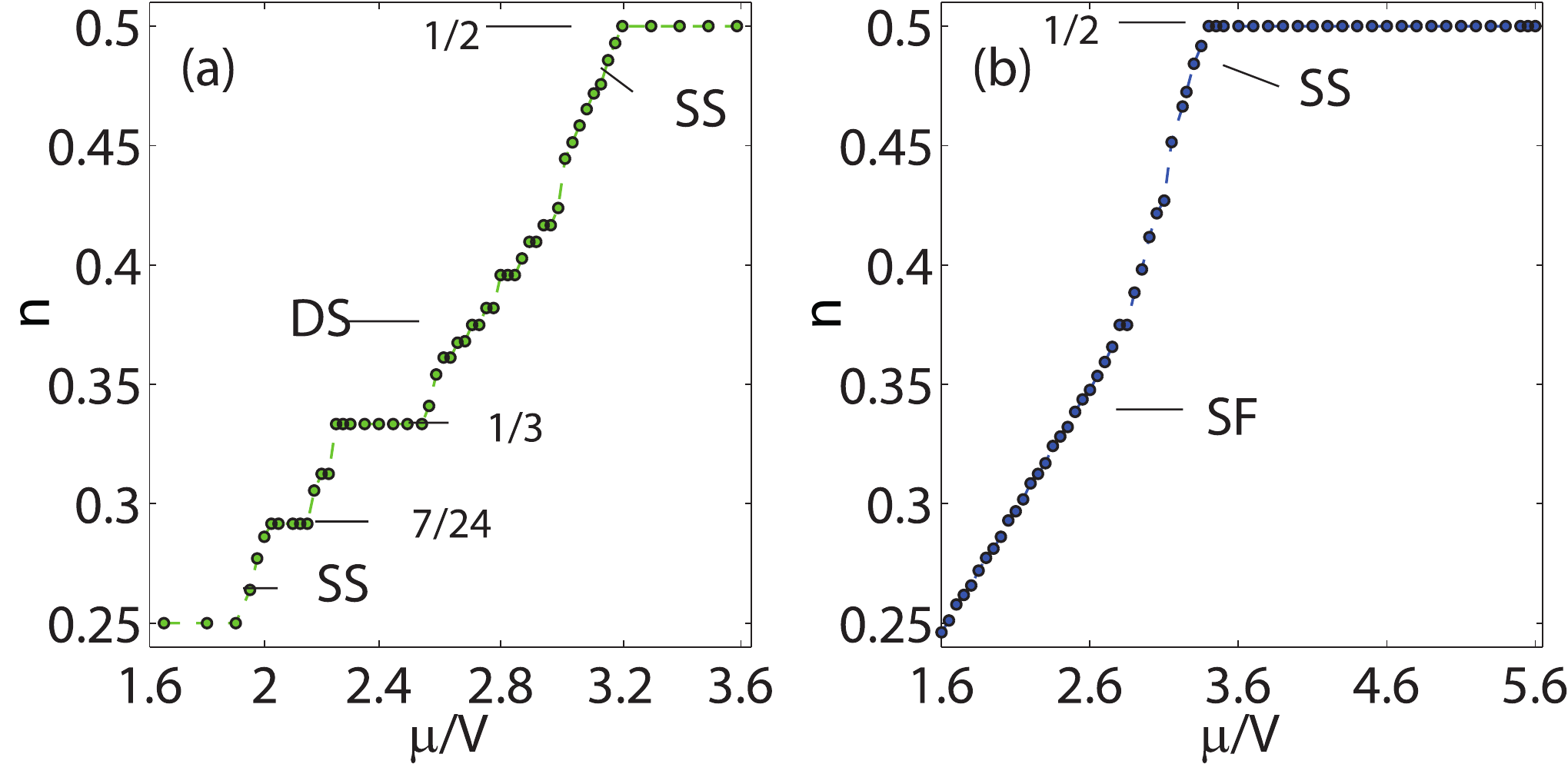}
\caption{(color online) $n$ vs. $\mu$. (a): Solids and SS for a system with
linear size $L=12$ and $J/V=0.05$. Some $n$ are indicated. (b): superfluid and
vacancy-SS for $L=16$ and $J/V=0.1$.}%
\label{fig:devil}
\end{figure}
\begin{figure}[t]
\includegraphics[width=0.55\columnwidth]{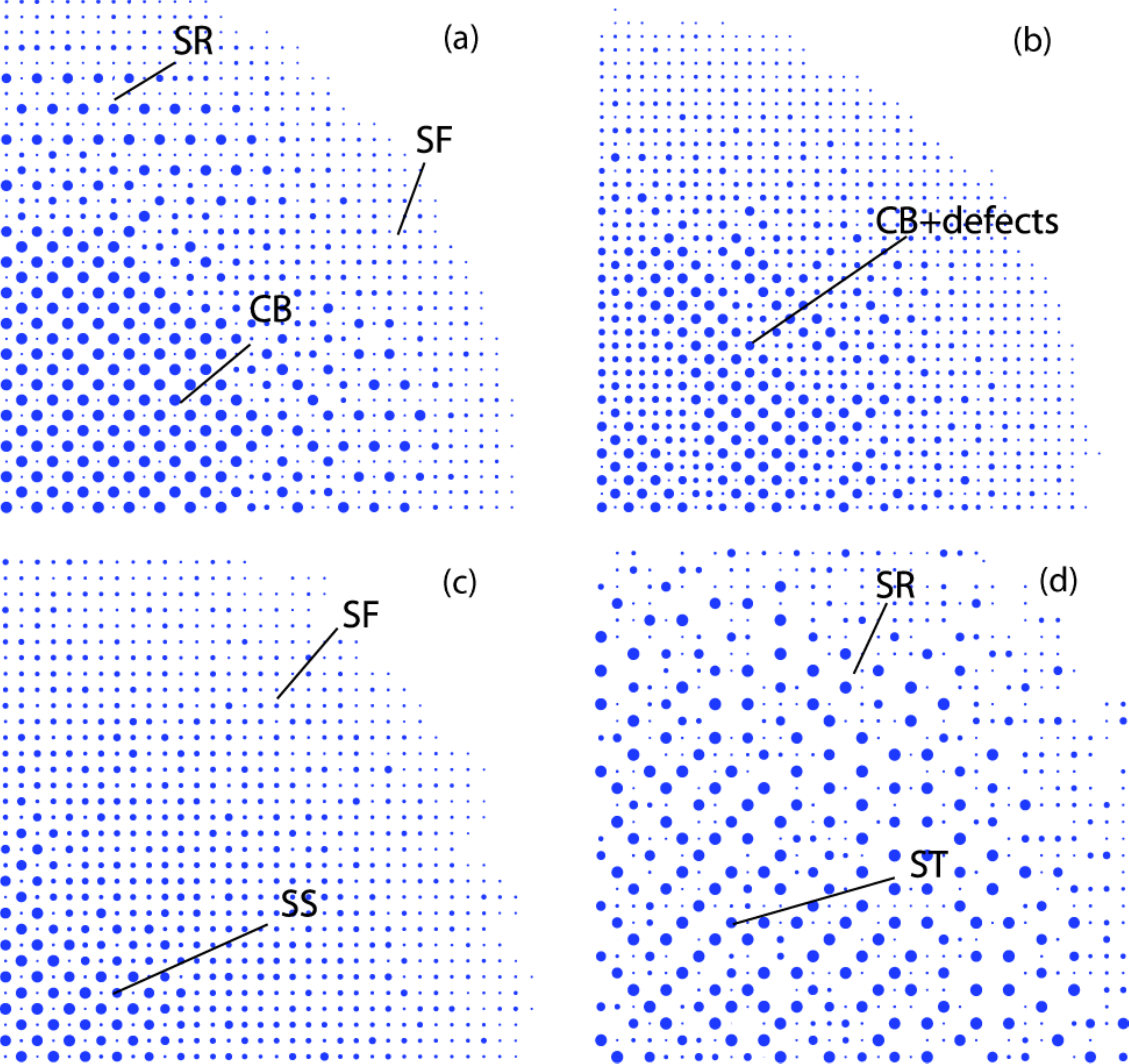}
\caption{(color
online) Spatial density profile in 2D for $N\simeq1000$ particles in a
harmonic potential. Phases are indicated (CB, SR, ST stand for checkerboard,
star, and stripe solids respectively). (a-b) $V/J=15$, $\mu/J=55$,
$\Omega/J=0.05$ and $T/J=0.0377$, with temperature annealing performed in
panel (a).; (c) $V/J=5$, $\mu/J=19$, $\Omega/J=0.01$ and $T/J=0.1$; (d)
$V/J=20$, $\mu/J=51$, $\Omega/J=0.04$ and $T/J=0.25$. [Adapted from
Ref.~\cite{Capogrosso2010}] }%
\label{trap}%
\end{figure}

The phase diagram of Eq.~\eqref{H} has been recently computed for
2D triangular~\cite{Pollet2010} and square~\cite{Capogrosso2010} lattices
using exact QMC methods and no cutoff in the range of interactions. The
zero-temperature phase diagram on a square lattice is shown in
Fig.~\ref{fig_lobes}. For large enough $J/V$, the low-energy phase is
superfluid, for all $\mu$. For finite $J$, three main solid Mott lobes emerge
with densities $n=1/2$, 1/3, and 1/4. These are the checkerboard (CB), stripe,
and star solids, respectively, the corresponding groundstate configurations
being sketched in panels (b-d). Analogous to the standard Mott insulating
phases of Fig.~\ref{fig:GuidoBHM}, these lattice solids at fractional filling
factor exist in some low-tunneling region of the $\mu$ vs. $J$ phase diagram.

Interestingly, it was found that a supersolid phase can be stabilized in a
broad range of parameters by doping the checkerboard and star Mott solids
\textit{either with vacancies} (removing particles) \textit{or interstitials}
(adding extra particles), in accordance with Andreev-Lifshitz's scenario of
defect-induced supersolidity~\cite{AndreevSS}. For example, a vacancy SS is
present for densities $0.5> n \gtrsim0.43$, roughly independent of the
interaction strength. In contrast to, e.g., case (iii) above, the long-range
interactions prevent phase separation from occurring below filling $n=1/2$ in
this microscopic model. This offers interesting prospects for the first
observation of this exotic phase using polar molecules.

Both the solid/supersolid and supersolid/superfluid quantum phase transitions
are second-order. By increasing $T$ the SS melts into a featureless normal
fluid via a two-step transition, the intermediate phase being a normal fluid
with finite density modulations, similar to a liquid
crystal~\cite{Capogrosso2010}.\newline

\emph{Devil's staircase and metastable many-body states:} The large Mott lobes
with $n=1/2$, 1/3, and 1/4 are robust in the presence of a confining harmonic
potential and moderate finite $T/J \sim1$, and thus are relevant to
experiments. Interestingly, for small-enough hopping $J/V \ll0.1$ the
low-energy phase is found to be incompressible ($\partial n / \partial\mu= 0$)
for most values of $\mu$. This parameter region is labeled as DS in
Fig.~\ref{fig_lobes} and it corresponds to a finite-$J$ version of the
classical \textit{Devil's staircase}. First discussed in the context of atomic
monolayers adsorbed on solids~\cite{Hubbard1978}, the latter consists of a
succession of incompressible ground states, dense in the interval $0 < n <1$,
with a spatial structure commensurate with the lattice for all rational
fillings~\cite{Hubbard1978,pokrovsky1978}. This behavior, which has no
analogue for shorter range interactions, is shown in Fig.~\ref{fig:devil}%
(a,b), where the particle density $n$ is plotted as a function of the chemical
potential $\mu$. In the figure, a continuous increase of $n$ as a function of
$\mu$ signals a compressible phase, while a solid phase is characterized by a
constant $n$. The main plateaux in panel (a) correspond to the Mott lobes of
Fig.~\ref{fig_lobes}, while the other steps are incompressible phases, with a
fixed, \textit{integer}, number of particles, indicating the Devil's-like
staircase. We will come back to this point in the discussion of
one-dimensional models.\newline

Determining the exact groundstate geometry and periodicity of the solid for a
given set of Hamiltonian parameters in the DS region is however a
computationally daunting task, since \textit{i)} for many rational fillings
[e.g. $n=7/24$ in Fig.~\ref{trap}(a)] it would require to consider system
sizes much larger than those accessible with reasonable numerical resources,
and \textit{ii)} the long-range interactions determine the presence of
numerous low-energy \textit{metastable} states~\cite{Menotti2007}, which for
finite $T$ can result in the presence of defects or in disordered structures.
The stability of low-energy metastable states has been thoroughly studied in
Refs.~\cite{Menotti2007,Trefzger2008}, where it is found that especially in
larger lattices, two metastable configurations might differ by the occupation
of just a few lattice sites. Thus, because of the presence of these metastable
states, in an experiment it will be very difficult to reach the ground state
or a given metastable configuration. This is directly reflected in the
numerics: Figure~\ref{trap} shows snapshots of the spatial density
distribution in the lattice in the presence of a realistic harmonic
confinement (shown is a single quadrant). Each circle corresponds to a
different site, and its radius is proportional to the local density. In panels
a) and b), $\mu$ is chosen such that particles at the trap center are in the
CB phase, with very small $T$. The density profile shows a wedding-cake
structure, with concentric Mott-lobes with density $n=1/2$ and 1/4, analogous
to the shells with contact interactions~\cite{Batrouni2002a,Batrouni2002b}.
However, while the system parameters are the same in both figures, panel (a)
shows regular CB and star patterns, while in panel (b) extended defects are
present in the CB phase and the star is barely visible. This is due to the
different preparation of the states in panels (a) and (b). In fact, in panel
(a) temperature annealing of the system prior to taking the snapshot was
performed in order to eliminate defects, while this was not done in panel (b).
Defects in (b) reflect the existence of a large number of low-energy
metastable states, which are a direct consequence of the long-range nature of
the interactions.\newline

\begin{figure}[h]
\centerline{\includegraphics[ width=0.5\columnwidth]{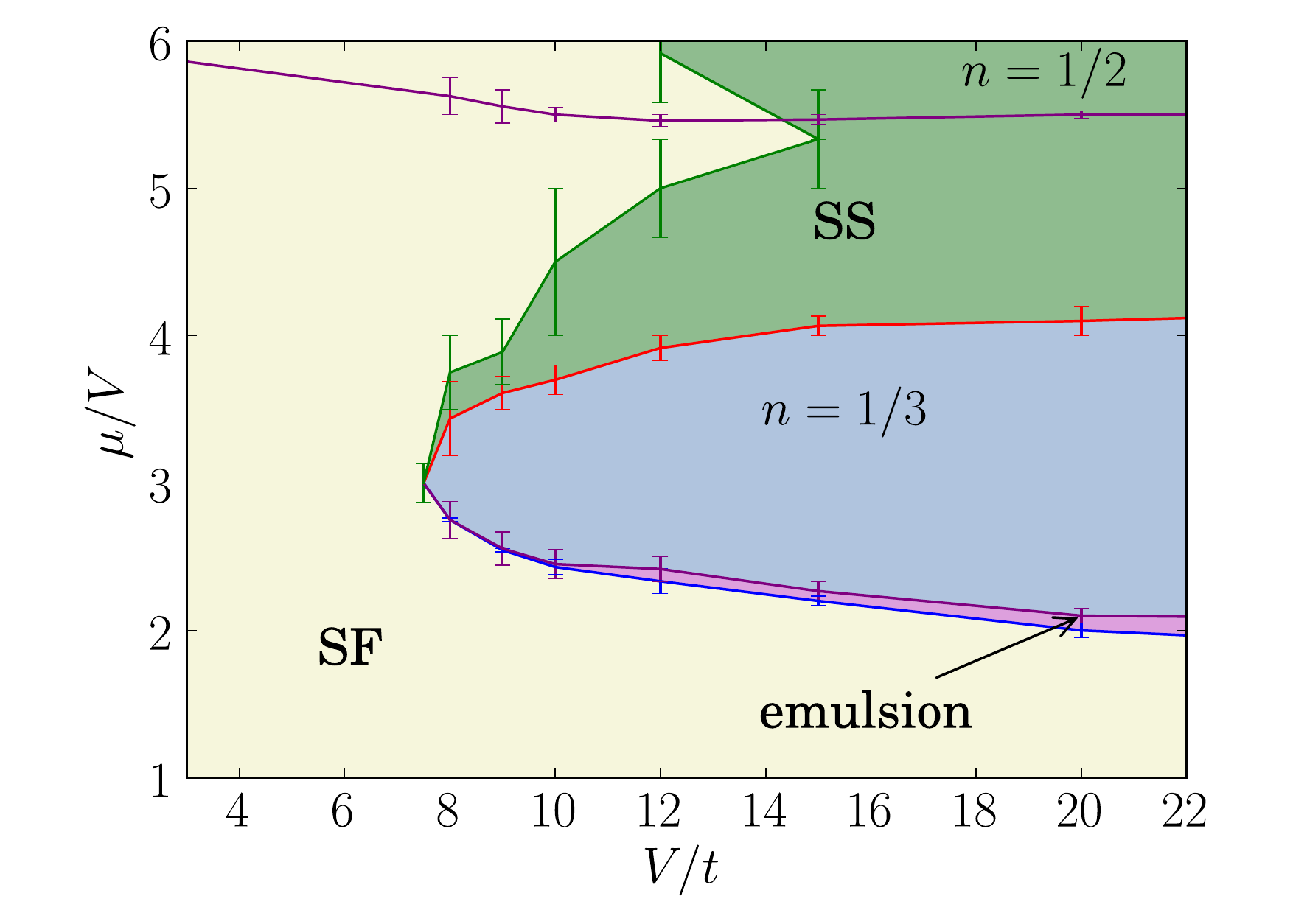}}
\caption{ Ground state phase diagram for the Hamiltionian Eq.~\eqref{H} on a
\textit{triangular} lattice in 2D, around $n=1/3$. The phases are a superfluid
`SF', supersolid `SS' and a commensurate solid at density $n=1/3$. A
transition region of the Spivak-Kivelson bubble type (emulsions) in indicated
with the double line, gradually going over to a region of incommensurate,
floating solids with increasing interaction strength. For large interaction
strength, and starting around half filling, the supersolid phase is suppressed
by emerging solid ordering (stripes at half filling and incommensurate,
floating solids at other fillings). [Adapted from Ref.~\cite{Pollet2010}]}%
\label{fig:DipolesTriangular2D}%
\end{figure}

\emph{Additional exotic phases:} Interestingly, the possibilities offered by
polarized dipoles in a 2D are not exhausted by the phases above. In
Ref.~\cite{Pollet2010} first numerical evidence has been proposed for a
\textit{bosonic} lattice version of the "microemulsion"
phase~\cite{Spivak2004} introduced in Sect.\ref{sec:secSelfAssembled} above,
for a system of polarized hard-core dipoles on a \textit{triangular} lattice.
This is shown in Fig.~\ref{fig:DipolesTriangular2D}, where an emulsion phase
is indicated as separating the Mott lobe at $n=1/3$ from a low-density
superfluid, suggesting a qualitative difference between the behavior of this
complex strongly-interacting system above and below the Mott lobe. In view of
future experiments with ultracold atomic gases with comparatively large
magnetic dipolar moments, such as Cr, Dy and Er atoms, it is an exciting
prospect to investigate how finite onsite interactions, which usually favor
supefluid and supersolid behavior, will modify this picture.\newline

\emph{Fermi gases:} A conclusive determination of the phase diagram of 2D
fermionic dipoles has been so-far hampered by the lack of exact theoretical
tools for the study of the strongly interacting regime. The experiments with
dipolar gases have motivated several theoretical works, which use different
approximate, often complementary, techniques. In analogy with Coulomb systems,
for dipolar gases polarized perpendicular to the 2D plane (e.g., with
isotropic in-plane interactions as discussed before) the existence of both
crystal-like~\cite{Mikelson2011,Sun2010,Yamaguchi2010} and quantum liquid
crystal phases~\cite{Lin2010,Lin2010b} (with interactions cut-off at NNN) has
been recently discussed. In particular, using a mean-field-theory approach
Ref.~\cite{Mikelson2011} has focused on determining the complex structure of
phases and phase-transitions of charge-density waves with different lattice
unit cells. References~\cite{Quintanilla2009,Fregoso2009} have discussed
quantum liquid crystal phases, which may be obtained for anisotropic
interactions. Analogous to the case of classical liquid
crystals~\cite{ChaikinBook}, these phases are classified as being
\textit{nematic} and \textit{smectic} according to their symmetry breaking
associated with the deformation of the Fermi surface as compared to the
isotropic case. In the nematic phase, the rotational symmetry is broken so the
typical Fermi surface has a cigar-like shape, i.e., it is stretched in one
direction and shrunk in other directions. In the smectic phase the system is
effectively in a reduced dimension, accordingly the Fermi surface is divided
into disconnected pieces. Experiments with atomic and molecular dipoles will
offer an enticing opportunity to test these predictions.

\subsubsection{Polarized dipoles on a bilayer optical lattice}

\begin{figure}[h]
\centerline{\includegraphics[ width=0.5\columnwidth]{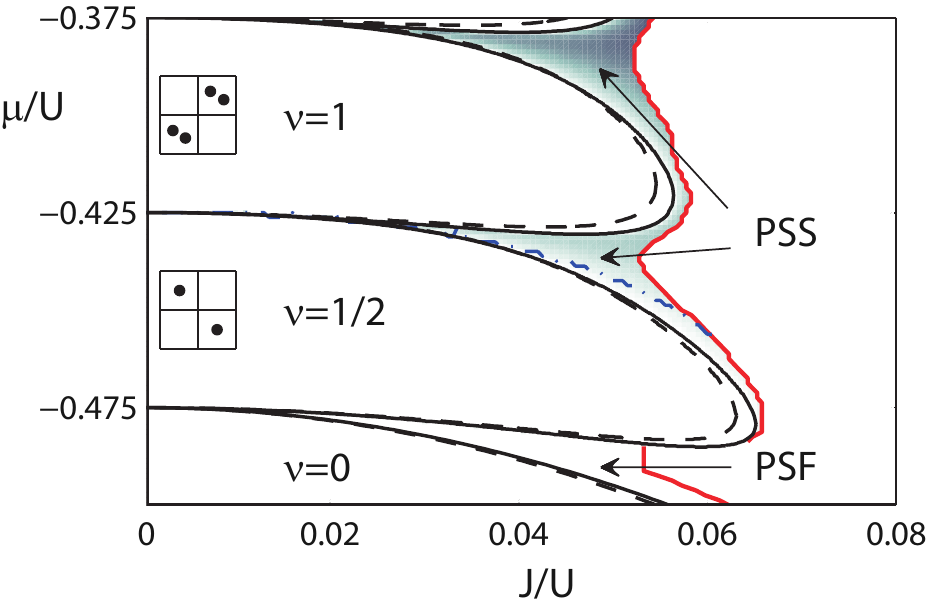}}
\caption{Phase diagram of the effective Hamiltonian of
Ref.~\cite{Trefzger2009a} describing bosonic dipoles trapped in a 2D two-layer
configuration. The model Hamiltonian has an extended Bose-Hubbard form with
$U_{\mathrm{i,i+1}}^{\sigma\sigma}= 0.025 U_{ii}^{\sigma\sigma}$,
$U_{ii}^{\sigma\sigma^{\prime}} = -0.95U_{ii}^{\sigma\sigma}$. Here, $\sigma$
and $\sigma^{\prime}$ label the two layers. The black full lines are
semi-analytic solutions indicating the boundaries of the insulating lobes for
the checkerboard ($n = 1/2$) and the doubly occupied checkerboard ($n = 1$).
The shaded area is the PSS phase predicted within a Gutzwiller mean-field
approach. PSS indicates a pair-supersolid-phase, and PSF a
pair-superfluid-phase, see text. [Adapted from Ref.~\cite{Trefzger2009a}]}%
\label{fig:figPSS}%
\end{figure}

As discussed in Sect.~\ref{s3} above, polarized dipoles trapped in a bilayer
configuration can form a paired superfluid phase (PSF), which originates from
the interlayer attraction due to dipole-dipole interactions, when tunneling
between the two layers is suppressed. The addition of an in-plane optical
lattice in the two layers makes this situation even more rich from a physical
point of view: in Ref.~\cite{Trefzger2009a} it was shown that the presence of
strong enough intra-layer repulsion and inter-layer attraction for bosonic
particles allows for the realization of a novel phase, named a pair-supersolid
phase (PSS), which is defined as a supersolid phase of composite particles.
Using a model Hamiltonian with finite on-site intra-layer interactions $U$ and
NN intra-layer interactions and within a mean-field analysis, it was shown
that the existence of the PSS phase relies on second order tunneling of
(composite) dipoles in parameter regimes close to insulating Mott phases of
(composite) dipoles, similar to the discussion of
Sect.~\ref{sec:2DLatticeMono} above. One example is given in
Fig.~\ref{fig:figPSS}. The existence of this exotic phase had been previously
discussed for anisotropic $t-J$ models~\cite{Boninsegni2008}, but no evidence
was found. It remains an open question whether exact (e.g., QMC) calculations
will confirm this prediction for realistic models with long-range interactions
and, e.g., onsite hard-core constraint. 

\subsection{Advanced Hamiltonian design with polar molecules}

The quantum phases of Sect.~\ref{sec:2DLatticeMono} are based on interactions
of the dipole-dipole type, see Eq.~\eqref{Vd}, and are thus present for any
dipolar gas. For polar molecules, the techniques described in Sect.~\ref{s2.2}
for modifying the \textit{shape} of interaction potentials allow for an
advanced engineering of microscopic Hamiltonians. This offers new
opportunities to explore exotic many-body phases in these systems. In the
following we review recent progress by discussing situations where pure
\textit{three-body} effective interactions can dominate the dynamics
(Sect.~\ref{sec:3body}), and the use of internal degrees of freedom for each
molecule in addition to rotational ones can provide toolboxes for the
simulation of condensed matter models of interest, including exotic lattice
spin models (Sect.~\ref{sec:secLattSpin}).

\subsubsection{Three-body interactions}

\label{sec:3body}

In Sect.~\ref{s2.2} techniques were introduced for the modification of the
shape of the effective interaction potentials using external DC electric and
AC microwave fields coupling rotational excitations of the molecules. In
particular, for a circularly polarized AC field the attractive time-averaged
interaction due to the rotating dipole moments of the molecules has been shown
to allow a strong reduction, and even a cancelation, of the total
dipole-dipole interaction. In a dense ensemble of molecules this can lead to
the realization of systems where the effective \emph{three-body} interaction
$W^{\mathrm{3D}}(\mathbf{r}_{i},\mathbf{r}_{j},\mathbf{r}_{k})$ of
Eq.~\eqref{eq:NbodyHamiltonian} dominates over the two-body term
$V^{\mathrm{3D}}(\mathbf{r}_{i}-\mathbf{r}_{j})$ and determines the properties
of the system in the ground state. We note that, as always, direct
particle-particle involve two particles only, while few-body interactions
emerge as an effective low-energy interaction after the high-energy degrees of
freedom are traced out.

Model Hamiltonians with strong three-body and many-body interactions are
strong candidates for exhibiting exotic ground state properties. This is
exemplified by the fractional quantum Hall states described by Pfaffian wave
functions which appear as ground states of a Hamiltonian with three-body
interactions \cite{Moore1991362,fradkin1998,cooper04}. These topological
phases support anyonic excitations with non-abelian braiding statistic. The
possibility of realizing a Hamiltonian where the two-body interaction can be
manipulated independently of the three-body term has been studied in
Ref.~\cite{Buchler:le}. There, it is shown that a stable system of particles
interacting via purely repulsive three-body potentials can be realized by
combining the setup above with a tight optical confinement provided by an
optical lattice. The latter ensures collisional stability of the setup and
defines a characteristic length scale (the lattice spacing) where an exact
cancelation of the two-body term can occur. The resulting extended
Hubbard-like Hamiltonian has the form 
\begin{eqnarray}
\label{Hubbard}H= - J \sum_{\langle i j \rangle} b^{\dag}_{i} b_{j} + \sum_{i
\neq j} \frac{U_{i j}}{2} n_{i} n_{j} + \sum_{i \neq j \neq k} \frac{W_{i j
k}}{6} n_{i} n_{j} n_{k},
\end{eqnarray}
where ${W_{i j k}}n_{i} n_{j} n_{k}$ is an offsite three-body term. The latter
is tunable independently of the two-body term $U_{i j} n_{i} n_{j}$, to the
extent that it can be made to dominate the dynamics and determine the
properties of the system in the ground state. This is in contrast to the
common approach to derive effective many-body terms from Hubbard models
involving two-body interactions, which are obtained in the $J\ll U$
perturbation limit, and are thus necessarily small~\cite{tewari06}.
\begin{figure}[t]
\centerline{\includegraphics[ width=0.75\columnwidth]{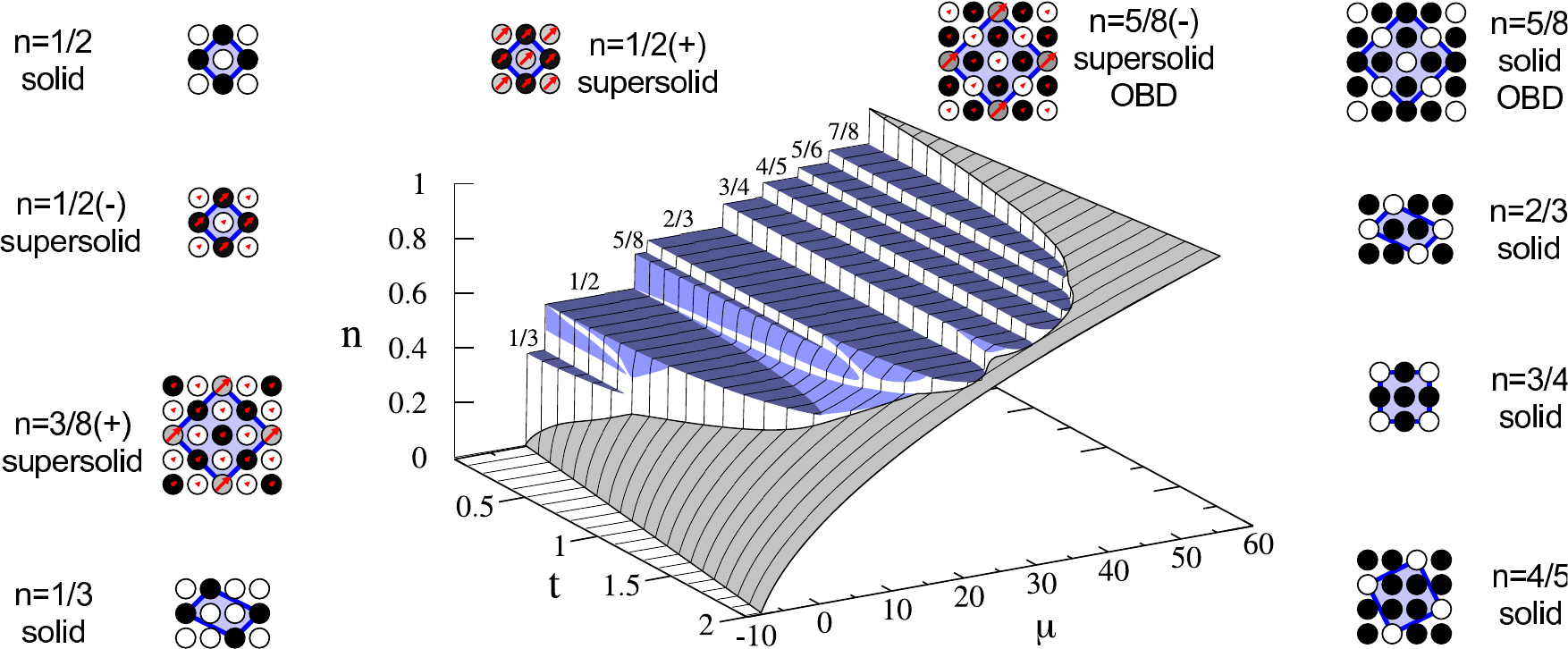}}
\caption{ Central plot: Phase diagram of the minimal model obtained within the
semiclassical approach as a function of the chemical potential $\mu$ and the
nearest-neighbor hopping amplitude $J$. Grey (white) regions are superfluid
(phase separated), and light (dark) blue denotes supersolids (solids).
Surrounding plots: schematic representation of the nature of some of the solid
and supersolid phases. The greyscale of the circles represents the filling
(white = empty, black = full), while the length and the direction of the red
arrows denotes the amplitude and the phase of the superfluid component. The
blue lines highlight the unit cell of the different structures. [Adapted from
Ref.~\cite{Schmidt2008}]}%
\label{fig:Lauchli}%
\end{figure}%

The phase diagram for bosonic particles on a 2D lattice with three-body
interactions has been recently investigated in Ref.~\cite{Schmidt2008}, where
a rich variety of superfluid, solid, supersolid and phase separated phases
have been found. In particular, several solid phases at fractional filling
factor are found to evolve upon doping into supersolid phases with complex
spatial structures. For example, the checkerboard supersolid at filling factor
$1/2$, which is unstable for hardcore bosons with nearest neighbor two-body
interaction, is found to be again stable in a wide range of tunneling
parameter, similar to the case of pure dipolar interactions discussed above.

The phase diagram for bosonic particles on a 1D lattice has been studied in
Ref.~\cite{capogrosso2009}, the corresponding phases being discussed in
Sect.~\ref{sec:1D} below.

\subsubsection{Lattice Spin models and quantum magnetism}

\label{sec:secLattSpin}

Lattice spin models are ubiquitous in condensed matter physics where they are
used to describe the characteristic behavior of complicated interacting
physical systems. Recent works have focused on realizing effective spin
Hamiltonian for the simulation of lattice spin models of fundamental interest
in condensed-matter, obtained by considering several internal states of each molecule.

\begin{figure}[ptb]
\includegraphics[width=0.5\columnwidth]{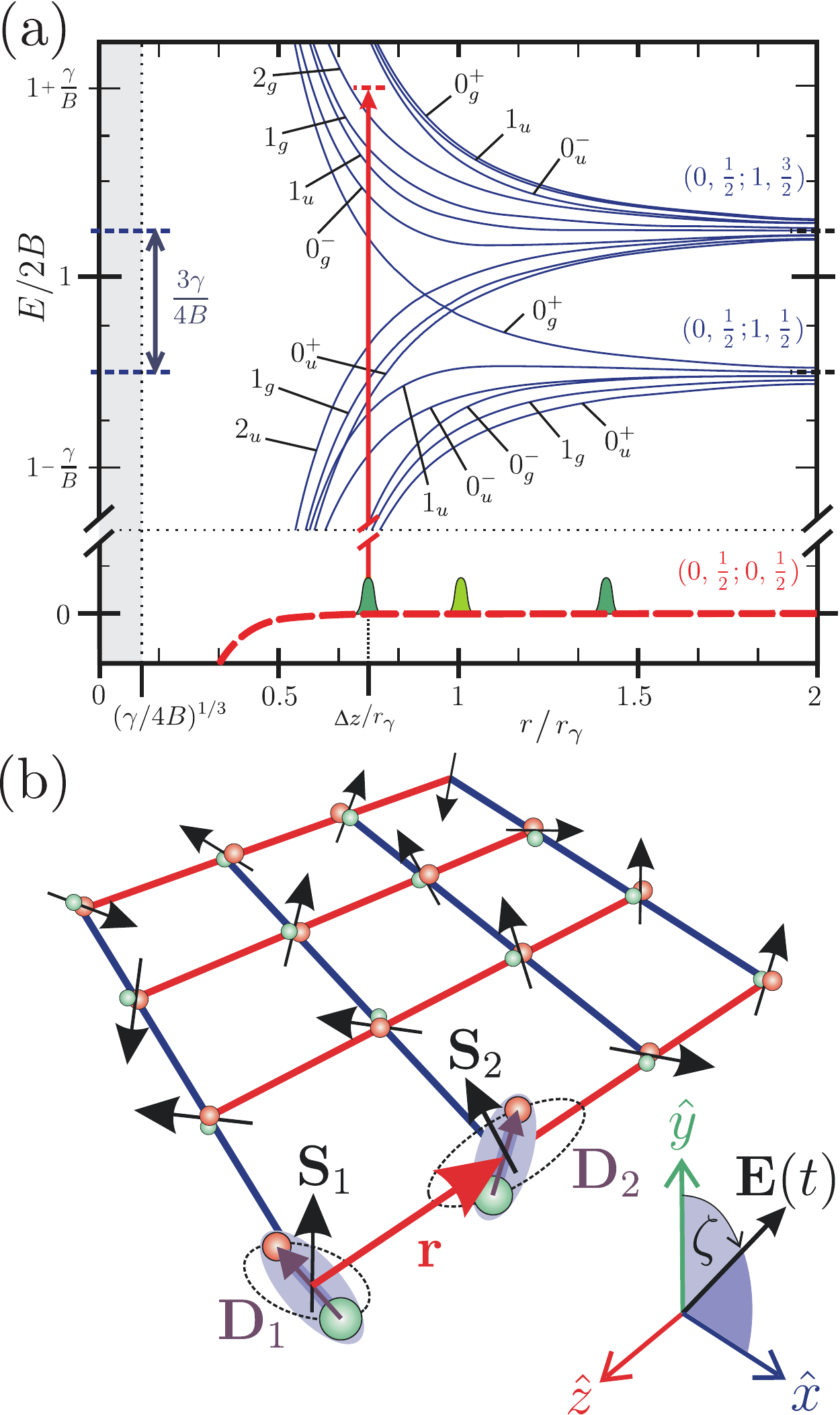}
\caption{(a) Potentials for a pair of molecules as a function of their
separation $r$. The symmetries $|Y|^{\pm}_{\sigma}$ of the excited manifolds
are indicated, as are the asymptotic manifolds $(N_{i},J_{i};N_{j},J_{j})$.
Here the quantum numbers are: $Y=M_{N}+M_{S}$ where $M_{N}=M_{N_{1}}+M_{N_{2}%
}$ and $M_{S}=M_{S_{1}}+M_{S_{2}}$ are the total rotational and spin
projections along the intermolecular axis; parity eigenvalues $\sigma=\pm1$
denoted as $g(u)$ for \textit{g}erade(\textit{u}ngerade); reflection symmetry
$R=\pm$ of all electronic and rotational coordinates through a plane
containing the intermolecular axis. (b) Example anisotropic spin models that
can be simulated with polar molecules trapped in optical lattices: Square
lattice in $2$D with nearest neighbor orientation dependent Ising interactions
along $\hat{x}$ and $\hat{z}$. Effective interactions between the spins
$\mathbf{S}_{1}$ and $\mathbf{S}_{2}$ of the molecules in their rovibrational
ground states are generated with a microwave field $\mathbf{E}(t)$ inducing
dipole-dipole interactions between the molecules with dipole moments
$\mathbf{D}_{1}$ and $\mathbf{D}_{2}$, respectively. [Adapted from
Ref.~\cite{Micheli:2006fh}] }%
\label{fig:1}%
\end{figure}

Reference~\cite{Micheli:2006fh} has shown that cold gases of polar molecules
can be used to construct in a natural way a \emph{complete toolbox} for any
permutation symmetric two spin-$1/2$ (qubit) interaction, based on techniques
of interaction engineering which are extensions of those discussed in
Sect.~\ref{s2.2}. The main ingredient of this (and related) proposal is the
dipole-dipole interaction: it couples strongly the rotational motion of the
molecules, it can be designed by means of microwave fields, and it can be made
spin-dependent, exploiting the spin-rotation splitting of the molecular
rotational levels. \newline


The basic building block in Ref.~\cite{Micheli:2006fh} is a system of two
polar molecules strongly trapped at given sites of an optical lattice, where
the spin-$1/2$ (or qubit) is represented by a single unpaired electron outside
a closed shell of a $^{2} \Sigma_{1/2}$ heteronuclear molecule in its
rotational ground state, as provided e.g. by alkaline-earth monohalogenides.
Rotational excitations are described by the Hamiltonian
\begin{eqnarray}
H_{\mathrm{m}} =B \mathbf{N^{2} + \gamma N\cdot S,}%
\end{eqnarray}
with $\mathbf{N}$ the rotational angular momentum of the nuclei, and
$\mathbf{S}$ the dimensionless electronic spin (assumed to be $S=1/2$ in the
following). Here $B$ denotes the rotational constant and $\gamma$ is the
spin-rotation coupling constant. The typical values of $B$ are a few tens of
GHz, and $\gamma$ is usually in the hundred MHz regime. The interaction
describing the internal degrees of freedom is $H_{\mathrm{in}}=V_{\mathrm{dd}%
}+\sum_{i=1}^{2} H_{\mathrm{m}}^{i}$, where $H_{\mathrm{dd}}$ is the
dipole-dipole interaction.

The molecules are assumed to be trapped in the optical lattice with a
separation $\Delta z\sim r_{\gamma}\equiv(2d^{2}/\gamma)^{1/3}$, where the
dipole dipole interaction is $d^{2}/r_{\gamma}^{3}=\gamma/2$. In this regime
the rotation of the molecules is strongly coupled to the spin and the excited
states are described by Hunds case (c) states in analogy to the dipole-dipole
coupled excited electronic states of two atoms with fine-structure. Thus,
while the two-particle ground states are essentially spin independent,
effective spin-dependent interactions in the groundstates can be obtained by
dynamically mixing dipole-dipole coupled excited states into the groundstates
using a microwave field $\mathbf{E}(\mathbf{x},t)$ with properly-chosen
polarization, frequency $\omega_{F}$ and Rabi-frequency $\Omega$, which is
tuned near resonance with the $N=0\rightarrow N=1$ transition.

The effective interactions in the lowest-energy states are obtained by
diagonalizing the $H_{\mathrm{BO}}=H_{\mathrm{m}} -\sum_{i=1,2}\mathbf{d}_{i}
\mathbf{E}$ potential, as described in Sect.~\ref{s2.2}. In second order
perturbation theory they read
\begin{eqnarray}
V_{\mathrm{eff}}^{\mathrm{3D}}(r)=\sum_{i,f}\sum_{\lambda(r)}\frac
{\langle{g_{f}}|H_{\mathrm{mf}}|{\lambda(r)}\rangle\langle{\lambda
(r)}|H_{\mathrm{mf}}|{g_{i}}\rangle}{\hbar\omega_{F}-E(\lambda(r))}|{g_{f}%
}\rangle\langle{g_{i}}|,\label{effHam1}%
\end{eqnarray}
where $\{{|{g_{i}}\rangle,|{g_{f}}\rangle}\}$ are ground states for two
molecules with $N_{1}=N_{2}=0$ and $\{|{\lambda(r)}\rangle\}$ are excited
eigenstates of $H_{\mathrm{in}}$ with $N_{1}+N_{2}=1$ and with excitation
energies $\{E(\lambda(r))\}$. The reduced interaction in the subspace of the
spin degrees of freedom is then obtained by tracing over the motional degrees
of freedom. For molecules trapped in the ground motional states of isotropic
harmonic wells with rms width $z_{0}$ the wave function is separable in center
of mass and relative coordinates, and the effective spin-spin Hamiltonian is
$H_{\mathrm{spin}}=\langle H_{\mathrm{eff}}(r)\rangle_{\mathrm{rel}}$.

The effective Hamiltonian Eq.~\eqref{effHam1} can in general be rewritten
as~\cite{Micheli:2006fh}
\begin{eqnarray}
H_{\mathrm{eff}}(r)=\frac{\hbar|\Omega|}{8}\sum_{\alpha,\beta=0}^{3}
\sigma^{\alpha}_{1}A_{\alpha,\beta}(r)\sigma^{\beta}_{2},\label{effHam2}%
\end{eqnarray}
where $\{\sigma^{\alpha}\}_{\alpha=0}^{3}\equiv\{\mathbf{ 1},\sigma^{x}%
,\sigma^{y},\sigma^{z}\}$ and $A$ is a real symmetric tensor.
Equation~\eqref{effHam2} describes a generic permutation symmetric two qubit
Hamiltonian. The components $A_{0,s}$ describe a pseudo magnetic field which
acts locally on each spin and the components $A_{s,t}$ describe two qubit coupling.

For a given field polarization, tuning the frequency near an excited state
induces a particular spin pattern on the ground states. These patterns change
as the frequency is tuned though multiple resonances at a fixed intermolecular
separation. For example, the anisotropic spin model $H_{XYZ}=\lambda_{x}
\sigma^{x}\sigma^{x}+\lambda_{y} \sigma^{y}\sigma^{y}+\lambda_{z}\sigma
^{z}\sigma^{z}$ can be simulated using three fields: one polarized along
$\hat{z}$ tuned to $0_{u}^{+}(3/2)$, one polarized along $\hat{y}$ tuned to
$0_{g}^{-}(3/2)$ and one polarized along $\hat{y}$ tuned to $0_{g}^{+}(1/2)$.
The strengths $\lambda_{j}$ can be tuned by adjusting the Rabi frequencies and
detunings of the three fields.

Of particular interest is the possibility of realizing highly anisotropic spin
models such as the following one
\begin{eqnarray}
H_{\mathrm{spin}}^{(\mathrm{I})}= \sum_{i=1}^{\ell-1}\sum_{j=1}^{\ell-1} J
(\sigma^{z}_{i,j}\sigma^{z}_{i,j+1}+\cos\zeta\sigma^{x}_{i,j}\sigma
^{x}_{i+1,j}),\label{Ioffe}%
\end{eqnarray}
which was first introduced by Dou\c{c}ot \textit{et al.} \cite{Duocot:05} in
the context of Josephson junction arrays. This model (for $\zeta\neq\pm\pi/2$)
admits a 2- fold degenerate ground subspace that is immune to local noise up
to $\ell$-th order and hence is a good candidate for storing a topologically
protected qubit for applications in quantum computing. Other Hamiltonians that
can be realized include the famous Kitaev model.\newline

References~\cite{Wall2010,Gorshkov2011a,Gorshkov2011b} have recently focused
on the realization of tunable Heisenberg-type models for unit filling of the
optical lattice. Effective spin degrees of freedom are encoded in internal,
e.g. rotational, states of the molecules, and molecular interactions are tuned
by a combination of DC and AC fields, using techniques similar to those of
Sect.~\ref{s2.2}. In Refs.~\cite{Gorshkov2011a,Gorshkov2011b} it is found that
doping the lattice with vacancies (that is, removing particles from a few
sites) leads to a tunable generalization of the so-called $t-J$ model, which
is relevant in the context of high-temperature
superconductivity~\cite{Ogata2008}. Typical Hamiltonian terms which are found
in addition to Eq.~\eqref{Hubbard10} are of the form
\begin{eqnarray}
H_{1}  & = &\frac{1}{2} \sum_{i \neq j} \left[  J_{z} S^{z}_{i} S^{z}_{j}+
\frac{J_{\perp}}{2} (S^{+}_{i} S^{-}_{j} + S^{-}_{i} S^{+}_{j}) \right.\\
&  +&  \left. V n_{i} n_{j} + W n_{i} S^{z}_{j} \right] /(\mathbf{r}_{i} -
\mathbf{r}_{j})^{3}, \label{eq:Magnetism}
\end{eqnarray}
where the operators $S_{j}^{z} = (n_{j, \sigma}-n_{j \sigma^{\prime}})/2$,
$S_{j}^{+} = c^{\dagger}_{j \sigma} c_{j \sigma^{\prime}}$, and $S_{j}^{-} =
(S_{j}^{+})^{\dagger}$ are spin-1/2 angular momentum operators on site $j$
describing a two-level dressed rotor degree of freedom $\sigma$ (e.g., the
lowest two $N_{z} = 0$ states of the molecule in the presence of a DC electric
field along $\hat{\mathbf{z}}$), and satisfying $[S_{j}^{z},S_{j}^{\pm}] = \pm
S_{j}^{\pm}$. By intuitively thinking of the ground state as a (classical)
dipole $\boldsymbol{\mu}_{0} = \mu_{0} \hat{\mathbf{z}}$ oriented along the DC
field (i.e.\ $\mu_{0} > 0$) and of the excited state as a dipole
$\boldsymbol{\mu}_{1} = \mu_{1} \hat{\mathbf{z}}$ oriented against the DC
electric field (i.e.\ $\mu_{1} < 0$), the terms $J_{z} = (\mu_{0} - \mu
_{1})^{2}$, $V = (\mu_{0} + \mu_{1})^{2}/4$, and $W = (\mu_{0}^{2} - \mu
_{1}^{2})/2$ derive from re-writing the direct dipole-dipole interaction,
while the term $J_{\perp}= 2\mu_{01}^{2}$ comes from the transition dipole
moment $\mu_{01}$, describing the exchange of a microwave excitation between
the molecules. One difference with the derivation of the effective spin models
of Ref.~\cite{Micheli:2006fh} described above is that dipole-dipole
interactions in Eq.~\eqref{eq:Magnetism} are used in first order, rather than
second order, which can allow for stronger interactions.

We note that the design of extended Hamiltonians with terms as in
Eq.~\eqref{eq:Magnetism} is in fact a hot topic of research in atomic and
molecular systems: for example, Refs.~\cite{Yu2009b,Schachenmayer2010} have
discussed the derivation of similar terms for trapped neutral atoms and
molecules at unit filling, while Refs.~\cite{Hauke2010,Porras2004} have
focused on ionic particles. The $J_{z} = 0$ case is also studied in the
context of molecules in Ref.~\cite{Barnett2006}. 
In Ref.~\cite{Herrera2010}, a similar Hamiltonian describes the dynamics 
of exciton-like interactions with impurities in gases of polar molecules.
There, the authors considered first a clean system with one particle per
site, subject to a very deep optical lattice potential in order to prevent
particle tunneling: in this regime, dipole-dipole interactions induce
collective excitonic modes, in sharp contrast with individual rotational
excitations of short-range interacting atoms. After substituting a small
number of initial components with a different kind of molecule still strongly
confined by the optical potential, the system displays an exciton-impurity
interactions between the original excitonic modes and the new particles,
which may be treated as effective impurities, leading to a disordered background
potential with both diagonal and off-diagonal contributions. Such a setup 
represents an ideal platform to investigate the effect of disorder and interaction
induced localization in excitonic gases in both 1, 2 and 3D.

Reference~\cite{Perez-Rios2010} has considered an ensemble of $^{2}\Sigma$
molecules in the rotationally ground state trapped on an optical lattice and
have shown that collective spin excitations can be controlled using external
electric and magnetic fields, in the context of the formation of Frenkel
excitons~\cite{Agranovich,Zoubi2005}. This system may be used for the quantum
simulation of spin excitation transfer in many-body crystals without phonons.
In Ref.~\cite{Rabl2007} a similar Hamiltonian has been studied in the context
of molecular self-assembled crystals for quantum memory applications. \newline


The inclusion of more internal states for the particles, such as hyperfine
levels, offers extensions to spin systems with larger
spin~\cite{Brennen2007,Wall2010,Gorshkov2011a,Gorshkov2011b}. For example, the
design of a large class of spin-1 interactions for polar molecules has been
shown in Ref.~\cite{Brennen2007}, which allows e.g. for the realization of a
generalized Haldane model in 1D, using a strategy of interaction-engineering
similar to those described above.
Recent work in Ref.~\cite{Wall2010} has started exploring the
non-equilibrium dynamics of molecules trapped in optical lattices in
the presence of several internal states, including, e.g., the rich hyperfine
structure often present in the ground state rotational manifolds.
In particular, it was shown that the number of effective degrees of freedom participating to the dynamics  may be indeed dynamically changed for each molecule using external electric and
magnetic fields, similar to the discussion above. This raises interesting prospects for the study of non-equilibrium dynamics in these complex systems.
In section Sec.~\ref{sec:1D}. below, we discuss some of the intriguing scenarios opened up in the many-body context by the accurate fine-tuning of interactions presented here, for the particular case of polar molecules trapped in 1D.

\subsubsection{Hubbard models in self-assembled dipolar lattices}

\label{sec:secFloat}

\begin{figure}[tbh]
\includegraphics[width=0.5\columnwidth]{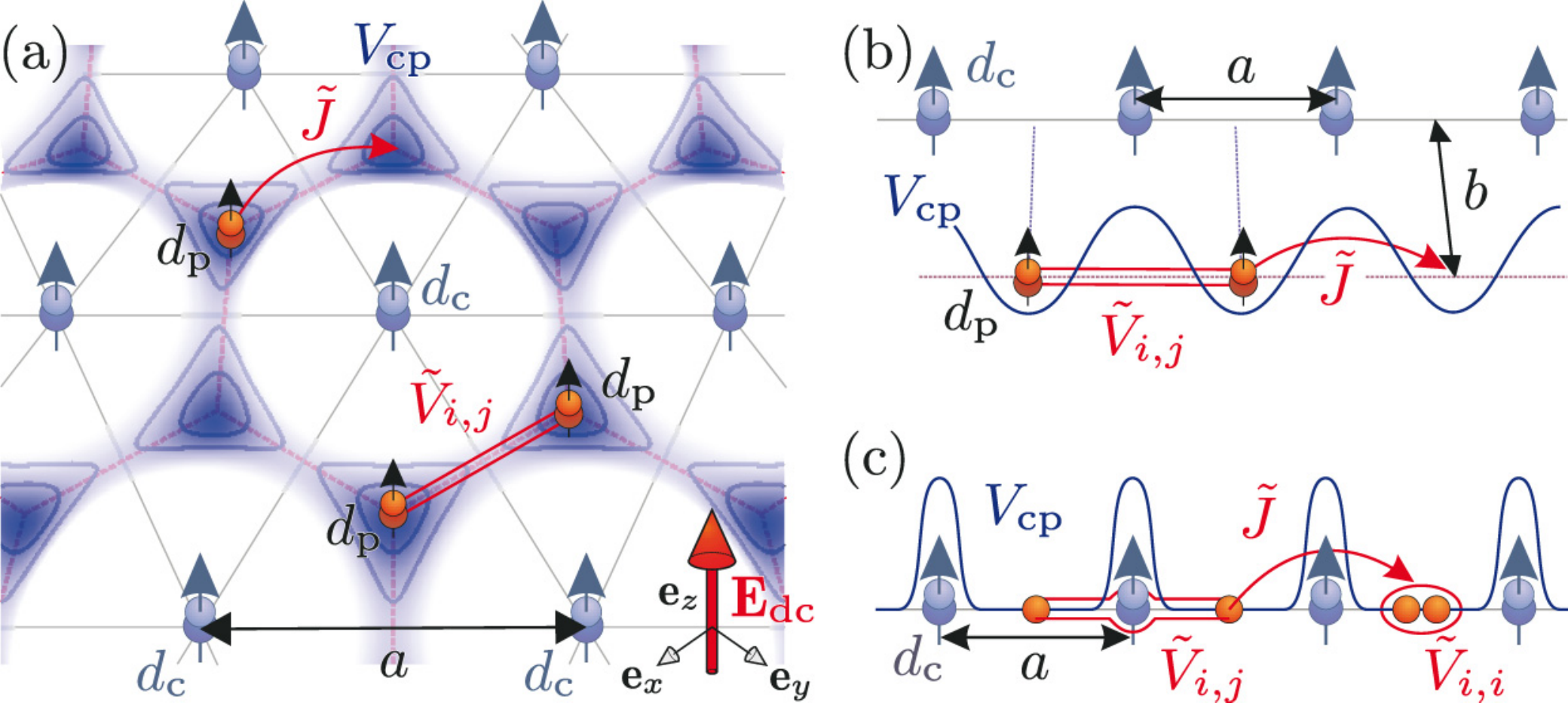}
\caption{Floating lattices of dipoles: A self-assembled crystal of polar
molecules with dipole moment $d_{\mathrm{c}}$ provides a 2D periodic honeycomb
lattice $V_{\mathrm{cp}}$ (darker shading corresponds to deeper potentials)
for extra molecules with dipole $d_{\mathrm{p}} \ll d_{\mathrm{c}}$ giving
rise to a lattice model with hopping $\tilde{J}$ and long-range interactions
$\tilde{V}_{i,j}$. }%
\label{figLattPhonons}%
\end{figure}

In Hubbard models with cold atoms or molecules in optical lattices there are
no phonon degrees of freedom corresponding to an intrinsic dynamics of the
lattice, since the back action on the optical potentials is typically
negligible. Thus, the simulation of polaronic materials~\cite{Alexandrov}
where the presence of crystal phonons affects strongly the Hubbard-like
dynamics of particles remains largely a challenge. References~
\cite{Pupillo08,Ortner09} have shown that a self-assembled floating lattice of
molecules as discussed in Sect.~\ref{sec:secSelfAssembled} can provide a
periodic trapping potential for extra atoms or molecules, whose dynamics can
be described in terms of a Hubbard model with phonons, the latter
corresponding to vibrations of the dipolar lattice.

The Hamiltonian for extra atoms or molecules in a self-assembled dipolar
lattice is
\begin{eqnarray}
H    =  -J\sum_{<i,j>}c_{i}^{\dagger}c_{j} +\tfrac{1}{2}\sum_{i,j}V_{ij}%
c_{i}^{\dagger}c_{j}^{\dagger}c_{j}c_{i}
  +  \sum_{q} \hbar\omega_{q} a_{q}^{\dagger}a_{q} +\sum_{q,j}M_{
q}e^{i\mathbf{q}\cdot\mathbf{R}_{j}^{0}}c_{j}^{\dagger}c_{j}(a_{q}
+a_{-q}^{\dagger}).\label{eq:eqSmallPolaron}%
\end{eqnarray}
Here, the first and second terms define a Hubbard-like Hamiltonian for the
extra-particles of the form of Eq.~\eqref{Hubbard10}, where the operators
$c_{i}$ ($c_{i}^{\dagger}$) are destruction (creation) operators of the
extra-particles. However, the third and fourth terms describe the acoustic
phonons of the crystal [$a_{q}$ destroys a phonon of quansimomentum
$\mathbf{q}$ in the mode $\lambda$] and the coupling of the extra-particles to
the crystal phonons, respectively. Tracing over these phonon degrees of
freedom in a strong coupling limit provides effective Hubbard models for the
extra-particles dressed by the crystal phonons
\begin{eqnarray}
\tilde{H}=-\tilde{J}\sum_{ <i,j>}c_{i}^{\dagger}c_{j}+\tfrac{1}{2} \sum
_{i,j}\tilde{V}_{ij}c_{i}^{\dagger}c_{j}^{\dagger}c_{j}c_{i}.\label{eq:Heff}%
\end{eqnarray}
The hopping of a dressed extra-particle between the minima of the periodic
potential occurs at a rate $\tilde{J}$, which is exponentially suppressed due
to the co-propagation of the lattice distortion, while offsite
particle-particle interactions $\tilde{V}_{i,j}$ are now a combination of
direct particle-particle interactions and interactions mediated by the
coupling to phonons, as given by polaronic dynamics~\cite{Mahan}.
Figure~\ref{figLattPhonons} shows two implementations of this idea in two- and
quasi-one-dimensional configurations, [panels (a) and (b-c), respectively].
The distinguishing features of this realization of lattice models are the
tunability of interactions among crystal dipoles, and of the particle-phonon
couplings. In addition, the lattice spacings are tunable with external control
fields, ranging from a $\mu$m down to the hundred nm regime, i.e. potentially
smaller than for optical lattices. Compared with optical lattices, for
example, a small scale lattice yields significantly enhanced hopping
amplitudes, which set the relevant energy scale for the Hubbard model, and
thus also the temperature requirements for realizing strongly correlated
quantum phases. Phonon-mediated interactions can be quite long-ranged,
decaying with distance as $1/r^{2}$ in 1D.\newline

\section{Dipolar gases in one- and quasi-one dimensional geometries}
\label{sec:1D}
In reduced dimensionality, the effects of quantum fluctuations are so relevant
that the standard Fermi-liquid picture breaks down due to the emergence of
several strongly correlated states of matter. The addition of interaction with
long-range, anisotropic tail such as the dipolar one leads to the
stabilization of very rich phase diagram in both purely 1D systems and coupled
ones such as ladders and planar arrays. In the following, we will first review
the basic physics of the single tube configuration and then illustrate recent
results on two-leg and multi-leg ladders.


\subsection{Dipolar gases in a single tube}

We now consider a system of dipolar particles confined in a one dimensional
(1D) geometry by a sufficiently deep optical lattice with frequency
$\omega_{\perp}$, so that their dynamics is purely 1D. In order to ensure
collisional stability, we will consider all dipole moments aligned
perpendicularly to the wire direction by an external field in such a way that
the inter-particle interaction is always repulsive. Defining $m$ the mass of
the particles and $C_{3}$ the strength of the dipolar interaction, a proper
description for, e.g., a gas of polar molecules in the long-distance regime
(see Sec.~\ref{s2}) is encoded in the following model Hamiltonian:
\begin{eqnarray}
\label{ham_1D}H   = \int dx \: \psi^{\dagger}(x)\left[ -\frac{\hbar^{2} }%
{2m}\partial_{x}^{2}\right] \psi(x)
 + \frac{1}{8 \pi} \int dx \:dx^{\prime}\psi^{\dagger}(x)\psi^{\dagger
}(x^{\prime})\frac{C_{3}}{|x-x^{\prime}|^{3}}\psi(x^{\prime})\psi
(x)\label{eq:eqHam}%
\end{eqnarray}
where $\psi(x),\psi^{\dagger}(x)$ are annihilation/creation operators with
bosonic or fermionic statistics. For a polar molecule gas, $C_{3}%
=d^{2}/\epsilon_{0}$, where $\epsilon_{0}$ is the vacuum permeability and $d$
the dipole moment induced by a DC electric field. After a proper rescaling of
all quantities, one can see that the only relevant parameter is the ration
between the dipolar length $R_{3}^{-1}=2\pi\hbar^{2}/(mC_{3})$ and the linear
density $\rho_{0}$ and, as such, we expect all thermodynamic properties to
depend only on $\rho_{0} R_{3}$. While statistical properties are expected not
to play a major role in this setup since bosons are subject to an effective
hard-core condition due to the infinite short-distance repulsion, some
properties of the system are still affected by statistics; thus, we will
at first consider a bosonic gas, and then comment on the fermionic case at the
end of the section. Qualitative features of Eq.~\eqref{ham_1D} are obtained
using the bosonization technique, which allows to map the original interacting
problem into a free one via the
identities~\cite{haldane1981,giamarchi_book,gogolin_book,giamarchi_review}:
\begin{eqnarray}
\psi^{\dagger}_{B/F}(x)   \simeq\sqrt{\rho_{0}+\frac{\partial_{x}%
\vartheta(x)}{\pi}}e^{i\varphi(x)}\times
\sum_{m}^{odd/even}\exp[im(\vartheta(x)+\pi\rho_{0}x)]
\end{eqnarray}
\begin{eqnarray}
 n(x)= \left( \rho_{0}+\frac{\partial_{x}\vartheta(x)}{\pi}\right) \left[ \sum_{m=-\infty}^{\infty}\exp[i2m(\vartheta(x)+\pi\rho_{0} x)]\right]
\end{eqnarray}
Here, the field $\varphi$ and $\vartheta$ represent density and phase
fluctuations of the original field, and satisfy bosonic commutation relations
$[\partial_{x}\vartheta(x),\varphi(y)]=i\pi\delta(x-y)$. As long as the
long-range tail of the interaction potential decays faster than Coulomb-like
interactions\cite{tsukamoto2000,inoue2006}, $1/|x-x^{\prime}|$, as in the
dipolar case, the Hamiltonian in Eq.~\eqref{ham_1D} is mapped into the
so-called Tomonaga-Luttinger liquid (TLL)\cite{haldane1981}:
\begin{eqnarray}
\label{wire_TLL}H=\frac{\hbar v}{2\pi}\int dx\left[ (\partial_{x}%
\vartheta(x))^{2}/K+K(\partial_{x}\varphi(x))^{2}\right]
\end{eqnarray}
which is a purely quadratic Hamiltonian with a linear dispersion relation at
small momenta, $\omega(k)\simeq vk$; the excitations in the system are sound
waves of the density with correspondent sound velocity $v$, whereas the factor
$K$, known as Tomonaga-Luttinger liquid parameter, is related to the
compressibility of the system $\mathcal{C}$ via $K=v\pi\mathcal{C}$, and in
Galilean invariant systems satisfies $K=v_{F}/v, v_{F}=\hbar\rho_{0}\pi/m$.
The TLL picture captures the entire low-energy physics of the original model,
and represents the substitute of the Landau-Fermi liquid scenario in 1D;
analogously, the parameters $v,K$ play a similar role to the Landau parameters
in higher dimensions. In addition, the long-distance decay of correlations
functions is completely captured by $K$, so that after the bosonization
mapping one gets:
\begin{eqnarray}
\label{sf_corr}B(x)=\langle\psi^{\dagger}(x)\psi(y)\rangle\simeq\frac
{1}{|x-y|^{1/2K}}%
\end{eqnarray}
for the one-body density matrix, also known as \textit{superfluid} (SF)
correlation, and
\begin{eqnarray}
\label{cdw_corr}D(x)= \langle n(x)n(y)\rangle- \rho_{0}^{2} \simeq\frac
{1}{|x-y|^{2}}+ \frac{\cos(2\pi(x-y)\rho_{0})}{|x-y|^{2K}}%
\end{eqnarray}
for the correlated part of density-density, or \textit{charge-density-wave}
(CDW), correlation function. The possibility of qualitatively describing the
asymptotic decay of correlations is remarkable. Since spontaneous symmetry
breaking is not allowed in many 1D models because of the
Mermin-Wagner-Hohenberg theorem~\cite{mermin1966,hohenberg1967}, correlation
functions encode the necessary information to distinguish between different
phases in low dimensional geometries; the slowest decaying correlation is
usually referred to as \textit{dominant order}, corresponding to the more
diverging susceptibility in the system~\cite{gogolin_book,giamarchi_book}.
From Eq.~\eqref{sf_corr} and~\eqref{cdw_corr}, we thus expect a transition
from dominant SF correlations to CDW ones at a precise value of the TLL
parameter, $K=1/2$, where the two decay exponents coincide.

While in principle all information regarding the quantum phases in 1D is
encoded in the LL parameters, the bosonization techniques sketched above do
not in allow in general to establish a relation between the microscopic
quantities in Eq.~\eqref{ham_1D} and $K$, as the latter has to be regarded as a
phenomenological parameter to be determined by comparisons with experiments or
numerical results. Before discussing the general case, let us comment on some
relevant situations. In the weakly interacting limit, $\rho_{0}R_{3}\ll1$, the
dipolar interaction is relevant only at very short distances, and thus the
system behaves very similarly to a Tonks-Girardeau
gas~\cite{tonks1936,girardeau1960}, for which $K_{TG}=1$; as such, we expect
$K^{(wc)}\simeq1$ in the weakly interacting regime. In the strongly
interacting case, $\rho_{0}R_{3}\gg1$, it has been noticed in
Ref.~\cite{citro2007,citro2008} that, approximating the system as a
classical crystal where particles are quasi-localized at equally spaced
distances, one has an energy per particle $e_{0}=\zeta(3)[\rho_{0}
R_{3}/(2m)]$ (where $\zeta$ is the Riemann zeta function), and deriving from
it the compressibility $\mathcal{C}$, the resulting TLL parameter in the
strong coupling regime reads:
\begin{eqnarray}
\label{wire_K_citro}K(\rho_{0}R_{3}\gg1)=\left( 0.73\rho_{0}R_{3} \right)
^{-1/2}%
\end{eqnarray}
Away from these two limits, the mapping of K into the microscopic parameters
of  Eq.~\eqref{ham_1D} has to be performed using non-perturbative techniques.
Numerical QMC simulations based on both reptation, diffusion and Worm
algorithm have been performed~\cite{citro2007,arkhipov2005,roscilde2010},
allowing to estimate the TLL parameter from the so called static structure
factor
\begin{eqnarray}
S(k)=\int_{0}^{L} dx\; e^{-ikx}\langle n(x)n(0)\rangle
\end{eqnarray}
and from the long-distance decay of $B(x)$ (see Fig.~\ref{fig:TLL_K}).
Moreover, QMC also allows to quantitatively estimate the momentum
distribution
\begin{eqnarray}
n(k)=\frac{1}{L}\int_{0}^{L} dx\;B(x)e^{ikx}%
\end{eqnarray}
which is directly measurable in experiments, and, according to TLL theory,
displays a clear signature of the SF/CDW transition as the typical peak of
quasi-condensation at $k=0$ disappears just across the phase boundary between
the SF and CDW phases. Typical QMC results for $n(k)$ are shown in
Fig.~\ref{fig:n_k_QMC} just across the phase boundary, as can be argued from
the sharp decrease of $n(0)$ when the interaction is increased. Moreover,
analytical results obtained through approximate methods lead to the following
expression~\cite{dalmonte2010}:
\begin{eqnarray}
\label{wire_K}K=\left( 1+ \frac{6\zeta(3)}{\pi^{2}}\rho
_{0}R_{3} \right) ^{-1/2}%
\end{eqnarray}
which compares favorably with numerical simulations, as can be seen in
Fig.~\ref{fig:TLL_K}, and recovers the limit in Eq.~\eqref{wire_K_citro} at
strong coupling.

\begin{figure}[t]
{\ }
\par
\begin{center}
\includegraphics[width=0.5\columnwidth]{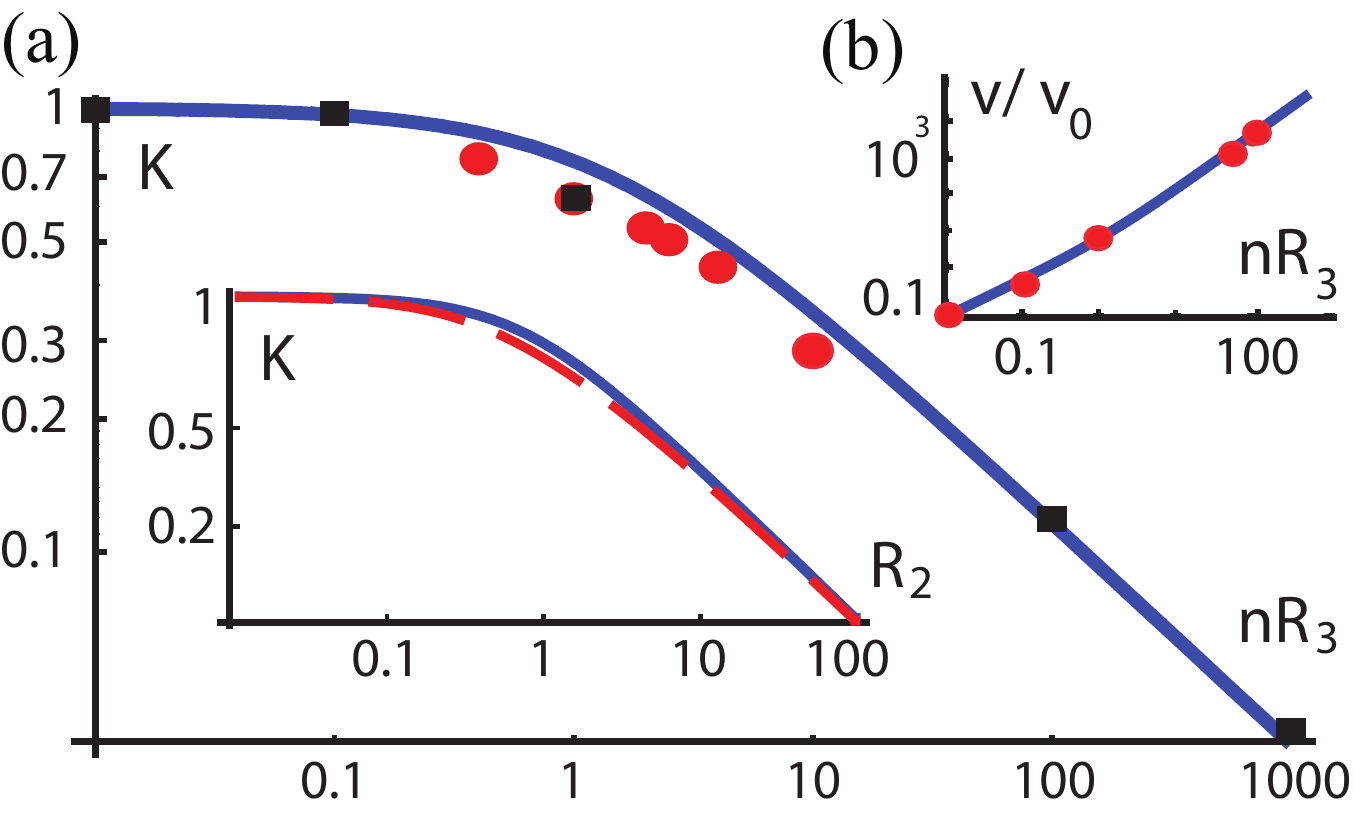}
\end{center}
\caption{Panel \textit{(a)}: TLL parameter $K$ in a dipolar wire as a function
of the dimensionless interaction strength $\rho_{0} R_{3}$ (here, $\rho
_{0}\equiv n$). The blue line is Eq.~\eqref{wire_K}, while red points and
black squares are QMC results from Ref. \cite{roscilde2010} and
\cite{citro2007} respectively The inset shows a similar prediction with
inverse-square interaction (blue line) compared with the exact result (red
dashed line). Panel \textit{(b)}: sound velocity to Fermi velocity ratio
$v/v_{F}$ ($v_{0}\equiv v_{F}$) as a function of $\rho_{0} R_{3}$; blue line
and red points are analytical and QMC results from
Ref.~\cite{dalmonte2010} and \cite{citro2007} respectively. Image
taken from Ref.~ \cite{dalmonte2010}.}%
\label{fig:TLL_K}%
\end{figure}\begin{figure}[tb]
\center{{\includegraphics[width= 0.5 \columnwidth]{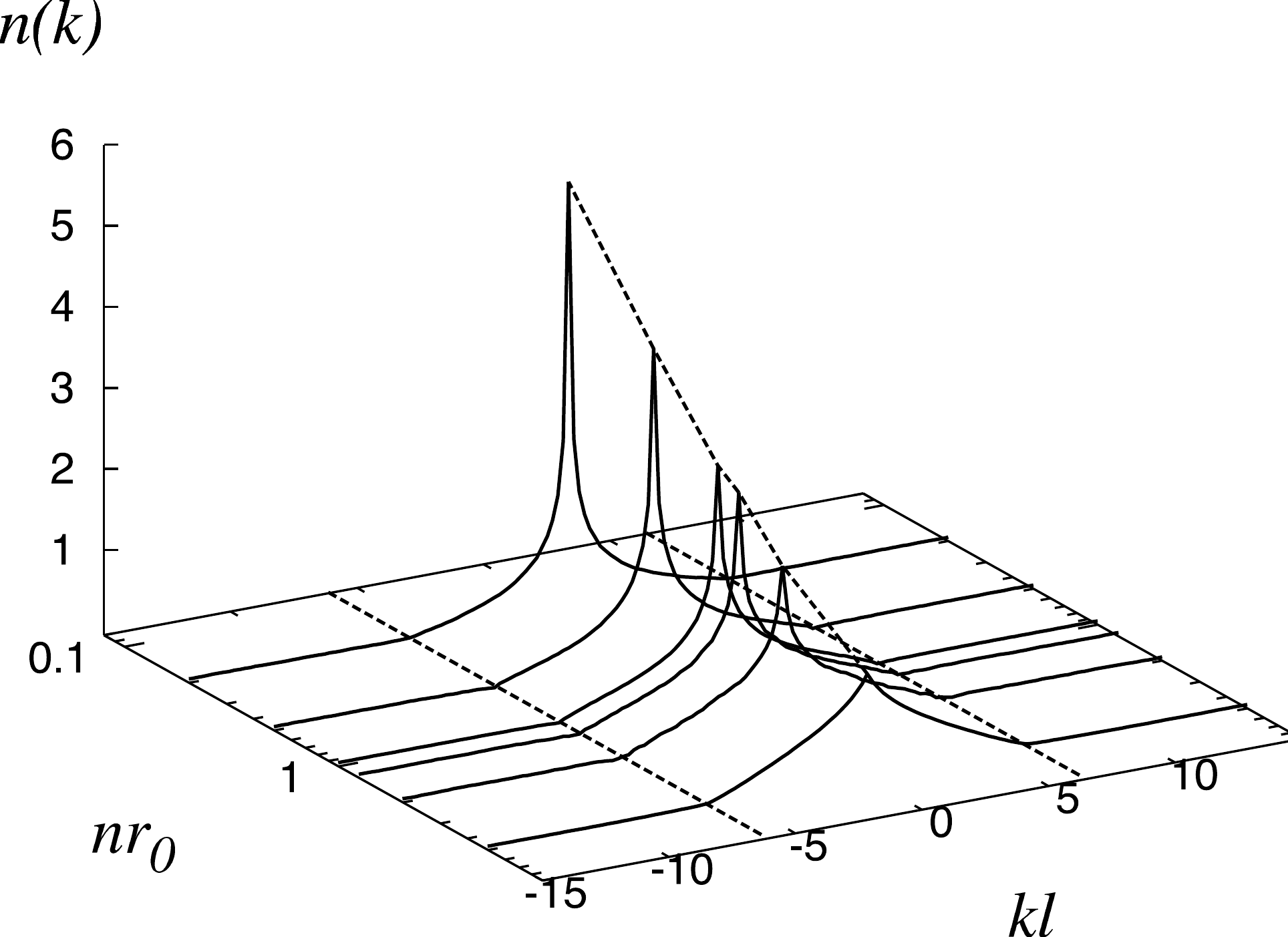}}\caption{Momentum distribution as a function of the interaction strength in a single tube as obtained from QMC simulation~\cite{roscilde2010}; the quasi-condensation peak at $k=0$ smoothens out with increasing repulsion according to TLL theory. Here, $n=\rho_0,r_0=R_3$ fix the notations, and $l$ is the system size. Image taken from Ref.\cite{roscilde2010}.}
\label{fig:n_k_QMC}}\caption{Panel \textit{(a)}: TLL parameter $K$ in a
dipolar wire as a function of the dimensionless interaction strength $\rho_{0}
R_{3}$ (here, $\rho_{0}\equiv n$). The blue line is Eq.~\eqref{wire_K}, while
red points and black squares are QMC results from Ref.
\cite{roscilde2010} and \cite{citro2007} respectively The inset
shows a similar prediction with inverse-square interaction (blue line)
compared with the exact result (red dashed line). Panel \textit{(b)}: sound
velocity to Fermi velocity ratio $v/v_{F}$ ($v_{0}\equiv v_{F}$) as a function
of $\rho_{0} R_{3}$; blue line and red points are analytical and QMC results
from Ref.~\cite{dalmonte2010} and \cite{citro2007} respectively.
Image taken from Ref.~ \cite{dalmonte2010}.}%
\label{fig:TLL_K}%
\end{figure}
Let us briefly discuss the relationship between a dipolar
interacting gas and a bosonic one with a contact repulsive interaction, the
Lieb-Liniger model~\cite{lieb1963}. Besides sharing the same low-energy
universality class (TLL) which allows to describe thermodynamical properties of
both systems on the same footing, there are quantitative and qualitative
differences between the two cases. The former is related to the very different
domain spanned by the TLL parameter, that is, $0<K<1$ for dipoles, whereas
$K\geq1$ for contact interactions. In addition to the momentum distribution,
such a feature can in principle be detected by investigating the so called
breathing mode, which, as shown in Ref.~\cite{pedri2008}, displays a
remarkably different behavior. Qualitative differences are not only captured
by to the fact that non-local interactions allow to cross the SF-CDW
transition; another very important feature is the response to a periodic
potential, which changes qualitatively when considering non-local repulsion,
as we will see in the next section. As a final remark, we resume the main
differences between bosonic and fermionic dipolar wires: while the TLL picture
remains quantitatively the same between the two, single particle correlation
functions are slightly different from Eq.~\eqref{sf_corr}, and, since no
superfluid instability may occur, the dominant correlation in the fermionic
case is always a CDW.

\subsubsection{Bosonic and fermionic gases in an shallow optical lattice}

\label{bos_shallow_lattice}

The effects of dipolar interactions become even more relevant in the presence
of external perturbations such as an underlying periodic potential, as
realized, e.g., by an optical lattice along the direction of the
tube~\cite{Bloch2008}. In this configuration, long-range interactions compete
with both kinetic energy and the periodic lattice potential, which in 1D has
the form~\cite{buechler2003}:
\begin{eqnarray}
H_{OL}=U\int\;dx\; \psi^{\dagger}(x) \sin^{2}(2\pi x/\lambda)\psi(x)
\end{eqnarray}
where $\lambda$ is the laser wavelength, $U_{L}=U/E_{R}$ the dimensionless
depth of the lattice potential, and $E_{R}=h^{2}/(2m\lambda^{2})$ the recoil
energy. In the limit of a very shallow lattice, $U_{L}\ll1$, the periodic
potential may be considered as a perturbation on the TLL Hamiltonian,
Eq.~\eqref{wire_TLL}, and its effects may be investigated using a sine-Gordon
description~\cite{dalmonte2010,buechler2003,gogolin_book}. The phase diagram,
and thus the influence of the underlying potential, sharply depends on the
ratio between the mean inter-particle distance $1/\rho_{0}$ and the lattice
spacing $\lambda/2$; when such ratio is an integer, $2/(\lambda\rho_{0}%
)=p\in\mathbb{N}$, and for large enough interactions, the system undergoes a
Berezinskii-Kosterlitz-Thouless (BKT)
transition~\cite{berezinskii1972,kosterlitz1973} from a gapless (e.g., SF or
CDW) to a crystalline phase, characterized by a gap in the excitation spectrum
and broken translational symmetry with one particle pinned every $p$ lattice
sites. On the contrary, if $2/(\lambda\rho_{0}) \notin\mathbb{N}$, the system
is always gapless regardless of the dipolar interaction strength. This
remarkable behavior suggests the following argument: when there is an allowed
configuration where particles can, at the same time, \textit{sit} on the
minima of the underlying potential and maintain a constant interparticle
distance, the dipolar interactions and the lattice potential can pin the gas,
as both lattice and dipolar potential energy contributions may be minimized by
the crystalline configuration. In the opposite case, the frustration between
the long-range repulsion, which tries to keep interparticle distance constant,
and the periodic potential does not allow for any stabilization of crystalline structure.

\begin{figure}[t]
\center{{\includegraphics[width= 0.5 \columnwidth]{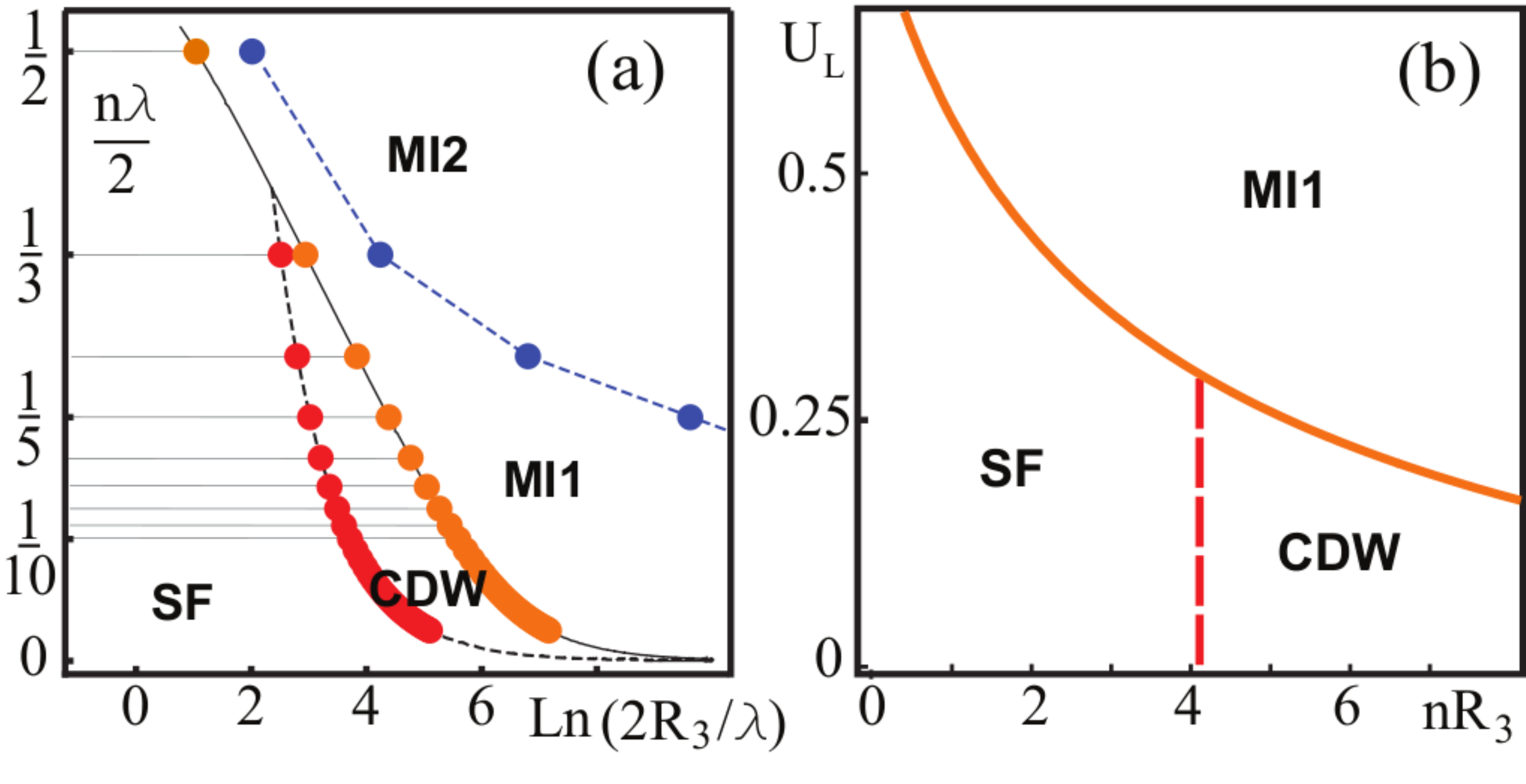}}\caption{Panel {\it (a)}: commensurate phase diagram for bosons with dipolar interactions in a shallow lattice with depth $U_L = 0.1$. Physical configurations correspond to commensurate fillings $\rho_0 \lambda/2 \equiv 1/p$, with $p \in \mathbb{N}$ (horizontal lines are guides to the eye for $p \leq 10$). Quantum phase transitions from a TLL to an insulating Mott insulator (MI) occur for each $1/p$ at the position of the dots on the continuous line, while red and blue dots on dashed lines indicate crossovers. MI1 and MI2 indicate MI with different low-energy spectra (see Ref.~\cite{dalmonte2010} for further details). Panel {\it (b)}: phase diagram at commensurate filling $1/p= 1/3$ in the $U_L$ vs $\rho R_3$ plane. Continuous line: quantum phase transition between a TLL and a MI. The phase diagram for fermions is identical to the one for bosons, except the TLL is always a CDW. In both panels, $n\equiv\rho_0$; image taken from Ref.~\cite{dalmonte2010}.}
\label{fig:sG_pd}}\end{figure}

The sine-Gordon model, combined with Eq.~\eqref{wire_K}, allows to make
quantitative predictions on the BKT transition in the regime of a shallow
lattice  potential where lattice bands are formed (which usually  corresponds
to $U/E_{R} \lesssim2$)~\cite{gogolin_book}. Typical phase diagrams for
bosonic gases are plotted in Fig.~\ref{fig:sG_pd}; the fermionic case is
obtained by substituting the superfluid phase with a CDW. A Luttinger
staircase, that is, a series of insulating states with commensurate density
satisfying $2/(\lambda\rho_{0})=p\in\mathbb{N}$, appears as a function of the
interaction strength at a fixed $U_{L}$~\cite{dalmonte2010}.


\subsubsection{Bosonic gases in a deep optical lattice: extended Bose-Hubbard
models}

As discussed in the previous section, when all energy scales involved in the
system dynamics are much smaller that the lattice bandwidth, both bosonic and
fermionic gases are properly described in the context of Hubbard-like models.
In the bosonic, single species case, the resulting extended Bose Hubbard
Hamiltonian (EBHH)
\begin{eqnarray}
\label{hdb}H_{db} =-J \sum_{i=1}^{L} (b^{\dagger}_{i}b_{i+1}+h.c.)+V\sum
_{i<j}\frac{n_{i}n_{j}}{|i-j|^{3}}
 +\sum_{i}\left[ \frac{U}{2}n_{i}(n_{i}-1)+\mu n_{i}\right]
\end{eqnarray}
contains kinetic energy ($J$), local ($U$) and dipolar ($V$) interaction
terms. The corresponding phase diagram has been studied in the
grand-canonical  ensemble in Ref.~\cite{burnell2009}, where a Devil's
staircase structure  has been shown to appear both in the hard-core (e.g.,
infinitely-large $U$)  and soft-core (e.g., finite $U$) limits. Moreover,
close to rational fillings, supersolid behavior has been suggested via both
numerical and analytical methods, in particular close to the half-filled case,
for both nearest-neighbor\cite{batrouni2006,mishra2009} and dipolar
interactions\cite{kumar2008}, whereas low and incommensurate filling fraction
are still described by TLLs with both CDW and SF dominant
orders\cite{dalmonte2011b}

\begin{figure}[t]
{\ \includegraphics[width= 0.5 \columnwidth]{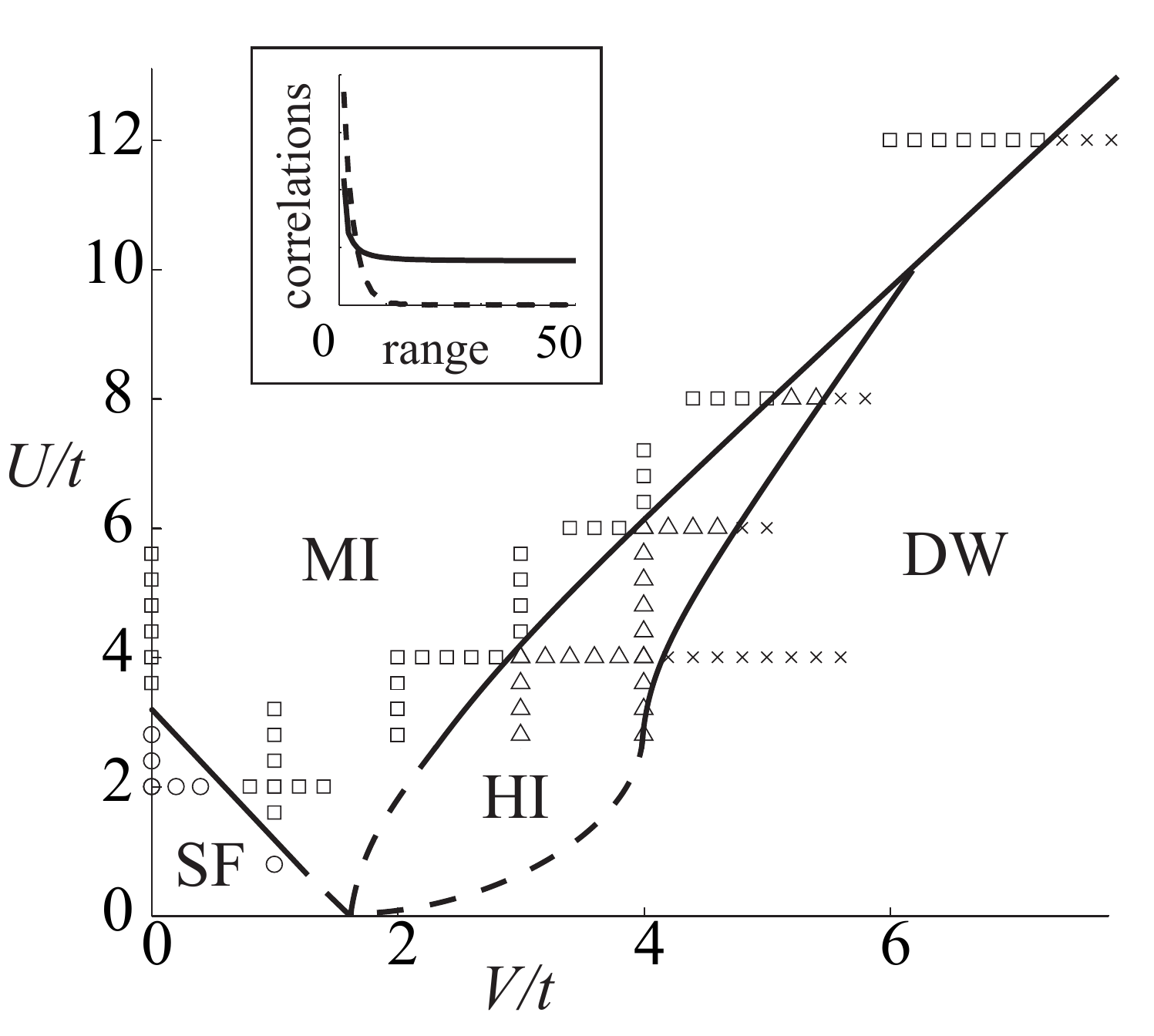}
\hspace{0.2mm} \includegraphics[width= 0.4 \columnwidth]{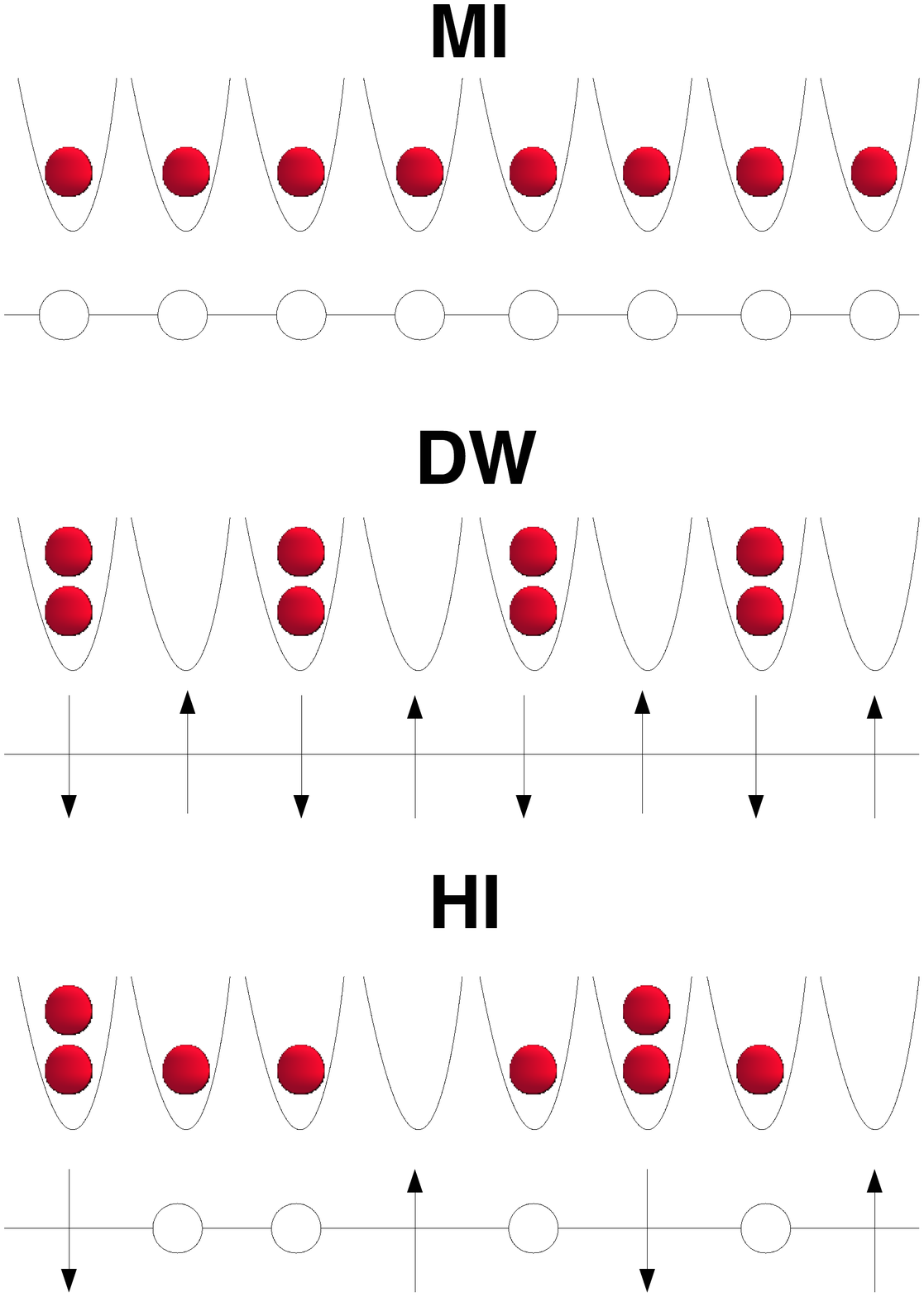}
}\caption{Left panel (image taken from Ref.~\cite{dallatorre2006}):
numerical phase diagram of dipolar bosons at unit filling: the Haldane
insulator is stable close to the $U\simeq2V$ line up to relatively strong
magnitude of $U$. Inset: string (solid line) and density wave (dashed line)
correlation functions in the HI phase as a function of $i-j$. Right panels:
cartoons of the magnetic phases at filling one, in both bosonic and spin-1
representation. }%
\label{fig:HI_pd}%
\end{figure}

\emph{Hidden order.} In the special case of unit filling, that is, when the
number of particles is equal to the number of lattice sites, an additional
instability may occur. When dipolar and local interactions are much larger
than the hopping rate $J\ll U, V$, the competition between $U$ and $V$ gives
rise to two different states of matter: a Mott insulator phase when
$U\gtrsim2V$, where double occupancies are suppressed, and a density wave (DW)
when $2V\gtrsim U$, where particles rearrange in a periodic structure in order
to minimize the non-local repulsion (see Fig.~\ref{fig:HI_pd}). While these
phases are separated by a first order phase transition line at strong
coupling, in the intermediate interaction regime it has been shown that an
additional state of matter, the so called \textit{Haldane insulator} (HI), can
emerge between the two~\cite{dallatorre2006}. The HI is gapped and characterized
in real space by a non-regular density
pattern, where doubly occupied and empty sites are spatially separated by
strings of singly occupied sites of uneven length (see the corresponding
cartoon in Fig.~\ref{fig:HI_pd}). The occurrence of such magnetic-like order
(between two doubly occupied sites there is always one and only one empty
site) together with positional disorder (the length of the strings is not
constant) is encoded into a \textit{string order parameter}:
\begin{eqnarray}
\mathcal{O}_{string}(j-k)=\langle\delta n_{j} \;\exp[i\pi\sum_{j<l<k}\delta
n_{l}]\; \delta n_{k} \rangle
\end{eqnarray}
which approaches a constant value at large distances, $|i-j|\gg1$; here
$\delta n_{j}\equiv n_{j}-1$. In addition, the density wave order parameter,
$\mathcal{O}_{DW}(j-k)=\langle\delta n_{j} \delta n_{k} \rangle$, which is
constant in the DW phase, decays exponentially in the HI. Numerical
simulations based on the density-matrix-renormalization-group
(DMRG)\cite{white1992,schollwock2005} algorithm have quantitatively determined
how this phase persists up to relatively large values of the interaction
strength $U\simeq8$ close to the $U\simeq2V$ line~\cite{dallatorre2006}. In
such a regime, where local occupations with $n_{j}\geq3$ are strongly
suppressed, the bosonic Hamiltonian can be mapped into a spin-1 problem where
the spin component along the $z$-direction satisfies $S^{z}_{j}=1-n_{i}$, and
the $J-, V-$ and $U-$terms play the role of spin-exchange, dipolar interaction
along the $z$-axis and on-site anisotropy. The system thus resemble the so
called Heisenberg model with single ion-anisotropy, or $\lambda-D$ model, the
only qualitative difference being the presence of a full dipolar tail instead
of nearest-neighbor interaction, establishing a strong connection between the
HI and the Haldane phase extensively studied in the context of magnetic
systems~\cite{tasaki1991,kennedy1992}. An exact correspondence between
bosonic and spin problem can be obtained in case of strong three-body
losses~\cite{daley2009}, where the on-site Hilbert space is truncated up to
doubly occupied states and the HI phase appears at smaller values of the
dipolar interaction~\cite{dalmonte2011a}. The existence of such HI has been
established even at higher integer fillings by mapping the problem to integer
spin chains with spin $s>1$\cite{amico2010}.

\begin{figure}[b]
{\ \includegraphics[width= 0.4 \columnwidth]{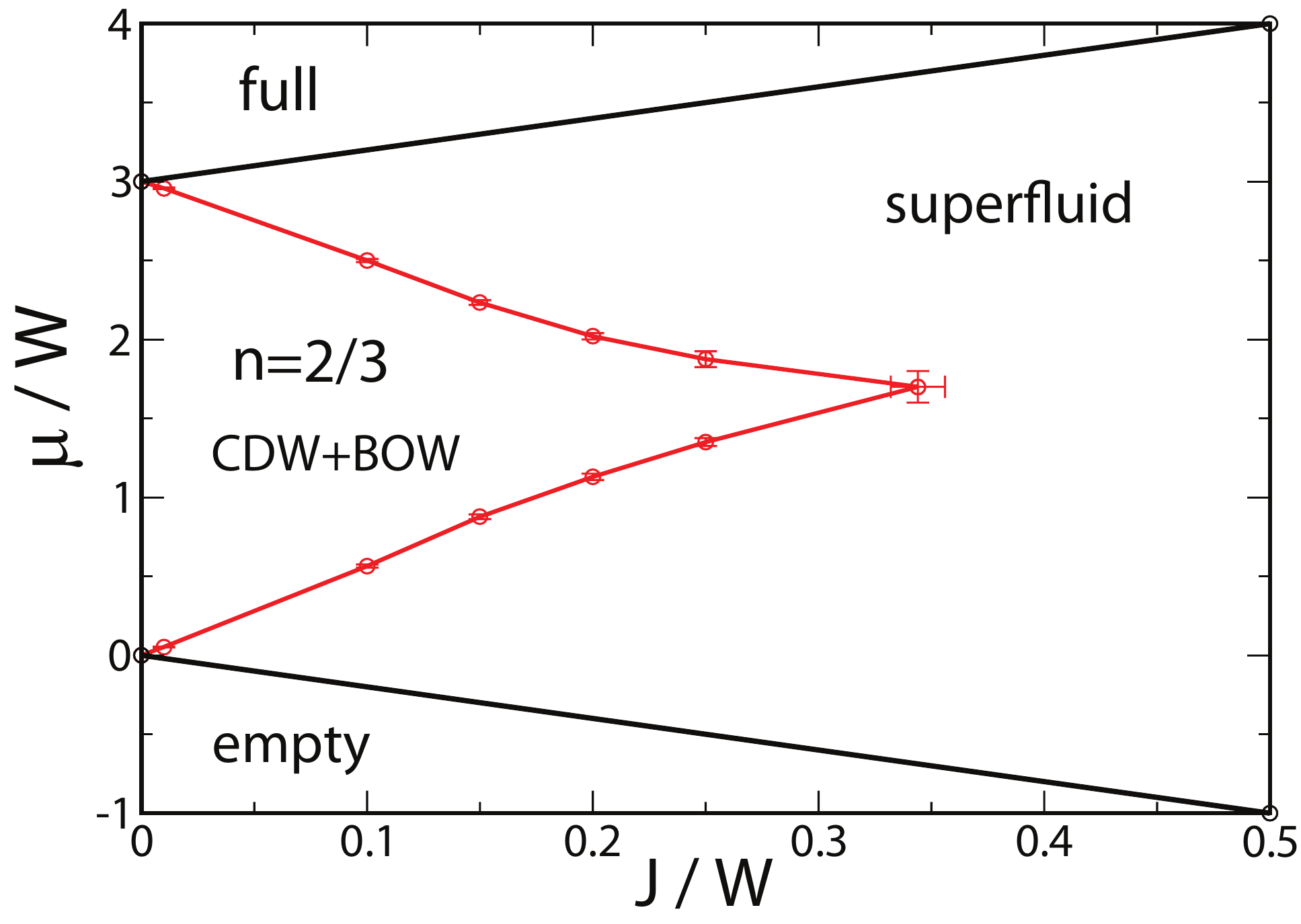}
}\caption{Phase diagram of hard-core bosons with three-body interactions in
the grand canonical ensemble. In the insulating phase at $J/W\lesssim0.3$, CDW
and bond order coexist. Image taken from Ref.~\cite{capogrosso2009}. }%
\label{fig:3body_pd}%
\end{figure}

\emph{Three-body interactions.} Additional interesting states of matter can be
found in 1D version of Hamiltonians already treated in 2D in the previous
section. As an example,it has been shown hot to engineer three-body
interactions with polar molecules, leading to the effective Hamiltonian in
Eq.~\eqref{Hubbard}. In 1D, the three-body term is responsible for the
stabilization of an insulating phase at filling $2/3$, which is driven by the
fact that, above a certain ratio of $W/J$, particles will minimize their
potential energy by sitting in a periodic patter at the expenses of the
kinetic one. Such phase transition belongs to the same universality class of
the pinning one described in the shallow lattice case, and the critical value
of the TLL parameter $K_{c}=2/9$ at the tip of the lobe is obtained from
general considerations~\cite{capogrosso2009}. The addition of a weak nearest
neighbor repulsion $V/J\simeq1$ can also stabilize a insulating state at
filling $1/2$ when $W/J\gg1$; remarkably, the further addition of a
next-nearest neighbor repulsion induces a competition between density wave and
bond order wave, as the BOW phase broadens thanks to the three-body repulsion
with respect to the $W=0$ case~\cite{schmitteckert2004}.

\subsection{Two-species mixtures}

The high degree of control over internal states of ultracold dipolar
gases\cite{ospelkaus2010}  has opened the way towards the theoretical
investigation of multispecies Hubbard-like models with additional dipolar
interactions. While such specific models were not considered in standard
condensed matter literature, Hamiltonians with nearest-neighbor interactions
(which is, in many respects, similar to a dipolar one close to half-filling)
such as the \textit{extended Hubbard model}~\cite{giamarchi_book} have been
shown to present a richer plethora of phases with respect to system with
contact interactions only. The theoretical advantage of multispecies 1D
systems with long-range interactions stems from the fact that one can
understand their basic physical properties employing the same low-energy
formalism used for models with contact interactions
only\cite{giamarchi_book,gogolin_book}. In the following, we will present an
overview of theoretical results on two-species models with dipolar
interactions, with both Fermionic and Fermi-Bose statistics.

\subsubsection{Fermi-Fermi mixtures}



\begin{figure}[b]
{\ \includegraphics[width= 0.35 \columnwidth]{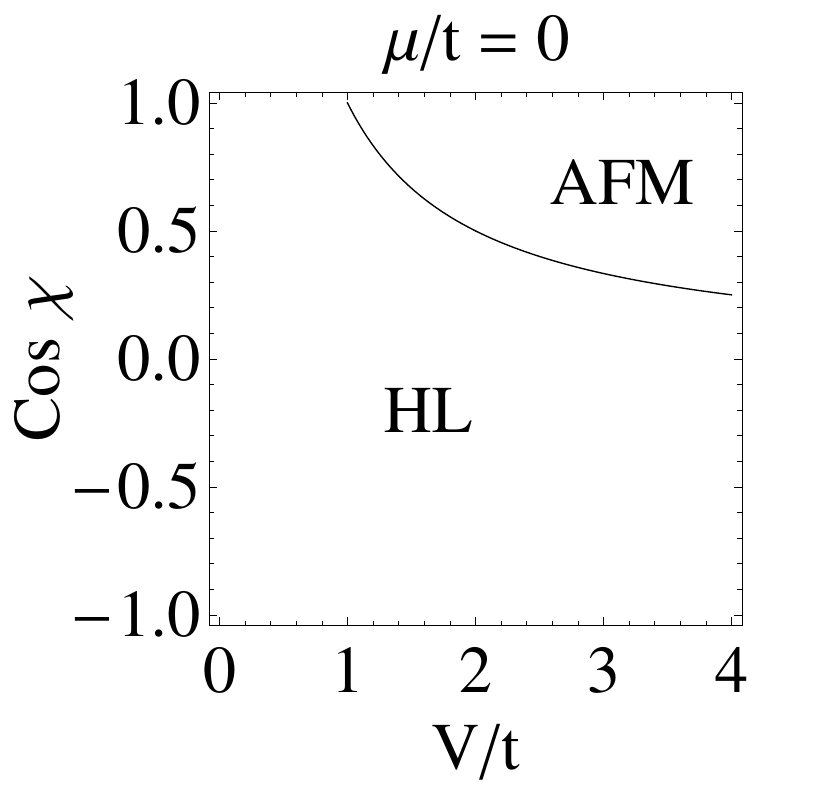}
\hspace{0.2mm} \includegraphics[width= 0.2 \columnwidth]{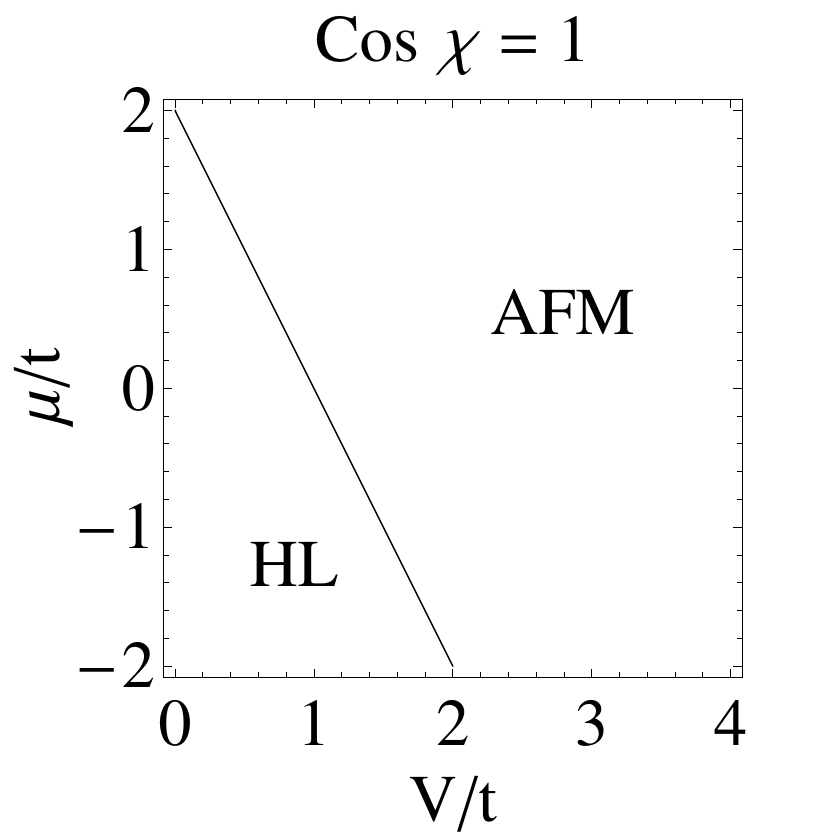}
}\caption{Mean field phase diagram for Eq.~\eqref{H_FF_eff} with
$\mu_{\uparrow}=\mu_{\downarrow}, V_{\uparrow\uparrow}=V_{\downarrow
\downarrow}=V=-V_{\uparrow\downarrow}/\cos(\chi)$. Image taken from
\cite{kestner2011}.}%
\label{fig:kestner_pd}%
\end{figure}

Contrary to their two-dimensional counterparts, fermionic dipolar mixtures in
1D may be theoretically investigated with the same accuracy as bosonic ones.
Defining as $c^{\dagger}_{i,\sigma},c_{i,\sigma}$ the creation/annihilation
operators at the site $i$ of the species $\sigma$ (we take here $\sigma
=\uparrow,\downarrow$), Fermi-Fermi mixtures in a deep optical lattice are
usually described by the following Hamiltonian:
\begin{eqnarray}
H_{FF}  =-t\sum_{i,\sigma}(c^{\dagger}_{i,\sigma}c_{i+1,\sigma}%
+h.c.)+U\sum_{i} n_{i,\uparrow}n_{i,\downarrow}+
 -\sum_{i,\sigma}\mu_{\sigma}n_{i,\sigma}+\frac{1}{2}\sum_{i\neq
j,\sigma,\delta}\frac{V_{\sigma\delta}}{|i-j|^{3}}n_{i,\sigma}n_{j,\delta}%
\end{eqnarray}
where the first two terms are hopping and on-site interaction, the third one
is a species-dependent chemical potential, and the last one represent the
dipolar contribution. When dealing with strong onsite two-body losses, the
effective Hamiltonian has to be projected onto a constrained Hilbert space
without double occupancies:
\begin{eqnarray}
\label{H_FF_eff}H^{(eff)}=\mathcal{P}H_{FF}\mathcal{P}%
\end{eqnarray}
where $\mathcal{P}=\prod_{i}(1-n_{i,\uparrow}n_{i,\downarrow})$. Such a
picture, which reduces the on-site Hilbert space to just three states, is
connected with the spin-1 chains (once $S^{z}_{i}=c^{\dagger}_{i,\uparrow
}c_{i,\downarrow}-c^{\dagger}_{i,\downarrow}c_{i,\uparrow}$ is defined)
already discussed in the context of the Haldane insulator phase for single
species dipolar bosons; however, differently with respect to spin models, the
key conserved quantity is $\sum_{i}(S^{z}_{i})^{2}$, as can be inferred by
performing a rigorous spin mapping~\cite{kestner2011}. The phase diagram of
Eq.~\eqref{H_FF_eff} has been investigated at the mean field level and via
numerical simulations based on the infinite time-evolving block decimation
(iTEBD)~\cite{vidal2007} algorithm in Ref.~\cite{kestner2011}
considering equal chemical potentials and intraspecies interactions,
$\mu_{\uparrow}=\mu_{\downarrow}, V_{\uparrow\uparrow}=V_{\downarrow
\downarrow}=V$, and varying the interspecies one $V_{\uparrow\downarrow
}=-V\cos(\chi)$. This setup can be realized by, e.g., applying a pair of ac
microwave fields close to the resonance of both internal
states\cite{kestner2011}. In the repulsive case, $V>0$, the ground state is
expected to show a dilute antiferromagnetic order of the form $..\uparrow
000\downarrow00\uparrow\downarrow00\uparrow..$, where, between molecules of
the same species, there is always one and only one molecule of the other
species, in addition to two strings of empty sites. Such configurations
display long-range order in the string correlation function
\begin{eqnarray}
\mathcal{O}^{z}_{string}(|k-j|)=-\langle S^{z}_{k} \exp[i\pi\sum_{k<l<j}%
S^{z}_{l}]S_{j}^{z}\rangle
\end{eqnarray}
and can be distinguished into two different cases: when $V\cos(\chi)>t-\mu/2$,
all sites are occupied, and the ground state is an antiferromagnetic insulator
(AFM); otherwise, the string order is dilute, the ground state is compressible
and the so called \textit{Haldane liquid} (HL) phase is stabilized. The mean
field phase diagram presenting such competition is shown in
Fig.~\ref{fig:kestner_pd}, and has been supported by numerical iTEBD results,
which show how, while both the AFM and the HL have long-range order in
$\mathcal{O}_{string}^{z}$, other quantities such as standard spin-spin
correlation functions have remarkably different behaviors~\cite{kestner2011},
and in the HL phase may be qualitatively described by mapping the low-energy
physics to a spinless TLL.

\begin{figure}[t]
{\ \includegraphics[width= 0.35 \columnwidth]{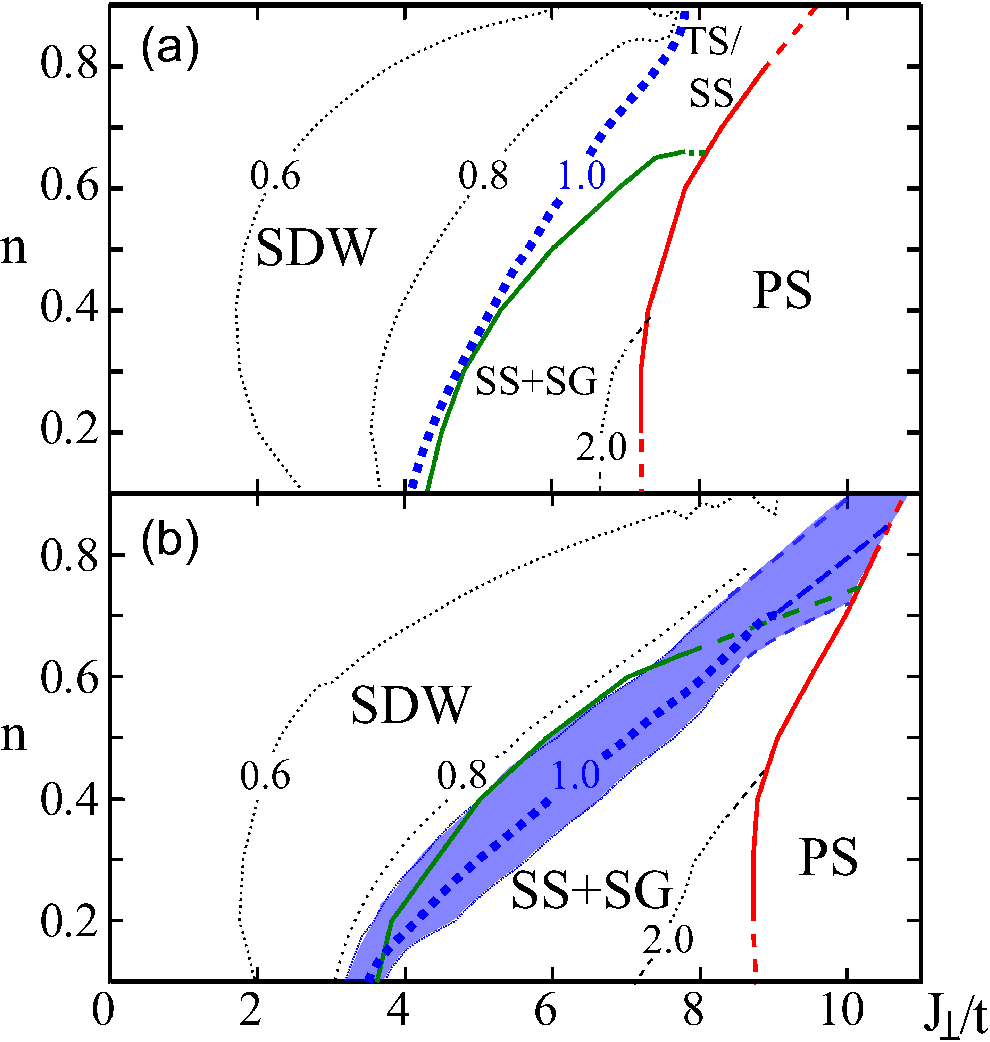}
}\caption{DMRG phase diagrams of the $t-J$ (upper panel) and $tJVW$ (for
$V=W=J_{z}=0$ , lower panel) models: blue dashed lines denotes $K_{\rho}=1$,
and green thick line separate spin gapped (SS+SG) and gapless (SDW or SS/TS)
regions. In the lower panel, the blue shaded region displays a charge
Luttinger parameter $K_{\rho}=1\pm0.15$ within numerical accuracy, and, as
such, may contain a gapless superfluid phase TS/SS. Image taken from
\cite{Gorshkov2011b}.}%
\label{fig:manmana_pd}%
\end{figure}

Using a combination of both DC and microwave external fields, as discussed in
Sec.~\ref{s6}, it is also possible to engineer $t-J$-like Hamiltonian with
dipolar spin-exchange interactions, as described by the $t-J-V-W$
Hamiltonian:
\begin{eqnarray}
\label{tJVW}H_{tJVW}=  -t\sum_{i,\sigma}(c^{\dagger}_{i,\sigma}c_{i+1,\sigma
}+h.c.)+\sum_{i\neq j} |R_{i}-R_{j}|^{-3}
\times\left[ \frac{J_{\perp}}{2}S^{+}_{i}S^{-}_{j}+
\frac{J_{z}}{2}S^{z}_{i}S^{z}_{j}+\frac{V}{2}n_{i}n_{j}+Wn_{i}S^{z}_{j}
\right]
\end{eqnarray}
where $S^{z}_{i}=(n_{i,\uparrow}-n_{i,\downarrow})$ and $S^{+}_{i}=c^{\dagger
}_{i,\uparrow}c_{i,\downarrow}$. A complete investigation of the phase diagram
of Eq.~\eqref{tJVW} is still lacking; nevertheless, the special case with
$V=W=J_{z}=0$, one of the simplest experimental realization, has been
numerically investigated in Ref.~\cite{Gorshkov2011b} by means of DMRG
calculations; in this case, the main difference with respect to the standard
$t-J$ model resides in the long-range nature of the spin-exchange
interactions, which are limited nearest-neighbor in the $t-J$ itself. At small
values of $J$, the ground state is a spin-density wave (SDW) at all densities,
resembling the ground state physics of the strongly repulsive Hubbard
model~\cite{giamarchi_book}. Then, increasing $J$, one enters first a region of
singlet superfluidity (SS) with a finite spin gap (SG), and finally phase
separation. In the SS region, inter-particle attraction is encoded into the
TLL parameter of the density (or \textit{charge}~\cite{giamarchi_book}) sector
$K_{\rho}>1$. The main difference with the phase diagram of the $t-J$ model,
which is reported in the upper panel of Fig.~\ref{fig:manmana_pd}, is the
presence of a larger gapped region; besides, the absence of a superfluid
region with no spin gap (denoted as TS/SS) cannot be completely ruled out, as
the broad blue region may indeed present such instability.

The realization of the $tJVW$ model with polar molecules presents two main
advantages for the purpose of exploring the physics of the $t-J$ model with
ultracold gases: first, the accessible regions are not limited to small
exchange couplings $J\ll t$, and second the spin gap in the superfluid region
is twice larger, making such phase more stable against thermal fluctuations~\cite{Gorshkov2011b}.

\begin{figure}[t]
 \includegraphics[width= 0.55\columnwidth]{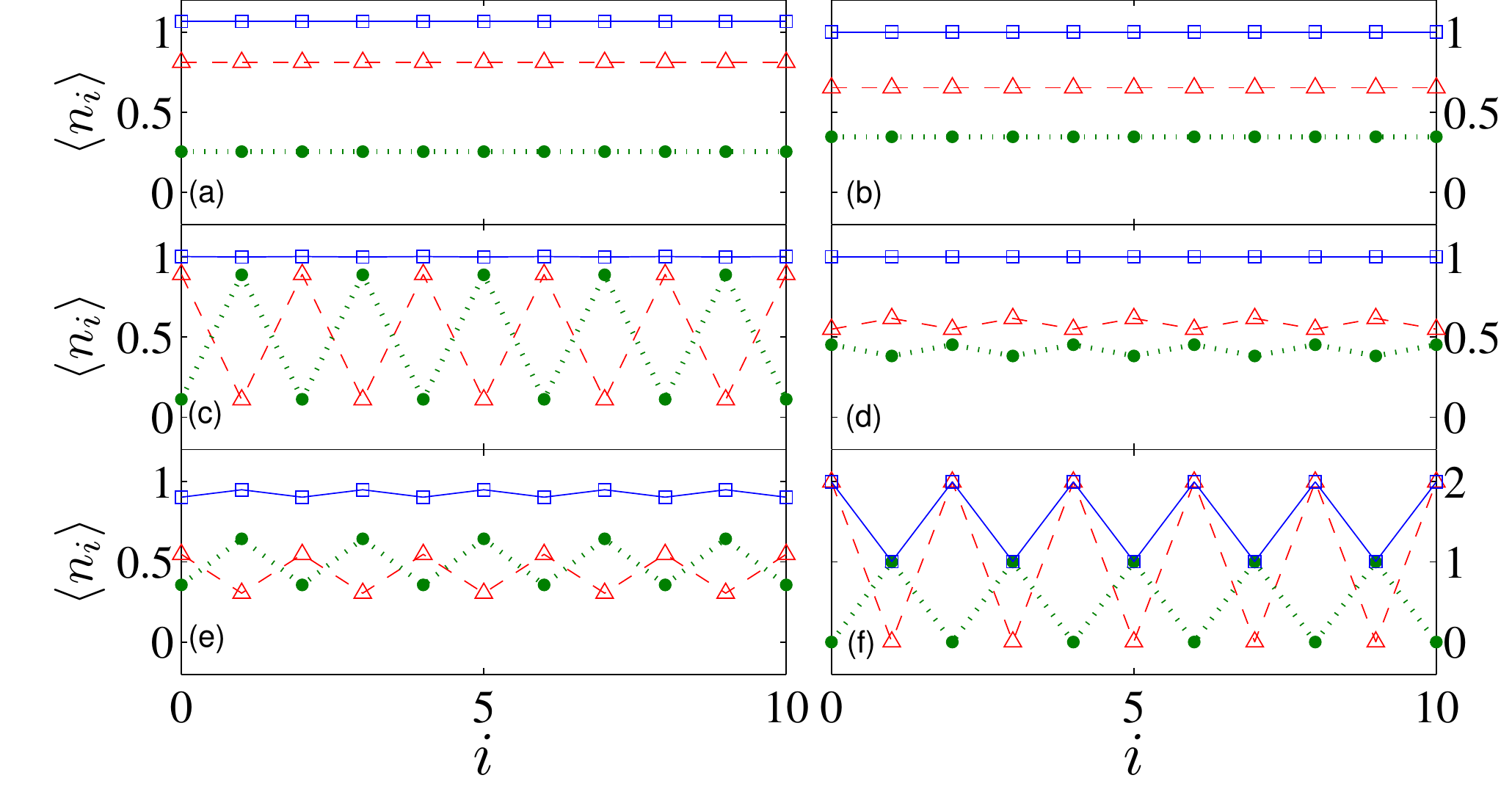}
\caption{Density distribution $n_{B}$ (triangles), $n_{F}$ (filled circles)
and $n_{B}+n_{F}$ (squares) for various phases as realized with dipolar
Bose-Fermi mixtures: \textit{(a)} BF liquid, \textit{(b)} BF Mott Insulator,
\textit{(c) and (f)} BF solids with different bosonic fillings, \textit{(d)}
density-wave BF Mott and \textit{(e)} density-wave BF liquid. Results have
been obtained by means of numerical iTEBD simulations. Image taken from Ref.~
\cite{wang2010}}%
\label{fig:wang_corr}%
\end{figure}

\subsubsection{Bose-Fermi mixtures}


Multispecies gases with different statistics, that is Bose-Fermi (BF)
mixtures, have been the subject of numerous theoretical and experimental
efforts in the field of cold atoms, and are currently being investigated even
in presence of dipolar interactions. In 1D, the effective Hamiltonian of such
systems when confined in a deep optical lattice is:
\begin{eqnarray}
\label{H_BFmix}H_{BF}  &=&-\sum_{i,\sigma=B/F}\left[ J_{\sigma}(a^{\dagger
}_{i,\sigma}a_{i+1,\sigma}+h.c.)-\mu_{\sigma}n_{i,\sigma}\right] +\nonumber\\
 &+&\sum_{i} n_{i,B}\left[ U_{BF}n_{F}+\frac{U_{BB}}{2}(n_{i,B}-1) \right]
 +\sum_{\sigma,\delta}\sum_{i<j}\frac{V_{\sigma\delta}}{|i-j|^{3}}%
n_{i,\sigma}n_{j,\delta}%
\end{eqnarray}
where $a^{\dagger}_{i,\sigma}$ are creation operators obeying
bosonic/fermionic statistics with $\sigma=B/F$ respectively, the first line
contains both hopping and chemical potential terms and the last two lines
onsite and off-site interactions. The insulating instabilities of this
Hamiltonian occurring close to half filling have been investigated
in Ref.~\cite{wang2010} by means of numerical simulations based on the
iTEBD algorithm by truncating the dipolar interactions up to nearest-neighbors
and considering several combination in the broad parameter space of
$\{J_{\sigma}, U_{\sigma\delta},V_{\sigma\delta}\}$. In addition to gapless
liquid and density-wave phases (a BF liquid, Fig.~\ref{fig:wang_corr} (a), and
a BF density wave, Fig.~\ref{fig:wang_corr} (e)), BF Mott phases with and
without density oscillations are present (a BF Mott insulator,
Fig.~\ref{fig:wang_corr} (b), and a BF Mott density wave,
Fig.~\ref{fig:wang_corr} (d)). Finally, in the special case of half filling
for both species, a BF solid, an incompressible phase with an alternating
density pattern (Fig.~\ref{fig:wang_corr} (c) and (f)), is also stable in the
strongly interacting regime $U_{\sigma\delta}\gg J_{\sigma}$. Such phase
displays true-long range order and, differently from the N\'eel
antiferromagnetic case realized with purely contact interactions, is stable
even when considering equal tunneling rates $J_{B}=J_{F}$; moreover, it
displays very different melting processes as the interaction are increased
depending on the relative value of $V_{BB}/V_{FF}$~\cite{wang2010}.





\subsection{Quasi-1D physics: coupling between tubes}

The long-range nature of dipolar interactions is well suited to create hybrid
systems with purely 1D dynamics but 2D interactions, such as arrays of coupled
tubes divided by very deep optical lattices preventing inter-tube tunneling
together with a proper tuning of the external DC electric fields in order to
manipulate the inter-tube interaction, as seen in Sec.\ref{s2}.  Such systems
establish a deep connection between the physics of several condensed matter
systems such as, e.g., spin compounds \cite{giamarchi_book} and ultracold
dipolar gases which cannot be explored with contact interactions only. In the
following, we first present recent results on the two-tube case, or
\textit{two-leg ladder}, and then consider the 2D limit where a large number
of tubes, a \textit{planar array}, is taken into account.



\subsubsection{Two-leg ladders}

Dipolar ladders allow to investigate rich physical phenomena by matching the
advantages of reduced loss rates typical of 1D confinement with the
possibility of considering multispecies physics, where the physical species
index is represented by the wire one~\footnote{The case of multispecies gases
confined in quasi-1D geometries, which allows to treat problems with even more
degrees of freedom, has attracted little attention so far.}. In this respect,
the special case of two-coupled tubes has been widely investigated. The
effective Hamiltonian of such systems, in case an optical lattice in the tube
direction is applied, is very similar to a two-species gas,
\begin{eqnarray}
H_{L} & =&-\sum_{\alpha,i}t_{\alpha}(c^{\dagger}_{\alpha,
i}c_{\alpha, i+1}+h.c.)+\mu\sum_{\alpha,i}n_{i,\alpha}\\
& +&\sum_{\alpha,\beta}\sum_{i,j}\mathcal{V}_{\alpha,\beta}(i,j)n_{i,\alpha
}n_{j,\beta}%
\label{H_ladder}
\end{eqnarray}
with the remarkable difference that here the anisotropic nature of the
dipole-dipole interaction plays a prominent role. Here, $\alpha=1,2$ is the
wire-index, and the specific shape of $\mathcal{V}_{\alpha\beta}(r)$ depends
on both the inter-wire distance $g$ and the angle between the dipole moments
and the plane the ladder lies on. Moreover, for bosonic particles, an
additional term $\sum_{i,\alpha}U n_{i,\alpha}(n_{i,\alpha}-1)/2$ denotes the
on-site interaction.

\begin{figure}[t]
{\ \includegraphics[width= 0.55 \columnwidth]{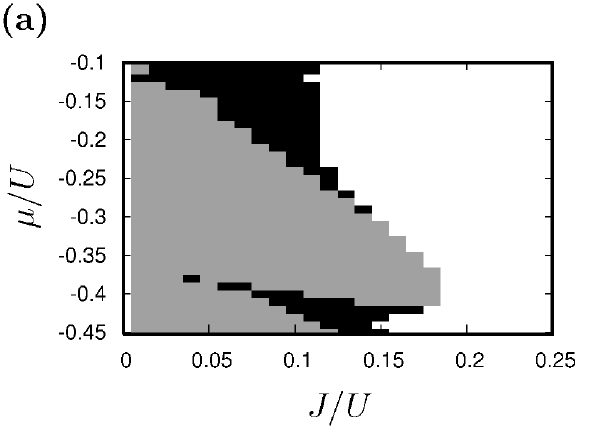}
}\caption{Numerical phase diagram of bosonic two-leg ladders with inter-wire
nearest-neighbor attraction $\mathcal{V}_{12}(0)/U=-0.75$: white, gray and
black regions represent SF, PSF and MI phases respectively. Image taken from
Ref.~\cite{arguelles2007}.}%
\label{fig:arguelles}%
\end{figure}

\begin{figure}[t]
{\ \includegraphics[width= 0.45 \columnwidth]{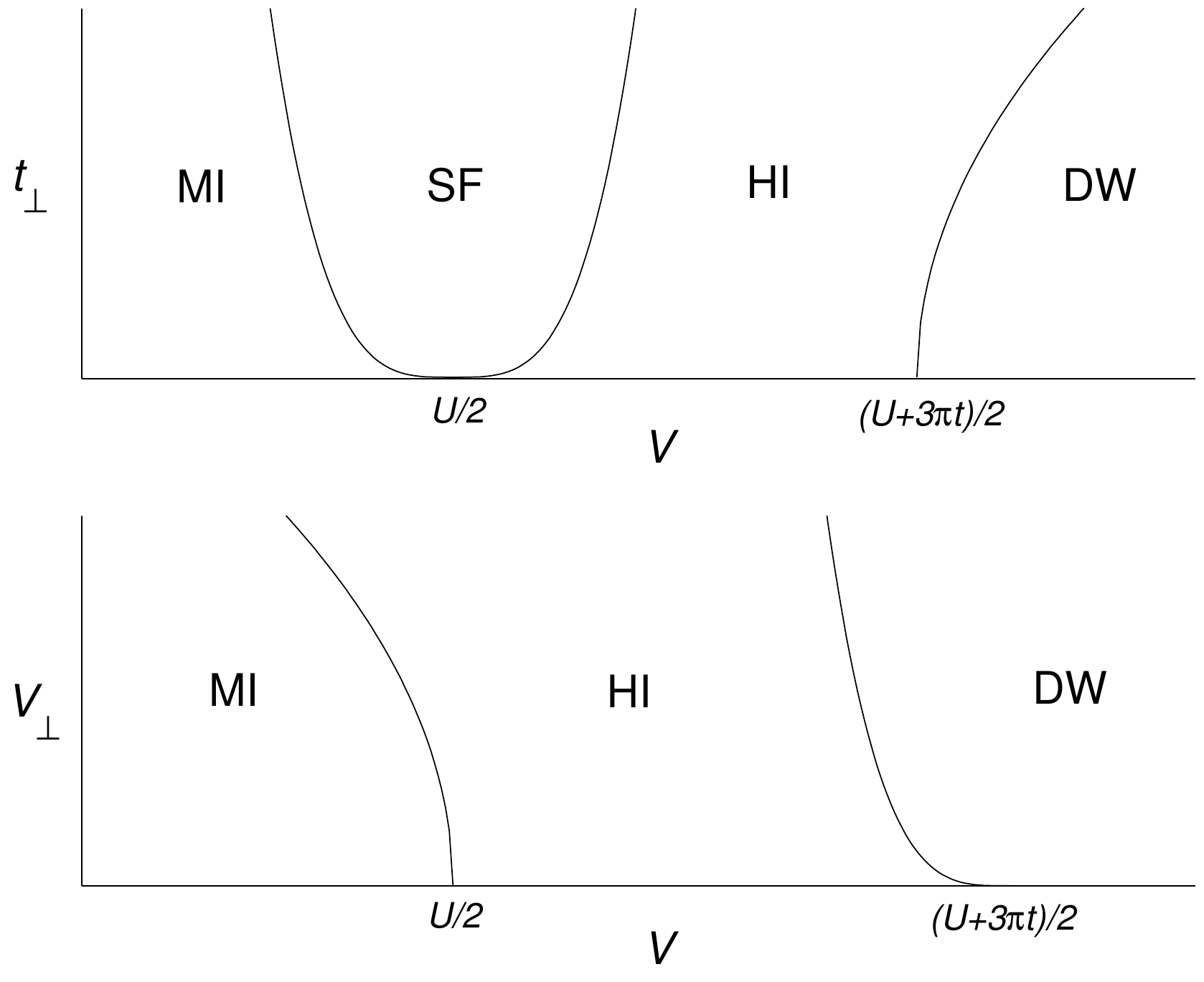}
}\caption{Qualitative changes in the phase diagram of two-coupled bosonic
chains at unit filling and fixed $U/J$ close to the HI phase. \textit{Upper
panel}: at finite interchain tunneling and no interchain interaction, a
superfluid phase appears between the DW and the HI. \textit{Lower panel}: at
finite interchain interaction and no interchain tunneling, the phase diagram
changes only quantitatively. Here, $V_{\perp}=\mathcal{V}_{12}(0),t=J$ and
$t_{\perp}=J_{\perp}$ fix the notations with respect to Ref.~
\cite{berg2008}, from which this image has been taken.}%
\label{fig:berg}%
\end{figure}

\textit{Bosonic ladders.} As a first step, it is worth considering if and how
phases typical of 1D setups such as Mott insulators, superfluids and Haldane
insulators are changed in ladder geometries. In case of interwire attraction,
both MI and SF phases appear in the limit of small ($J/U\ll1$)and large
($J/U\gg1$) intrawire tunneling respectively, and a pair superfluid phase
(PSF), that is, a superfluid ground state of composite particles made of one
boson on each tube, is stabilized at intermediate values of $U/J$%
~\cite{arguelles2007}. The Mott lobes close to the PSF phase have remarkably
different excitations, that is, particle-hole are substituted by
creation/annihilation of composite particles, further increasing the
re-entrant shape of the Mott lobes, as can be seen in Fig.~\ref{fig:arguelles}%
. The evolution of the HI phase in ladder setups in indeed more tantalizing,
as in the case of spin-1 chains it has been shown that the string order of the
Haldane phase is unstable towards weak antiferromagnetic interchain exchange
perturbations~\cite{anfuso2007}. In the case of bosonics coupled tubes,
combining TLL theory with renormalization group arguments, it has been
shown~\cite{berg2008} that, while a small intertube repulsion $\mathcal{V}%
_{12}(0)/J\ll1$ changes only quantitatively the shape of the phase diagram,
interchain tunneling terms of the type:
\begin{eqnarray}
\label{Jperp}H_{\perp}=J_{\perp}\sum_{i} (c^{\dagger}_{i,1}c_{i,2}+h.c.)
\end{eqnarray}
open an additional gapless, superfluid phase between the two: a schematic view
of such changes is presented in Fig.~\ref{fig:berg}, and has been confirmed
numerically via DMRG simulations~\cite{berg2008,berg2011}.

\begin{figure}[t]
\includegraphics[width= 0.55 \columnwidth]{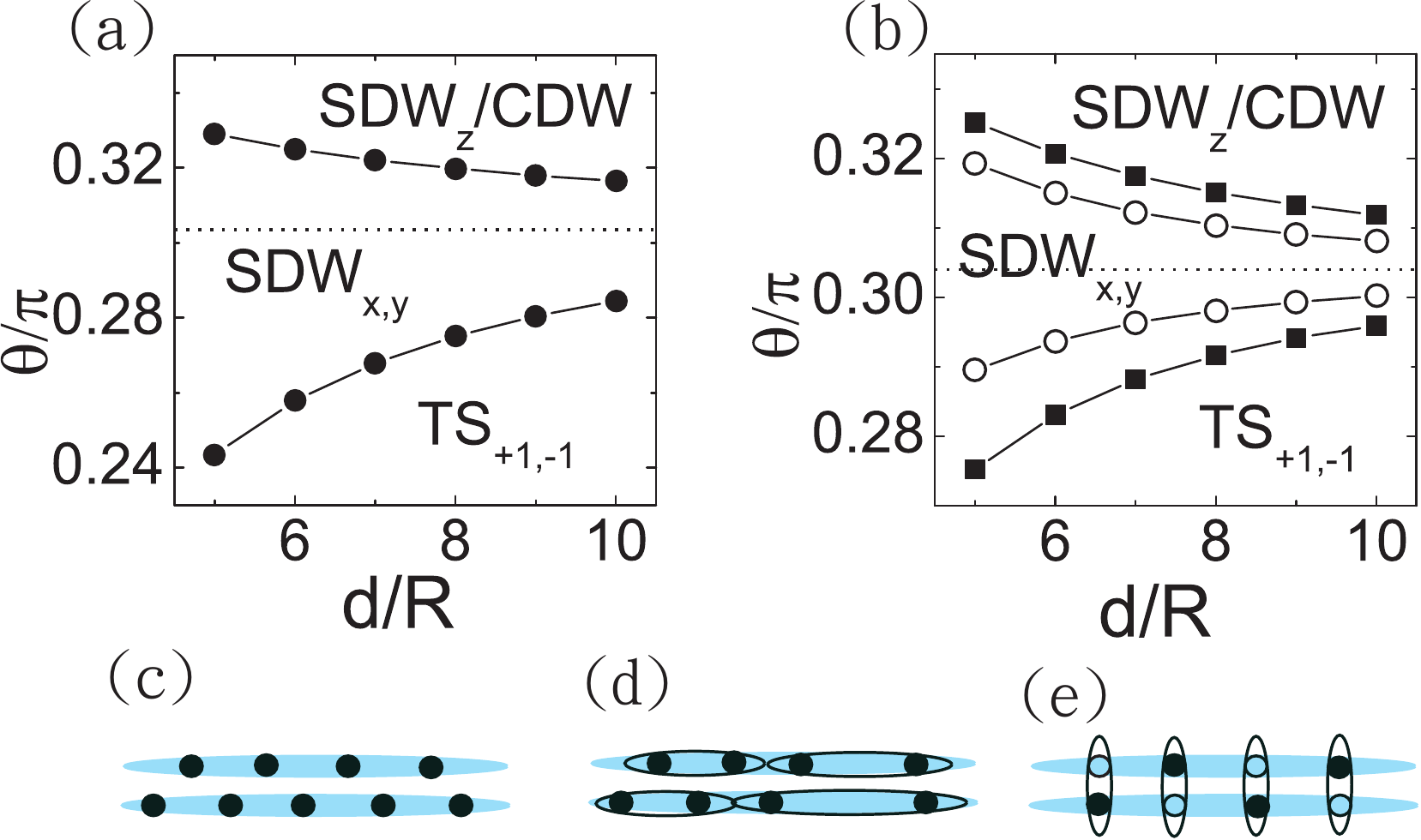}
\caption{Panels \textit{(a), (b)}: phase diagram of fermionic two-leg ladders
with inter-wire nearest-neighbor attraction repulsion as a function of the
interwire distance-to-wire width ratio $d/R$ and the angle $\theta$ (see
Fig.~\ref{fig:Set2D}); $k_{F}R=0.1, 0.2$ in the left and right panel,
respectively, where $k_{F}$ is the Fermi momentum of the single tube. Panel
\textit{(c), (d), (e)}: cartoons of the SDW$_{z}$, TS and SDW$_{x,y}$ phases
discussed in the text; horizontal ellipses denote pairing between particles in
the same tube, while intertube correlation is indicated by vertical ellipses.
Image taken from Ref.~\cite{chang2009}.}%
\label{fig:wang}%
\end{figure}

\textit{Fermionic ladders.} Fermionic ladders present more instabilities with
respect to the single tube, single species case. In absence of a lattice, the
weak-coupling phase diagram has been investigated using the bosonization
formalism~\cite{giamarchi_review,gogolin_book} in various
setups~\cite{chang2009,dalmonte2011}. In case of short range interwire
repulsion, the interplay between the longitudinal confinement length $R$, the
interwire distance $d$ and the angle $\theta$ (see Fig.~\ref{fig:Set2D}) is
responsible for the stabilization of three main phases. In case of intrawire
attraction, $\theta\lesssim0.3\pi$, pairing on the single wire is favored,
giving rise to a \textit{triplet superfluid} phase (TS); in the opposite case
of intrawire repulsion, $\theta\gtrsim0.3\pi$, a spin density wave with
alternating order (SDW$_{z}$) appears, while in the intermediate regime
$\theta\simeq0.3 \pi$ the dominant order is an in-plane spin-density wave
(SDW$_{x,y}$)~\cite{chang2009}. In the case of a deep lattice,
Eq.~\eqref{H_ladder} becomes an extension of the so called anisotropic
extended Hubbard model (AEHM)~\cite{otsuka2000} with both large anisotropy and
long-range, dipolar terms; while all the previous phases are expected to
appear in the low-density limit, the possibility of tuning independently
inter- and intra-wire repulsion, making the former stronger than the latter,
leads to a spontaneous breaking of the $\mathbb{Z}_{2}$ symmetry associated
with the ladder geometry, stabilizing a fully \textit{polarized} ground state
where all particles stay on the same tube to minimize the inter-tube
repulsion. Moreover, the addition of inter-wire tunneling terms
(Eq.~\eqref{Jperp}) may stabilize additional phases~\cite{chang2009}.

When inter-wire interactions are turned attractive, a different physical
picture arises depending on the population ratio between the wires,
$n_{1}/n_{2}$. When $n_{1}=n_{2}$, pairing between the tubes is always
favored, and the system behaves like a quantum liquid of composite bosonic
particles made of one boson per wire; such a liquid can be both a superfluid
(that is, a PSF), or a CDW depending on both the interwire distance and dipole
strength~\cite{dalmonte2011}. If a shallow optical lattice commensurate with
the particle density is introduced along the tube direction, that is its
lattice wavelength $\lambda$ satisfies $2/n_{1}\lambda\in\mathbb{N}$, the
system can undergo a BKT transition to a composite crystal as a function of
the lattice depth, in analogy with what happens in the single tube bosonic
case (see Sec.~\ref{bos_shallow_lattice}): in a grand canonical ensemble, this
picture evolves establishing a Luttinger staircase of dimer
crystals~\cite{dalmonte2011}. In the deep lattice case, these predictions have
been checked by means of DMRG simulations.

\begin{figure}[tb]
\includegraphics[width= 0.55 \columnwidth]{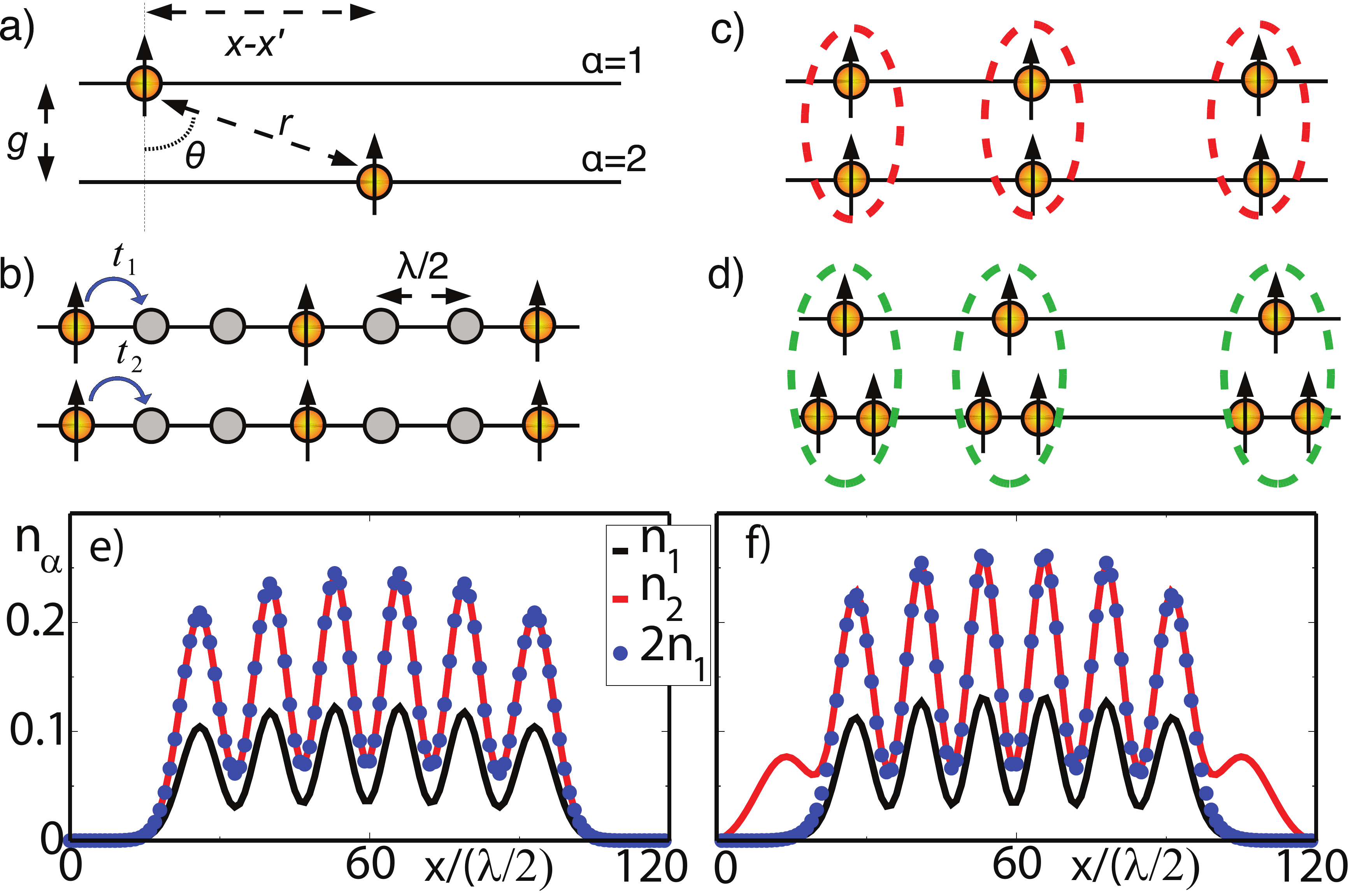}
\caption{Panel \textit{(a)}: typical configuration of two-leg ladder with
attractive short-distance interaction between the wires. Panel \textit{(b-d)}:
cartoons of a dimer crystal (\textit{(b)}), dimer liquid (or PSF,
\textit{(c)}) and a trimer liquid (\textit{(d)}). Panel \textit{(e-f)}:
typical density distribution of strongly imbalanced wires with $2n_{1}\lesssim
n_{2}$ (see Ref.~\cite{dalmonte2011} for technical details). In both
cases, the density match $n_{2}=2n_{1}$ in the central part of the system
indicate the presence of a trimer liquid. Image taken from
Ref.~\cite{dalmonte2011}. }%
\label{fig:trimers}%
\end{figure}

\emph{Imbalanced ladders.} A completely different phenomenon occurs in
presence of a density imbalance between the tubes. For interwire distances $g$
larger than the inverse particle density, the long-range nature of the
dipolar interaction creates an effective fixed range interwire attraction,
thus favoring pairing not only between two, but also between many particles.
In the special case of commensurate densities between the wires, $n_{1}%
/n_{2}=p/q\in\mathbb{Q}$, a gas of composite particles, or \textit{multimers},
composed by $p(q)$ particles in the first (second) wire respectively can be
stabilized for sufficiently low densities~\cite{dalmonte2011}. Such a multimer
liquid picture has a description in terms of low-energy field
theory~\cite{burovski2009}, and allows to study generalized pairing mechanism
beyond two-particle ones. In the special case of $p=1, q=2$, the stability of
\textit{trimer liquids} has been confirmed numerically for sufficiently large
tunneling and dipole imbalance between the tubes. When also a trapping
potential and a non-ideal population ratio is considered, such trimer liquids
(which are not stable in homogeneous setups with unmatched densities) appear
in the center of the system, and are surrounded by both dimer liquids or
singles species ones depending on the population ration, forming a
pancake-like structure of composite liquids (see Fig.~\ref{fig:trimers}) as
shown in recent DMRG simulations~\cite{dalmonte2011}.

The possibility of further tuning the numerous parameters of the ladder
Hamiltonian Eq.~\eqref{H_ladder} may indeed allow to identify additional
exotic state of matter, as the complete phase diagram of these system is still
largely unexplored. Moreover, before turning our attention to systems composed
by a large number of tubes, it is worth noticing that ladder systems with more
than two tubes and mixtures in quasi-1D geometries are in principle good
candidates to study new model Hamiltonians in connection to many-body problems
with many degrees of freedom.

\subsubsection{Planar array of tubes}

Planar arrays of tubes of dipolar gases constitute a hybrid setup where one
can investigate 2D states of matter with physical properties typical of 1D
systems such as, e.g., quasi long range order of correlation functions. Even
in this case, the effective Hamiltonian:
\begin{eqnarray}
H_{plane}  =\sum_{\alpha,j}\int dx\; \psi_{\alpha,j}^{\dagger}(x)(-\frac
{\hbar^{2}}{2m}\partial_{x}^{2})\psi_{\alpha,j}(x)+
 + \sum_{\alpha,\beta, j, l}\int dx\;dx^{\prime}\rho_{\alpha,j}(x)
\rho_{\beta,l}(x^{\prime}) \mathcal{V}_{\alpha\beta}(x,x^{\prime})\nonumber
\end{eqnarray}
has many controllable parameters: here, $\psi_{\alpha}^{\dagger},\psi_{\alpha
}$ are creation/annihilation operator on the tube $\alpha$ (sums over Greek
indices denote sums over the tube index, while Latin indices denote particles
along each tube), $\rho_{\alpha}=\psi^{\dagger}_{\alpha}\psi_{\alpha}$, the
first term represents the kinetic energy over each tube and the second one the
dipolar interactions, which depend on both strength and orientation of the
dipole moments. In the zero-density limit, where only one bosonic particle per
wire is considered and inter-wire interaction are attractive, a
\textit{quantum rough chain} forms with the off-diagonal long range order in
the $M$-body correlation function, where $M$ is the number of
wires~\cite{capogrosso2011}.

\begin{figure}[t]
{\
\includegraphics[width= 0.55 \columnwidth]{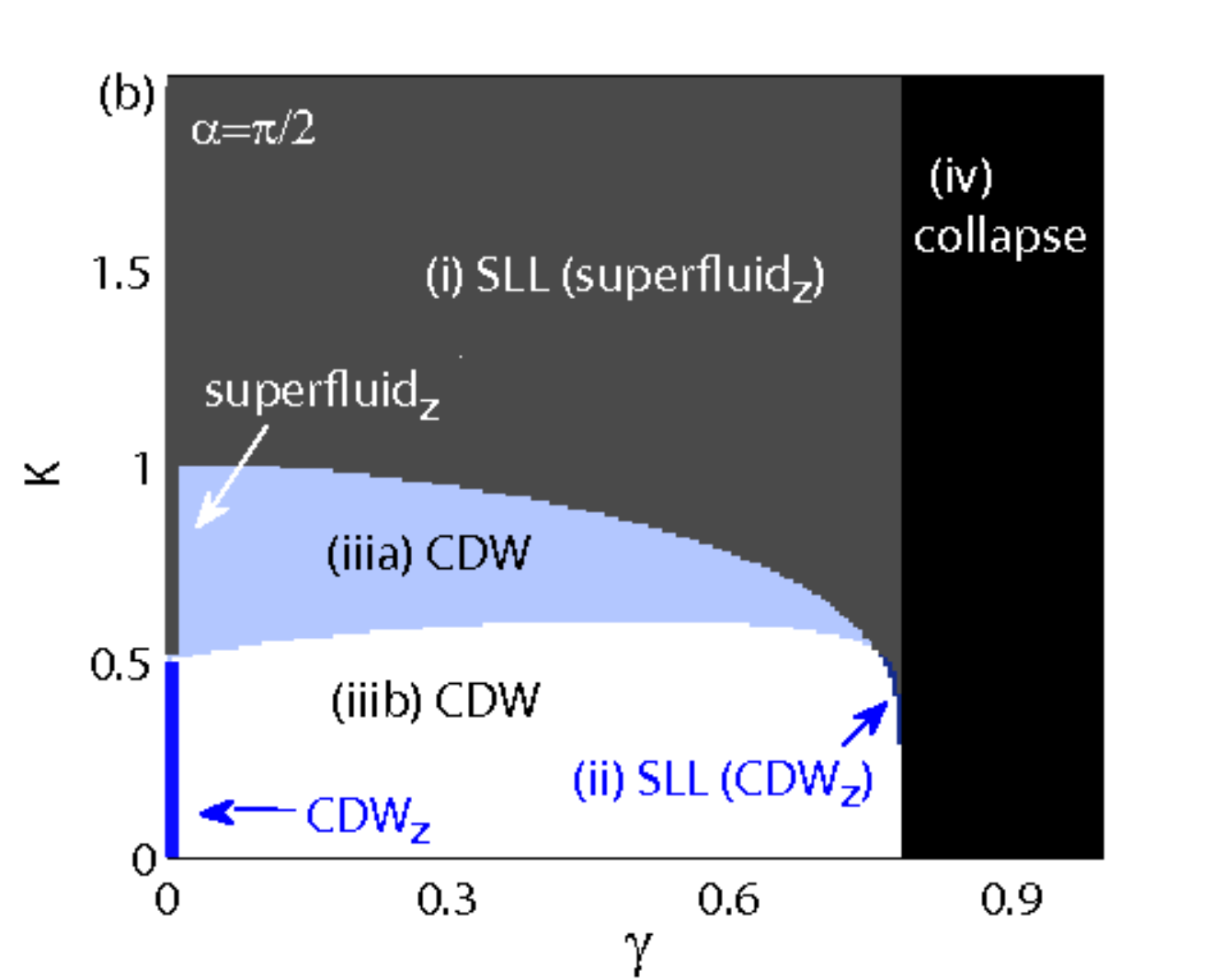}
}\caption{Phase diagram of a planar array of dipolar bosons as a function of
the single wire Luttinger parameter $K$ and the ration between the dipolar
interaction and the scattering length $\gamma=4V_{0}K/(a^{2} U)$ (see text).
Here, the angle $\theta$ ($\alpha$ according to notations in
Ref.~\cite{kollath2008}) is fixed at $\pi/2$. Image taken from
Ref.~\cite{kollath2008}.}%
\label{fig:kollath}%
\end{figure}
In the finite density regime, the many body physics displays very different
effects. In the bosonic case with intrawire dipolar repulsion, the competition
between local interactions and long-range ones is encoded into a dimensionless
parameter $\gamma=4V_{0}K/(ua^{2})$, where $u$ and $K$ are the sound velocity
and the TLL parameter of a single tube, $V_{0}$ is the strength of the dipolar
interactions and $a$ the intertube spacing. By varying both $K$ and $\gamma$,
one can span a broad parameter region with respect to the dipolar interaction,
the density of particles and the $s$-wave scattering length $a_{1D}%
$~\cite{kollath2008}. In addition to a series of density wave states driven by
the strong dipolar repulsion, which may lead to both a stripe and checkerboard
order depending on the relative strength between inter- and intra-tube
repulsion, a gapless phase (denoted as (i) in Fig.~\ref{fig:kollath}) with
dominant superfluid correlations along the tubes is present for large values
of $K$ (that is, for weak intratube interactions)~\cite{kollath2008}. Such a
phase, which despite being effectively two-dimensional still preserves an
algebraic decay of correlation functions typical of the quasi-long range order
of 1D systems, is known as sliding Luttinger liquid (SLL), and has been
investigated even in several fermionic models related to stripe physics and
high-T$_{c}$ superconductivity~\cite{kivelson2003}. Moreover, the SLL phase
may also have dominant density-density correlation along the tube in a tiny
region at intermediate $\gamma$ (denoted as (ii)). In the strongly interacting
regime, a collapsed regime take place, whose precise shape depends on the sign
of the intertube interactions. The complete phase diagram for a fixed value of
the angle $\theta$ is shown in Fig.~\ref{fig:kollath}.

\begin{figure}[t]
{\ \includegraphics[width= 0.55 \columnwidth]{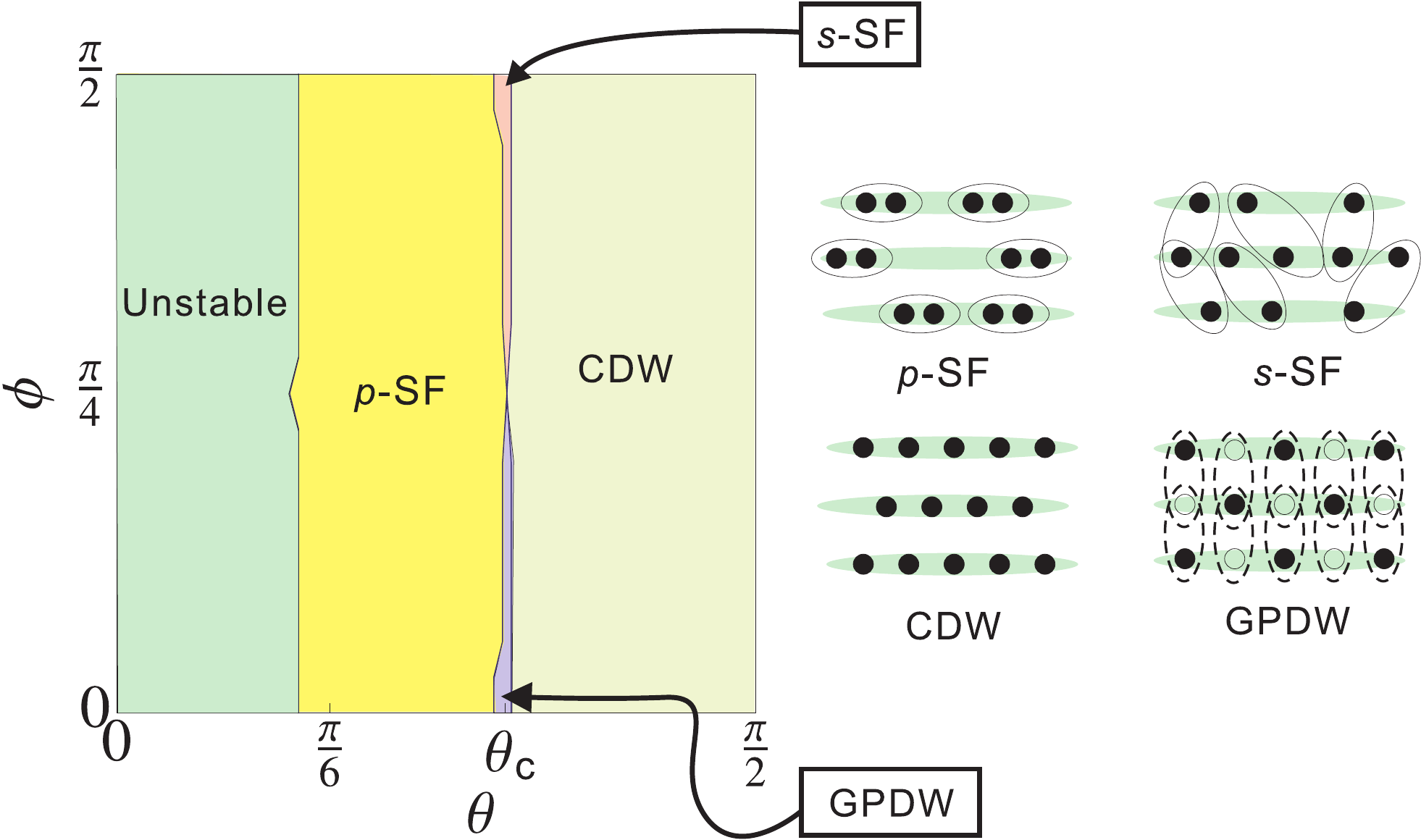}
}\caption{Left panel: schematic phase diagram of a planar array of fermionic
dipolar tubes as a function of the angles $\theta$ and $\phi$; all phases are
discussed in the text, and their cartoons are presented in the right panel,
where solid ellipses denote (intra- or inter-tube) pairing and shaded ones
indicate inter-tube coherence. Image taken from Ref.~ \cite{huang2009}.}%
\label{fig:huang}%
\end{figure}

The planar array of fermionic dipoles displays a different phase diagram,
albeit some features such as the density wave patterns emerge independently on
the statistics. The phase diagram of this system as a function of the angles
$\theta$ and $\varphi$ (see Fig.~\ref{fig:Set2D}) have been derived in the
framework of a generalized TLL theory~\cite{huang2009} by comparing the
long-distance decay of several correlation functions, as reported in
Fig.~\ref{fig:huang}. When both inter- and intra-tube interactions are
strongly repulsive, the ground state exhibits CDW (checkerboard) order,
similarly to the bosonic case. Close to the magic angle $\theta=\theta
_{c}=0.3\pi$, where the intrawire interaction turns from repulsive to
attractive, an intertube superfluid ($s$-SF) phase together with a density
wave with broken particle number conservation along the tubes (or gauge-phase
density wave, GPDW) are present. Finally, for attractive intratube
interactions, a $p$-wave superfluid ($p$-SF) is the precursor of an unstable
phase toward intratube collapse.

\section{Conclusions and outlook}

\label{s6}

The many-body systems discussed in this review are examples of the variety of
physical properties which originate from the anisotropy and long-range
character of dipole-dipole interactions, in combination with their tunability
with external fields. In many physical situations quantum dipolar gases behave
qualitatively different when compared to atomic gases with short-range van der
Waals interactions, and provide us with a large number of novel quantum
systems with unique physical properties. Understanding the many-body behavior
of these systems is a very interesting and challenging problem with several
potentially significant consequences for fundamental science and practical applications.

In the present review we have focused on dipolar quantum gases represented by
polar molecules in the rovibrational ground state. An alternative realization
of a dipolar quantum gas, although in a completely different regime, is a gas
of laser excited Rydberg atoms \cite{saffman2010quantum} or molecules
\cite{merkt1997molecules,yamakita2004deflection}. Highly excited Rydberg atoms
and molecules interact via remarkably strong electric dipole moments or Van
der Waals interactions. This leads to the phenomenon of a \emph{dipole
blockade} and formation of \emph{superatoms} where within a given blockade
radius only a single Rydberg atom can be excited, and, for example, crystals
of these superatoms can be formed \cite{saffman2010quantum}. In view of the
finite life time of Rydberg states these many body phases will only exist for
a comparatively short time in the so-called \emph{frozen gas} regime where
there is no atomic motion. However, as discussed in
Ref.~\cite{Pupillo2010}, the large Rydberg dipoles can also be admixed
weakly to the ground state by off-resonant laser excitation which provides a
situation loosely reminiscent of the polar molecule case, although decoherence
due to spontaneous emission remains always an issue.

Finally, we remark that the tools for manipulating interactions in dipolar
systems, as described in the present review, also provide promising
ingredients for controlled entanglement between polar molecules, and thus
possible new scenarios for quantum computing
\cite{PhysRevLett.88.067901,Jelin,Ortner2011}. Loading exactly one atom or
molecules per lattice site via a Mott insulator transition in an optical
lattice provides us with an array of qubits. In the case of polar molecules,
the qubits can he represented by long-lived rotational or spin degrees of
freedom. Single site addressing, as developed in
Refs.~\cite{bakr2009,sherson2010single} for atoms, allows to both
manipulate as well as read out the single qubit. Entanglement of qubits can be
achieved via the strong and long range dipolar interactions between molecules
\cite{PhysRevLett.88.067901,Jelin,Ortner2011}, or in the atomic case via
dipolar Rydberg interactions\cite{saffman2010quantum}.

\section{Acknowledgements}

We would like to thank 
I. Bloch,
D. Blume,
J. Bohn, 
H. P. B\"uchler,
L. D. Carr,
N. R. Cooper,  
R. C\^ot\'e,
D. DeMille, 
E. Demler,
J. Doyle, 
F. Ferlaino
A. Gorshkov,
M. Lewenstein,
M. Lukin,
D. Jin,
P. Julienne,
R. Krems, 
B. Lev,
G. Mejier,
H. C. N\"agerl,
D. Petrov, 
T. Pfau, 
G. Qu\'em\'ener, 
L. Santos,
G. Shlyapnikov,
J. Ye, 
S. Yelin,
D.-W. Wang and
M. Weidem\"uller
for many interesting discussion on the topics 
treated in this review. Work supported by the Austrian Science Fund, EU grants AQUTE, COHERENCE and
NAME-QUAM, and by MURI, AFOSR, ISIS, EOARD, IPCMS and the RTRA Foundation.

\newpage

\section{Biographies}

\begin{figure}[t]
{ \includegraphics[width= 0.35 \columnwidth]{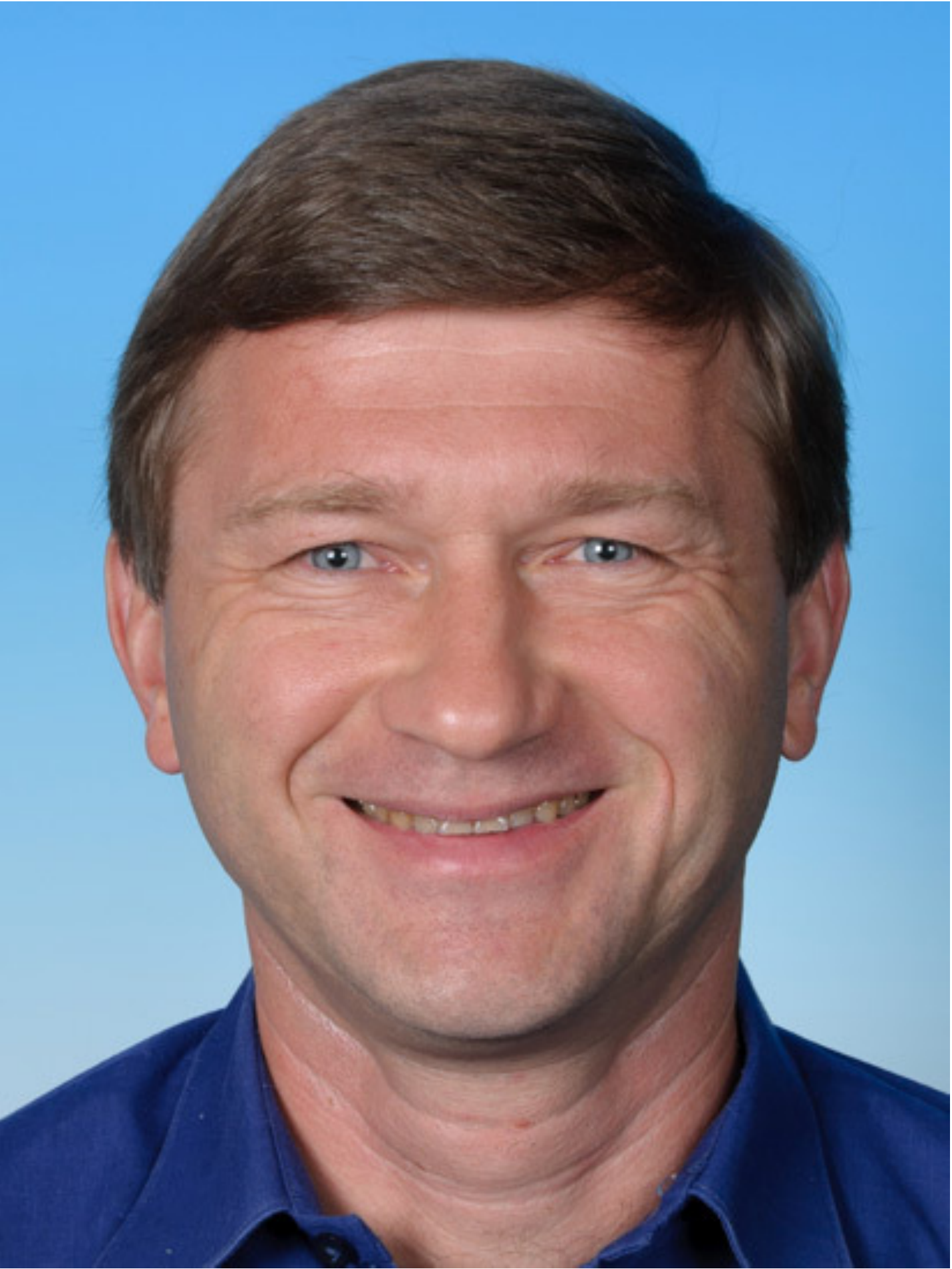}
}\caption{Mikhail Baranov is a senior researcher at the Austrian Academy of
Sciences in Innsbruck. He received his M.S. in Theoretical Physics in 1984
from the Moscow Institute of Engineering and Physics Russia, and PhD in
Physics and Mathematics in 1987 in the same Institute. In 2008 he joined Prof.
Zoller's group in Innsbruck. Scientific Interests are many-body systems,
strongly correlated states, ultracold atomic and molecular gases. }%
\label{fig:baranov_picture}%
\end{figure}

\begin{figure}[t]
{ \includegraphics[width= 0.35 \columnwidth]{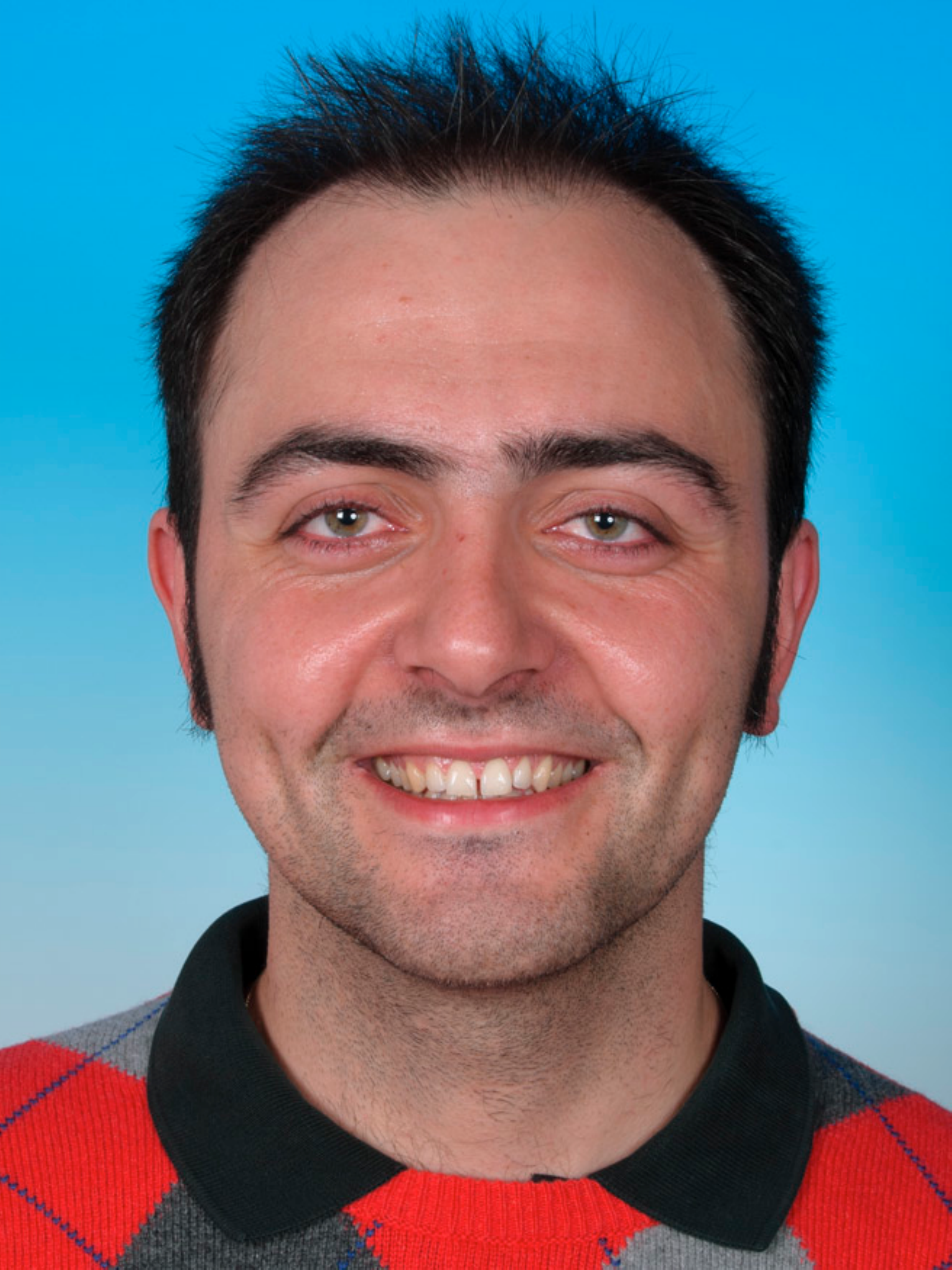}
}\caption{Marcello Dalmonte is Junior Scientist at the Austrian Academy of
Sciences in Innsbruck. He received both his M.S. (2007) and Ph.D (April 2011)
from the Department of Physics of the University of Bologna (Italy), working
under the supervision of Dr. Elisa Ercolessi. During his studies, he spent
several months as a visiting student at the Institute for Theoretical Physics
in Innsbruck. His main research interests include different aspects of
many-body theories in one-dimensional geometries, with particular focus on
quantum phenomena and strong correlations in ultracold atomic and molecular
gases.}%
\label{fig:dalmonte_picture}%
\end{figure}

\begin{figure}[t]
{ \includegraphics[width= 0.35 \columnwidth]{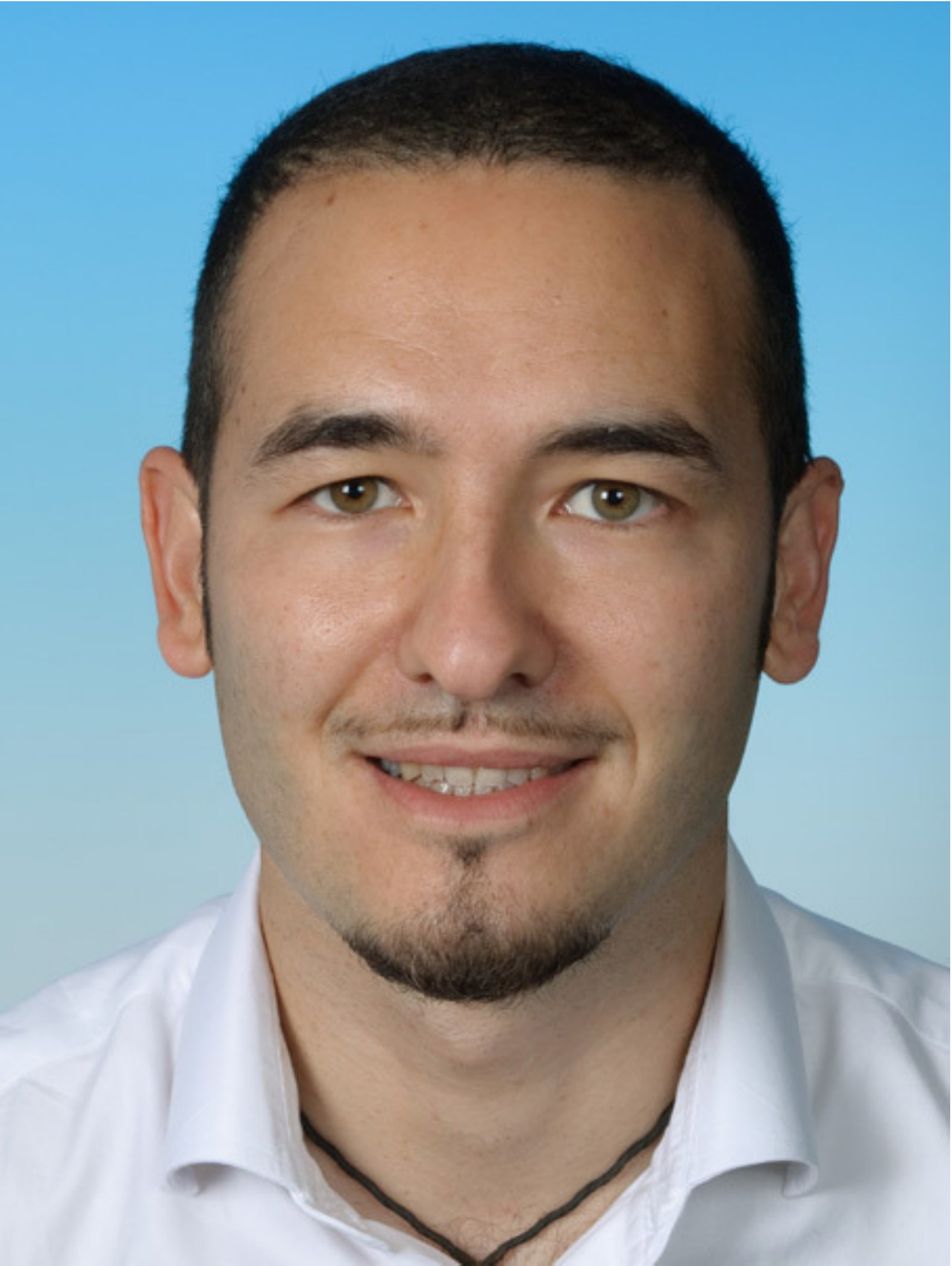}
}\caption{Guido Pupillo is professor at the University of Strasbourg and director of the Laboratory of Quantum Physics at the Institut de Science d'Ing\'enierie Supramol\'eculaires and the Institut de Physique et Chimie des Mat\'eriaux in Strasbourg. He received
his Ph.D. in Physics from the University of Maryland in the USA, for research
conducted at the National Institute of Standard and Technology
(Gaithersburg). In 2005 he joined Prof. Zoller's group at the
University of Innsbruck, where he got the Habilitation for
professorship in theoretical physics in 2011. His research interests focus on the
physics of strongly correlated atomic and molecular gases.}%
\label{fig:pupillo}%
\end{figure}

\begin{figure}[t]
{ \includegraphics[width= 0.35 \columnwidth]{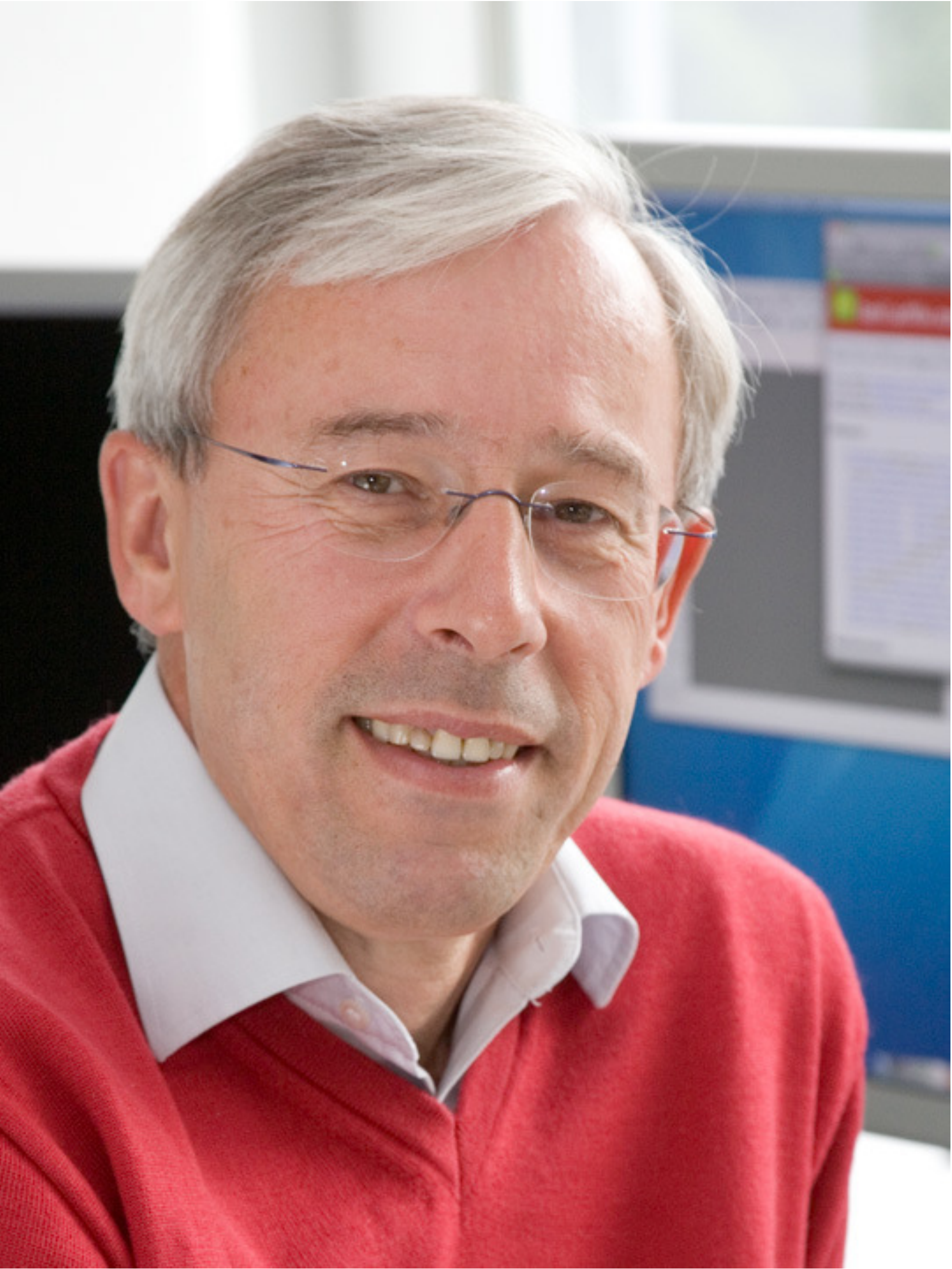}
}\caption{Peter Zoller is professor at the University of Innsbruck and works
on quantum optics and quantum information. He has obtained his Doctorate from
the University of Innsbruck in 1977, and the Habilitation for professorship in
1981. After spending several years as visiting fellow, from 1991 to 1994 he
was professor of physics and JILA Fellow at JILA and at the Physics Department
of the University of Colorado, Boulder USA. Back in Innsbruck, from 1995 to
1999 he headed the Institute of Theoretical Physics, and was vice-dean of
studies from 2001 to 2004. He was Loeb lecturer in Harvard, Boston, MA (2004),
Yan Jici chair professor at the University of Science and Technology of China,
Hefei, and chair professor at Tsinghua University, Beijing (2004), as well as
Lorentz professor at the University of Leiden in the Netherlands (2005). Since
2003, Peter Zoller has also held the position of Scientific Director at the
Institute for Quantum Optics and Quantum Information (IQOQI) of the Austrian
Academy of Sciences. He is the recipient of several international prizes,
including the Benjamin Franklin Medal in Physics (2010), the Dirac Medal
(2006), the Niels Bohr/UNESCO Gold Medal (2005), and the Max Planck Medaille
2005. He is best known for his research on quantum computing and quantum
communication and for bridging quantum optics and solid state physics.}%
\label{fig:zoller_picture}%
\end{figure}

\bibliography{chemrev_bibliography}
\end{document}